\documentclass[aps,prb,notitlepage,nofootinbib]{revtex4-2}
\usepackage{graphicx}
\usepackage[dvipsnames]{xcolor}
\usepackage{amsfonts}
\usepackage{amssymb}
\usepackage{amsmath}
\usepackage{mathtools}
\usepackage{afterpage}
\usepackage[mathscr]{euscript}
\usepackage{braket}
\usepackage{subfigure}
\usepackage{multirow}
\usepackage{float}
\usepackage{qcircuit}
\usepackage{enumitem}
\usepackage{color}   
\usepackage{hyperref}
\hypersetup{
    colorlinks=true, 
    linktoc=all,     
    linkcolor=blue,  
}

\usepackage[utf8]{inputenc} 
\usepackage{yfonts}
\usepackage{dsfont}
\usepackage{cancel}
\usepackage[scr=boondox]{mathalpha}

\usepackage{pst-all}
\usepackage{leftidx}
\usepackage{epstopdf}
\usepackage{booktabs}
\usepackage[off]{auto-pst-pdf}

\usepackage[bottom]{footmisc}

\newcommand{\coho}[1]{\textswab{#1}}

\newcommand{\act}[2]{{}^{{\bf #1}} #2}
\newcommand{\acts}[3]{{}^{{\bf #1}_{#2}} #3}

\newcommand\restr[2]{{
  \left.\kern-\nulldelimiterspace 
  #1 
  \vphantom{\big|} 
  \right|_{#2} 
  }}

\begin{document}
	\def\U{\mathrm{U}(1)}
	\def\SU{\mathrm{SU}}
	\def\SO{\mathrm{SO}}
	\def\Pin{\mathrm{Pin}}
	\def\Spin{\mathrm{Spin}}
	\def\H{\mathcal{H}}
	\def\F{\mathcal{F}}
	\def\E{\mathbb{E}}
	\def\TT{\mathsf{T}}
	\def\A{\mathcal{A}}
	\def\L{\mathcal{L}}
	\def\C{\mathcal{C}}
	
\newcommand{\Z}{\mathbb Z}
\newcommand{\R}{\mathbb R}
    \renewcommand{\L}{\mathcal L}
	\renewcommand{\H}{\mathcal H}
	
\newcommand{\commentrk}[1]{\textcolor{blue}{[RK: #1]}}
	   
\newcommand{\commentdb}[1]{\textcolor{red}{[DB: #1]}}

\newcommand{\commentst}[1]{\textcolor{olive}{[ST: #1]}}
\newcommand{\commentmb}[1]{\textcolor{purple}{[MB: #1]}}

\newcommand\curvearrowbotright{\rotatebox[origin=c]{180}{$\curvearrowleft$}}
\newcommand\curvearrowbotleft{\rotatebox[origin=c]{180}{$\curvearrowright$}}

\def\Sq{\mathop{\mathrm{Sq}}\nolimits}
\newcommand{\mathbbm}[1]{\text{\usefont{U}{bbm}{m}{n}#1}}

        \title{Anomalies in (2+1)D fermionic topological phases and (3+1)D path integral state sums for fermionic SPTs}
        \author{Srivatsa Tata$^1$}
        \author{Ryohei Kobayashi$^2$}
        \author{Daniel Bulmash$^1$}
        \author{Maissam Barkeshli$^1$}
        \affiliation{$^1$Condensed Matter Theory Center and Joint Quantum Institute, Department of Physics, University of Maryland, College Park, Maryland 20472 USA}
        \affiliation{
        $^2$Institute for Solid State Physics, University of Tokyo, Kashiwa, Chiba 277-8581, Japan
        }

    \begin{abstract}
        Given a (2+1)D fermionic topological order and a symmetry fractionalization class for a global symmetry group $G$, we show how to construct a (3+1)D topologically invariant path integral for a fermionic $G$ symmetry-protected topological state ($G$-FSPT), in terms of an exact combinatorial state sum. This provides a general way to compute anomalies in (2+1)D fermionic symmetry-enriched topological states of matter.
        Equivalently, our construction provides an exact (3+1)D combinatorial state sum for a 
        path integral of any FSPT that admits a symmetry-preserving gapped boundary, which includes the (3+1)D topological insulators and superconductors in class AII, AIII, DIII, and CII that arise in the free fermion classification. Our construction proceeds by using the fermionic topological order (characterized by a super-modular tensor category) and symmetry fractionalization data to define a (3+1)D path integral for a bosonic theory that hosts a non-trivial emergent fermionic particle, and then condensing the fermion by summing over closed $3$-form $\Z_2$ background gauge fields. This procedure involves a number of non-trivial higher-form anomalies associated with Fermi statistics and fractional quantum numbers that need to be appropriately canceled off with a Grassmann integral that depends on a generalized spin structure. We show how our construction reproduces the $\Z_{16}$ anomaly indicator for time-reversal symmetric topological superconductors with ${\bf T}^2 = (-1)^F$. Mathematically, with some standard technical assumptions, this implies that our construction gives a combinatorial state sum on a triangulated 4-manifold that can distinguish all $\Z_{16}$ $\Pin^+$ smooth bordism classes. As such, it contains the topological information encoded in the eta invariant of the pin$^+$ Dirac operator, thus giving an example of a state sum TQFT that can distinguish exotic smooth structure. 
    \end{abstract}

        \maketitle
        
\twocolumngrid
\tableofcontents
\onecolumngrid
        
        \section{ Introduction}

        Topological phases of matter in (2+1) space-time dimensions can be partially characterized by the braiding and fusion properties of topologically non-trivial finite-energy quasiparticles, referred to as anyons. For bosonic systems, the algebraic data that describes the anyons mathematically defines a unitary modular tensor category (UMTC) \cite{moore1989b,wang2008,nayak2008,Bonderson07b} $\mathcal{C}$, while for fermionic systems $\mathcal{C}$ is taken to be a super-modular tensor category \cite{bruillard2017a}. It is believed that, ignoring any global symmetries, (2+1)D gapped phases of matter can be fully characterized by $\mathcal{C}$ together with the chiral central charge $c_-$ of the potentially gapless (1+1)D edge theory. 
        
        In the presence of a global symmetry group $G$, topological phases of matter are further characterized by the braiding and fusion properties of symmetry defects combined with the anyons \cite{barkeshli2019}. Part of the defining data involving symmetry defects is the symmetry fractionalization pattern, which specifies how anyons carry fractional quantum numbers under the symmetry. In terms of a (2+1)D space-time path integral, the symmetry fractionalization pattern can be characterized by changes in the amplitude for anyon worldlines passing through junctions of codimension-1 symmetry defects \cite{barkeshli2019}. 
        
        Certain patterns of symmetry fractionalization can be anomalous, which means that there is an obstruction to defining a consistent set of braiding and fusion data \cite{barkeshli2019,barkeshli2019rel,Chen2014} for the symmetry defects together with the anyons. Over the past several years, it has been shown in various examples how the (2+1)D surface of certain (3+1)D symmetry-protected topological states (SPTs) can host symmetry-preserving gapped boundaries with non-trivial topological order and anomalous symmetry fractionalization patterns \cite{vishwanath2013,wang2013b,MetlitskiPRB2013,MetlitskiPRB2015,wang2013b,chen2014b,Chen2014,fidkowski2013,bonderson2013sto,wang2014,metlitski2014,ChoPRB2014,kapustin2014b,seiberg2016gapped,hermele2016,YangPRL2015,song2017}.
        An SPT \cite{chen2013,senthil2015} is a topological phase of matter that is potentially non-trivial only in the presence of symmetry; a mathematical classification of SPTs has been conjectured in terms of invertible TQFTs and bordism groups \cite{freed2016,kapustin2014,kapustin2014c,Kapustin:2014dxa}.
        Recently it has been shown by explicit construction that any given symmetry fractionalization class for a (2+1)D bosonic topological phase, including anomalous ones, can be realized at the surface of a (3+1)D SPT \cite{bulmash2020}. 
        The anomaly of a (2+1)D topological phase is then characterized by the specific (3+1)D SPT that hosts the given (2+1)D topological phase at its surface. More generally, the modern understanding of anomalies in quantum field theory is in terms of invertible TQFTs in one higher dimension \cite{freed2014}.
        
        For bosonic topological phases, there is now a completely general theory to compute both relative anomalies (i.e. the difference in anomalies between two theories with different symmetry fractionalization classes) \cite{barkeshli2019rel} and the full absolute anomaly given the UMTC $\mathcal{C}$ and the algebraic data that characterizes the symmetry fractionalization class \cite{bulmash2020}. In particular, the absolute anomaly can be computed by using the UMTC and symmetry fractionalization data to define a (3+1)D path integral in terms of a state sum for an SPT that hosts the given symmetry-enriched topological order at its surface \cite{bulmash2020,barkeshli2019tr}. Evaluating the path integral on the appropriate $G$-bundles then gives anomaly indicators that can fully characterize the SPT phase and thus the anomaly. 
        Moreover, the path integral state sum can also be used to develop a commuting projector Hamiltonian for the (3+1)D SPT and also a symmetric surface termination that hosts the given (2+1)D symmetry-enriched topological order \cite{bulmash2020}. 
        
        From a broader perspective, these results show how to compute anomalies of topological quantum field theories from the intrinsically quantum data that defines the TQFT -- e.g. the UMTC -- rather than through the examination of a  classical Lagrangian in the presence of background gauge fields. This provides a step forward in eliminating the crutch of classical Lagrangians in describing quantum field theories \cite{seiberg2015}. It is especially useful when a classical Lagrangian description may not be known or cumbersome to work with. 
        
        The main purpose of this paper is to generalize the results of Ref.~\cite{bulmash2020} to the case of fermionic topological phases. The fermionic problem is richer in a number of ways. First, the classification of fermionic SPTs (FSPTs), and equivalently of anomalies of fermionic systems, is richer than that of bosonic SPTs. While bosonic SPTs with symmetry group $G$ in $4$ space-time dimensions are almost entirely classified by group cohomology \cite{chen2013,kapustin2014,kapustin2014c} $\mathcal{H}^{4}(G, \U)$, FSPTs have a richer classification involving first, second, third, and fourth cohomologies \cite{wang2020}, which are related to certain $4$-dimensional generalized $\Spin$ and $\Pin$ bordism groups \cite{Kapustin:2014dxa}. Secondly, the topological path integrals for SETs of bosons depends on a $G$-bundle, whereas for fermionic topological phases the topological path integrals depend, in addition, on a certain generalization of spin or pin structures. 
        
        In this paper we show how, given a (2+1)D fermionic topological phase described by a super-modular tensor category and a symmetry fractionalization class, one can define a (3+1)D topologically invariant path integral in terms of a state sum for a FSPT. Since the path integral describes a FSPT, it depends on a choice of spin structure when evaluated on trivial $G$ bundles, and on a type of generalized spin or pin structure appropriate for fermions coupled to a $G$-gauge field. The (3+1)D FSPT defined in this way is expected to host the given (2+1)D SET at its surface. This therefore gives an explicit method to compute the anomaly of the (2+1)D theory, by identifying the (3+1)D FSPT through the evaluation of the topological path integral on the appropriate space-time 4-manifolds and $G$-bundles. 
        
        Our construction is in a sense a combination of the construction of \cite{bulmash2020} applied to super-modular categories, combined with a generalization of the prescription of Gaiotto and Kapustin \cite{Gaiotto:2015zta} for condensing fermions within a path integral framework. Our generalization of the latter allows us to consider symmetry groups which are not a direct product of fermion parity with a bosonic symmetry group. In particular, this gives us enough power to construct 
        an explicit topological path integral state sum for all fermion topological insulators and superconductors that appear in the free fermion classification and which admit a symmetry-preserving surface topological order \cite{wang2013b,MetlitskiPRB2015,wang2013b,bonderson2013sto,wang2014}. The existence of an explicit state sum for the topological path integral of (3+1)D FSPTs that admit symmetry-preserving gapped boundaries suggests that there should also exist corresponding commuting projector Hamiltonians, tensor network descriptions, and explicit quantum circuits that prepare these FSPTs. In particular, we note that the existence of a commuting projector Hamiltonian for the generator of the $\Z_{16}$ classification in class DIII has been an open question; our work suggests that such a Hamiltonian indeed exists. We leave it to future work to explicitly develop these Hamiltonian, tensor network, and quantum circuit descriptions. 
        
        As an explicit example, we study the case of time-reversal symmetry where ${\bf T}^2 = (-1)^F$, which corresponds to the case of class DIII in the free fermion periodic table of topological insulators and superconductors. (3+1)D FSPTs with this symmetry have a $\Z_{16}$ classification \cite{kitaev2011,fidkowski2013,metlitski2014,wang2014}, which corresponds to the $\Pin^+$ bordism group in 4-dimensions \cite{Kapustin:2014dxa}. By using an appropriate anomalous surface topological order, our construction can be used to obtain an explicit combinatorial state sum that, given a triangulation of a $4$-manifold and a pin$^+$ structure on it, can distinguish which $\Z_{16}$ class in the bordism group it belongs to. This is particularly interesting as it gives perhaps the first example of a combinatorial state sum for a TQFT which can detect exotic smooth structure \footnote{We note that by ``exotic smooth structure," we are specifically referring to the existence of 4-manifolds that are homeomorphic but not diffeomorphic.}. There exist two closed 4-manifolds, $\mathbb{RP}^4$ and ``fake $\mathbb{RP}^4$'' , which are homeomorphic but not diffeomorphic to each other \cite{cappell1976,stolz1988}. These two manifolds correspond to the $\pm 1$ and $\pm 9$ elements of the $\Z_{16}$ $\Pin^+$ smooth bordism group (with the sign determined by the choice of pin$^+$ structure) \cite{stolz1988,kirby_taylor_1991}. Therefore a combinatorial state sum for a TQFT that can distinguish this $\Z_{16}$ implies that it can also distinguish these two classes of smooth structure given a triangulation and pin structure on the manifold. 
        More generally, the eta invariant of the pin$^+$ Dirac operator is also known to be an invariant of smooth Pin$^+$ bordisms and can distinguish all $\Z_{16}$ classes \cite{stolz1988}; therefore our path integral in this case can distinguish the same exotic smooth structure as the eta invariant. This includes additional examples such as exotic smooth structures on $\mathbb{RP}^4 \# S$, where $S$ is the connected sum of 11 copies of $S^2 \times S^2$\cite{stolz1988}. We note that recently, a no-go theorem \cite{reutter2020} has been proven that rules out oriented semi-simple 4-dimensional TQFTs from being able to distinguish smooth structure; our construction circumvents this no-go theorem as it is an unoriented TQFT \footnote{We note that in these examples, the manifolds in question, such as $\mathbb{RP}^4$ vs. fake $\mathbb{RP}^4$, are also not stably diffeomorphic to each other (under connected sum with $S^2 \times S^2$). Thus our results are consistent with the main theorem of \cite{reutter2020} if the assumption of orientability is removed.}. 
        
        We expect similar results to hold for all spatial dimensions $\geq 3$: it is natural to conjecture that all bosonic and fermionic SPT phases in any dimensions that admit symmetry-preserving gapped boundaries also admit commuting projector Hamiltonians and state sums for topological path integrals\footnote{We note that previous studies \cite{fidkowski2013} utilized commuting projector Walker-Wang \cite{walker2012} Hamiltonians as an intermediate step in constructing Hamiltonians for fermionic topological superconductors; these constructions also require an additional step to condense fermions, which spoils the commuting projector nature of the full Hamiltonian.}. At a mathematical level, this suggests a general way of obtaining invariants of smooth bordism classes of $G_b$ bundles with generalized spin structures in general dimensions. 
        
        \subsection{ Summary of main results}
        
        We consider a system of fermions with global symmetry group $G_f$. $G_f$ contains a central subgroup $\Z_2^f$, generated by fermion parity $(-1)^F$. The subgroup that acts on bosonic operators is then $G_b = G_f/\Z_2^f$. In general, $G_f$ is a $\Z_2$ central extension of $G_b$, where the extension is specified by $[\omega_2] \in \mathcal{H}^2(G_b, \Z_2)$. Physically, the existence of a non-trivial $[\omega_2]$ can be understood as the physical fermion carrying fractional quantum numbers under $G_b$. The specific $2$-cocycle $\omega_2({\bf g}, {\bf h})$ is determined by the action of $G_f$ on local fermion operators in the microscopic theory \cite{bulmashSymmFrac,aasen21ferm}. 
        
        In this paper, we assume we are given the data of a super-modular tensor category \cite{bruillard2017a,bruillard2017b,bonderson2018} $\mathcal{C}$, which is a unitary braided fusion category that contains a single invisible object $\psi$ that physically describes the fermion. In addition to $\mathcal{C}$, we are given the data that specifies symmetry fractionalization for the symmetry group $G_b$. This data consists of a group homomorphism $\rho : G_b \rightarrow \text{Aut}(\mathcal{C})$, which specifies how the symmetry permutes the anyons, and a set of $\U$ phases $\{\eta_a({\bf g}, {\bf h})\}$ for each anyon $a$, which must obey certain consistency equations and which are defined up to certain gauge transformations. We review this data in Sec.~\ref{sec:symmFracReview}. In particular, $\omega_2$ specifies $\eta_\psi$: $\eta_\psi({\bf g}, {\bf h}) = \omega_2({\bf g},{\bf h})$.
        
        Given $\mathcal{C}$ and the symmetry fractionalization data, we show how to construct a path integral $Z(M^4,A_b,\xi_{\mathcal{G}})$, where $M^4$ is a 4-manifold, $A_b$ is a flat $G_b$ gauge field, which specifies a flat $G_b$ bundle over $M^4$, and $\xi_{\mathcal{G}}$, referred to as a $\mathcal{G}_f$-structure, is a $\Z_2$-valued $1$-cochain.  $\xi_{\mathcal{G}}$ satisfies
        \begin{align}
            \delta \xi_{\mathcal{G}} = w_2 + w_1^2 + A_b^* \omega_2,
        \end{align}
        where $w_1$ and $w_2$ are the first and second Stiefel-Whitney classes of $M^4$, and $A_b^*$ denotes the pullback of the gauge field $A_b$, which we can, with some abuse of notation, also view as a map $A_b : M^4 \rightarrow BG_b$, where $BG_b$ is the classifying space of $G_b$. As we will explain, the group $\mathcal{G}_f$ is essentially a lift of the space-time symmetry group that acts on bosonic local operators to one that acts on fermionic operators. 
        
        In the case where $[\omega_2]$ is trivial, so that $G_f = G_b \times \Z_2^f$, $\xi_{\mathcal{G}}$ is simply a spin structure on orientable manifolds and a pin$^-$ structure on non-orientable manifolds. 
        
        The path integral $Z(M^4,A_b,\xi_{\mathcal{G}})$ is constructed by first taking the super-modular category and symmetry fractionalization data through the construction of Ref.~\cite{bulmash2020}, which is a symmetry-enriched generalization of the Crane-Yetter-Walker-Wang construction. This gives a (3+1)D
        path integral $Z_b(M^4,A_b)$, for a bosonic topological phase which contains a single non-trivial point-like excitation, which is a fermion. We then condense this fermion, by generalizing
        the prescription of Ref.~\cite{Gaiotto:2015zta} to include non-trivial $\omega_2$. In particular, the existence of a single non-trivial point-like excitation implies that the theory has a $\Z_2$ $2$-form symmetry \cite{gaiotto2014}. Therefore we can turn on a coupling to a background $3$-form $\Z_2$ gauge field $f_3$, thus yielding a path integral $Z_b((M^4,T), A_b, f_3)$. Here $T$ is a triangulation of $M^4$ equipped with a branching structure (local ordering of vertices) and $\pm$ assignments to $d$-simplices, and $f_3$ is a $\Z_2$-valued $3$-cocycle on the triangulation. Condensing the fermion corresponds to ``gauging'' the $2$-form symmetry by summing over all possible background configurations of $f_3$.
        
        The Fermi statistics of $\psi$ implies on general grounds that the $\Z_2$ $2$-form symmetry has an 't Hooft anomaly \cite{Gaiotto:2015zta}. Moreover, as we explain in Section \ref{generalTheorySec}, the existence of a non-trivial class $[\omega_2] \in \mathcal{H}^2(G_b, \Z_2)$ implies on general grounds that there is a mixed anomaly between the global $G_b$ symmetry and the $\Z_2$ $2$-form symmetry. One significant technical result of this work is a careful accounting for this mixed anomaly in the fermion condensation procedure. These anomalies imply that, when $f_3$ is non-zero, $Z_b$ is not invariant under gauge transformations of $A_b$ or $f_3$ or under changes of triangulation, branching structure, and $\pm$ assignment of simplices.  
        
        In order to obtain a topologically invariant path integral by condensing the fermion, we need to appropriately compensate for these anomalies by including an additional factor. The full fermionic path integral then takes the form:
        \begin{align}
            Z((M^4,T), A_b, \xi_{\mathcal{G}}) = \frac{1}{\sqrt{|H^2(M^4,\Z_2)|}} \sum_{[f_3] \in H^3(M^d,\Z_2)} Z_b((M^4,T), A_b, f_3) z_c( (M^4, T), f_3, \xi_{\mathcal{G}}) ,
        \end{align}
        where
        \begin{align}
            z_c( (M^4, T), f_3, \xi_{\mathcal{G}}) = \sigma( (M^4, T), f_3 ) (-1)^{\int_{M^4} \xi_{\mathcal{G}} \cup f_3}
        \end{align}
        
        Here $\sigma( (M^4, T), f_3 )$ is the so-called Grassmann integral, for which we review two alternate definitions in Sec.~\ref{sec:zpsi}, and $H^2(M^4,\Z_2)$ is a singular cohomology group of $M^4$ with $\Z_2$ coefficients. One definition of $\sigma$ is based on an integral over Grassmann variables that decorate the triangulation; for orientable manifolds this is given by the construction of Gu and Wen \cite{gu2014}, while the generalization to non-orientable manifolds was given in \cite{Kobayashi2019pin}. An alternate definition in terms of a certain winding number of loops associated to $f_3$ is also given, which was first presented for two-dimensional orientable manifolds in \cite{Gaiotto:2015zta} and later generalized to higher dimensions and non-orientable manifolds in \cite{tata2020}. In Appendix~\ref{sec:windingDefOfSigma} we demonstrate the equivalence between these two definitions.
        
        The full fermionic path integral $Z$ is retriangulation-invariant and also invariant under changes of $A_b$ and $\xi_{\mathcal{G}}$ by gauge transformations (i.e. by coboundaries). (A precise formulation of retriangulation-invariance in the presence of $A_b$ and $\xi_{\mathcal{G}}$ is given in Secs.~\ref{sec:anomaliesOfShadow},\ref{sec:fermStateSum}.) On the other hand, $Z_b$ and $z_c$ are not individually retriangulation or gauge invariant when $f_3$ is non-zero. One of the main technical contributions of our paper is to give in Sec.~\ref{sec:craneyettershadow} a precise and general definition of the bosonic theory $Z_b((M^4,T), A_b, f_3)$, which generalizes the construction of \cite{bulmash2020} to the case of a non-zero $f_3$, and to show that the resulting state sum possesses the expected pure and mixed anomalies under gauge transformations of $f_3$ and $A_b$ and under retriangulations. 
        
        Next, in Sec.~\ref{sec:fermStateSum}, we argue that our resulting path integral $Z(M^4,\xi_{\mathcal{G}}, A_b)$ describes an invertible TQFT, and therefore defines a (3+1)D FSPT, which implies that it should be a bordism invariant. As we will discuss, while the bordism invariance is strongly expected on general grounds, a fully general and mathematically rigorous proof is beyond the scope of this paper. 

        Another result of this paper is a new derivation of the $\Z_{16}$ anomaly indicator for $G_f = \Z_4^{{\bf T}, f}$ \cite{wang2017,tachikawa2016b,Kobayashi2019pin}, which corresponds to the case of time-reversal symmetry with ${\bf T}^2 = (-1)^F$, appropriate to the surface topological order of (3+1)D class DIII topological superconductors. We show in Sec.~\ref{sec:examples} by explicit computation that 
        \begin{align}
        \label{ZRP4eq}
            Z( \mathbb{RP}^4 ) = \frac{1}{\sqrt{2}\mathcal{D}} \left(\sum_{x | x = {\,^{\bf T}}x} d_x \theta_x \eta^{\bf T}_x + i \sum_{x | x = {\,^{\bf T}}x \times \psi} d_x \theta_x \eta^{\bf T}_x\right)
        \end{align}
        where 
        \begin{equation} 
            \eta_a^{\bf T} := \begin{cases}
            \eta_a({\bf T},{\bf T}) & \,^{\bf T}a=a\\
            \eta_a({\bf T},{\bf T})U_{\bf T}(a,\psi;a\psi)F^{a,\psi, \psi} & \,^{\bf T}a = a \times \psi 
            \end{cases} 
        \end{equation}
        is a gauge-invariant quantity defined by the symmetry fractionalization data which can be thought of as the action of ${\bf T}^2$ on the anyon $x$. Here $\eta_a$ and $U_{\bf T}$ are part of the data that characterizes symmetry fractionalization and $F$ is the $F$-symbol of the anyon theory, as reviewed in Section \ref{sec:symmFracReview}. We note that Eq.~\eqref{ZRP4eq} makes a choice of one of the two possible pin$^+$ structures on $\mathbb{RP}^4$; using the other choice would complex conjugate Eq.~\eqref{ZRP4eq}. 
        
        Finally, we note that our construction gives a state sum for an unoriented $\Pin^+$ TQFT that can detect certain exotic smooth structure. Applying our construction to the case of $G_f = \Z_4^{{\bf T}, f}$, we obtain a topological invariant $Z((M^4,T), \xi)$ in terms of an exact state sum for a triangulation $T$ of $M^4$, where $\xi$ is a pin$^+$ structure. Since $Z((M^4,T), \xi)$ describes an FSPT, it is expected to be a $\Pin^+$ bordism invariant. Using as input the super-modular category that describes the anyon content in $\SO(3)_3$ Chern-Simons theory and a particular symmetry fractionalization class, we conclude that our construction gives a $\Z_{16}$ smooth bordism invariant. This implies that $Z((M^4,T), \xi)$ should be able to distinguish the same exotic smooth structure as the eta invariant of the pin$^+$ dirac operator, which is also a $\Z_{16}$ smooth bordism invariant \cite{stolz1988}. An example is real and fake $\mathbb{RP}^4$, which are homeomorphic but not diffeomorphic to each other \cite{cappell1976,stolz1988,kirby_taylor_1991}. In fact, in this example the bosonic shadow 
        $Z_b( (M^4,T), f = 0 )$ can also distinguish real vs. fake $\mathbb{RP}^4$. This is because the bosonic shadow $Z_b$ can be obtained from $Z$ by gauging fermion parity, which mathematically corresponds to averaging over the pin$^+$ structures, which preserves the distinction between real and fake $\mathbb{RP}^4$ in the $\Pin^+$ bordism group. 

        This paper is organized as follows. In Sec.~\ref{generalTheorySec}, we explain the general theory of fermion condensation to produce path integrals for fermionic topological phase for general $G_f$ by starting with a bosonic shadow and summing over closed $(d-1)$-form $\Z_2$ gauge fields. In particular we give a general discussion of the  higher-form anomalies that appear in theories hosting fermionic particles. In Sec.~\ref{sec:geometricChoices}, we establish notation for various geometric objects and discuss a number of geometric choices related to triangulated manifolds that we need to set up our fermionic path integral. In Sec.~\ref{sec:zpsi}, we define in two ways, one geometric and one algebraic, the factor $z_c$ which cancels the anomalies of the bosonic shadow. We then review symmetry fractionalization, in particular its application to fermionic topological phases, in Sec.~\ref{sec:symmFracReview}. We construct the bosonic shadow for our fermionic SPT and demonstrate that it has the desired anomalies in Sec.~\ref{sec:craneyettershadow}, then construct and analyze the full fermionic state sum in Sec.~\ref{sec:fermStateSum}. We evaluate the state sum for some example cases in Sec.~\ref{sec:examples}, and close with some discussion of open questions in Sec.~\ref{sec:Discussion}. Many highly technical discussions and detailed calculations are relegated to appendices. In particular, we highlight a few which may be of general interest: Appendix \ref{topologicalPreliminaries} includes geometrical descriptions of the first and Stiefel-Whitney classes, spin and pin structures, and how such objects are determined by branching structures of a triangulation and orientation of $d$-simplices; Appendix \ref{twistedSpinSec} includes an explanation of spin, pin, and $\mathcal{G}_f$ structures from an algebraic perspective of lifting transition functions of bundles; Appendix \ref{appPachnerLemmas} contains mathematical results about Pachner moves in the presence of branching structures and background gauge fields which do not appear to have been discussed previously in the literature.

               \section{ General theory of path integrals for fermionic topological phases via fermion condensation}
               \label{generalTheorySec}
        
        Here we provide a general overview for constructing a path integral for a TQFT describing a fermionic topological phase by starting with a path integral for a bosonic topological phase in $d$ space-time dimensions and condensing an emergent fermion by gauging an anomalous $(d-2)$-form symmetry. 
        That is, we wish to define a topologically invariant path integral $Z(M^d, A_b, \xi_{\mathcal{G}})$ defined on a $d$-dimensional space-time manifold $M^d$. The manifold further is equipped with a flat $G_b$ gauge field $A_b$ together with a generalized spin structure $\xi_{\mathcal{G}}$, which we refer to as a $\mathcal{G}_f$-structure. 
        
        In the case where $G_f = \Z_2^f \times G_b$, where $G_b$ is a unitary bosonic symmetry and $\Z_2^f$ is fermion parity, then $\mathcal{G}_f = \Spin(d)$, and $\xi_{\mathcal{G}}$ is a spin structure. This case was studied in detail by Gaiotto-Kapustin \cite{Gaiotto:2015zta}, building on Gu-Wen \cite{gu2014} and our discussion below follows closely that of Gaiotto-Kapustin. In the more general case where $[\omega_2]$ is non-trivial, we obtain a generalization of the Gaiotto-Kapustin results to more general fermionic symmetry groups $G_f$. 
        
        It is well-known that one can describe fermionic topological phases by starting with a bosonic topological phase, which is referred to as a bosonic shadow, that possesses a topologically non-trivial point-like fermionic excitation, $\psi$, and then condensing the fermion.\footnote{This procedure has also been made mathematically precise in the context of braided fusion categories \cite{aasen2019}. } By bosonic topological phase, we mean a phase of matter whose topological path integral can be well-defined without a choice of spin structure. Physically, the fermion condensation procedure can be carried out by first stacking the bosonic system with a trivial fermionic insulator, which contains a topologically trivial fermionic excitation $c$, and then condensing the bound state $\psi c$. This effectively converts the topologically non-trivial fermion of the bosonic phase into a topologically trivial fermion of a fermionic phase.\footnote{This basic idea has been used to propose experiments to probe emergent fermions in quantum spin liquids and fractional quantum Hall states \cite{barkeshli2014sledge,barkeshli2015fqhsc,aasen2020}.} This is simply the reverse of the procedure where we start with a fermionic system and gauge fermion parity; in the field theory path integral, gauging fermion parity corresponds to summing over spin structures, while in categorical language, gauging fermion parity corresponds to taking a modular extension.\cite{bruillard2017a}
        
        Gaiotto and Kapustin provided an explicit method to carry out the fermion condensation procedure in a path integral setting. Since the bosonic shadow possesses a fermionic particle, it follows that it must have a $(d-2)$-form symmetry \cite{gaiotto2014}, generated by the Wilson loops for the fermionic particle. Therefore the theory can be coupled to a $(d-1)$-form $\Z_2$ gauge field $f_{d-1} \in Z^{d-1}(M^d, \Z_2)$, where $Z^k(M^d, \Z_2)$ denotes $\Z_2$-valued $k$-cocycles on $M^d$. Physically, $f_{d-1}$ defines, through Poincar\'e duality, a 1-dimensional submanifold along which a fermionic particle propagates in space-time. Therefore, we have a bosonic path integral $Z_b(M^d, A_b, f_{d-1})$, where $A_b$ is a flat $G_b$ connection, which can be thought of as the amplitude for a fermion to propagate along the loop determined by $f_{d-1}$. Summing over distinct $(d-1)$-form gauge fields $f_{d-1}$ is equivalent to summing over all possible closed paths for the fermions, which by definition condenses the fermion. 
        
         Importantly, since the $(d-2)$-form symmetry is associated with a fermionic particle, it possesses an 't Hooft anomaly that reflects the non-trivial Fermi statistics. Moreover, a non-trivial $[\omega_2]$ implies that the fermion carries fractional quantum numbers under the $G_b$ symmetry and implies the existence of a mixed anomaly between the $(d-2)$-form $\Z_2$ symmetry and the global $0$-form symmetry $G_b$. Therefore, condensing the fermion to obtain a topologically invariant, well-defined fermionic TQFT requires addressing some additional complications associated with these anomalies. 
         
         Below, we begin by briefly reviewing some geometric structures that appear when discussing worldlines of fermionic particles. Subsequently, we describe a series of anomalies that are expected on general grounds when working with theories that contain fermions and which show up in various parts of our constructions. Finally, we discuss the specific anomalies expected of the bosonic shadow $Z_b(M^d, A_b, f_{d-1})$ and how to cancel the anomalies to implement a well-defined fermion condensation procedure. 
         
         \subsection{ Geometric structures and fermion loops}
         
          Let us consider some general path integral $\mathcal{Z}(M^d, A_b, f_{d-1},\cdots)$, which can be thought of as a quantum amplitude for a fermion propagating in space-time along a loop that corresponds to the Poincar\'e dual of $f_{d-1}$, which we denote $f^\vee_{d-1}$. This fermion propagates in the presence of a flat $G_b$ gauge field $A_b$, which in turn determines a network of codimension-1 $G_b$ domain walls. $\mathcal{Z}$ may also depend on some additional structure indicated by the $\cdots$, like a spin structure, although for the current discussion this is not important.
         
          A loop in space-time along which a fermion can propagate is associated with a choice of fermion boundary conditions. Physically, such boundary conditions are related to the fact that there is some fermion field $\Psi(x)$ on the curve $x \in [0,L] / \{0 \sim L\}$ which may have either periodic or anti-periodic boundary conditions, such that $\Psi(0) = \pm \Psi(L)$. Mathematically, this is encoded by a choice of spin structure on the curve. The periodic (anti-periodic) boundary conditions are referred to geometrically as non-bounding (bounding) spin structures. 
          
          The spin structures on the curves can arise geometrically in a number of ways. If an orientable manifold $M$ is equipped with a spin structure, then a framing of the curve can be used to define an induced spin structure on the curve. However, even if the orientable manifold $M$ is not equipped with a spin structure, we can define an induced spin structure on generic curves that lie on the $1$-skeleton of $M$ as follows. Consider a set of $d$ vector fields ${v}_1,\cdots, {v}_d$ defined on $M$, which are linearly independent on the 1-skeleton of $M^d$. We refer to this set of $d$ vector fields as a ``background framing'' of $M$. Given a loop, we can use the tangent of the loop together with the first $(d-2)$ vector fields and the orientation of $M$ to define a framing of the loop, which we refer to as the tangent framing, which is a set of $d$ vector fields along the loop. The difference between the background and tangent framings along the loop defines a loop in $\SO(d)$, and thus a unique path in $\Spin(d)$ starting at the identity. The endpoint of this path exactly tells us the induced spin structure and is related to whether the loop in $\SO(d)$ was contractible. In our specific construction as discussed in the Appendix \ref{w1PmAssignments_AndVectorFields}, we are given a triangulation, branching structure, and assignment of $\pm$ signs to simplices, in which case there are canonical constructions of such a set of $d$ vector fields $\{{v}_i\}$. Since the first $(d-2)$ vector fields in the background and tangent framing can be taken to be shared, the induced spin structure can be computed by projecting the first $(d-2)$ vectors away, before computing the relative winding, as depicted in Figs. \ref{fig:inducedSpinStruct2D},\ref{fig:inducedSpinStruct3D}. This winding number, mod $2$, defines the induced spin structure on the loop.
          
          A crucial aspect of this spin structure is that when the framing of the curve twists by a unit, the induced spin structure is also twisted by a unit. Physically, this is the familiar statement that a $2\pi$ rotation of the fermion gives a minus sign, which is essentially the spin-statistics theorem of quantum field theory. 
          
          The above discussion can be generalized to non-orientable manifolds by considering induced pin structures on the fermion loops, as we explain in Appendix \ref{inducedPinStruct}. 
          
          A key property of fermionic systems is the existence of a fermion parity operator $(-1)^F$, which gives the fermion parity of a quantum state of fermions. It is well-known in QFT that the $(-1)^F$ operator lets one `toggle' between different fermionic boundary conditions. \footnote{See, e.g., Appendix A.2 of~\cite{book:polchinski1998_1} for an exposition on inserting $(-1)^F$ on a spatial slice changing boundary conditions in the $S^1$ time direction of a manifold $M^d \times S^1$. In general, $(-1)^F$ inserted on a codimension-1 slice of space-time is the same as twisting the spin structure along that slice.} Associated to this, there is the possibility of a fermion parity flux, which mathematically is a codimension-2 submanifold in space-time where continuously deforming a fermion loop past it twists the fermion boundary conditions by a unit and changes the path integral $\mathcal{Z}(M^d, A_b, f_{d-1},\cdots)$ by a minus sign. 
         
         \subsection{ Possible anomalies of $\Z_2$ $(d-2)$-form symmetries in theories containing fermions}
        
        In general we can consider a gauge transformation $f_{d-1} \rightarrow f_{d-1}' = f_{d-1} + \delta \lambda_{d-2}$, which physically corresponds to deforming the fermion loop to another topologically equivalent loop. We can also consider a gauge transformation $A_b \rightarrow A_b' = A_b + \delta \lambda_0$, which corresponds to moving the domain walls defined by $A_b$. 
        
        The path integral will not be gauge invariant, meaning that the quantum amplitude defined by $\mathcal{Z}(M^d, A_b, f_{d-1})$ can change as the fermion loop and $G_b$ domain walls are deformed. The gauge non-invariance is controlled by an 't Hooft anomaly, which means that 
        \begin{align}
        \mathcal{Z}(M^d, A_b', f_{d-1}') = \mathcal{Z}(M^d, A_b, f_{d-1}) e^{2\pi i S_{d+1}(M^d \times I, \tilde{A}_b, \tilde{f}_{d-1})} ,
        \end{align}
        where $S_{d+1}$ is a topological action for an invertible TQFT defined on the $(d+1)$-manifold $M^d \times I$, and where $\tilde{A}_b$, $\tilde{f}_{d-1}$ are defined on $M^d \times I$ and reduce to $A_b$, $f_{d-1}$ on one copy of the boundary $M^d$ and $A_b'$, $f_{d-1}'$ on the other copy. The dual of $\tilde{f}_{d-1}$, denoted $\tilde{f}^\vee_{d-1}$, defines a 2-dimensional worldsheet in $M^d \times I$ that describes how the fermion loop evolves in the extra $(d+1)^{\text{th}}$ direction. 
        
        When the particle propagating along the $f_{d-1}$ loop has Fermi statistics and also fractional quantum numbers under $G_b$, we can on general grounds expect the following set of anomalies may appear. 
        
        \subsubsection{$\Sq^2$, framing anomaly}
        
        The most important anomaly associated with a fermionic particle is the so-called Steenrod square anomaly, for which we have
        \begin{align}
            S_{d+1}[W^{d+1}, \tilde{f}_{d-1}] = \frac{1}{2} \int_{W^{d+1}} \text{Sq}^2(\tilde{f}_{d-1}). 
        \end{align}
        For a general $\mathbb{Z}_2$ $(d-1)$-cocycle, the Steenrod square can be defined in terms of the higher cup product (see Appendix \ref{prelimSecB} for a discussion of higher cup products):
        \begin{align}
        \Sq^2(\tilde{f}_{d-1}) = \tilde{f}_{d-1} \cup_{d-3} \tilde{f}_{d-1} . 
        \end{align}    
        
\begin{figure}[h!]
        \centering
        \includegraphics[width=\linewidth]{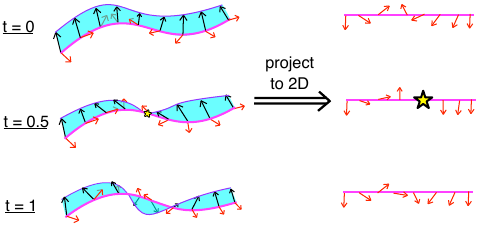}
         \caption{Illustration in $d=3$ of a movie from $t = 0 \to 1$ of how the $\Sq^2$ anomaly corresponds to a `framing anomaly' of fermion loops. $f$ (pink line) and a shifted version of $f$ (purple lines) shifted by the black vectors correspond to a framing (blue sheet) of the fermion line. Under the bordism movie, $f$ and its shifted version may intersect, which corresponds to contributions to $\int f \cup_{d-3} f = \int f \cup f$ at $t=0.5$ here. Projecting away the framing, the winding of this extra red vector field around $f$'s tangent line around the loop determines the induced spin structure on $f$, as in Sections \ref{sec:inducedSpinStructOnCurve},\ref{inducedPinStruct}. For every intersection, this winding jumps by $\pm 2\pi$ and will twist the spin structure by a unit. This can be thought of as a framing anomaly. The star indicates the point of intersection, where the red vector field degenerates after the projection. }
          \label{fig:sq2FramingPic}
        \end{figure}

        We can understand the origin of the Steenrod square anomaly as a framing anomaly. As discussed above, the fermion loop is essentially a framed loop, and changing the framing by one unit gives a minus sign due to the spin of the fermion. We will see here that this change of framing is accounted for by the Steenrod square.  See Fig.~\ref{fig:sq2FramingPic} for an illustration of this.
        
        Mathematically, the term $\Sq^2(\tilde{f}_{d-1}) = \tilde{f}_{d-1} \cup_{d-3} \tilde{f}_{d-1}$ can be understood geometrically by a theorem of Thom~\cite{thom1950}. Suppose that the manifold $M^d$ is equipped with a background framing, i.e. a generic collection of $d$ vector fields. Then we thicken the $2$-dimensional sheet defined by $\tilde{f}_{d-1}^\vee$ along the first $d-3$ vector fields, and then shift it along the $(d-2)^{\text{nd}}$ vector field, to define a $(d-1)$-dimensional submanifold $\tilde{f}^\vee_{\text{thickened,shifted}}$. Then $\Sq^2(\tilde{f}_{d-1})$ counts the number of points in the intersection $\tilde{f}^\vee_{\text{thickened,shifted}} \cap \tilde{f}^\vee$, modulo $2$. 
        
        In the special case where we are given a triangulation and branching structure, one can define $d$ linearly independent vector fields along the dual $1$-skeleton as discussed in Appendix \ref{w1PmAssignments_AndVectorFields}. One can then further define $\Sq^2$ at the cochain level in terms of a geometric interpretation of higher cup products, developed in Ref.~\cite{tata2020}, which we explain in Appendix \ref{prelimSecB}, in terms of the thickening and shifting prescription above.
        
        The description of $\Sq^2$ given above can be related to a change of framing of the fermion curve. 
        The $(d-3)$ thickening directions and $(d-2)^{\text{nd}}$ shifting direction, together with the tangent of the curve and the orientation of $M$, define $d$ vector fields along the curve, which constitute a framing of the fermion line in $d$ dimensions. This framing of the loop, together with the $d$ vector fields constituting the background framing of $M$, induce a spin structure on the loop. The bordism $W^{d+1}$ from $M^d \to M^d$ can be thought of as a movie that at $t = 0$ starts off with a given collection of framed fermion worldlines. As time increases, the framing of the fermion worldlines can change; this occurs precisely when the $(d-2)$ vector fields become degenerate at certain space-time points.
        
        In our setting of a triangulated bordism, these changes of induced framings correspond to the higher cup product, since the $\cup_{d-3}$ above computes the number of intersections, which is exactly the number of framing twists. Therefore the Steenrod square term implies that the space-time path integral $\mathcal{Z}(M^d, A_b, f_{d-1})$ should change by a minus sign each time the framing of the fermion line twists by a unit, giving us one of our anomaly mechanisms.
        See Fig.~\ref{fig:sq2FramingPic} for an illustration. 
        
        \subsubsection{$w_2$ anomaly}
        
        A closely related but geometrically distinct anomaly action involves the second Stiefel-Whitney class, which is represented by a $\Z_2$-valued $2$-cocycle $w_2$. 
    
\begin{figure}[h!]
            \centering
            \begin{minipage}{0.42\textwidth}
                \centering
                \includegraphics[width=\linewidth]{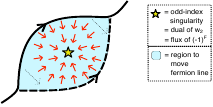}
                \end{minipage}
            \begin{minipage}{0.52\textwidth}
                 \centering
                 \includegraphics[width=\linewidth]{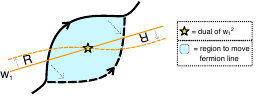}
            \end{minipage}
            \caption{(Left) $w_2$ is dual to odd-index singularities of a frame of $(d-1)$ generic vector fields (single red vector field here). It can be thought of as a flux of $(-1)^F$ because changing a curve's path through it twists the induced spin structure / fermionic boundary conditions on the curve. (Right) A dual of $[w_1]$ and a perturbation determine the dual of $w_1^2$. Physically, the perturbing direction determines an action of ${\bf R}$ (as opposed to ${\bf R}^{-1}$) on the loop. 
            }
            \label{fig:anomalyPics}
        \end{figure}

        The origin of this anomaly action is due to the fact that, in the absence of a spin structure, $w_2$ defines a  codimension-$2$ submanifold, which we denote as $w_2^\vee$, associated with the location of fermion parity vortices; that is, a codimension-$2$ submanifold around which a fermion worldline acquires a minus sign relative to the absence of the submanifold. In Appendix \ref{sec:inducedSpinStructOnCurve},\ref{inducedPinStruct} and \ref{twistedSpinSec}, we provide two perspectives on how to understand the relationship between $w_2$ and sources of fermion parity flux in the absence of spin structure. Geometrically, this relative minus sign arises from the fact that the induced spin structure on the loop dual to $f$ gets twisted by a unit upon every crossing with the dual of $w_2$. This is illustrated in Fig.~\ref{fig:anomalyPics}. 
        
        Therefore, one expects a sign whenever the loop $f^\vee_{d-1}$ is deformed to cross $w_2^\vee$. This leads to an anomaly action that counts the number of $0$-dimensional crossings in the intersection $w_2^\vee \cap \tilde{f}^\vee_{d-1}$ in $W^{d+1}$, which we denote $\# \mathrm{int}\big(w_2^\vee,\tilde{f}^\vee_{d-1} | W^{d+1}\big)$. 
        Let $w_2$ be a $2$-cocycle defined on the dual cellulation of a triangulation of the space-time manifold and $\tilde{f}_{d-1}$ to be a $(d-1)$-cocycle on the triangulation of $W^{d+1}$. 
        We will denote the pairing between $w_2$ and $\tilde{f}$ as $w_2(\tilde{f})$, and thus we write:\footnote{We note that in previous work, this anomaly is often written as $w_2 \cup \tilde{f}_{d-1}$, however there is no canonical definition of such a term at the cochain level in the continuum. On a triangulation such a term is well-defined if $w_2$ is a $2$-cocycle on the triangulation, however in our constructions we naturally have a $2$-cocycle on the dual cellulation.}
        \begin{align}
            S_{d+1}[W^{d+1}, \tilde{f}_{d-1}] = \frac{1}{2} \int_{W^{d+1}} w_2(\tilde{f}) = \frac{1}{2}\cdot\# \mathrm{int}\big(w_2^\vee,\tilde{f}_{d-1}^\vee | W^{d+1}\big).
        \end{align}
        
        As discussed in Appendix \ref{prelimSecB} and \ref{twistedSpinSec}, a spin structure $\xi$ is additional data that essentially cancels out the fermion parity fluxes defined by $w_2$. From the geometrical discussion in Appendix \ref{app:SpinStructReview}, we see that the spin structure on the manifold is a specification for fixing the $d$ vector fields along the $1$-skeleton of the manifold so that they can be extended over the $2$-skeleton, which essentially trivializes $w_2$. From the algebraic discussion in Appendix \ref{twistedSpinSec}, we see that the spin structure specifies the lift of the transition functions of an $\SO(d)$ bundle to $\Spin(d)$, in such a way that the cocycle on three overlapping charts is trivial. 
        
        Therefore, we expect that in a theory of fermions that appropriately depends on a spin structure, the $w_2$ anomaly above should vanish. In contrast, a fermionic theory which does not depend on a spin structure, but in which the fermions are sensitive to the background geometry of the manifold, should possess the above $w_2$ anomaly; the relevant example in this paper is the Gu-Wen Grassmann integral $\sigma$, which is defined in Secs.~\ref{windingGrassDefSec},\ref{grassIntegral} and Appendix \ref{sec:windingDefOfSigma}. 
        
        \subsubsection{$w_1^2$ anomaly}
        
        On a non-orientable manifold, the first Stiefel-Whitney class, represented by a $1$-cocycle $w_1$, is non-vanishing. The dual of $w_1$ is a codimension-$1$ submanifold $w_1^\vee$ across which the local orientation gets flipped. Therefore a fermion crossing $w_1^\vee$ is effectively acted on by a reflection. In general we can consider two classes of reflection action: ${\bf R}^2 = (-1)^F$, or ${\bf R}^2 = 1$. These descend from Wick-rotating corresponding time-reversal symmetries ${\bf T}^2 = 1$ or ${\bf T}^2 = (-1)^F$ respectively.\footnote{See, e.g., Ref.~\cite{Kapustin:2014dxa,witten2016rmp} for discussions of this.}  Here we consider the effects of the former case, where \textit{time-reversal} and fermion parity do not mix. The latter case is handled in the next subsection about $A_b^* \omega_2$.
        
        In the case where ${\bf R}^2 = (-1)^F$, it turns out that $w_1^2$, dual to the self-intersection $(w_1^2)^\vee$ of the orientation-reversing wall, also acts a source of fermion parity flux, which leads to the anomaly action
        \begin{align}
           S_{d+1}[W^{d+1}, \tilde{f}_{d-1}] = \frac{1}{2} \int_{W^{d+1}} w_1^2(\tilde{f}_{d-1}) = \frac{1}{2}\cdot\# \mathrm{int}\big((w_1^2)^\vee,\tilde{f}_{d-1}^\vee | W^{d+1}\big). 
        \end{align}
        
        We can heuristically understand the origin of the above term as follows. ${\bf R}^2 = (-1)^F$ implies ${\bf R} = {\bf R}^{-1} (-1)^F$. The symmetry action defines a perturbation of the orientation-reversing wall by demanding that crossing along the perturbation acts as ${\bf R}$, while crossing against the perturbation acts as ${\bf R}^{-1}$. 
        As such, crossing the orientation-reversing wall in one direction differs from crossing in the other direction by a factor of $(-1)^F$. Also, the perturbation defines a codimension-2 submanifold consisting of where $w_1^\vee$ and its perturbation intersect each other. Encircling this codimension-2 intersection is equivalent to being acted upon by ${\bf R}^2 = (-1)^F$, which thus explains why $w_1^2$ should be thought of as sourcing fermion parity flux. This is sketched in Fig.~\ref{fig:anomalyPics}.
        
        At a mathematical level, $w_1^2$ gives a codimension-2 submanifold across which the induced pin$^{-}$ structure of the fermion loop changes by a sign, as discussed in Sec.~\ref{inducedPinStruct}. 
        
        We note that on closed manifolds, we have that
        \begin{align}
            \int_{W^{d+1}} \text{Sq}^2 (\tilde{f}_{d-1}) 
            &= \int_{W^{d+1}} (w_2 + w_1^2)( \tilde{f}_{d-1}). 
        \end{align}
        However, the two sides of the equation are distinct when $\partial W^{d+1}$ is non-trivial, and there is no known local formula for the difference in terms of a coboundary. As we will see, the distinction between these terms plays an important role in our constructions.
        
        As in the discussion of the $w_2$ anomaly, we expect the above $w_1^2$ anomaly may possibly appear whenever a fermionic theory is defined on a non-orientable manifold and the fermions are sensitive to the background geometry of the manifold, but a pin$^-$ structure has not been specified. The Gu-Wen Grassmann integral $\sigma$ provides an example of such a fermionic theory. 
        
        \subsubsection{$A_b^* \omega_2(f)$, mixed anomaly}
        
        The last important anomaly is a mixed anomaly between the $0$-form $G_b$ symmetry and the $(d-2)$-form $\Z_2$ symmetry, which occurs in the case where $[\omega_2] \in \H^2(G_b, \Z_2)$ is non-trivial. In this case, the fermions possess fractional $G_b$ quantum numbers; therefore condensing the fermions will in general break the $G_b$ symmetry, which is a signature of the anomaly.  
        \begin{figure}[h!]
         \centering
         \includegraphics[width=0.5\linewidth]{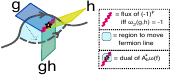}
        \caption{$G_b$ domain walls labeled by ${\bf g},{\bf h}$ meeting at a trijunction with a ${\bf gh}$ domain wall will trap a $(-1)^F$ flux if and only if $\omega_2({\bf g},{\bf h}) = -1 \in \Z_2$. The $A_b^*\omega_2(f)$ anomaly term tells us that the induced spin structure / fermionic boundary conditions be twisted by a unit upon passing through such trijunctions with $\omega_2({\bf g},{\bf h}) = -1$. This gauge-field dependence on fermionic boundary conditions signals the need for a $G_b$-twisted spin structure on the manifold.}
         \label{fig:anomalyPic_AbOmega}
        \end{figure}
        
        The anomaly gives an account of the following physical phenomena. Recall that a flat $G_b$ gauge field can be thought of as a network of codimension-1 domain walls labeled by elements of $G_b$. The statement of the symmetry fractionalization is that as a fermion line passes through a trijunction of a ${\bf g}$, ${\bf h}$, and ${\bf gh}$ domain wall, it should pick up a sign $(-1)^{\omega_2({\bf g}, {\bf h})}$. 
        See Fig.~\ref{fig:anomalyPic_AbOmega} for an illustration. 

        This can be encoded in the following $(d+1)$-dimensional action~\footnote{Note that later on when describing our construction, we use different notation, either $f \cup A_b^* \omega_2 = (f_\infty A_b^* \omega_2)(f)$, to encode this, due to technicalities of working on a triangulation. See Sec.~\ref{sec:geometricChoices}.}:
        \begin{align}
        \label{mixedAnomalyAction}
            S_{d+1} = \frac{1}{2} \int_{W^{d+1}} \tilde{A}_b^* \omega_2 ( \tilde{f}_{d-1}) = \frac{1}{2}\cdot\# \mathrm{int}\big((\tilde{A}_b^* \omega_2)^\vee,\tilde{f}_{d-1}^\vee | W^{d+1}\big). 
        \end{align}
        Here, a flat gauge field on $M^d$ can be defined in terms of a map $A_b: M^d \rightarrow BG_b$, where $BG_b$ is the classifying space of $G_b$. $\tilde{A}_b: W^{d+1} \rightarrow BG_b$ restricts to $A_b$ on $\partial W^{d+1} = M^d$, and $\tilde{A}_b^*$ is the pullback which maps the cohomology class $[\omega_2] \in H^2(BG_b, \Z_2) \rightarrow H^2(W^{d+1}, \Z_2)$. Thus $A_b^*\omega_2$ is a representative 2-cocycle in $H^2(M^d, \Z_2)$ and $\tilde{A}_b^* \omega_2$ gives a representative 2-cocycle in $H^2(W^{d+1}, \Z_2)$. If we consider the gauge field $A_b(ij)$ on links $\braket{ij}$ of a simplicial complex (we use $A_b$ to refer both to the map and gauge field), we have $A_b^* \omega_2(ijk) = \omega_2\big(A_b(ij), A_b(jk)\big)$ on the $3$-simplex $\braket{ijk}$. This cocycle is Poincaré dual to a codimension-2 submanifold $(\tilde{A}_b^* \omega_2)^\vee$.
            
        Note that the bundle associated to the anti-unitary part of $G_b$ is given by the orientation bundle of $M^d$: orientation-reversing loops correspond to loops with flux of time-reversal symmetry.
        
        For example, in the case $G_b = \Z_2^{\bf T}$, we set $A_b = w_1(M^d) \in Z^1(M^d, \Z_2)$, where $w_1(M^d)$ is a 1-cochain representative of the first Stiefel-Whitney class $[w_1]$. In the dual-picture, this means that the `time-reversal' domain wall is exactly the orientation-reversing wall. If $[\omega_2]$ is non-trivial, in which case $G_f = \Z_4^{{\bf T},f}$, then $A_b^*\omega_2 = w_1^2$, since $\omega_2(A_b(01), A_b(12)) = \omega_2(w_1(01), w_1(12)) = w_1(01) w_1(12)$. Extending these to the bulk manifold $W^{d+1}$, we have $\tilde{A}_b^* \omega_2 = w_1^2$, where now $w_1 = w_1(W^{d+1})$ is a 1-cochain representative of $[w_1(W^{d+1})] \in H^1(W^{d+1}, \Z_2)$. The mixed anomaly is therefore governed by:
        \begin{align}
         S_{d+1} = \frac{1}{2} \int_{W^{d+1}} w_1^2( \tilde{f}_{d-1}) =\frac{1}{2}\cdot \#\mathrm{int}\big((w_1^2)^\vee,\tilde{f}_{d-1}^\vee | W^{d+1}\big). 
        \end{align}
        As such, in this example of ${\bf T}^2 = (-1)^F \leftrightarrow {\bf R}^2 = 1$, we have that the $A_b^*\omega$ term is equivalent to the $w_1^2$ term. Looking forward, this is the reason why $\Z_4^{{\bf T},f}$ requires $\mathcal{G}_f = \text{pin}^+$ structures that trivialize $w_2 + w_1^2 + A_b^* \omega_2 = w_2 + w_1^2 + w_1^2 = w_2$.
    
        \subsection{ Anomalies of the bosonic shadow $Z_b(M^d, A_b, f_{d-1})$} \label{sec:anomaliesOfShadow}

        Now let us consider a bosonic theory which contains an emergent fermion with fractional quantum numbers under $G_b$ defined by $[\omega_2]$. In this case, $Z_b(M^d, A_b, f_{d-1})$ must have the Steenrod square and mixed anomalies discussed above, which arise directly because of the Fermi statistics and fractional $G_b$ quantum numbers of the fermion. However, importantly, we do not expect an emergent fermion to have the $w_2 + w_1^2$ anomaly. If we have both the $\text{Sq}^2$ anomaly and the $w_2 + w_1^2$ anomaly, then the anomaly would be cohomologically trivial as we would essentially be double-counting the effect of the Fermi statistics. At the cochain level we expect $\text{Sq}^2$ as opposed to $w_2 + w_1^2$. Since the emergent fermion is fundamentally a non-local disturbance of a bosonic theory, we not expect it to interact directly with any particular representative of $w_2 + w_1^2$; its Fermi statistics manifests under changes of framing of its worldline, which is directly taken into account by $\text{Sq}^2$. 
        As we will see, this is indeed borne out in the specific construction of $Z_b$ that we present in Section \ref{sec:craneyettershadow}. 
        
        Therefore, on general grounds we expect that the anomaly of the bosonic shadow is given by 
        \begin{align}
        S_{d+1,b} = \frac{1}{2} \int_{W^{d+1}} \left( \text{Sq}^2(\tilde{f}_{d-1}) + \tilde{A}_b^* \omega_2 ( \tilde{f}_{d-1}) \right).
        \end{align}
        
        The path integral $Z_b$ will, in the cases that we study, be presented in terms of a triangulation. In this case, $Z_b$ will not be retriangulation invariant in the presence of non-zero background $f_{d-1}$ due to the above anomalies. This makes perfect sense because retriangulations will require changing $f_{d-1}$, which will change the path integral. The change in phase can be understood in terms of the $(d+1)$-dimensional action. Given two triangulations $T$ and $T'$ of $M^d$, we consider a triangulation of $M^d \times [0,1]$ which reduces to $T$ and $T'$ on the two boundaries. We also consider extending the definition of $A_b$ and $f_{d-1}$ to $\tilde{A}_b$, $\tilde{f}_{d-1}$, which are defined on the triangulation of $M^d \times [0,1]$, and which reduce to their values on the initial and final triangulations. Then the change in phase of $Z_b$ is given by
        \begin{align}
            Z_b( (M^d, T), A_b, f_{d-1}) = Z_b( (M^d, T'), A_b', f_{d-1}' ) e^{2 \pi i S_{d+1,b}(M^d \times [0,1], \tilde{A}_b, \tilde{f}_{d-1})}.
        \end{align}
        Here we have explicitly written the dependence on the triangulation $T$ in the path integral. Note that we could also consider $(M^d)'$ on the right-hand side instead, in which case we consider a bordism from $M^d$ to $(M^d)'$. 
        
        \subsection{ Anomaly cancellation and fermion condensation}
        \label{anomalyCancelOverview}
        
        In order to obtain a retriangulation invariant, anomaly-free path integral, we must therefore consider ``stacking'' the bosonic theory with another theory which transforms in exactly the same way as to cancel these anomalies. From the discussion above, it is clear that this other theory should be an intrinsically fermionic theory. An important requirement of such a fermionic theory is that it should depend on a choice of a $\mathcal{G}_f$-structure $\xi_{\mathcal{G}}$, which is required to specify how to couple the theory with $G_b$ internal symmetry to fermions. Thus, we consider the product:
        \begin{align}
        \label{}
        Z_b( (M^d, T), A_b, f_{d-1}) z_c( (M^d, T), f_{d-1}, \xi_{\mathcal{G}}) . 
        \end{align}
        Here $z_c$ is the path integral of a fermionic theory, i.e. which contains physical fermions, and will be defined in Section \ref{sec:zpsi}. Note that $z_c$ depends on the $G_b$-connection $A_b$ implicitly through the $\mathcal{G}_f$ structure $\xi_{\mathcal{G}}$ which we will define in Section \ref{sec:geometricChoices} and Appendix \ref{twistedSpinSec}. When $A_b = 0$ or $[\omega_2]$ is trivial, $\xi_{\mathcal{G}}$ is a spin or pin structure on the manifold, which justifies why we refer to $z_c$ as a fermionic theory. When $z_c$ is defined appropriately, the product above will be retriangulation invariant, anomaly-free, and independent of a change of $\xi_{\mathcal{G}}$ by a 1-coboundary. 

\begin{figure}[h!]
        \centering
        \includegraphics[width=\linewidth]{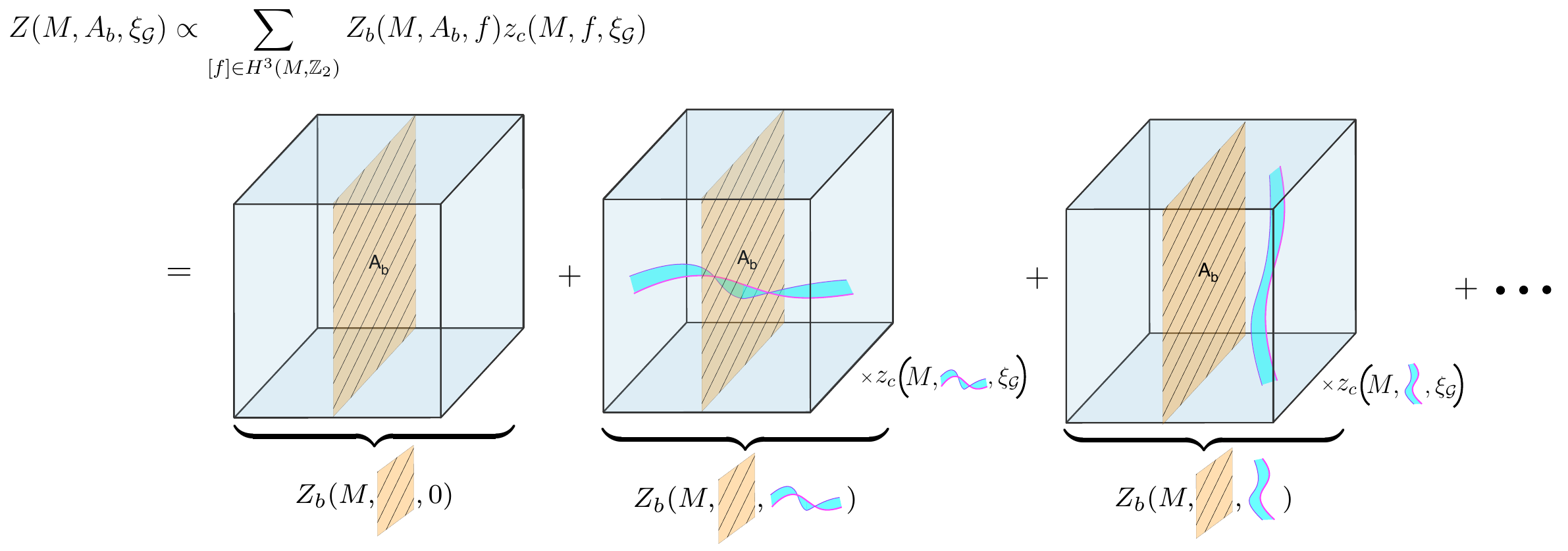}
        \caption{Cartoon of our state sum. We are given 4-manifold $M$ (depicted here in $d=3$) and a flat $G_b$ gauge field $A_b$ represented by a closed network of codimension-1 domain walls specifying holonomies. The full state sum $Z$ depends on a $\mathcal{G}_f$ structure, $\xi_\mathcal{G}$. It is a sum over all cohomology classes $[f] \in H^3(M,\Z_2)$ which are represented here by their Poincaré dual loops. For each $[f]$ and a representative loop $f$ (pink curve), we add the contribution of the `shadow-theory' $Z_b(M,A_b,f)$ in the background of $f$, times a factor $z_c(f,\xi_\mathcal{G})$. The curve dual to $f$ is endowed with a framing (purple curve and blue sheet) depending on the background geometry of $M$, and the factor $z_c(f,\xi_\mathcal{G})$ gives one of $\{\pm 1, \pm i\}$ depending on the induced spin structure (i.e. fermionic boundary conditions) on the framed curve from the $\xi_{\mathcal{G}}$.} 
        \label{fig:stateSum_cartoon}
        \end{figure}

        Finally, the topologically invariant path integral for our fermionic theory can then be obtained by condensing the fermions of the bosonic shadow by summing over all inequivalent choices of $f_{d-1}$:
        \begin{align}
        \label{pathIntegral}
        Z(M^d, A_b, \xi_{\mathcal{G}}) \propto \sum_{[f_{d-1}] \in H^{d-1}(M^d, \Z_2)} Z_b((M^d,T), A_b, f_{d-1}) z_c((M^d,T),f_{d-1},\xi_{\mathcal{G}}).
        \end{align}
        This is illustrated schematically in Fig.~\ref{fig:stateSum_cartoon}. 
        
        A careful and detailed definition of $Z(M^d, A_b, \xi_{\mathcal{G}})$ in our (3+1)D constructions will be given in subsequent sections.

        \section{ Setting up the construction: geometric choices on triangulations}
        \label{sec:geometricChoices}

        The construction of the fermionic path integral $Z(M^4, A_b, \xi_\mathcal{G})$ requires a number of important geometric choices as we now briefly describe. While $Z$ itself will eventually only depend on $M^4$, $A_b$, and $\xi_{\mathcal{G}}$, the quantities $Z_b$ and $z_c$ arising in the construction do explicitly depend on the additional geometric choices described below.
        
        First, we pick a triangulation of $M$ (we often drop the notation labeling the dimension of $M$). Following Appendix~\ref{topologicalPreliminaries}, this triangulation defines a dual cellulation $M^\vee$ of $M$. For both $M$ and $M^\vee$, there are $\Z_2$ chain spaces $Z_k(M,\Z_2),Z_k(M^\vee,\Z_2)$ with boundary operators $\partial$ and $\Z_2$ cochain spaces $Z^k(M,\Z_2),Z^k(M^\vee,\Z_2)$ with coboundary operators $\delta$. Poincaré duality on the chain-level is the statement that $\left( Z_k(M,\Z_2) , \partial \right) \equiv \left( Z^{d-k}(M^\vee,\Z_2) , \delta \right)$ in the sense that both spaces are canonically equal and their respective $\partial$,$\delta$ operators are canonically equal. As such, we will often abuse notation in referring to objects as living in either space, as with their respective (co)boundary operators.  Similarly, we have $\left( Z^k(M,\Z_2) , \delta \right) \equiv \left( Z_{d-k}(M^\vee,\Z_2) , \partial \right)$.
        
        Now equip $M^d$ with a branching structure, which is a local ordering of vertices. As we explain in Appendix \ref{prelimSecB}, the branching structure immediately determines a canonical representative $w_2 \in Z^2(M^\vee, \Z_2)$ for the second Stiefel-Whitney class; that is, $w_2$ is a 2-cocycle on the dual cellulation. 
        Next, we pick a flat $G_b$ gauge field $A_b$ which defines a map from $1$-simplices of the triangulation into $G_b$. Importantly, to ensure a retriangulation-invariant path integral, holonomies of $G_b$ through non-contractible cycles should correspond to anti-unitary (resp. unitary) symmetry group elements if the cycles are orientation-reversing (resp. preserving). To make this more precise, let $s : G_b \rightarrow \Z_2$ be the $\Z_2$ grading on $G_b$ which determines which group elements are anti-unitary. Then, the pullback $A_b^\ast s \in Z^1(M, \Z_2)$ defines a flat $\Z_2$ gauge field on the triangulation, i.e. a $\Z_2$ bundle over $M$, which should coincide with the orientation bundle. 
        
        The $\Z_2$ gauge field $A_b^\ast s$ determines a collection of $1$-simplices across which the local orientation is reversed. We will instead want a collection of dual 1-cells across which the local orientation is reversed, or equivalently an \textit{orientation-reversing wall} consisting of a closed submanifold of $(d-1)$-simplices, which lives in $Z_{d-1}(M,\Z_2) \equiv Z^1(M^\vee,\Z_2)$, and which we will denote as $w_1$. 
        As we explain in Appendix~\ref{app:FInfty}, one can define~\cite{Thorngrenthesis} a map $f_\infty$ which maps a $k$-cocycle on the original triangulation to a $(d-k)$-cycle on the original triangulation. Thus, $f_\infty(A_b^*s) \in Z_{d-1}(M, \Z_2)$ defines the desired orientation-reversing wall, comprised of $(d-1)$-simplices of the original triangulation. The Poincaré dual of this orientation-reversing wall defines a representative $w_1$ of the first Stiefel-Whitney class as a $1$-cochain on the dual cellulation, in $Z^1(M^\vee,\Z_2)$.
        
        Finally, for each $d$-simplex $\Delta_d$, we choose an assignment $\epsilon(\Delta_d) = \pm$, which defines an orientation for each $d$-simplex, subject to a compatibility requirement. As shown in Appendix~\ref{w1PmAssignments_AndVectorFields}, a branching structure together with a collection of $\pm$ signs for $d$-simplices defines a representative for $w_1$ as a cochain on the dual cellulation. We require that the $\pm$ signs are chosen such that $w_1 = f_\infty (A_b^\ast s)$. 
        
        Not only does $A_b$ define a representative $w_1$ as discussed above, but as we explain in Appendix~\ref{fInftyAndPerturbationOf_w1}, it also defines a perturbation of the orientation-reversing wall associated with $w_1$. This perturbation can be used to define a self-intersection of the orientation-reversing wall, which has support on $(d-2)$-simplices, i.e. on the dual $2$-cells. This defines a representative $2$-cocycle for $w_1^2$ defined on the dual cellulation.
        
        In what follows, we will use $T$ to denote the triangulation, branching structure, and assignment of $\pm$ signs. $w_2$, $w_1$, and $w_1^2$ will be cochains on the dual cellulation and will be assumed to arise from the choices of $M^d$, $T$, and $A_b$ discussed above.
        
        Our definition of $z_c$ below will depend on a $\mathcal{G}_f$ structure, which is a $1$-cochain defined on the dual cellulation, with the property that
        \begin{align}
            \delta \xi_\mathcal{G} = w_2 + w_1^2 + f_\infty A_b^* \omega_2 . 
        \end{align}
        The $\mathcal{G}_f$-structure $\xi_{\mathcal{G}}$ allows us to specify a lift from the full bosonic symmetry (including space-time symmetries) of the field theory to the fermionic theory. We describe it mathematically in detail in Appendix \ref{twistedSpinSec}. 
        
        In our discussion of anomalies, we will be working in a five-dimensional retriangulation geometry $W^5$. There, we denote $w_2(f),w_1^2(f)$ to refer to the cochain-chain pairing between $w_2,w_1^2 \in Z^2(W^\vee,\Z_2) = Z_3(W,\Z_2)$ and the fermion lines $f \in Z_2(W^\vee,\Z_2) = Z^3(W,\Z_2)$. Notationally it makes just as much sense to write $f(w_2), f(w_1^2)$, but we choose not to do this to avoid confusion with $f_\infty$ and to conceptually treat the fermion lines $f$ as the \textit{input}. For the $A_b^* \omega_2$ anomaly, we will often see $(f \cup A_b^* \omega_2)(012345)=f(0123)A_b^* \omega_2(34,45)$ which by definition is equivalent on the cochain-level to $\big(f_\infty A_b^* \omega_2\big)(f)(012345)$. We will also sometimes express the cochain-chain pairing as an integration over some corresponding chain: for example, $(-1)^{\int_{w_2} \lambda} := (-1)^{\int_M w_2(\lambda)} = (-1)^{\int {\lambda(w_2)}} =: (-1)^{\int_{\lambda} w_2}$. 
        
        \section{ Definition of $z_c$}
        \label{sec:zpsi}
        
        Here we define the factor $z_c$, which is an important ingredient in canceling the anomalies of $Z_b$ in order to implement the fermion condensation procedure by summing over the closed $(d-1)$-form field $f_{d-1}$. 
        
        \subsection{ Factorization of $z_c$ and anomaly cancellation}
        
        A natural idea to define the fermionic path integral $z_c$ is to set up a Grassmann integral that describes an integral over the worldlines of fermionic particles. Such a path integral, which we denote $\sigma(M^d, f_{d-1})$, was defined by Gu and Wen \cite{gu2014}. However, we will see that $\sigma(M^d, f_{d-1})$ actually does not cancel the anomalies of $Z_b$ discussed in the preceding section, fundamentally due to its lack of dependence on a $\mathcal{G}_f$ structure; it can cancel the Steenrod square anomaly at the cost of introducing a distinct set of terms in the $(d+1)$-dimensional anomaly action that are cohomologically equivalent to the Steenrod square term. The additional terms can then be canceled by the introduction of one extra simple factor in $z_c$ that depends on a a $\mathcal{G}_f$ structure.
        
        According to the above discussion, the fermionic path integral $z_c( (M^d, T), f_{d-1}, \xi_{\mathcal{G}})$ can be split into two pieces by extracting the dependence on the $\mathcal{G}_f$ structure, $\xi_{\mathcal{G}}$:
        \begin{align}
            z_c( (M^d, T), f_{d-1}, \xi_{\mathcal{G}}) = \sigma( (M^d, T), f_{d-1}) e^{\pi i \int_{M^d} \xi_{\mathcal{G}}(f_{d-1})} . 
        \end{align} 
        Here we use the notation introduced in the previous section, where $\xi_{\mathcal{G}}(f_{d-1})$ counts the number of intersections (mod 2) between the $(d-1)$-dimensional manifold defined by the dual of the 1-cochain $\xi_{\mathcal{G}}$ and the loop $f_{d-1}^\vee$. 
        
        The Gu-Wen Grassmann integral $\sigma( (M^d, T), f_{d-1})$ will be defined in two distinct but equivalent ways in Sections \ref{windingGrassDefSec} and \ref{grassIntegral}.
        
        The anomalies of $z_c$ can be understood in terms of the anomalies of the two factors. The anomaly for the factor $e^{i \pi \int_{M^d} \xi_{\mathcal{G}} (f_{d-1})} $ is controlled by the $(d+1)$-dimensional action:
        \begin{align}
            S_{d+1; \xi} = \frac{1}{2}\int_{W^{d+1}} (\delta\tilde{\xi}_{\mathcal{G}})(\tilde{f}_{d-1}).
        \end{align}
        Here $\tilde{\xi}_{\mathcal{G}}$ restricts to $\xi_{\mathcal{G}}$ on $\partial W^{d+1} = M^d$ and is a trivialization, satisfying $\delta \tilde{\xi}_{\mathcal{G}} = w_2 + w_1^2 + f_\infty \tilde{A}_b^* \omega_2$, extended into the bulk retriangulation geometry $W^{d+1}$.

        As we explain in subsequent sections, the Gu-Wen Grassmann integral $\sigma((M^d,T), f_{d-1})$ is defined in such a way that its anomalies are controlled by the following $(d+1)$-dimensional action:
        \begin{align} \label{sigmaGrAnomalyAction}
            S_{d+1,\sigma} = \frac{1}{2} \int_{W^{d+1}} \left( (w_2 + w_1^2)(\tilde{f}_{d-1}) + \Sq^2(\tilde{f}_{d-1}) \right).
        \end{align}
        On a \textit{closed} $W^{d+1}$, the `Lagrangian' $(w_2 + w_1^2)(\tilde{f}_{d-1}) + \Sq^2(\tilde{f}_{d-1})$ is cohomologically trivial due to the Wu relation~\cite{ManifoldAtlasWu}, but survives as a cochain in general.
        Remarkably, the two anomalies above conspire together to cancel off the pure and mixed anomalies of $Z_b$ at the cochain level. Indeed, we can write the total $(d+1)$-dimensional action for the combined system:
        \begin{align}
            S_{d+1, b} + S_{d+1, \xi} + S_{d+1, \sigma} &= 
            \frac{1}{2}\int_{W^{d+1}} [\Sq^2(\tilde{f}) + \tilde{f} \cup \tilde{A}_b^* \omega_2] + [(w_2 + w_1^2 + f_\infty\tilde{A}_b^* \omega_2) (\tilde{f})] + [(w_2+w_1^2) (\tilde{f}) + \Sq^2(\tilde{f})] 
            \nonumber \\
            &= 0 \text{ mod } 1,
        \end{align}
        where $(f_\infty \tilde{A}_b^* \omega_2)(\tilde{f}) := \tilde{f} \cup \tilde{A}_b^* \omega_2$ by definition.
        
        Finally, we note that an important property of $z_c(M^d, f_{d-1}, \xi_{\mathcal{G}})$ is that it is a quadratic refinement of a higher cup product pairing:
        \begin{equation} \label{eq:quadRefinement_zC}
        z_c(M^d, f_{d-1} + f_{d-1}',\xi_{\mathcal{G}}) = z_c(M^d, f_{d-1}, \xi_{\mathcal{G}})z_c(M^d, f'_{d-1}, \xi_{\mathcal{G}}) e^{i \pi \int_{M^d} f_{d-1} \cup_{d-2} f_{d-1}' } .
        \end{equation}
        Since the $\xi_{\mathcal{G}}$ factor is linear in $f$, this quadratic property arises from the quadratic refinement of the function $\sigma$, which we define below.
        
        \subsection{ Geometric definition of Gu-Wen $\sigma(f)$ in terms of winding numbers}
        \label{windingGrassDefSec}
        
        The Gu-Wen Grassmann integral $\sigma( (M^d, T), f_{d-1})$ was given a geometric formulation for oriented $d = 2$ manifolds by Gaiotto and Kapustin \cite{Gaiotto:2015zta}, following a geometric formulation of two-dimensional spin structures studied previously in the math literature, e.g. ~\cite{atiyah1971,johnson1980}. This geometric formulation also appeared in a combinatorial context in earlier work on the dimer model in statistical mechanics on oriented~\cite{cimasoniReshetikhin2007} and unoriented~\cite{cimasoniNonorientable2009} surfaces.
        Recently, \cite{tata2020} gave a generalization of this geometric formulation to $d > 2$, including non-orientable manifolds, which we briefly review here. 
        
        An important subtlety in the definition of the Grassmann integral is that it also implicitly depends on a choice of perturbation of the orientation-reversing wall determined by $w_1$. As mentioned in Section~\ref{sec:geometricChoices} and explained in Appendix~\ref{usingFInftyOrientationWall}, this perturbation of $w_1$ is determined by $A_b$. Therefore, $\sigma( (M^d, T), f_{d-1})$ does have an implicit dependence on $A_b$ through its specification of both $w_1$ and the perturbation of $w_1$ defining $w_1^2$. In what follows, we will often for ease of notation drop the explicit dependence of $\sigma( (M^d, T), f_{d-1})$ on $(M^d,T)$ and keep it implicit. 
        
\begin{figure}[h!]
            \centering
            \includegraphics[width=0.9\linewidth]{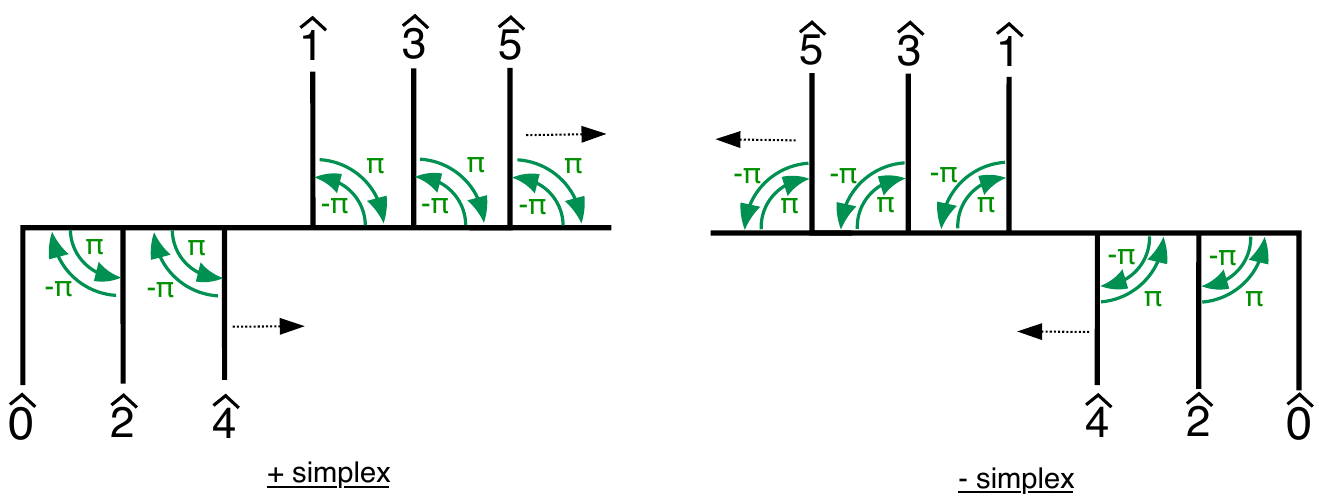}
            \caption{Trivalent resolution of the dual 1-skeleton of a $d$-simplex, and partial windings at the resolved junctions. Here $\hat{i}$ specifies a 1-cell in the dual cellulation, which is dual to the $(d-1)$-simplex obtained by removing the $i$th vertex in the $d$-simplex $\langle 0 \cdots d\rangle$. 
            Note that this resolution is different than the one in~\cite{tata2020}, but still works using the same methods.}
            \label{trivalentResolutionsAndWindings}
        \end{figure}

        The definition of $\sigma(f_{d-1})$ proceeds as follows. First, note that the closed $\Z_2$ $(d-1)$-cochain $f_{d-1}$ defined on the triangulation defines a closed $\Z_2$-valued $1$-chain on the dual cellulation. We can therefore use a canonical trivalent resolution of the dual $1$-skeleton in order to unambiguously split this $1$-chain into a set of distinct loops on the dual $1$-skeleton
        \footnote{Not every trivalent resolution will be sufficient; we need to pick one to be consistent with quadratic refinement Eq.~\eqref{quadRefinementEq} with the standard definition of higher cup product (see Appendix~\ref{topologicalPreliminaries}). Among such possible choices, we choose one that is related to the order of Grassmann variables in the alternate definition of $\sigma(f)$ given in Section \ref{grassIntegral}, which helps prove equivalence of the definitions. This choice differs from~\cite{tata2020}, but the same methods in that paper show that the one used here works.}. 
        Let us denote this set of loops as $\{L_i\}$. The precise trivalent resolution that we use is depicted in Fig.~\ref{trivalentResolutionsAndWindings}. 
        
        Next, as we explain in Appendix \ref{prelimSecB}, the branching structure, together with the choice of $\pm$ assignments on $d$-simplices, defines $d$ ``background'' vector fields ${v}_1,\cdots, {v}_d$ along the dual $1$-skeleton, which gives a trivialization of the tangent bundle $TM$ along the dual $1$-skeleton. Using these background vector fields, given a $(d-1)$-cochain $f_L$ dual to a single loop $L$ on the dual $1$-skeleton, we can define a winding number $\sigma(f_L)$, which essentially determines how the tangent of the curve winds with respect to the background vector fields. The total winding can be computed locally along the path by summing up partial contributions along each of the segments within a 4-simplex. Thus, in specifying the trivalent resolution, we should also specify the partial winding angles experienced by the loops traversing their respective junctions. These partial winding angles, which are $\pm \pi$, are also indicated in Fig.~\ref{trivalentResolutionsAndWindings}. 
        
        Physically, the winding number $\sigma(f_L)$ determines the boundary conditions for a fermion traversing the loop. For orientation-preserving loops, the winding determines whether the fermion boundary conditions are periodic or anti-periodic along the loop, which is equivalent to an induced spin structure on the loop. For orientation-reversing loops, $\sigma(f_L) = \pm i$, which is equivalent to an induced pin$^-$ structure on the loop. The appearance of the $i$ implies that a fermion traversing the orientation-reversing loop twice acquires a $-1$ sign, which is closely tied to the fact that reflection squares to $-1$ in $\Pin^-(d)$.  In particular, we have, with $f_L$ dual to the loop $L$,
        \begin{align}
        \label{sigmaLbc}
            \sigma(f_L) = \begin{cases}
            1 \text{ if $L$ has anti-periodic boundary conditions} \\
            -1 \text{ if $L$ has periodic boundary conditions} \\
            \pm i \text{ if $L$ is orientation-reversing}
            \end{cases}
        \end{align}
        The precise method to compute $\sigma(f_L)$ is given in Appendix \ref{prelimSecB} and \ref{sec:windingDefOfSigma}.  
        
        The full Grassmann integral $\sigma(f_{d-1})$ is then given as
        \begin{align}
            \sigma( f_{d-1}) = \prod_{i} \sigma(f_{L_i}) . 
            \label{eqn:GrassmannAsProductOfLoops}
        \end{align}

     On elementary loops of the dual $1$-skeleton, we have
        \begin{equation}
        \label{sigmaElLoop}
            \sigma(\delta \lambda) = (-1)^{\int_\lambda (w_2 + w_1^2)} . 
        \end{equation}
    Here $\lambda$ is a $(d-2)$-cochain which is non-trivial on a single $(d-2)$-simplex. The dual of $\lambda$ is then a single $2$-cell in the dual cellulation, while $\delta \lambda$ is an elementary loop which is the boundary of a single $2$-cell in the dual cellulation. We slightly abuse notation and also use $\lambda$ to denote the dual $2$-cell, so that 
    $\int_\lambda (w_2 + w_1^2) = \int_M (w_2 + w_1^2)(\lambda)$. 
     
    Furthermore, as mentioned above, $\sigma$ has the quadratic refinement property: 
        \begin{equation}
        \label{quadRefinementEq}
            \sigma(f)\sigma(f') = \sigma(f+f') (-1)^{\int f \cup_{d-2} f'}
        \end{equation}
    In~\cite{tata2020}, quadratic refinement is shown by bootstrapping two-dimensional pictures into higher dimensions using the characterization of $\cup_i$ by thickening and shifting.
    Actually, Eqs.~(\ref{sigmaElLoop},\ref{quadRefinementEq}) together imply that $\sigma$ has the anomaly property Eq.~\eqref{sigmaGrAnomalyAction} under retriangulation. In particular, Eq.~\eqref{sigmaElLoop} gives the $\int (w_2 + w_1^2)(f)$ term and Eq.~\eqref{quadRefinementEq} gives the $\Sq^2(f)$ term. First, quadratic refinement implies that the anomaly action is $\Sq^2$ up to a linear term, and the values on the elementary loops $\delta \lambda$ give the linear factor. The reader can refer to~\cite{Gaiotto:2015zta,Kobayashi2019pin} for a derivation.
    
    Finally, we note that Eqs.~\eqref{sigmaElLoop} and \eqref{quadRefinementEq} actually define the Grassmann integral for all cases where $f_3$ is cohomologically trivial, since the choice of triangulation, branching structure, and $\pm$ signs defines a representative $w_2 + w_1^2$, as explained in Appendix \ref{prelimSecB}.     
        
\subsection{ Algebraic definition of Gu-Wen $\sigma(f)$ in terms of Grassmann variables}
        \label{grassIntegral}
        
       Here, we recall the algebraic construction of the Grassmann integral $\sigma((M^d,T), f_{d-1})$ directly in terms of an integral over Grassmann variables that decorate the triangulation, following \cite{gu2014, Gaiotto:2015zta,Kobayashi2019pin}. The formulation of the construction in relation to spin and pin structures was given in~\cite{Gaiotto:2015zta, Kobayashi2019pin} but was formulated on barycentrically-subdivided triangulations. Here we use the original definition in~\cite{gu2014} for arbitrary orientable triangulations and extend the formulation in~\cite{Kobayashi2019pin} to arbitrary non-orientable triangulations. 
       
       Throughout this section we will refer to the Grassmann variable based definition as $\sigma^\text{gr}$. In Appendix \ref{grassEquivsec}, we prove that $\sigma^\text{gr}$ matches the winding definition presented above, both for orientable and non-orientable manifolds.
        
        \subsubsection{Definition on orientable manifolds}
        
        We first work on an orientable spin manifold $(M^d, T)$. Also, we will use $T^k$ to denote the set of $k$-simplices of $T$. 
        
        Here $T$ is a triangulation of $M^d$ equipped with a branching structure and consistent assignments $\epsilon(\Delta_d)$ of $\pm$ signs to $d$-simplices $\Delta_d$ such that the branching structure is compatible with their local orientations. See Appendix \ref{w1PmAssignments_AndVectorFields} for discussion about how exactly this is accomplished. 
        
        For a given $(d-1)$-form field $f_{d-1}\in Z^{d-1}(M,\mathbb{Z}_2)$, we assign a pair of Grassmann variables $\theta_{\Delta_{d-1}}, \overline{\theta}_{\Delta_{d-1}}$ on each $(d-1)$-simplex $\Delta_{d-1}$ of $M^d$ such that $f_{d-1}(\Delta_{d-1}) = 1$ (note that in general $f_{d-1}(\Delta_{d-1}) \in \{0,1\}$). We associate $\theta_{\Delta_{d-1}}$ on one side of ${\Delta_{d-1}}$ contained in one of the $d$-simplices neighboring ${\Delta_{d-1}}$ (which will be specified later), $\overline{\theta}_{\Delta_{d-1}}$ on the other side.
        
Define the Grassmann integral as 
\begin{equation}
    \sigma^\text{gr}((M^d,T), f_{d-1})=\int\prod_{{\Delta_{d-1}}|f_{d-1}({\Delta_{d-1}})=1}d\theta_{\Delta_{d-1}} d\overline{\theta}_{\Delta_{d-1}} \prod_{\Delta_{d} \in T^d} u(\Delta_d),
    \label{sigmadef}
\end{equation}
where $\Delta_{d}$ denotes any $d$-simplex, and $u(\Delta_{d})$ is a product of Grassmann variables contained in $\Delta_{d}$, which will be defined below.
For instance, for $d=2$, $u(\Delta_{2}) = \vartheta_{12}^{f(12)}, \vartheta_{01}^{f(01)}, \vartheta_{02}^{f(02)}$ for $\Delta_{2}=(012)$. Here, $\vartheta$ denotes $\theta$ or $\overline{\theta}$ depending on the choice of the assigning rule, which will be discussed later. The order of Grassmann variables in $u(\Delta_d)$ will also be defined shortly.
We note that $u(\Delta_{d})$ is ensured to be Grassmann-even since $f_{d-1}$ is closed. 

Due to the fermionic sign of Grassmann variables, $\sigma(f_{d-1})$ is a quadratic function of $f_{d-1}$, whose quadratic property depends on the order of Grassmann variables in $u(\Delta_d)$. We will adopt the order used in Gaiotto-Kapustin \cite{Gaiotto:2015zta}, which is defined as follows. 

\begin{figure}[htb]
\centering
\includegraphics[width=0.5\linewidth]{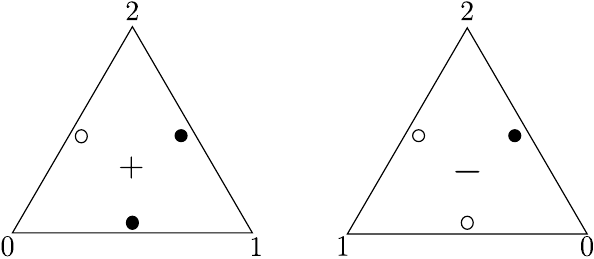}
\caption{Assignment of Grassmann variables on 1-simplices in the case of $d=2$. $\theta$ (resp.~$\overline{\theta}$) is represented as a black (resp.~white) dot.}
\label{fig:Grassmann}
\end{figure}

\begin{itemize}
\item
For $\Delta_{d}=\braket{0 1\dots d}$, we label a $(d-1)$-simplex $\braket{01\dots\hat{i}\dots d}$ (i.e., a $(d-1)$-simplex given by omitting a vertex $i$) as $\hat{i}$. 
\item Then, the order of $\vartheta_{\hat{i}}$ appearing in $u(\Delta_{d})$ for a $d$-simplex $\Delta_{d} = \braket{01...d}$ with $\epsilon(\Delta_d)=+$ is defined by first assigning even $(d-1)$-simplices in ascending order, then odd simplices in ascending order again:
\begin{equation}
    \hat{0} \to \hat{2} \to \hat{4}\to \cdots \to \hat{1} \to \hat{3} \to \hat{5} \to \cdots
\end{equation}
\item For $-$ $d$-simplices, the order is defined in opposite way:
\begin{equation}
    \cdots \to \hat{5} \to \hat{3} \to \hat{1} \to \cdots \to \hat{4} \to \hat{2} \to \hat{0}.
\end{equation}
\end{itemize}
Note that this is similar to the trivalent resolution of Fig.~\ref{trivalentResolutionsAndWindings} used in the preceding section, a fact which we use in proving the equivalence between the two definitions in Appendix \ref{grassEquivsec}. 
For example, for $d=2$, $u(012)=\vartheta_{12}^{f(12)}\vartheta_{01}^{f(01)}\vartheta_{02}^{f(02)}$ when $\epsilon(\braket{012})=+$, 
and $u(012)=\vartheta_{02}^{f(02)}\vartheta_{01}^{f(01)}\vartheta_{12}^{f(12)}$ when $\epsilon(\braket{012})=-$. 
Then, we choose the assignment of $\theta$ and $\overline{\theta}$ on each $\Delta_{d-1}$ such that, if $\epsilon(\Delta_{d})=+$ (resp.~$-$) simplex with vertices labeled $012 \cdots d$ according to the branching structure, then $u(\Delta_{d})$ includes $\overline{\theta}_{\hat{i}}$ (resp. $\theta_{\hat{i}}$) when $i$ is an odd (resp.~even) number. It is convenient to represent the choices of $\theta_{\Delta_{d-1}}$ versus $\overline{\theta}_{\Delta_{d-1}}$ that appear in $u(\Delta_{d})$ as illustrated in Fig.~\ref{fig:Grassmann}.

So, for $\epsilon(\Delta_d)=+$, we have

\begin{align*}
u({\Delta_{d}}) = \theta_{\hat{0}}^{f(\hat{0})} \theta_{\hat{2}}^{f(\hat{2})} \theta_{\hat{4}}^{f(\hat{4})} \cdots \overline{\theta}_{\hat{1}}^{f(\hat{1})} \overline{\theta}_{\hat{3}}^{f(\hat{3})} \overline{\theta}_{\hat{5}}^{f(\hat{5})} \cdots
\end{align*}

whereas for $\epsilon(\Delta_d)=-$ we have

\begin{align*}
u({\Delta_{d}}) = \cdots \theta_{\hat{5}}^{f(\hat{5})} \theta_{\hat{3}}^{f(\hat{3})} \theta_{\hat{1}}^{f(\hat{1})} \cdots \overline{\theta}_{\hat{4}}^{f(\hat{4})} \overline{\theta}_{\hat{2}}^{f(\hat{2})} \overline{\theta}_{\hat{0}}^{f(\hat{0})} 
\end{align*}

We note that the above definition is well-defined and does not depend on any ordering of the $d$-simplices ${\Delta_{d}}$. This is because $u({\Delta_{d}})$ are Grassmann-even, since $f_{d-1}$ is a cocycle, $\delta f_{d-1} = 0$. 
        
        \subsubsection{Definition on non-orientable manifolds} \label{sigmaNonOrientableGrassmannDef}
        
Now let us construct the Grassmann integral $\sigma^\text{gr}(M, f)$ on a $d$-manifold $M^d$ which might be non-orientable, following~\cite{Kobayashi2019pin}. 
We construct an non-orientable manifold by picking locally oriented patches, and then gluing them along codimension one loci by transition functions. The locus where the transition functions are orientation reversing constitutes a representative of the dual of the first Stiefel-Whitney class $w_1$. We will sometimes call the locus an orientation-reversing wall. As discussed in Appendix \ref{prelimSecB}, an assignment of $\pm$ signs to $d$-simplices gives, along with the branching structure, an assignment of representative of $w_1$. 

For the oriented case, the factor $\prod_{\Delta_{d} \in T^d} u({\Delta_{d}})$ contains a pair of Grassmann variables $\theta_{\Delta_{d-1}}$ and $\overline{\theta}_{\Delta_{d-1}}$ for each ${\Delta_{d-1}}$ such that $f_{d-1}({\Delta_{d-1}}) = 1$. Whether $u({\Delta_{d}})$ for a given ${\Delta_{d}}$ contains $\theta_{\Delta_{d-1}}$ or $\overline{\theta}_{\Delta_{d-1}}$ is determined by the assignment $\epsilon(\Delta_{d})=\pm$ and whether ${\Delta_{d-1}}$ is an odd or even numbered $(d-1)$-simplex within ${\Delta_{d}}$. This rule ensures that neighboring $d$-simplices always produce $\theta_{\Delta_{d-1}}$ and $\overline{\theta}_{\Delta_{d-1}}$ in pairs. 

In the non-orientable case, the assigning rule fails when ${\Delta_{d-1}}$ lies on the orientation-reversing wall, $w_1$. In this case, we would naively have to assign Grassmann variables of the same color on both sides of ${\Delta_{d-1}}$ (i.e., both are black ($\theta$) or white ($\overline{\theta}$)).
This would mean that $\prod_{\Delta_{d} \in T^d} u({\Delta_{d}})$ would include two factors of $\theta_{\Delta_{d-1}}$ or $\overline{\theta}_{\Delta_{d-1}}$, which would then imply $\prod_{\Delta_{d} \in T^d} u({\Delta_{d}}) = 0$. 

For later convenience, we define the orientation of each $(d-1)$-simplex on the orientation-reversing wall in terms of the mismatched colors of the Grassmann variables on the wall. Namely, a $(d-1)$-simplex ${\Delta_{d-1}}$ on the wall has $+$ orientation if the naive assigning rule puts two $\theta$'s on ${\Delta_{d-1}}$, and $-$ orientation for ${\Delta_{d-1}}$ with two $\overline{\theta}$'s. 

Hence, we need to slightly modify the construction of the Grassmann integral on the orientation-reversing wall. To do this, instead of specifying a canonical rule to assign Grassmann variables on the wall, we could place a pair $\theta_{\Delta_{d-1}}$, $\overline{\theta}_{{\Delta_{d-1}}}$ on the wall in an arbitrary fashion. The order of the Grassmann variables will still be the same: for $\epsilon(\Delta_d)=+$,

\begin{equation*}
\vartheta_{\hat{0}}^{f(\hat{0})} \vartheta_{\hat{2}}^{f(\hat{2})} \cdots \vartheta_{\hat{1}}^{f(\hat{1})} \vartheta_{\hat{3}}^{f(\hat{3})} \cdots
\end{equation*}
and for $\epsilon(\Delta_d)=-$ 

\begin{equation*}
\cdots \vartheta_{\hat{3}}^{f(\hat{3})} \vartheta_{\hat{1}}^{f(\hat{1})} \cdots \vartheta_{\hat{2}}^{f(\hat{2})} \vartheta_{\hat{0}}^{f(\hat{0})}
\end{equation*}

\begin{figure}[htb]
\centering
\includegraphics[width=0.25\linewidth]{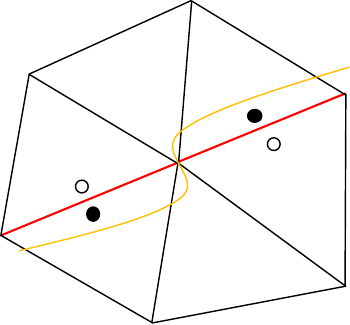}
\caption{The signs of $d$-simplices near the orientation-reversing wall, which is represented as a red line. The assignment of Grassmann variables on the wall specifies a deformation of the wall that intersects the wall transversally at $(d-2)$-simplices.
}
\label{fig:wall}
\end{figure}

As before, for every $\hat{i}$ that is \textit{not} on the orientation-reversing wall $w_1$, $\vartheta_{\hat{i}}$ will again be $\theta$ (resp. $\overline{\theta}$) for $i$ even (resp. $i$ odd) for $\epsilon(\Delta_d)=+$, and $\overline{\theta}$ (resp. $\theta$) for $i$ even (resp. $i$ odd) for $\epsilon(\Delta_d)=-$. For those Grassmann variables on $w_1$, we may choose the variables arbitrarily and obtain a non-vanishing Grassmann integral. However, this choice encodes a representative of $w_1^2$; as in Fig.~\ref{fig:wall}, a choice of the assignment of $\theta_{\Delta_{d-1}}$, $\overline{\theta}_{\Delta_{d-1}}$ on the orientation-reversing patch corresponds to choosing a deformation of $w_1$ in the direction from $\overline{\theta} \to \theta$, which gives a representative of $w_1^2$. 

As explained in Section~\ref{sec:geometricChoices} and our discussion of the winding definition, the choice of branching structure, $A_b$, and assignment of signs $\epsilon(\Delta_d)=\pm$ to $d$-simplices also determines a choice of $w_1$, a perturbation of the orientation-reversing wall, and $w_1^2$; these choices are the same as those made in the winding definition. To correctly obtain the anomaly cancellation, we need to make sure that all of these choices agree with each other. Therefore, the choice of Grassmann variables along $w_1$ is fixed by the perturbation of the orientation-reversing wall determined by the branching structure and $A_b$.

Armed with the choice of perturbation of the orientation-reversing wall, we define the Grassmann integral as
\begin{equation}
    \sigma^\text{gr}(M, f)=\int\prod_{{\Delta_{d-1}}|f({\Delta_{d-1}})=1}d\theta_{\Delta_{d-1}} d\overline{\theta}_{\Delta_{d-1}} \prod_{\Delta_{d} \in T^d} u({\Delta_{d}})\prod_{{\Delta_{d-1}}|\mathrm{wall}}(\pm i)^{f({\Delta_{d-1}})},
    \label{sigmadefpin}
\end{equation}
where the $\prod_{{\Delta_{d-1}}|\mathrm{wall}}(\pm i)^{f({\Delta_{d-1}})}$ term assigns weight $(+i)^{f(e)} \in \{1,i\}$ (resp.~$(-i)^{f(e)} \in \{1,-i\}$) on each $(d-1)$-simplex ${\Delta_{d-1}}$ on the orientation-reversing wall, when ${\Delta_{d-1}}$ is assigned $+$ orientation (resp. $-$ orientation). 
This factor makes the Grassmann integral a $\Z_4$ valued function.

\section{ Symmetry fractionalization}
\label{sec:symmFracReview}

In this section, we briefly review the mathematical characterization of symmetry fractionalization and its application in fermionic topological phases of matter. The material presented in this section is mainly review from \cite{barkeshli2019,bulmash2020,bulmashSymmFrac,aasen21ferm}. 

\subsection{ Review of BTC notation} \label{sec:btcReview}

Here we briefly review the notation that we use to describe BTCs. For a more comprehensive review
of the notation that we use, see, e.g., Ref.~\cite{barkeshli2019}. The topologically 
non-trivial quasiparticles of a (2+1)D topologically ordered state are equivalently referred to
as anyons, topological charges, and quasiparticles. In the category theory terminology, they correspond
to isomorphism classes of simple objects of the BTC. 

A BTC $\mathcal{C}$ contains splitting spaces $V_{c}^{ab}$, and their dual fusion spaces, $V_{ab}^c$,
where $a,b,c \in \mathcal{C}$ are anyons. These spaces have dimension 
$\text{dim } V_{c}^{ab} = \text{dim } V_{ab}^c = N_{ab}^c$, where the fusion coefficients $N_{ab}^c$ determine the fusion rules. In particular, the fusion rules of the anyons are written as $a \times b = \sum_c N_{ab}^c c$, so that fusion from $a \times b \to c$ is possible if and only if $N_{ab}^c \ge 1$. If $N_{ab}^c > 1$, then each fusion corresponds to a higher dimensional vector space with more possible `fusion outcomes'. 

The fusion spaces are depicted graphically as: 

\begin{equation}
\left( d_{c} / d_{a}d_{b} \right) ^{1/4}
\raisebox{-0.5\height}{\begin{pspicture}[shift=0.5](-0.1,-0.2)(1.5,1.2)
  \small
  \psset{linewidth=0.9pt,linecolor=black,arrowscale=1.5,arrowinset=0.15}
  \psline{-<}(0.7,0)(0.7,-0.35)
  \psline(0.7,0)(0.7,-0.55)
  \psline(0.7,-0.55) (0.25,-1)
  \psline{-<}(0.7,-0.55)(0.35,-0.9)
  \psline(0.7,-0.55) (1.15,-1)	
  \psline{-<}(0.7,-0.55)(1.05,-0.9)
  \rput[tl]{0}(0.4,0){$c$}
  \rput[br]{0}(1.4,-0.95){$b$}
  \rput[bl]{0}(0,-0.95){$a$}
 \scriptsize
  \rput[bl]{0}(0.85,-0.5){$\mu$}
  \end{pspicture}
  }
=\left\langle a,b;c,\mu \right| \in
V_{ab}^{c} ,
\label{eq:bra}
\end{equation}

\begin{equation}
\left( d_{c} / d_{a}d_{b}\right) ^{1/4}
\raisebox{-0.5\height}{\begin{pspicture}[shift=-0.65](-0.1,-0.2)(1.5,1.2)
  \small
  \psset{linewidth=0.9pt,linecolor=black,arrowscale=1.5,arrowinset=0.15}
  \psline{->}(0.7,0)(0.7,0.45)
  \psline(0.7,0)(0.7,0.55)
  \psline(0.7,0.55) (0.25,1)
  \psline{->}(0.7,0.55)(0.3,0.95)
  \psline(0.7,0.55) (1.15,1)	
  \psline{->}(0.7,0.55)(1.1,0.95)
  \rput[bl]{0}(0.4,0){$c$}
  \rput[br]{0}(1.4,0.8){$b$}
  \rput[bl]{0}(0,0.8){$a$}
 \scriptsize
  \rput[bl]{0}(0.85,0.35){$\mu$}
  \end{pspicture}
  }
=\left| a,b;c,\mu \right\rangle \in
V_{c}^{ab},
\label{eq:ket}
\end{equation}
where $\mu=1,\ldots ,N_{ab}^{c}$, $d_a$ is the quantum dimension of $a$, 
and the factors $\left(\frac{d_c}{d_a d_b}\right)^{1/4}$ are a normalization convention for the diagrams. 

Diagrammatically, inner products come from connecting the fusion/splitting spaces' lines as:
\begin{equation}
  \begin{pspicture}[shift=-0.95](-0.2,-0.35)(1.2,1.75)
  \small
  \psarc[linewidth=0.9pt,linecolor=black,border=0pt] (0.8,0.7){0.4}{120}{240}
  \psarc[linewidth=0.9pt,linecolor=black,arrows=<-,arrowscale=1.4,
    arrowinset=0.15] (0.8,0.7){0.4}{165}{240}
  \psarc[linewidth=0.9pt,linecolor=black,border=0pt] (0.4,0.7){0.4}{-60}{60}
  \psarc[linewidth=0.9pt,linecolor=black,arrows=->,arrowscale=1.4,
    arrowinset=0.15] (0.4,0.7){0.4}{-60}{15}
  \psset{linewidth=0.9pt,linecolor=black,arrowscale=1.5,arrowinset=0.15}
  \psline(0.6,1.05)(0.6,1.55)
  \psline{->}(0.6,1.05)(0.6,1.45)
  \psline(0.6,-0.15)(0.6,0.35)
  \psline{->}(0.6,-0.15)(0.6,0.25)
  \rput[bl]{0}(0.07,0.55){$a$}
  \rput[bl]{0}(0.94,0.55){$b$}
  \rput[bl]{0}(0.26,1.25){$c$}
  \rput[bl]{0}(0.24,-0.05){$c'$}
 \scriptsize
  \rput[bl]{0}(0.7,1.05){$\mu$}
  \rput[bl]{0}(0.7,0.15){$\mu'$}
  \endpspicture
=\delta _{c c ^{\prime }}\delta _{\mu \mu ^{\prime }} \sqrt{\frac{d_{a}d_{b}}{d_{c}}}
  \pspicture[shift=-0.95](0.15,-0.35)(0.8,1.75)
  \small
  \psset{linewidth=0.9pt,linecolor=black,arrowscale=1.5,arrowinset=0.15}
  \psline(0.6,-0.15)(0.6,1.55)
  \psline{->}(0.6,-0.15)(0.6,0.85)
  \rput[bl]{0}(0.75,1.25){$c$}
  \end{pspicture}
  ,
\end{equation}
This is a way of phrasing topological charge conservation. In addition, we have the usual `resolution of the identity' in a UMTC, phrased diagrammatically:
\begin{equation} \label{resOfIdentity}
\raisebox{-0.5\height}{
\begin{pspicture}[shift=-0.65](-0.1,-0.2)(1.0,1.2)
  \small
  \psset{linewidth=0.9pt,linecolor=black,arrowscale=1.5,arrowinset=0.15}
  \psline{->}(0.25,0)(0.25,0.6)
  \psline(0.25,0)(0.25,1.0)
  \psline{->}(0.7,0)(0.7,0.6)
  \psline(0.7,0)(0.7,1.0)
  \rput[br]{0}(0.15,0.5){$a$}
  \rput[bl]{0}(0.8,0.5){$b$}
 \end{pspicture}
 }
=\sum_{c} \sqrt{\frac{d_c}{d_a d_b}}
\raisebox{-0.5\height}{\begin{pspicture}[shift=-0.65](-0.4,-0.2)(1.5,1.3)
  \small
 \psset{linewidth=0.9pt,linecolor=black,arrowscale=1.5,arrowinset=0.15}
  \psline{->}(0.7,0.25)(0.7,0.7)
  \psline(0.7,0.25)(0.7,0.8)
  \psline(0.7,0.8) (0.25,1.25)
  \psline{->}(0.7,0.8)(0.3,1.2)
  \psline(0.7,0.8) (1.15,1.25)	
  \psline{->}(0.7,0.8)(1.1,1.2)
  \psline{->}(0.25,-0.3)(0.6,0.15)
  \psline(0.25,-0.3)(0.7,0.25)
  \psline{->}(1.15,-0.3)(0.8,0.15)
  \psline(1.15,-0.3)(0.7,0.25)
  \rput[bl]{0}(0.4,0.5){$c$}
  \rput[br]{0}(1.4,1.05){$b$}
  \rput[bl]{0}(0,1.05){$a$}
  \rput[bl]{0}(0,-0.2){$a$}
  \rput[br]{0}(1.4,-0.2){$b$}
  \end{pspicture}
  },
\end{equation}
implicitly assuming $N_{ab}^c \leq 1$ for all $a,b,c$. 

We denote $\bar{a}$ as the topological charge conjugate of $a$, for which
$N_{a \bar{a}}^1 = 1$, i.e.
\begin{align}
a \times \bar{a} = 1 +\cdots
\end{align}
Here $1$ refers to the identity particle, i.e. the vacuum topological sector, which physically describes all 
local, topologically trivial bosonic excitations. 

The $F$-symbols are defined as the following basis transformation between the splitting
spaces of $4$ anyons:
\begin{equation}
\raisebox{-0.5\height}{\begin{pspicture}[shift=*](0,-0.45)(1.8,1.8)
  \small
  \psset{linewidth=0.9pt,linecolor=black,arrowscale=1.5,arrowinset=0.15}
  \psline(0.2,1.5)(1,0.5)
  \psline(1,0.5)(1,0)
  \psline(1.8,1.5) (1,0.5)
  \psline(0.6,1) (1,1.5)
   \psline{->}(0.6,1)(0.3,1.375)
   \psline{->}(0.6,1)(0.9,1.375)
   \psline{->}(1,0.5)(1.7,1.375)
   \psline{->}(1,0.5)(0.7,0.875)
   \psline{->}(1,0)(1,0.375)
   \rput[bl]{0}(0.05,1.6){$a$}
   \rput[bl]{0}(0.95,1.6){$b$}
   \rput[bl]{0}(1.75,1.6){${c}$}
   \rput[bl]{0}(0.5,0.5){$e$}
   \rput[bl]{0}(0.9,-0.3){$d$}
 \scriptsize
   \rput[bl]{0}(0.3,0.8){$\alpha$}
   \rput[bl]{0}(0.7,0.25){$\beta$}
\end{pspicture}
}
= \sum_{f,\mu,\nu} \left[F_d^{abc}\right]_{(e,\alpha,\beta)(f,\mu,\nu)}
 \raisebox{-0.5\height}{\begin{pspicture}[shift=-1.0](0,-0.45)(1.8,1.8)
  \small
  \psset{linewidth=0.9pt,linecolor=black,arrowscale=1.5,arrowinset=0.15}
  \psline(0.2,1.5)(1,0.5)
  \psline(1,0.5)(1,0)
  \psline(1.8,1.5) (1,0.5)
  \psline(1.4,1) (1,1.5)
   \psline{->}(0.6,1)(0.3,1.375)
   \psline{->}(1.4,1)(1.1,1.375)
   \psline{->}(1,0.5)(1.7,1.375)
   \psline{->}(1,0.5)(1.3,0.875)
   \psline{->}(1,0)(1,0.375)
   \rput[bl]{0}(0.05,1.6){$a$}
   \rput[bl]{0}(0.95,1.6){$b$}
   \rput[bl]{0}(1.75,1.6){${c}$}
   \rput[bl]{0}(1.25,0.45){$f$}
   \rput[bl]{0}(0.9,-0.3){$d$}
 \scriptsize
   \rput[bl]{0}(1.5,0.8){$\mu$}
   \rput[bl]{0}(0.7,0.25){$\nu$}
  \end{pspicture}
  }
.
\end{equation}

To describe topological phases, these are required to be unitary transformations, i.e.

\begin{eqnarray}
\left[ \left( F_{d}^{abc}\right) ^{-1}\right] _{\left( f,\mu
,\nu \right) \left( e,\alpha ,\beta \right) }
&= \left[ \left( F_{d}^{abc}\right) ^{\dagger }\right]
  _{\left( f,\mu ,\nu \right) \left( e,\alpha ,\beta \right) }
  \nonumber \\
&= \left[ F_{d}^{abc}\right] _{\left( e,\alpha ,\beta \right) \left( f,\mu
,\nu \right) }^{\ast }
.
\end{eqnarray}

Anyon lines may be ``bent'' using the $A$ and $B$ symbols, given diagrammatically by

\vspace{-5pt}
\begin{align}
\raisebox{-0.5\height}{
\begin{pspicture}[shift=-0.65](0.0,-0.2)(1.5,1.2)
  \small
  \psset{linewidth=0.9pt,linecolor=black,arrowscale=1.5,arrowinset=0.15}
  \psline{->}(0,0)(0,0.55)
  \psline(0,0)(0,0.85)
  \psarc{->}(0.3,0.85){0.3}{25}{115}
  \psarc{-}(0.3,0.85){0.3}{25}{180}
  \psline{->}(1,0)(1,0.45)
  \psline(1,0)(1,0.55)
  \psline(1,0.55) (0.55,1)
  \psline{->}(1,0.55)(0.6,0.95)
  \psline(1,0.55) (1.45,1)	
  \psline{->}(1,0.55)(1.4,0.95)
  \rput[bl]{0}(0.7,0){$c$}
  \rput[br]{0}(1.7,0.8){$b$}
  \rput[bl]{0}(0.4,0.7){$a$}
  \rput[bl]{0}(0.15,0.1){$\bar{a}$}
 \scriptsize
  \rput[bl]{0}(1.15,0.35){$\mu$}
  \end{pspicture}
  }
  &= \sum_{\nu} \left[A^{ab}_c\right]_{\mu \nu}
\raisebox{-0.5\height}{\begin{pspicture}[shift=0.5](-0.1,-0.2)(1.5,1.2)
  \small
  \psset{linewidth=0.9pt,linecolor=black,arrowscale=1.5,arrowinset=0.15}
  \psline{-<}(0.7,0)(0.7,-0.35)
  \psline(0.7,0)(0.7,-0.55)
  \psline(0.7,-0.55) (0.25,-1)
  \psline{-<}(0.7,-0.55)(0.35,-0.9)
  \psline(0.7,-0.55) (1.15,-1)	
  \psline{-<}(0.7,-0.55)(1.05,-0.9)
  \rput[tl]{0}(0.4,0){$b$}
  \rput[br]{0}(1.4,-0.95){$c$}
  \rput[bl]{0}(0,-0.95){$\bar{a}$}
 \scriptsize
  \rput[bl]{0}(0.85,-0.5){$\nu$}
  \end{pspicture}
  },
  \label{eqn:Abend}
\end{align}
 \vspace{-10pt}
\begin{align}
\raisebox{-0.5\height}{\begin{pspicture}[shift=-0.65](-0.1,-0.2)(2.15,1.2)
  \small
  \psset{linewidth=0.9pt,linecolor=black,arrowscale=1.5,arrowinset=0.15}
  \psline{->}(2,0)(2,0.55)
  \psline(2,0)(2,0.85)
  \psarc{<-}(1.7,0.85){0.3}{65}{155}
  \psarc{-}(1.7,0.85){0.3}{0}{155}
  \psline{->}(1,0)(1,0.45)
  \psline(1,0)(1,0.55)
  \psline(1,0.55) (0.55,1)
  \psline{->}(1,0.55)(0.6,0.95)
  \psline(1,0.55) (1.45,1)	
  \psline{->}(1,0.55)(1.4,0.95)
  \rput[bl]{0}(0.7,0){$c$}
  \rput[br]{0}(1.55,0.65){$b$}
  \rput[bl]{0}(0.4,0.7){$a$}
  \rput[br]{0}(1.8,0.1){$\bar{b}$}
 \scriptsize
  \rput[bl]{0}(1.15,0.35){$\mu$}
  \end{pspicture}
  }
  &= \sum_{\nu} \left[B^{ab}_c\right]_{\mu \nu}
\raisebox{-0.5\height}{\begin{pspicture}[shift=0.5](-0.1,-0.2)(1.5,1.2)
  \small
  \psset{linewidth=0.9pt,linecolor=black,arrowscale=1.5,arrowinset=0.15}
  \psline{-<}(0.7,0)(0.7,-0.35)
  \psline(0.7,0)(0.7,-0.55)
  \psline(0.7,-0.55) (0.25,-1)
  \psline{-<}(0.7,-0.55)(0.35,-0.9)
  \psline(0.7,-0.55) (1.15,-1)	
  \psline{-<}(0.7,-0.55)(1.05,-0.9)
  \rput[tl]{0}(0.4,0){$a$}
  \rput[br]{0}(1.4,-0.95){$\bar{b}$}
  \rput[bl]{0}(0,-0.95){$c$}
 \scriptsize
  \rput[bl]{0}(0.85,-0.5){$\nu$}
  \end{pspicture}
  }.
  \label{eqn:Bbend}
\end{align}
In fact, these bending operators are unitary~\cite{Bonderson07b}, so the corresponding equations for the diagrams reflected vertically are Hermitian conjugates of the above. They can be expressed in terms of $F$-symbols by
\begin{align}
\left[A_c^{ab}\right]_{\mu \nu} &= \sqrt{\frac{d_a d_b}{d_c}}\varkappa_a \left[F^{\bar{a}ab}_b\right]^{\ast}_{1,(c,\mu, \nu)}\\
\left[B^{ab}_c\right]_{\mu \nu} &= \sqrt{\frac{d_a d_b}{d_c}}\left[F^{ab\bar{b}}_a\right]_{(c,\mu, \nu),1}
\end{align}
where the phase $\varkappa_a = \pm 1$ is the Frobenius-Schur indicator
\begin{equation}
\varkappa_a = d_a F^{a\bar{a}a}_{a11} .
\end{equation}
The $R$-symbols define the braiding properties of the anyons, and are defined via the the following
diagram:
\begin{equation}
\raisebox{-0.5\height}{\begin{pspicture}[shift=-0.65](-0.1,-0.2)(1.5,1.2)
  \small
  \psset{linewidth=0.9pt,linecolor=black,arrowscale=1.5,arrowinset=0.15}
  \psline{->}(0.7,0)(0.7,0.43)
  \psline(0.7,0)(0.7,0.5)
 \psarc(0.8,0.6732051){0.2}{120}{240}
 \psarc(0.6,0.6732051){0.2}{-60}{35}
  \psline (0.6134,0.896410)(0.267,1.09641)
  \psline{->}(0.6134,0.896410)(0.35359,1.04641)
  \psline(0.7,0.846410) (1.1330,1.096410)	
  \psline{->}(0.7,0.846410)(1.04641,1.04641)
  \rput[bl]{0}(0.4,0){$c$}
  \rput[br]{0}(1.35,0.85){$b$}
  \rput[bl]{0}(0.05,0.85){$a$}
 \scriptsize
  \rput[bl]{0}(0.82,0.35){$\mu$}
  \end{pspicture}
  }
=\sum\limits_{\nu }\left[ R_{c}^{ab}\right] _{\mu \nu}
\raisebox{-0.5\height}{\begin{pspicture}[shift=-0.65](-0.1,-0.2)(1.5,1.2)
  \small
  \psset{linewidth=0.9pt,linecolor=black,arrowscale=1.5,arrowinset=0.15}
  \psline{->}(0.7,0)(0.7,0.45)
  \psline(0.7,0)(0.7,0.55)
  \psline(0.7,0.55) (0.25,1)
  \psline{->}(0.7,0.55)(0.3,0.95)
  \psline(0.7,0.55) (1.15,1)	
  \psline{->}(0.7,0.55)(1.1,0.95)
  \rput[bl]{0}(0.4,0){$c$}
  \rput[br]{0}(1.4,0.8){$b$}
  \rput[bl]{0}(0,0.8){$a$}
 \scriptsize
  \rput[bl]{0}(0.82,0.37){$\nu$}
  \end{pspicture}
  }
  .
\end{equation}
Under a basis transformation, $\Gamma^{ab}_c : V^{ab}_c \rightarrow V^{ab}_c$, the $F$ and $R$ symbols change:
\begin{align} \label{eq:vertexBasisTransformation}
  F^{abc}_{def} &\rightarrow \check{F}^{abc}_d = \Gamma^{ab}_e \Gamma^{ec}_d F^{abc}_{def} [\Gamma^{bc}_f]^\dagger [\Gamma^{af}_d]^\dagger
  \nonumber \\
  R^{ab}_c & \rightarrow \check{R}^{ab}_c = \Gamma^{ba}_c R^{ab}_c [\Gamma^{ab}_c]^\dagger .
\end{align}
  where we have suppressed splitting space indices and dropped brackets on the $F$-symbol for shorthand.
  These basis transformations are referred to as vertex basis gauge transformations. Physical quantities correspond to gauge-invariant combinations
  of the data. 

Furthermore, we will fix all gauge transformations to be of the form
\begin{equation}
    \Gamma^{a 1}_a = \Gamma^{1 a}_a = +1
\end{equation}
so that there is a canonical choice of the basis vector $\ket{a,1;a}$ for the fusion spaces $V^{a 1}_a$. This requirement ensures that one can arbitrarily add identity anyon lines to a diagram and not alter its evaluation, and to make sure all diagrams that differ by additions of identity lines transform in the same way under gauge transformations.
  
The topological twist $\theta_a$ is defined via the diagram:
\begin{equation}
\theta _{a}=\theta _{\bar{a}}
=\sum\limits_{c,\mu } \frac{d_{c}}{d_{a}}\left[ R_{c}^{aa}\right] _{\mu \mu }
= \frac{1}{d_{a}}
\raisebox{-0.5\height}{\begin{pspicture}[shift=-0.5](-1.3,-0.6)(1.3,0.6)
\small
  \psset{linewidth=0.9pt,linecolor=black,arrowscale=1.5,arrowinset=0.15}
  \psarc[linewidth=0.9pt,linecolor=black] (0.7071,0.0){0.5}{-135}{135}
  \psarc[linewidth=0.9pt,linecolor=black] (-0.7071,0.0){0.5}{45}{315}
  \psline(-0.3536,0.3536)(0.3536,-0.3536)
  \psline[border=2.3pt](-0.3536,-0.3536)(0.3536,0.3536)
  \psline[border=2.3pt]{->}(-0.3536,-0.3536)(0.0,0.0)
  \rput[bl]{0}(-0.2,-0.5){$a$}
  \end{pspicture}
  }
.
\end{equation}
Finally, the modular, or topological, $S$-matrix, is defined as
\begin{equation}
S_{ab} =\mathcal{D}^{-1}\sum
\limits_{c}N_{\bar{a} b}^{c}\frac{\theta _{c}}{\theta _{a}\theta _{b}}d_{c}
=\frac{1}{\mathcal{D}}
\raisebox{-0.5\height}{\begin{pspicture}[shift=-0.4](0.0,0.2)(2.6,1.3)
\small
  \psarc[linewidth=0.9pt,linecolor=black,arrows=<-,arrowscale=1.5,arrowinset=0.15] (1.6,0.7){0.5}{167}{373}
  \psarc[linewidth=0.9pt,linecolor=black,border=3pt,arrows=<-,arrowscale=1.5,arrowinset=0.15] (0.9,0.7){0.5}{167}{373}
  \psarc[linewidth=0.9pt,linecolor=black] (0.9,0.7){0.5}{0}{180}
  \psarc[linewidth=0.9pt,linecolor=black,border=3pt] (1.6,0.7){0.5}{45}{150}
  \psarc[linewidth=0.9pt,linecolor=black] (1.6,0.7){0.5}{0}{50}
  \psarc[linewidth=0.9pt,linecolor=black] (1.6,0.7){0.5}{145}{180}
  \rput[bl]{0}(0.1,0.45){$a$}
  \rput[bl]{0}(0.8,0.45){$b$}
  \end{pspicture}
  }
,
\label{eqn:mtcs}
\end{equation}
where $\mathcal{D} = \sqrt{\sum_a d_a^2}$.

We also denote by $\mathcal{A}$ the Abelian group corresponding to fusion of Abelian anyons, for which each $a \in \mathcal{A}$ satisfies $d_a = 1$ and $a \times b$ has a unique fusion product for any $b \in \mathcal{C}$.

The double braid, or mutual statistics, of anyons $a$ and $b$ is defined
\begin{equation}
  \raisebox{-0.5\height}{\begin{pspicture}[shift=-0.6](0.0,-0.05)(1.1,1.45)
  \small
  \psarc[linewidth=0.9pt,linecolor=black,border=0pt] (0.8,0.7){0.4}{120}{225}
  \psarc[linewidth=0.9pt,linecolor=black,arrows=<-,arrowscale=1.4,
    arrowinset=0.15] (0.8,0.7){0.4}{165}{225}
  \psarc[linewidth=0.9pt,linecolor=black,border=0pt] (0.4,0.7){0.4}{-60}{45}
  \psarc[linewidth=0.9pt,linecolor=black,arrows=->,arrowscale=1.4,
    arrowinset=0.15] (0.4,0.7){0.4}{-60}{15}
  \psarc[linewidth=0.9pt,linecolor=black,border=0pt]
(0.8,1.39282){0.4}{180}{225}
  \psarc[linewidth=0.9pt,linecolor=black,border=0pt]
(0.4,1.39282){0.4}{-60}{0}
  \psarc[linewidth=0.9pt,linecolor=black,border=0pt]
(0.8,0.00718){0.4}{120}{180}
  \psarc[linewidth=0.9pt,linecolor=black,border=0pt]
(0.4,0.00718){0.4}{0}{45}
  \rput[bl]{0}(0.1,1.2){$a$}
  \rput[br]{0}(1.06,1.2){$b$}
  \end{pspicture}
  }
= M_{ab}
\raisebox{-0.5\height}{\begin{pspicture}[shift=-0.6](-0.2,-0.45)(1.0,1.1)
  \small
  \psset{linewidth=0.9pt,linecolor=black,arrowscale=1.5,arrowinset=0.15}
  \psline(0.3,-0.4)(0.3,1)
  \psline{->}(0.3,-0.4)(0.3,0.50)
  \psline(0.7,-0.4)(0.7,1)
  \psline{->}(0.7,-0.4)(0.7,0.50)
  \rput[br]{0}(0.96,0.8){$b$}
  \rput[bl]{0}(0,0.8){$a$}
  \end{pspicture}
  }
.
  \label{doubleBraid}
\end{equation}
 and is a phase if either $a$ or $b$ is an Abelian anyon.

Also, we will eventually make use of the `ribbon identity' 
\begin{equation} \label{ribbonId}
R^{a b}_c R^{b a}_c = \frac{\theta_c}{\theta_a \theta_b} \mathbbm{1}
\end{equation}

\subsection{ Super-modular tensor categories} \label{sec:superModularCategories}

        Physically realizable bosonic topological orders are described by modular tensor categories, which have the property that the $S$-matrix is unitary. This means that braiding is non-degenerate, that is, every anyon $a$ can be detected by its non-trivial mutual statistics with some other anyon $b$.
        
        Fermionic topological phases have local fermions, and locality requires that these fermions have trivial mutual statistics with all other excitations. One way to keep track of the anyon fusion and braiding properties in a fermionic topological phase is to use a super-modular tensor category $\mathcal{C}$. A super-modular tensor category is a unitary braided fusion category where there are exactly two transparent anyons: the identity $1$ and a fermion $\psi$.\footnote{Mathematically speaking, $\mathcal{C}$ is super-modular if its M{\"u}ger center is equivalent to the unitary symmetric fusion category sVec of super-vector spaces.} That is, $\psi$ has $\theta_{\psi}=-1$ and trivial mutual braiding with all other particles. As such, the braiding is degenerate.
        
        In a super-modular tensor category, the anyons (simple objects) \it as a set \rm form the structure $\{1,a,b, ...\} \times \{1,\psi\}$.
        The $S$ matrix factorizes as
        \begin{align}
          S = \tilde{S} \otimes \frac{1}{\sqrt{2}} \left(\begin{matrix} 1 & 1 \\ 1 & 1 \end{matrix} \right) ,
        \end{align}
        where $\tilde{S}$ is unitary.

        The $S$ and $T$ matrices do not form a representation of $SL(2;\Z)$. Rather, one has a representation of a subgroup of $SL(2;\Z)$ which keeps the spin structure invariant. Collecting the Hilbert spaces on the 2-torus for \textit{all} spin structures realizes a representation of the ``metaplectic group" $Mp_1(\Z)$, which is the non-trivial $\Z_2$ extension of $SL(2;\Z)$\cite{delmastro2021}. 
        
        There is a convenient, canonical gauge-fixing available for a super-modular category. In particular,
        \begin{align} 
            F^{\psi, \psi, a} &\rightarrow \Gamma^{\psi \psi}_1 F^{\psi, \psi, a} (\Gamma^{\psi a})^{\ast} (\Gamma^{\psi, \psi \times a})^{\ast} \label{eq:Fpsipsia_transform}\\
            F^{a,\psi,\psi} &\rightarrow \Gamma^{\psi,a} \Gamma^{a\times \psi, \psi} F^{a,\psi,\psi} \left(\Gamma^{\psi \psi}_1\right)^{\ast}  \label{eq:Fapsipsi_transform}
        \end{align}
        under a vertex basis transformation $\Gamma$. One can check that this is sufficient gauge freedom to gauge fix 
        \begin{align} 
            F^{\psi,\psi,a}&=1 \label{eqn:Fpsipsia}\\
            F^{a,\psi,\psi}&=1 \label{eqn:Fapsipsi}
        \end{align}
        $F^{\psi,\psi,a}=1$ for all $a \in \mathcal{C}$, and likewise gauge-fix $\Gamma^{a,\psi}$ such that $F^{a, \psi,\psi}=1$ for all $a \in \mathcal{C}$. Diagrammatically, in this canonical gauge, fermion lines may be bent freely and slid past each other.

\subsection{ Topological symmetry and braided auto-equivalence}

An important property of a BTC $\mathcal{C}$ is the group of ``topological symmetries,'' which are related to ``braided auto-equivalences'' in the mathematical literature, although the former contains anti-unitary symmetries as well. They are associated with the symmetries of the emergent BTC description, irrespective of any global symmetries of the microscopic model from which $\mathcal{C}$ emerges as the description of the universal properties of the anyons. 

The topological symmetries consist of the invertible maps
\begin{align}
\varphi: \mathcal{C} \rightarrow \mathcal{C} .
\end{align}
The different $\varphi$, modulo equivalences known as natural isomorphisms, form a group, which we denote as Aut$(\mathcal{C})$.\cite{barkeshli2019}

The symmetry maps can be classified according to a $\mathbb{Z}_2$ grading corresponding to whether $\varphi$ has a unitary or anti-unitary action on the category:
\begin{align} \label{defnOfS_anti-unitary}
s(\varphi) = \left\{
\begin{array} {ll}
+1 & \text{if $\varphi$ is unitary} \\
*  & \text{if $\varphi$ is anti-unitary} \\
\end{array} \right.
\end{align}
where $*$ refers to complex conjugation. We note that one can consider a more general $\mathbb{Z}_2\times \mathbb{Z}_2$ grading by considering separately transformations that correspond to time-reversal and spatial reflection symmetries.\cite{barkeshli2019,barkeshli2019rel} Here we will not consider spatial parity reversing transformations and thus do not consider this generalization. 
Thus the topological symmetry group can be decomposed as
\begin{align}
\text{Aut}(\mathcal{C}) = \text{Aut}_{0}(\mathcal{C}) \sqcup \text{Aut}_{1}(\mathcal{C})
\end{align}
Aut$_{0}(\mathcal{C})$ is therefore the subgroup corresponding to topological symmetries that are unitary (this is referred to in the mathematical literature as the group of ``braided auto-equivalences''), and $\text{Aut}_{1}(\mathcal{C})$ are the anti-unitary ones.

The maps $\varphi$ may permute the topological charges:
\begin{align}
\varphi(a) = a' \in \mathcal{C}, 
\end{align}
subject to the constraint that 
\begin{align}
N_{a'b'}^{c'} &= N_{ab}^c
\nonumber \\
S_{a'b'} &= S_{ab}^{s(\varphi)},
\nonumber \\
\theta_{a'} &= \theta_a^{s(\varphi)},
\end{align}
The maps $\varphi$ have a corresponding action on the $F$- and $R$- symbols of the theory, as well as on the fusion and splitting spaces, which we will discuss in the subsequent section. 

\subsection{ Global symmetry and symmetry fractionalization}
\label{globsym}

We now consider a system which has a global symmetry group $G$. The global symmetry acts on the anyons and the topological state space through the action of a group homomorphism
\begin{align}
[\rho] : G \rightarrow \text{Aut}(\mathcal{C}) . 
\end{align}
We use the notation $[\rho_{\bf g}] \in \text{Aut}(\mathcal{C})$ for a specific element ${\bf g} \in G$. The square brackets indicate the equivalence class of symmetry maps related by natural isomorphisms, which we define below. $\rho_{\bf g}$ is thus a representative symmetry map of the equivalence class $[\rho_{\bf g}]$. We use the notation
\begin{align}
\,^{\bf g}a \equiv \rho_{\bf g}(a). 
\end{align}
We associate a $\mathbb{Z}_2$ grading $s({\bf g})$ by defining
\begin{equation}
s({\bf g}) \equiv s( \rho_{\bf g})
\end{equation}
where we think of the complex conjugation $*$ as representing the $-1$ element of $\Z_2 = \{\pm 1\}$.

Each $\rho_{\bf g}$ has an action on the fusion/splitting spaces:
\begin{align}
\rho_{\bf g} : V_{ab}^c \rightarrow V_{\,^{\bf g}a \,^{\bf g}b}^{\,^{\bf g}c} .
\end{align}
This map is unitary if $s({\bf g}) = 1$ and anti-unitary if $s({\bf g}) = *$. We choose a basis $\ket{a,b;c,\mu}$ for $V_{ab}^c$ and write the action of $\rho_{\bf g}$ on the basis states as
\begin{align}
\rho_{\bf g} |a,b;c, \mu\rangle = \sum_{\nu} [U_{\bf g}(\,^{\bf g}a ,
  \,^{\bf g}b ; \,^{\bf g}c )]_{\mu\nu} \ket{\,^{\bf g} a, \,^{\bf g} b; \,^{\bf g}c,\nu},
  \label{eqn:rhoStates}
\end{align}
and the action of $\rho_{\bf g}$ defined on the rest of the fusion/splitting spaces by (anti-)linearity\footnote{For anti-unitary ${\bf g}$, Eq.~\eqref{eqn:rhoStates} looks different from much of the literature, e.g.~\cite{barkeshli2019}, where a complex conjugation operator typically appears on the right-hand side. The interpretation of that complex conjugation operator is very delicate; naive interpretation of the complex conjugation operator's action leads to erroneous complex conjugations in the consistency equations if they are derived from the literature version of Eq.~\eqref{eqn:rhoStates}. We find it simpler and unambiguous to write the equation in the above form on the basis states and simply declare the action of $\rho_{\bf g}$ to be extended to the whole state space via anti-linearity.}. Here $U_{\bf g}(\,^{\bf g}a , \,^{\bf g}b ; \,^{\bf g}c ) $ is a $N_{ab}^c \times N_{ab}^c$ matrix.

\begin{figure}
    \centering
    \subfigure[\label{fig:symmFrac}]{\includegraphics[width=0.4\linewidth]{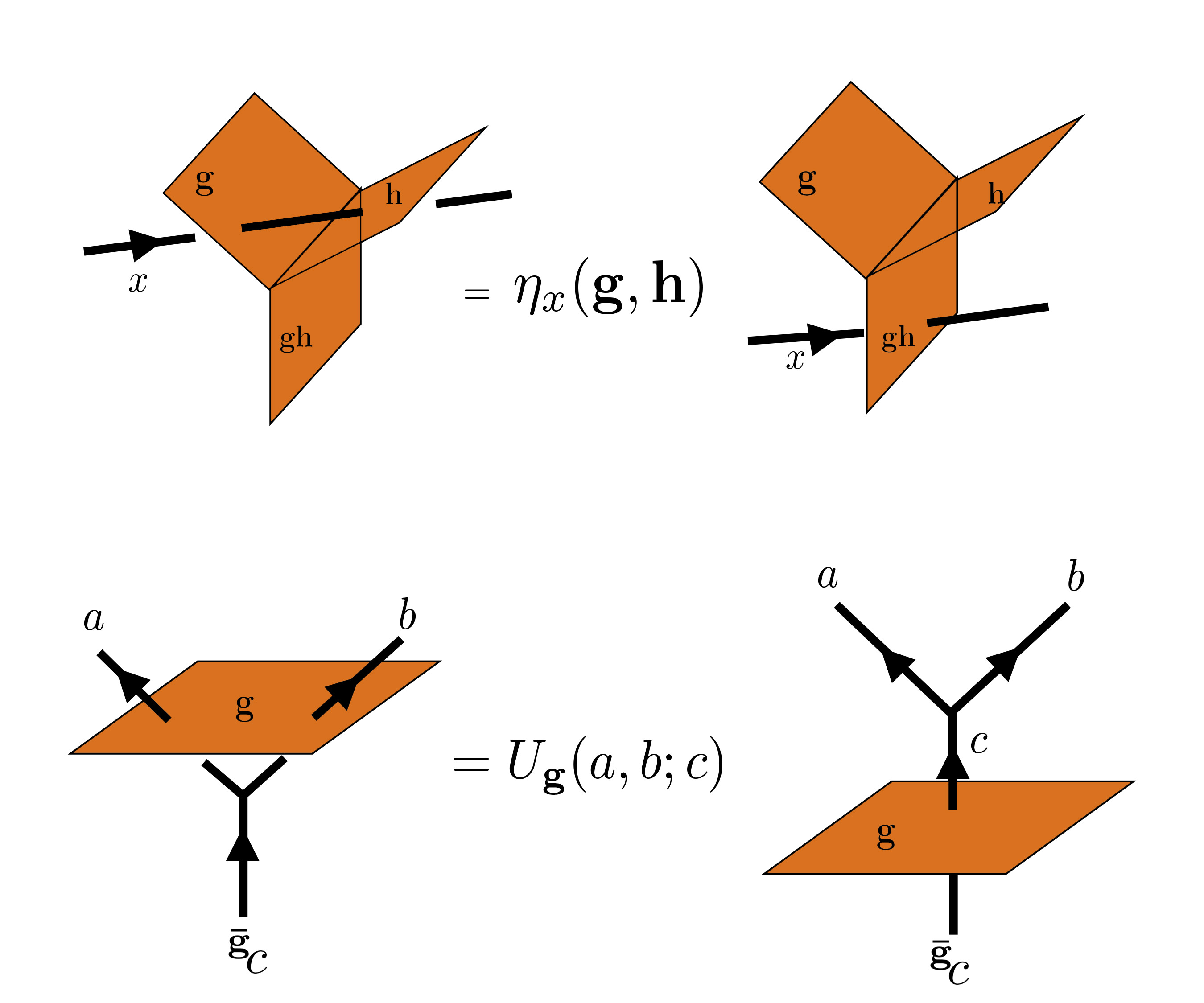}} \hspace{0.1\linewidth}
    \subfigure[\label{fig:graphicalCalculus_defs}]{\includegraphics[width=0.8\linewidth]{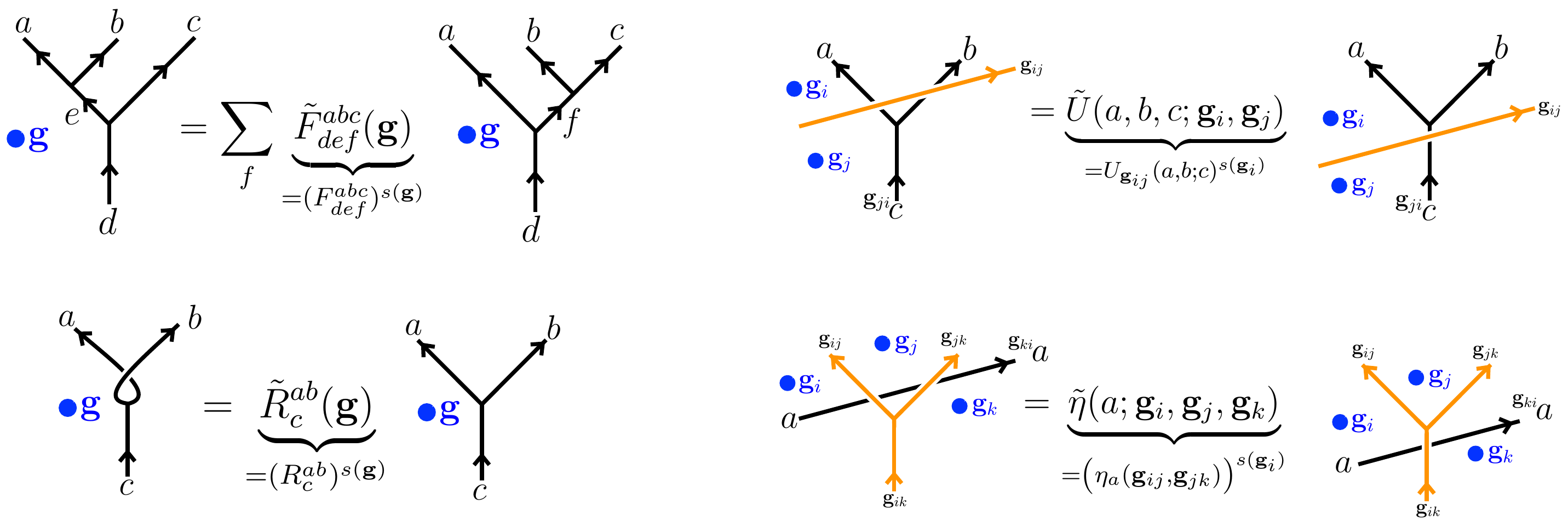}}
    \caption{(a) Anyon lines (black) passing through branch sheets (orange) and graphical definitions of the $U$ and $\eta$ symbols. (b) ``Tilded'' graphical calculus conventions. Orange lines are symmetry domain walls and black lines are anyons. The directions of the arrow on the ${\bf g}_{ij}={\bf g}_i{\bf g}_{j}^{-1}$ lines correspond to the direction in which the ${\bf g}_{ij}$ twists are applied. }
\end{figure}

In the presence of domain walls of $G$, a graphical calculus can be developed for the action of symmetry domain walls on anyon data. The basic pictures defining the graphical calculus are given in Fig.~\ref{fig:symmFrac}. However, if $G$ contains anti-unitary symmetries, then the anti-linearity of $\rho_{\bf g}$ is difficult to track within the usual BTC graphical calculus. We will therefore use a modified graphical calculus described in Refs.~\cite{bulmash2020,barkeshli2019rel}.

A key point from those works is that in the presence of anti-unitary symmetry domain walls, the local orientation of space reverses across the domain wall. Recall that the particular orientation chosen in space is important in defining the BTC data. Reflecting a BTC diagram (say, without any $G$ defects) across the vertical axis corresponds to a Hermitian conjugation of $F$- and $R$-symbols. The fact that space should reverse orientation after an anti-unitary domain-wall sweep means roughly that $F$- and $R$-symbols should be complex conjugated after sweeping the domain wall across.

The proposal to deal with such situations was to imagine that regions of space are labeled by group elements ${\bf g} \in G$ in which fusion spaces and their duals gain an extra label by group elements. For example, the space $V^{ab}_c$ is replaced with $\tilde{V}^{ab}_{c}({\bf g})$. Graphically this is depicted as
\begin{equation}
\left( d_{c} / d_{a}d_{b}\right) ^{1/4}
\begin{pspicture}[shift=-0.65](-0.1,-0.2)(1.5,1.2)
  \small
  \psset{linewidth=0.9pt,linecolor=black,arrowscale=1.5,arrowinset=0.15}
  \psline{->}(0.7,0)(0.7,0.45)
  \psline(0.7,0)(0.7,0.55)
  \psline(0.7,0.55) (0.25,1)
  \psline{->}(0.7,0.55)(0.3,0.95)
  \psline(0.7,0.55) (1.15,1)	
  \psline{->}(0.7,0.55)(1.1,0.95)
  \rput[bl]{0}(0.4,0){$c$}
  \rput[br]{0}(1.4,0.8){$b$}
  \rput[bl]{0}(0,0.8){$a$}
  \rput[bl]{0}(0.2,0.35){$\textcolor{blue}{{\bf g}}$}
 \scriptsize
  \rput[bl]{0}(0.85,0.35){$\mu$}
  \end{pspicture}
=\left| a,b;c,\mu \right\rangle_{{\bf g}} \in
\tilde{V}_{c}^{ab}({\bf g}),
\label{eq:ketWithG}
\end{equation}
where the tildes are meant to distinguish these spaces from the previous ones, and similarly for the dual vertices. In a region of space labeled by ${\bf g}$ we will have $\tilde{F}$ and $\tilde{R}$ symbols $\tilde{F}^{abc}_{def}({\bf g}), \tilde{R}^{ab}_c({\bf g})$. Similarly, there is now a ``tilded'' symmetry action
\begin{equation}
    \tilde{\rho}_{\bf h} : V_c^{ab}({\bf g}) \to V_{\,^{\bf h}c}^{\,^{\bf h}a \,^{\bf h}b}({\bf hg})
\end{equation}
which defines a ``tilded'' $U$ symbol
\begin{equation}
    \tilde{\rho}_{\bf h}\ket{a,b;c}_{\bf g} = \tilde{U}(\,^{\bf h}a,\,^{\bf h}b,\,^{\bf h}c;{\bf hg},{\bf g}) \ket{\,^{\bf h}a,\,^{\bf h}b;\,^{\bf h}c}_{\bf hg}
\end{equation}
where the ${\bf g}$ subscript on the state indicates that the state is in $V_c^{ab}({\bf g})$. Importantly, $\tilde{\rho}_{\bf h}$ is always unitary. The anti-unitarity of any symmetry action is tracked by the group element labels. The graphical representations of these tilded data are shown in Fig.~\ref{fig:graphicalCalculus_defs}.

The unitary or anti-unitarity of the group element labels play the role of the local orientation of space. This manifests itself in relating the $\tilde{F},\tilde{R},\tilde{U}$ to the original $F,R$ symbols
\begin{equation}
\begin{split}
\tilde{F}^{abc}_{def}({\bf g}) &= (F^{abc}_{def})^{s({\bf g})} \\
\tilde{R}^{ab}_c({\bf g})      &= (R^{ab}_{c})^{s({\bf g})}\\
\tilde{U}(a,b,c;{\bf g}_i,{\bf g}_j) &= \left(U_{{\bf g}_{ij}}(a,b;c)\right)^{s({\bf g}_i)},
\end{split}
\label{eqn:tildeToUntilded}
\end{equation}
where ${\bf g}_{ij}={\bf g}_i{\bf g}_j^{-1}$.

There is a similar equation for vertex basis transformations
\begin{equation}
    \tilde{\Gamma}_a^{bc}({\bf g}) = \left(\Gamma_a^{bc}\right)^{s({\bf g})}.
\end{equation}

We will develop the rest of the formalism interchangeably using the usual and tilded notation; they can always be related using Eq.~\eqref{eqn:tildeToUntilded}.

Under the map $\rho_{\bf g}$, the $F$ and $R$ symbols transform as well:
\begin{widetext}
\begin{align}
\rho_{\bf g}[ F^{abc}_{def}] &= U_{\bf g}(\,^{\bf g}a, \,^{\bf g}b; \,^{\bf g}e) U_{\bf g}(\,^{\bf g}e, \,^{\bf g}c; \,^{\bf g}d) F^{\,^{\bf g}a \,^{\bf g}b \,^{\bf g}c }_{\,^{\bf g}d \,^{\bf g}e \,^{\bf g}f} 
U^{-1}_{\bf g}(\,^{\bf g}b, \,^{\bf g}c; \,^{\bf g}f) U^{-1}_{\bf g}(\,^{\bf g}a, \,^{\bf g}f; \,^{\bf g}d) =  \left(F^{abc}_{def}\right)^{s({\bf g})}
\nonumber \\
\rho_{\bf g} [R^{ab}_c] &= U_{\bf g}(\,^{\bf g}b, \,^{\bf g}a; \,^{\bf g}c)  R^{\,^{\bf g}a \,^{\bf g}b}_{\,^{\bf g}c} U_{\bf g}(\,^{\bf g}a, \,^{\bf g}b; \,^{\bf g}c)^{-1} = \left(R^{ab}_c\right)^{s({\bf g})},
\label{eqn:UFURConsistency}
\end{align}
\end{widetext}
where we have suppressed the additional indices that appear when $N_{ab}^c > 1$. 

We demand that composition of $\rho_{\bf g}$ obey the group multiplication law up to a natural isomorphism $\kappa_{\bf g, h}$
\begin{align}
\kappa_{{\bf g}, {\bf h}} \circ \rho_{\bf g} \circ \rho_{\bf h} = \rho_{\bf g h} ,
\end{align}
where the action of $\kappa_{ {\bf g}, {\bf h}}$ on the fusion / splitting spaces is defined as
\begin{align}
\kappa_{ {\bf g}, {\bf h}} ( |a, b;c,\mu \rangle) = \sum_\nu [\kappa_{ {\bf g}, {\bf h}} ( a, b;c )]_{\mu\nu} | a, b;c,\nu \rangle
\end{align}
and, being a natural isomorphism, obeys by definition  
\begin{align}
[\kappa_{ {\bf g}, {\bf h}} (a,b;c)]_{\mu \nu} = \delta_{\mu \nu} \frac{\beta_a({\bf g}, {\bf h}) \beta_b({\bf g}, {\bf h})}{\beta_c({\bf g}, {\bf h}) },
\end{align}
where $\beta_a({\bf g}, {\bf h})$ are some (not-unique) choice of $\U$ phases. The above definitions imply that
\begin{widetext}
\begin{align} \label{kappaU}
\kappa_{ {\bf g}, {\bf h}} ( a, b;c ) = U_{\bf g}(a,b;c)^{-1} \left(U_{\bf h}( \,^{\bar{\bf g}}a, \,^{\bar{\bf g}}b; \,^{\bar{\bf g}}c  )^{-1}\right)^{s({\bf g})}  U_{\bf gh}(a,b;c )
\end{align}
\end{widetext}
where $\bar{\bf g} \equiv {\bf g}^{-1}$. 

After defining a symmetry action on the topological state space, we may define symmetry localization on a full quantum many-body state of the system. Conceptually, this amounts to splitting the symmetry action into a piece that acts as $\rho_{\bf g}$ on the topological degrees of freedom and a local piece $U_{\bf g}^{(i)}$ that acts near the $i$th anyon. For a precise definition, see, e.g.~\cite{barkeshli2014}; we will not need a full discussion for our purposes.

There is in general an obstruction to defining symmetry localization; see~\cite{barkeshli2019,fidkowski2015,barkeshli2018} for the bosonic case and~\cite{bulmashSymmFrac,aasen21ferm} for the fermionic case. We assume this obstruction is trivial so that we can define symmetry fractionalization.
Symmetry fractionalization is specified by a set of phases $\eta_{a}({\bf g,h})$, which satisfy certain consistency relations which we will discuss shortly. The data $\{U,\eta\}$ characterize a \textit{symmetry fractionalization class} and give us information about how the group symmetries fractionalize onto the different anyons. The $U_{\bf g}(a,b;c)$ relates the symmetry action on fusion vertices $c \to a,b$ to the fusion vertices $^{\bf g}c \to  \,^{\bf g}a, \,^{\bf g}b$, while the data $\eta_a({\bf g}, {\bf h})$ characterize the difference in phase obtained when acting ``locally'' on an anyon $a$ by ${\bf g}$ and ${\bf h}$ separately, as compared with acting on $a$ by the product ${\bf gh}$. The $\eta$ symbols also have a ``tilded'' version, graphically defined in Fig.~\ref{fig:graphicalCalculus_defs} and related to the usual symbols by
\begin{equation}
    \tilde{\eta}(a;{\bf g}_i,{\bf g}_j,{\bf g}_k) = \eta_a({\bf g}_{ij},{\bf g}_{jk})^{s({\bf g}_i)}.
\end{equation}

\begin{figure}[h!]
    \centering
    \includegraphics[width=\linewidth]{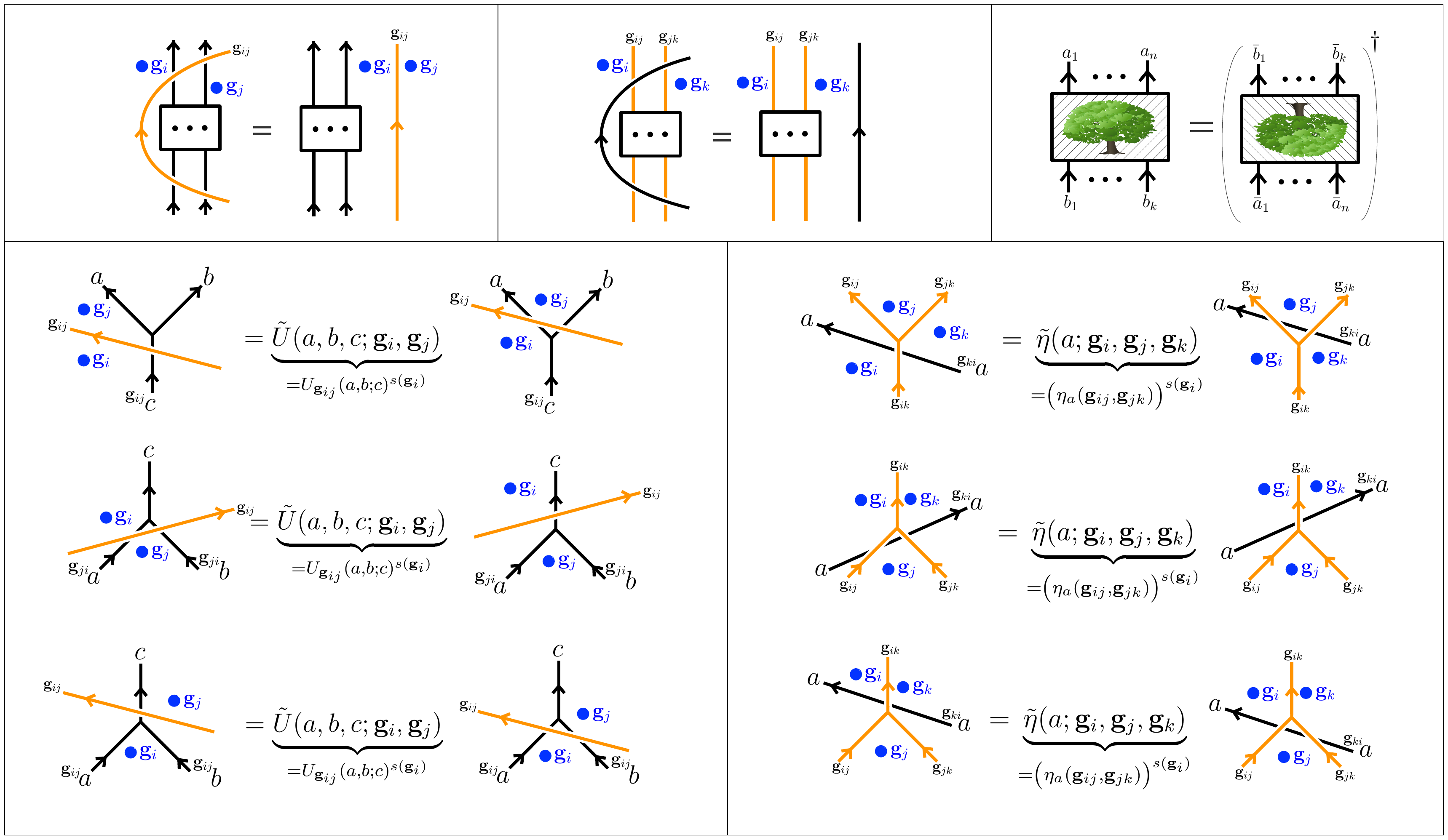}
    \caption{Other moves in the graphical calculus can be derived using isotopy invariance of sliding vertical strands (top-left and top-middle) and noting that reflecting the diagrams vertically corresponds to Hermitian conjugation (top-right).}
    \label{fig:graphicalCalculus_more}
\end{figure}

The diagrammatic rules and definitions in Fig.~\ref{fig:graphicalCalculus_defs} can be used to derive more rules as in Fig.~\ref{fig:graphicalCalculus_more}. One can derive them by demanding isotopy invariance under sliding vertical strands and noting that reflecting a partial diagram vertically amounts to Hermitian conjugation of relevant matrices.

Lastly, there are several consistency conditions that need to be imposed on the $U,\eta,F,R$ symbols in order for diagrammatic evaluations to be consistent under different orders of moves. In the case of $N_{ab}^c \le 1$ that we deal with in this paper, they can be written as:
\begin{widetext}
\begin{align}
\tilde{F}^{fcd}_{egl}({\bf g})\tilde{F}^{abl}_{efk}({\bf g}) &= \sum_h \tilde{F}^{abc}_{gfh}({\bf g})\tilde{F}^{ahd}_{egk}({\bf g})\tilde{F}^{bcd}_{khl}({\bf g}) \label{eqn:tildePentagon}\\
\tilde{R}^{ca}_e({\bf g})\tilde{F}^{acb}_{deg}({\bf g})\tilde{R}^{cb}_g({\bf g}) &= \sum_f \tilde{F}^{cab}_{def}({\bf g})\tilde{R}^{cf}_d({\bf g})\tilde{F}^{abc}_{dfg}({\bf g})\\
(\tilde{R}^{ac}_e({\bf g}))^{-1}\tilde{F}^{acb}_{deg}({\bf g})(\tilde{R}^{bc}_g({\bf g}))^{-1} &= \sum_f \tilde{F}^{cab}_{def}({\bf g})(\tilde{R}^{fc}_d({\bf g}))^{-1}\tilde{F}^{abc}_{dfg}({\bf g})\\
\tilde{U}(\acts{g}{12}{a},\acts{g}{12}{b} ;\acts{g}{12}{e},{\bf g}_1,{\bf g}_2)\tilde{U}(\acts{g}{12}{e},\acts{g}{12}{c} ;\acts{g}{12}{d},{\bf g}_1,{\bf g}_2)&\tilde{F}^{\acts{g}{12}{a}\act{12}{b}\act{12}{c}}_{\act{12}{d}\act{12}{e}\act{12}{f}}({\bf g}_1) \times \nonumber \\
\times \tilde{U}^{-1}(\acts{g}{12}{b},\acts{g}{12}{c} ;\acts{g}{12}{f},{\bf g}_1,{\bf g}_2)&\tilde{U}^{-1}(\acts{g}{12}{a},\acts{g}{12}{f} ;\acts{g}{12}{d},{\bf g}_1,{\bf g}_2) = \tilde{F}^{abc}_{def}({\bf g}_2)\\
  \tilde{U}(\acts{g}{12}{b},\acts{g}{12}{a} ;\acts{g}{12}{c},{\bf g}_1,{\bf g}_2)\tilde{R}^{\act{12}{a}\act{12}{b}}_{\act{12}{c}}({\bf g}_1)\tilde{U}^{-1}(\acts{g}{12}{a},\acts{g}{12}{b} ;\acts{g}{12}{c},{\bf g}_1,{\bf g}_2) &= \tilde{R}^{ab}_c({\bf g}_2) \label{eqn:tildeURConsistency}
\\
\tilde{U}(\acts{g}{21}{a},\acts{g}{21}{b} ;\acts{g}{21}{c} ,{\bf g}_2,{\bf g}_3)\tilde{U}(a,b;c,{\bf g}_1, {\bf g}_2) &= \tilde{U}(a, b;c, {\bf g}_1, {\bf g}_3)\frac{\eta_c({\bf g}_1, {\bf g}_2, {\bf g}_3)}{\eta_a({\bf g}_1, {\bf g}_2, {\bf g}_3)\eta_b({\bf g}_1, {\bf g}_2, {\bf g}_3)}
\label{UUoverU_equals_EtaOverEtaEta}
\\
  \tilde{\eta}_{{}^{{\bf g}_{21}}\!a}({\bf g}_2,{\bf g}_3,{\bf g}_4) \tilde{\eta}_a({\bf g}_1, {\bf g}_3, {\bf g}_4)
                                                             &= \tilde{\eta}_a({\bf g}_1,{\bf g}_2,{\bf g}_3) \tilde{\eta}_a({\bf g}_1, {\bf g}_2, {\bf g}_4)
                                                              \label{EtaEta_equals_EtaEta}
\end{align}
\end{widetext}
The top three are just the standard pentagon and hexagon equations from BTCs without symmetry. The next two ensure the symmetry action is compatible with the $F$- and $R$-symbols. The next ensures that the symmetry action and symmetry fractionalization are consistent with each other, and the last one is a generalized associativity condition for the $\eta$ symbols.

These data are subject to an additional class of gauge transformations, referred to as symmetry action gauge transformations, which arise by changing $\rho$ by a natural isomorphism: \cite{barkeshli2019}
\begin{align}
  U_{\bf g}(a,b;c) &\rightarrow \frac{\gamma_{a}({\bf g}) \gamma_b({\bf g})}{ \gamma_c({\bf g}) } U_{\bf g}(a,b;c)
\nonumber \\
  \eta_a({\bf g}, {\bf h}) & \rightarrow \frac{\gamma_a({\bf g h}) }{(\gamma_{\,^{\bf g} a}({\bf h}))^{s({\bf g})} \gamma_a({\bf g}) } \eta_a({\bf g}, {\bf h})
\end{align}
These symmetry gauge transformations may also be performed in the tilded notation:
\begin{equation}
    \tilde{\gamma}(a;{\bf g}_i, {\bf g}_j) = \gamma_a({\bf g}_{ij})^{s({\bf g}_i)}
\end{equation}
We note that $U$ also changes under a vertex basis gauge transformation according to
\begin{equation} \label{eq:U_change_vertex_basis}
\check{U}_{{\bf g}}(a,b,c)_{\mu \nu} = \sum_{\mu', \nu'} [\Gamma^{\act{\bar{g}}{a} \act{\bar{g}}{b}}_{\act{\bar{g}}{c}}]_{\mu, \mu'} U_{{\bf g}}(a,b,c)_{\mu' \nu'}\left[(\Gamma^{ab}_c)^{-1}\right]^{s({\bf g})}_{\nu'\nu},
\end{equation}
with the shorthand $\bar{{\bf g}}={\bf g}^{-1}$. Different gauge-inequivalent choices of $\{\eta\}$ and $\{U\}$ characterize distinct symmetry fractionalization classes.\cite{barkeshli2019} In this paper we will always fix the gauge
\begin{align}
  \eta_1({\bf g},{\bf h})=\eta_a({\bf 1},{\bf g}) = \eta_a({\bf g},{\bf 1})&=1
                                                                             \nonumber \\
  U_{\bf g}(1,b;c)=U_{\bf g}(a,1;c)&=1.
  \end{align}
  
One can show that symmetry fractionalization forms a torsor over $\mathcal{H}^2_{\rho}(G, \mathcal{A})$ in the bosonic case and $\mathcal{H}^2_{\rho}(G,\mathcal{A}/\{1,\psi\})$ in the fermionic case. That is, different possible patterns of symmetry fractionalization can be related to each other by elements of $\mathcal{H}^2_{\rho}(G, \mathcal{A})$ or $\mathcal{H}^2_{\rho}(G,\mathcal{A}/\{1,\psi\})$. In particular, given an element $[\coho{t}] \in \mathcal{H}^2_{\rho}(G, \mathcal{A})$ (or $\mathcal{H}^2_{\rho}(G,\mathcal{A}/\{1,\psi\})$), we can change the symmetry fractionalization class as
\begin{align}
\eta_a({\bf g}, {\bf h}) \rightarrow \eta_a({\bf g}, {\bf h}) M_{a \mathfrak{t}({\bf g},{\bf h})},
\end{align}
where $\coho{t}({\bf g},{\bf h}) \in \mathcal{A}$ is a representative 2-cocycle for the cohomology class $[\coho{t}]$ and $M_{ab}$ is the double braid in Eq.~\eqref{doubleBraid}.

\subsection{ Fermionic symmetry fractionalization}
\label{fermSymFracSec}

        We presently sketch a framework for symmetry fractionalization in fermionic topological phases, summarizing results from ~\cite{bulmashSymmFrac, aasen21ferm}.

        The above formalism for symmetry actions, localization, and fractionalization does not rely on whether $\C$ is modular, super-modular, or neither, but it does rely on the symmetry group being bosonic. To extend the above formalism to construct symmetry actions of $G_f$ on the super-modular tensor category $\C$, then, rather than directly plugging $G_f$ into the above formalism, we will instead consider $G_f$ as a central extension of $G_b$ and construct a symmetry action of $G_b$ on $\C$. 
        
        All of the above formalism for $G_b$ applies as usual. We expect that $G_f$, and in particular $G_b$ cannot permute $\psi$; it is easy to see that this follows from the consistency conditions and super-modularity (super-modularity is important to ensure that the only transparent fermion is $\psi$).
        
        However, we need to account for the locality of the fermions and for the fact that $G_f$ is present and may be a non-trivial extension of $G_b$. We discuss how to do this below.
        
        Locality of the fermion imposes a key constraint 
        \begin{equation}
            U_{\bf g}(\psi, \psi;1)=1.
            \label{eqn:Upsi1}
        \end{equation}
        Physically, this arises from the idea that symmetry localization separates the action of a symmetry into a product of its local action on anyons and its action on the topological state space. In the case of the fermion, the symmetry transformation rules of the local fermion operators are set entirely by the local Hilbert space, that is, the action of the symmetry on states containing only fermions must be determined entirely by the local action of the symmetry. Thus the action on the topological state space given by $U_{\bf g}(\psi,\psi;1)$ must be trivial. 
        This also means that there is a constraint on symmetry gauge transformations
        \begin{equation}
            \gamma_{\psi}({\bf g})=1.
            \label{eqn:gammaPsiConstraint}
        \end{equation}
        This is because a symmetry gauge transformation changes the local action of the symmetry operators on anyons but, as discussed above, the local action of symmetries on fermions is fixed from the outset by the microscopic Hilbert space because fermions are local excitations. Therefore, such gauge transformations are not allowed. We will also see later that we should constrain the vertex basis transformation
        \begin{equation}
            \Gamma^{\psi \psi}_1=+1.
            \label{eqn:GammaPsiPsi1Constraint}
        \end{equation}
        Although we do not have a microscopic justification for Eq.~\eqref{eqn:GammaPsiPsi1Constraint}, we expect it to also arise from the locality of the fermion; we will see later that relaxing this constraint leads to unphysical ambiguities in our path integral in the case of anti-unitary symmetries. 

        Finally, we must include the fact that $G_f$ may be a non-trivial central extension of $G_b$. We incorporate this by demanding that
        \begin{eqnarray}
        \eta_{\psi}({\bf g,h}) = \omega_2({\bf g,h})
        \label{eqn:etaPsiOmega2Constraint}
        \end{eqnarray}
        where $\omega_2 \in Z^2(G_b,\Z_2)$ specifies the central extension. Eq.~\eqref{eqn:etaPsiOmega2Constraint} is gauge-invariant under the restricted set of gauge transformations which obey Eq.~\eqref{eqn:gammaPsiConstraint}. 
        
        The constraint Eq.~\eqref{eqn:etaPsiOmega2Constraint} requires some physical and mathematical explanation. Physically, the fermion cannot carry fractional quantum numbers of $G_f$ since it is a local excitation. However, if ${\bf g,h}\in G_b$ and, if considered in $G_f$, ${\bf gh} = (-1)^F {\bf k}$ for some ${\bf k} \in G_b$, from the perspective of $G_b$ the fermion appears to carry a fractional quantum number because it picks up a minus sign from the action of $(-1)^F$. Hence we can reinterpret minus signs from fermion parity, that is, $\omega_2$, as symmetry fractionalization of $G_b$ on the fermion.

        Consider, for example, $G_b = \Z_2^{\bf T}$. Then $\eta_{\psi}({\bf T,T})$ tells us whether the local action of ${\bf T}^2$ on a fermion is the same as the identity, i.e., if ${\bf T}^2 = 1$, or if there is an additional minus sign. On the other hand, if $\omega_2({\bf T,T}) = -1$, then by definition ${\bf T}^2 = (-1)^F$, which has a local action of $-1$ on a fermion. In this sense, $\eta_{\psi}$ and $\omega_2$ encode the same information and should be equated.
        
        Mathematically, our requirement that $\eta_{\psi}$ must be a $\Z_2$ cocycle is actually necessary, as follows.  Since $\rho_{\bf g}$ cannot permute $\psi$, the consistency equation Eq.~\eqref{EtaEta_equals_EtaEta} for $\eta_{\psi}$ becomes a 2-cocycle condition. Furthermore, using the constraint Eq.~\eqref{eqn:Upsi1} the $U$-$\eta$ consistency condition Eq.~\eqref{UUoverU_equals_EtaOverEtaEta} forces $\eta_{\psi}({\bf g,h})\in \Z_2$. Hence $\eta_{\psi} \in Z^2(G_b,\Z_2)$ and it makes sense to set it equal to $\omega_2$. 
        
        The constraints discussed in this section, as we will see, are crucial to ensuring that the bosonic shadow path integral we define in Sec.~\ref{sec:craneyettershadow} has the appropriate higher form anomalies required for the fermion condensation procedure. In particular, they ensure that the bosonic shadow path integral transforms appropriately when decorating the theory with fermion worldlines by coupling the theory to a background 3-form gauge field.

        \section{ Bosonic Shadow: Symmetry-enriched Crane-Yetter path integral with 3-form gauge field}
        \label{sec:craneyettershadow}
        
        \subsection{ Definition of the path integral}
        
        In this section, we construct a path integral $Z_b( (M^4,T) , A_b, f_3)$ that corresponds to the bosonic shadow discussed in the previous sections. The construction is a generalization of the construction of \cite{bulmash2020} to take as input a super-modular category $\mathcal{C}$ and symmetry fractionalization data, and then to couple the theory to a non-zero background flat $3$-form $\Z_2$ gauge field $f_3$. 
        
        Our construction takes as input a super-modular tensor category $\mathcal{C}$ describing the fermionic topological order, the symmetry group $G_f$, and symmetry fractionalization data $U$ and $\eta$ for $G_b$ on $\mathcal{C}$. It associates an amplitude $Z_b( (M^4,T) , A_b, f_3)$ to a triangulated 4-manifold $M^4$ with branching structure, background flat $G_b$ gauge field $A_b$, and background flat $3$-form $\mathbb{Z}_2$ gauge field $f_3$.
        
        We denote the transparent fermion of $\mathcal{C}$ by $\psi$. We demand, as discussed in Sec.~\ref{sec:symmFracReview}, that 
        \begin{align}
            U_{\bf g}(\psi,\psi;1) &= 1\\
            \eta_{\psi}({\bf g}, {\bf h}) &= \omega_2({\bf g}, {\bf h})
        \end{align}
        where $[\omega] \in \H^2(G_b,\mathbb{Z}_2)$ is the cohomology class specifying $G_f$ as a group extension of $G_b$ by $\mathbb{Z}_2$. Subject to the first constraint, the consistency condition Eq.~\eqref{UUoverU_equals_EtaOverEtaEta} between $\eta$ and $U$ implies that $\eta_{\psi}({\bf g,h}) = \pm 1$, which turns the consistency condition for $\eta_{\psi}$ Eq.~\eqref{EtaEta_equals_EtaEta} into an (untwisted) $\Z_2$-valued cocycle condition, so the second condition makes sense.

        Given a triangulation and branching structure $T$ of $M^4$, we assign the following fixed background data:
        \begin{itemize}
            \item A cocycle $f \in Z^3((M^4,T),\Z_2)$ representing the background fermion line, which consists of an assignment $f_{ijkl}$ on each 3-simplex satifying $f_{0123} + f_{0124} + f_{0134} + f_{0234} + f_{1234} = 0 \text{ (mod 2)}$ on each 4-simplex $\braket{01234}$ (using the additive $\Z_2$ notation). 
            \item A flat gauge field $A_b$ which consists of group elements ${\bf g}_{ij} \in G_b$ on each edge $\braket{ij}$ satifying ${\bf g}_{ij}{\bf g}_{jk} = {\bf g}_{ik}$
        \end{itemize}
        The state sum consists of a summation over all possible assignments of the following data:
        \begin{itemize}
            \item To each 2-simplex $\braket{ijk}$, assign a simple object (anyon) $a_{ijk}  \in \mathcal{C}$
            \item To each 3-simplex $\braket{ijkl}$, assign an anyon $b_{ijkl} \in \C$ and an element of the vector space $V^{\psi^{f_{ijkl}}\times b}_{ijl,jkl} \otimes V_b^{ikl,\act{lk}{ijk}}$ .
        \end{itemize}
        
        Where it does not lead to ambiguity, we will often ignore the distinction between a simplex and the anyon or group element data assigned to it, i.e. simply write ${\bf g}_{ij}$ as $ij$, $a_{ijk}$ as $ijk$, or $b_{ijkl}$ as $ijkl$. We will always refer to the 3-form gauge field as $f_{ijkl}$.
        
       We will also need to assign to each $4$-simplex $\Delta_4$ of $M$ an orientation $\epsilon(\Delta_4) = \pm$. If $M$ is orientable, this simply amounts to choosing a global orientation of $M$. If $M$ is non-orientable, there must be an important compatibility between these choices and the choice of $A_b$, as discussed in Sec.~\ref{sec:geometricChoices}.

\begin{figure}
           \centering
           \includegraphics[width=0.7\columnwidth]{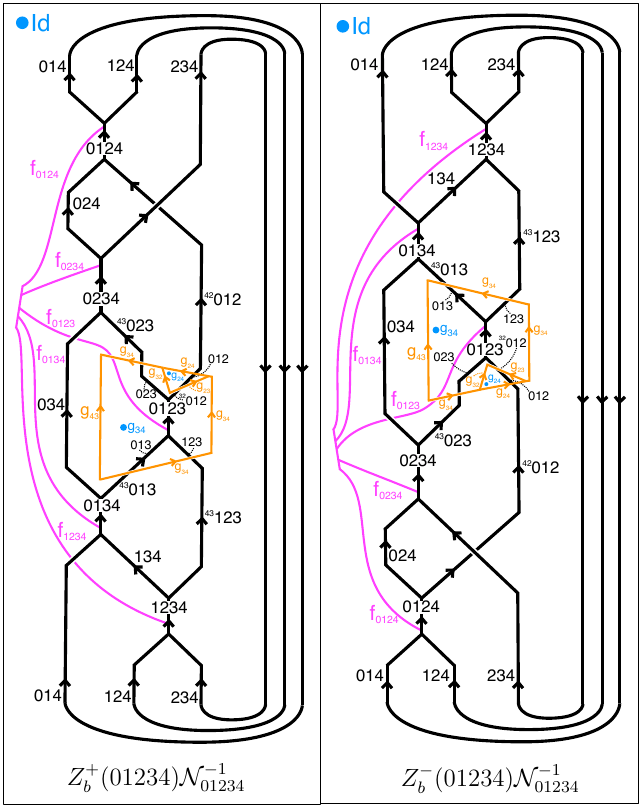}
           \caption{Amplitudes associated to a positively (left) and negatively (right) oriented 4-simplex 01234, where the branching structure induces the order on the labels. Orange are symmetry domain walls, black are anyon lines taking values in the input super-modular braided tensor category $\mathcal{C}$, and pink lines are either the identity or the transparent fermion of $\mathcal{C}$ according to the value of the background $3$-form gauge field $f$ on the 3-simplex in question. The blue dots and corresponding group elements label regions of space as required for the graphical calculus in Sec.~\ref{sec:symmFracReview}. Here Id is the identity in $G_b$.
           }
           \label{fig:15jSymbols}
       \end{figure}

       We then assign an orientation-dependent amplitude $Z_b^{\epsilon(\Delta_4)}(\Delta_4)$ to each 4-simplex $\Delta_4$ of $M$. This amplitude is given diagrammatically in Fig.~\ref{fig:15jSymbols}, with a normalization factor
       \begin{equation} \label{eq:normalizationFactor_15j}
           \mathcal{N}_{01234} = \sqrt{\frac{\prod_{\Delta_3 \in \text{3-simplices}} d_{b_{\Delta_3}}}{\prod_{\Delta_2 \in \text{2-simplices}} d_{a_{\Delta_2}}}} .
       \end{equation}
       These diagrams are fermion-decorated versions of the diagrams in \cite{bulmash2020}, or alternatively symmetry-decorated versions of the diagrams in \cite{kapustinThorngren2017FermionSPT}, both of which are decorated versions of the Crane-Yetter 15j symbol.  We will discuss the origin of the diagram shortly.
       
       With the canonical gauge-fixing Eqs.~(\ref{eqn:Fpsipsia},\ref{eqn:Fapsipsi}), (we will discuss relaxing this gauge-fixing in Sec.~\ref{subsubsec:shadowChoices}) we can explicitly evaluate the diagrams:
       \begin{equation}
       \begin{split}
           Z_b^+(01234) =& \frac{1}{\eta_{012}(23,34)^{s(24)}} \Big( \frac{U_{34}(013,123 ; 0123 \times f_{0123})}{U_{34}(023, {^{32}}012 ; 0123 ) U_{34}(f_{0123},0123 ; 0123 \times f_{0123})} \Big)^{s(34)} \\
           &\cdot 
           \Big( F^{f_{0123} \, , \,  {^{43}}013 \, , \,  123}_{0123 \times f_{0123} \, , \,  {^{43}}013 \times f_{0123} \, , \,  0123} \,\,\,
                  \big(F^{034 \, , \, f_{0123} \, , \,  {^{43}}013 \times f_{0123}}_{0134 \times f_{0134} \, , \,  034 \times f_{0123} \, , \,  {^{43}}013} \big)^* \,\,\, 
                  R^{f_{0123} \, , \,  034}_{034 \times f_{0123}} \Big) \\
           &\cdot
           \left( F^{f_{0234} \, , \,  034 \, , \,  {^{43}}023}_{0234 \, , \,  034 \times f_{0234} \, , \,  0234 \times f_{0234}} \right) \\
           &\cdot
           \Big( F^{f_{0124} \, , \,  024 \, , \,  {^{42}}012}_{0124 \, , \,  024 \times f_{0124} \, , \,  0124 \times f_{0124}} \,\,\,
                 F^{f_{0124} \, , \,  024 \times f_{0124} \, , \,  234}_{0234 \, , \,  024 \, , \,  0234 \times f_{0124}} \,\,\,
                 F^{f_{0124} \, , \,  034 \times f_{0234} \, , \,  {^{43}}023}_{0234 \times f_{0124} \, , \,  034 \times f_{0234} \times f_{0124} \, , \,  0234} \Big) \\
           &\cdot 
           \Big( \big(F^{f_{0134} \, , \,  034 \times f_{0123} \, , \,  {^{43}}013 \times f_{0123}}_{0134 \, , \,  034 \times f_{0123} \times f_{0134} \, , \,  0134 \times f_{0134}} \big)^* \Big) \\
           &\cdot 
           \Big( \big(F^{f_{1234} \, , \,  134 \, , \,  {^{43}}123}_{1234 \, , \,  134 \times f_{1234} \, , \,  1234 \times f_{1234}} \big)^* \,\,\,
                  F^{014 \, , \,  f_{1234} \, , \,  134 \times f_{1234}}_{0134 \, , \,  014 \times f_{1234} \, , \,  134} \,\,\,
                  \big( R^{f_{1234} \, , \,  014}_{014 \times f_{1234}} \big)^* \\
                  &\quad\quad
                  \big(F^{f_{1234} \, , \,  014 \, , \,  134 \times f_{1234}}_{0134 \, , \,  014 \times f_{1234} \, , \,  0134 \times f_{1234}} \big)^* \,\,\,
                  \big(F^{f_{1234} \, , \,  034 \times f_{0123} \times f_{0134} \, , \,  {^{43}}013 \times f_{0123}}_{0134 \times f_{1234} \, , \,  034 \times f_{0123} \times f_{0134} \times f_{1234} \, , \,  0134} \big)^* \,
           \Big) \\
           &\cdot \Bigg( 
           \sum_{d,a \in \mathcal{C}} 
           F^{024 \times f_{0124} \, , \,  234 \, , \,  {^{42}}012}_{d \, , \,  0234 \times f_{0124} \, , \,  a} \,\,
           R^{{^{42}}012 \, , \,  234}_a \,\,
           \big(F^{024 \times f_{0124} \, , \,  012 \, , \,  234}_{d \, , \, 0124 \, , \,  a} \big)^* \,\,
           F^{014 \, , \,  124 \, , \,  234}_{d \, , \,  0124 \, , \,  1234} \\
           &\quad\quad\quad\quad\quad
           \big(F^{014 \, , \,  134 \times f_{1234} \, , \,  {^{43}}123}_{d \, , \,  0134 \times f_{1234} \, , \,  1234} \big)^* \,\,
           F^{034 \times f_{0234} \times f_{0124} \, , \,  {^{43}}013 \times f_{0123} \, , \,  {^{43}}123}_{d \, , \,  0134 \times f_{1234} \, , \,  {^{43}}0123 \times f_{0123}} \,\,
           \big(F^{034 \times f_{0234} \times f_{0124} \, , \,  {^{43}}023 \, , \,  {^{42}}012}_{d \, , \,  0234 \times f_{0124} \, , \,  {^{43}}0123 \times f_{0123}} \big)^*  \,\, d_d 
           \Bigg)
       \end{split}
       \end{equation}
       where we mean $f_{ijkl} = (\psi)^{f(ijkl)}$, $s(ij) = s({\bf g}_{ij})$, ${^{lk}}ijk = {^{{\bf g}_{lk}}a_{ijk}} = \rho_{{\bf g}_{lk}}(a_{ijk})$, etc. The top row is obtained from contracting the $G_b$ domain walls using the graphical calculus of Sec.~\ref{globsym}. Each large parentheses-set in the next five rows comes from $F$-, $R$-moves bringing the fermions $f_{0123},f_{0234},f_{0124},f_{0134},f_{1234}$ onto the $034$ edge, often using the free bending of fermion lines, discussed in Sec.~\ref{sec:superModularCategories}. The last row of summing over $d,a$ is from first resolving the identity on $234,012$ to give a sum over $a$, then resolving the identity on $024,a$ to give a sum over $d$, then doing the various $F$-, $R$-moves and taking inner products that enforce various charge conservations as in Sec.~\ref{sec:btcReview}, then finally removing a $d$-loop giving a factor $d_d$. Note that by definition, we took away various other quantum-dimension factors from the normalization $\mathcal{N}_{01234}$. On a $-$-simplex, the expression is simply the complex conjugate.
       \begin{equation}
           Z_b^-(01234) = (Z_b^+(01234))^*
       \end{equation}
      
       Finally, we can define the full path integral. Let $\chi$ be the Euler characteristic of $M$, and let $T^k$ be the set of $k$-simplices of $M$. Then define 
       \begin{equation}
           Z_b(M,A_b,f)=\mathcal{D}^{2(N_0-N_1)-\chi}\sum_{\lbrace a,b \rbrace} \frac{\prod_{\Delta_2\in T^2} d_{a_{\Delta_2}} \prod_{\Delta_4 \in T^4}Z_b^{\epsilon(\Delta_4)}(\Delta_4)}{\prod_{\Delta_3 \in T^3} d_{b_{\Delta_3}}}
       \end{equation}
       where $N_k$ is the number of $k$-simplices of $M$.
     
       We expect that the path integral described above, with $\mathcal{C}$ a super-modular category, describes a topological phase of matter with a single topologically non-trivial point-like particle, which is a fermion. That is, the above path integral gives a $G_b$ symmetry-enriched $\Z_2$ gauge theory coupled to fermionic matter. We will discuss in Section \ref{equivToZ2GaugeTheory} the evidence for this. 
 
       We note that the construction given above differs somewhat from the construction of \cite{bulmash2020}, as the latter also decorated the vertices ($0$-simplices) with group elements and then included in $Z_b$ a sum over all possible group elements at the vertices. The purpose of such a decoration is so that the path integral, wave function on the boundary of $\partial M^4$ and related Hamiltonian construction have a locally generated $G_b$ symmetry. This decoration is more subtle in the fermionic case; for simplicity we omit these vertex labels here and will discuss how to include them in Section \ref{GbSymVertexSec}. 
       
       The normalization, apart from the factor of $\mathcal{D}^{-\chi}$ and factors of quantum dimension, are chosen to ensure the correct anomaly under retriangulation of $M$. If $f=0$, then this amounts to retriangulation invariance; the factors of $d_a$ and $d_b$ are required to ensure retriangulation invariance under $3-3$ Pachner moves, while the $\mathcal{D}^2$ normalization is also required for $2-4$ and $1-5$ Pachner moves. The factor of $\mathcal{D}^{-\chi}$ is topologically invariant and is chosen to ensure that $Z[S^4,0,0]=1$.
       
       \subsubsection{Origin of the diagrams}
       
\begin{figure}
           \centering
           \includegraphics[width=\linewidth]{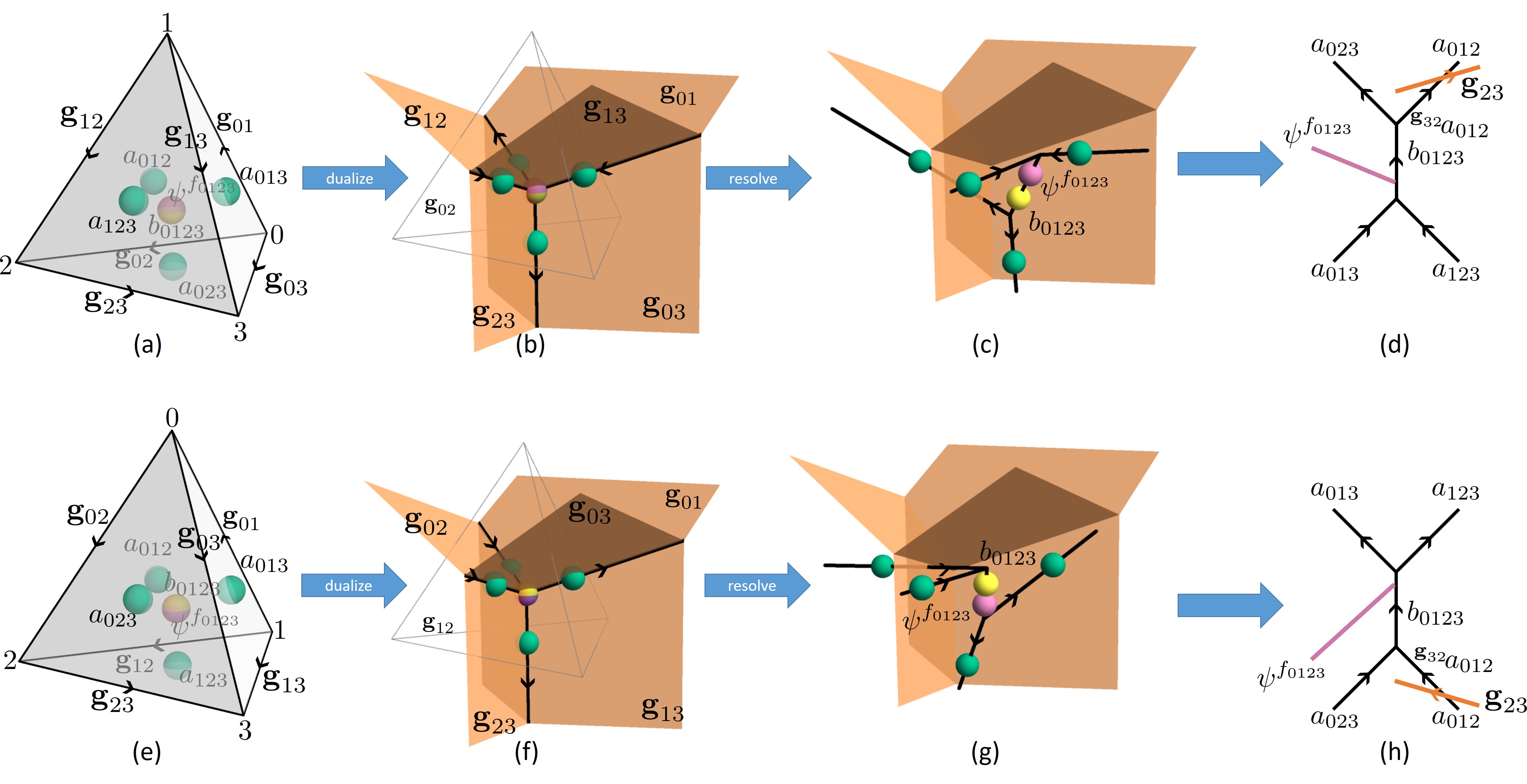}
           \caption{Turning state sum data for the bosonic shadow into graphical calculus for a 3-simplex with (a)-(d) positive relative orientation and (e)-(h) negative relative orientation. (a) Data associated to 2- and 3-simplices in the state sum. Anyons $a_{ijk}$ (green spheres) are associated to 2-simplices, while an anyon $b_{ijkl}$ (yellow sphere) and either a fermion $\psi$ or the identity (purple sphere), depending on $f_{ijkl}$, are associated to 3-simplices. (b) After dualizing in 3 dimensions, four anyons $a$ are associated to dual 1-cells, all fusing at the dual 0-cell where $b$ and $\psi$ live. The dual 2-cells are domain walls carrying a $G_b$ ${\bf g}_{ij}$ element from the background $A_b$ gauge field. The anyons all live on domain wall junctions. (c) Resolving the fusion channels and symmetry action; fusion is resolved trivalently, including fusion with the fermion, and the anyons are pushed towards a 0-simplex. One anyon line, $a_{012}$, passes through a domain wall before it reaches a fusion vertex. (d) Representation of (c) in graphical calculus. Parts (e)-(h) are the same process as (a)-(d) for a negatively oriented 3-simplex.}
           \label{fig:dualizingDiagrams}
       \end{figure}

       The diagrams in Fig.~\ref{fig:15jSymbols} arise naturally once the data on the 3-simplices are combined into a braided fusion diagram with symmetry defects, as discussed in \cite{bulmash2020}. Here there is an additional modification to include the fermions arising from $f$ defined on the $3$-simplices. Below we briefly comment how the above process works. The procedure is illustrated in Fig.~\ref{fig:dualizingDiagrams}. 
       
       Consider the dual of the 3-simplex in three dimensions. Since 2-simplices are assigned anyon lines, each dual 1-cell is now an anyon line, as in Fig.~\ref{fig:dualizingDiagrams}(b). Four such anyon lines fuse at the dual 0-cell. Ignoring the group elements and fermions for the moment, the data at the dual 0-cell (original 3-simplex) is a choice of resolution of those four anyon lines into trivalent fusion vertices, as in Fig.~\ref{fig:dualizingDiagrams}(c). This dual picture can now be interpreted as a sensible object in the graphical calculus, as in Fig.~\ref{fig:dualizingDiagrams}(d).
       
       Next we add in the symmetry defects. Since 1-simplices carry group elements ${\bf g}_{ij}$ from the background gauge field $A_b$, that group element is associated to the dual 2-cells, which we can think of as domain walls separating different $G_b$ domains, as in Fig.~\ref{fig:dualizingDiagrams}(b). When an anyon line crosses the domain wall, it is acted on by the corresponding group element. However, anyon lines, which live on dual 1-cells, naturally live on domain wall trijunctions. To disambiguate the symmetry action, we deform all of the data on a $k$-simplex towards the highest-numbered 0-simplex (dual 3-cell) according to the branching structure, as in Fig.~\ref{fig:dualizingDiagrams}(c). The symmetry action in the 3-simplex data then appears naturally and can be interpreted in the graphical calculus as in Fig.~\ref{fig:dualizingDiagrams}(d).
       
       Next, we add the fermion lines. The 3-form gauge field $f$ naturally lives on 3-simplices. Suppose that $\Delta_3$ is a 3-simplex for which $f(\Delta_3)\neq 0$. On a closed 4-manifold, $\Delta_3$ is on the boundary of exactly two 4-simplices $\Delta_4^a$ and $\Delta_4^b$. Then we should interpret $f(\Delta_3)\neq 0$ as saying that a fermion worldline passes between $\Delta_4^a$ and $\Delta_4^b$ through $\Delta_3$. This is incorporated by fusing a fermion line into the diagram for $\Delta_3$.
       
       Finally, for each 4-simplex, we can consider its boundary 3-simplices as a triangulation of $S^3$ and project the corresponding diagram on the dual cellulation of $S^3$ onto a plane to obtain the diagram in Fig.~\ref{fig:15jSymbols}. More informally, we can lay out all of the diagrams corresponding to each 3-simplex and connect corresponding anyon lines and domain walls to form a closed diagram. By flatness $\delta f=0$, the five lines $f_{ijkl}$ in the 4-simplex fuse to the identity, so we simply choose a trivalent resolution of that fusion process. 
       
       In our graphical calculus, a ${\bf g}_{ij}$ domain wall separates regions labeled with (blue) $G_b$ elements that differ by ${\bf g}_{ij}$. Hence all group element labelings are determined by the outermost region's group element ${\bf g}$, which we take to be the identity for the following reason.
       It suffices to show that we should set ${\bf g}$ to be unitary because if this is the case, then multiplying all group elements by $\overline{\bf g}$ leaves the diagram invariant, so we may as well set ${\bf g}=1$. The interpretation of the (anti-)unitarity of ${\bf g}$ is that it tracks the local orientation of space-time. However, as discussed in Sec.~\ref{sec:geometricChoices}, this information is already encoded in the choice of $\pm$ signs assigned to each 4-simplex, and this assignment is inconsistent with a global orientation only at the orientation-reversing wall $w_1$. Choosing all of the ${\bf g}$ to be unitary means that we have implicitly cut open $M$ on $w_1$, assigned a global orientation to the cut manifold, and then glued the manifold back together with an orientation twist. Only the 3-simplices which lie on the cut are affected by the orientation twist. Furthermore, as explained in Appendix~\ref{fInftyAndPerturbationOf_w1}, the orientation-reversing wall can be thought of as `perturbed' into exactly one of the 4-simplices $\braket{01234}$, the unique one for which the domain wall surrounding $\braket{0123} \subset \braket{01234}$ is anti-unitary. As such, an anti-unitary region around a $\braket{0123}$ means that the local orientation on $\braket{0123}$ is twisted relative to the rest of the diagram.
       Our choice of cochain representative $w_1=f_{\infty}(A_b^*s)$ ensures that the orientation twist is correctly encoded via the domain walls.

       \subsubsection{Choices in the definition}
       \label{subsubsec:shadowChoices}
       
       The definition described above involves a large number of choices, which we now highlight. Some of these choices are arbitrary, while others are made to be consistent with choices made in defining the Grassmann integral. We refer the reader to the Appendix of~\cite{bulmash2020} for an explanation of the independence of the path integral $Z_b$ under the arbitrary choices when $f = 0$. 
       
       One choice is the resolution of the five objects (four anyon lines and a fermion line) that meet at a 3-simplex into a trivalent resolution. On an orientable manifold, it is straightforward to see that this choice is arbitrary: different options are related by $F$- and $R$-moves which cancel out on closed manifolds, as each 3-simplex appears exactly twice and with opposite induced orientation. The choice we have made is the most aesthetic and matches the trivalent resolution that would appear in the 3$d$ Grassmann integral.\footnote{We note that it is also the same trivalent resolution that appears in a different context in \cite{qingrui}.} On an non-orientable manifold, it is still true that the choice is arbitrary, but this is more subtle to show because any 3-simplex on the orientation-reversing wall $w_1$ will appear twice in the \textit{same} induced orientation. In order for the $F$- and $R$-moves relating different resolutions to cancel, such a 3-simplex must appear surrounded by a domain wall for an anti-unitary group element exactly once. In Appendix~\ref{usingFInftyOrientationWall}, we prove that this indeed occurs due to the choice of cochain representative of $w_1$ that we made in Sec.~\ref{sec:geometricChoices}.
       
       Similarly, a change of basis for fusion vertices $\Gamma^{ab}_c$ and symmetry gauge transformation $\gamma_a({\bf g})$ cancels out on a closed manifold, provided we maintain the constraints Eqs.~(\ref{eqn:gammaPsiConstraint},\ref{eqn:GammaPsiPsi1Constraint}). In particular, our results are invariant under relaxing the gauge-fixing Eqs.~(\ref{eqn:Fapsipsi},\ref{eqn:Fpsipsia}), so long as we do so in a way which maintains Eq.~\eqref{eqn:GammaPsiPsi1Constraint}. Given the aforementioned result from Appendix~\ref{usingFInftyOrientationWall}, the proof of these statements is essentially identical to those for the case without fermions given in~\cite{bulmash2020}, and is discussed in the Appendix~\ref{app:GammaPsiPsi_1}. We will discuss the effect of transformations which violate Eqs.~(\ref{eqn:gammaPsiConstraint},\ref{eqn:GammaPsiPsi1Constraint}) shortly in Sec.~\ref{subsubsec:fermionLocality}.

       As discussed in~\cite{bulmash2020}, subject to a mild technical assumption, changing the over-crossing of the anyon lines $234$ and ${}^{42}012$ to an under-crossing amounts to an overall complex conjugation of the path integral and thus does not affect the required cancellation of the higher-form anomalies associated with gauge transformations of $f$. The choice of over/under-crossing of fermion lines with anyon lines is unimportant since those choices differ by the mutual statistics of the fermion with the anyon, which is trivial as long as the input category is super-modular. From these two facts, mirroring the 3-simplices about the vertical axis produces the same result as in the theory without fermions, i.e. it complex conjugates the path integral.
       
       Another choice is the order in which we fuse the fermion lines together. This order is chosen to be consistent with the trivalent resolutions shown in Fig.~\ref{trivalentResolutionsAndWindings} used in the Grassmann integral and ensures proper anomaly cancellation as discussed in Section \ref{anomalyCancelOverview}.
       
       As discussed above, we have chosen ``deformations'' of the data on the 2- and 3-simplices to ensure that the anyon lines stay away from junctions of domain walls. As shown in an appendix of~\cite{bulmash2020} for the case $f = 0$, one can make other choices as well without affecting the overall path integral; one can verify that the same is true when $f \neq 0$. 
        
       Finally, the choices of triangulation and branching structure of $M$, gauge for $A_b$, gauge for $f$ and, if appropriate, orientation-reversing wall (representative of $w_1$) all need to be made consistently in our path integral and in the Grassmann integral. As we will show shortly, our path integral is not invariant under changes of these choices but is anomalous as described in Section \ref{sec:anomaliesOfShadow} when $f \neq 0$. When $f = 0$, $Z_b(M^4,A_b,f=0)$ is indeed independent of these choices and therefore is a topological invariant.  
     
       \subsubsection{On the fermion locality constraints} \label{subsubsec:fermionLocality}
       
       If $f=0$, then the bosonic shadow is exactly invariant under symmetry action gauge transformations, and vertex basis transformations, regardless of whether the symmetry fractionalization data and gauge transformations obey the fermion locality constraints Eqs.~(\ref{eqn:Upsi1}-\ref{eqn:etaPsiOmega2Constraint}). This is unsurprising; the shadow is perfectly legitimate as a bosonic theory, so we may relax these constraints and no problems should appear unless we attempt to condense the fermion, i.e., gauge the 2-form symmetry. 
       
       If $f \neq 0$, then as discussed above, the bosonic shadow is invariant under gauge transformations that obey the fermion locality constraints, and as we will see in the next subsection, it also possesses the correct anomalies for the fermion condensation procedure to work. However, we show in Appendix~\ref{app:GammaPsiPsi_1} that under a non-trivial vertex basis transformation $\Gamma^{\psi \psi}_1$ which preserves the constraint Eq.~\eqref{eqn:Upsi1}, the path integral transforms as 
         \begin{equation}
      Z_b[M^4,A_b,f] \rightarrow Z_b[M^4,A_b,f] \times \left(-1\right)^{\int_{M_4} f \cup w_1}.
       \label{eqn:gammapp1shift}
       \end{equation}
       For $G_b=\Z_2^{\bf T}$, Eq.~\eqref{eqn:gammapp1shift} also gives the transformation of the path integral under $\gamma_\psi({\bf T})=-1$ \footnote{More generally, if we chose $\gamma_\psi({\bf g}) = -1$, then essentially the same argument of~\cite{bulmash2020} shows $Z_b$ gains an extra $(-1)$ for every crossing of $f$ with a ${\bf g}$ domain wall. For example, the gauge transformation $\gamma_\psi({\bf T}) = -1$ for the $G_b = \Z_2^{\bf T}$ time-reversal symmetry changes $Z_b(\cdots) \to Z_b(\cdots) (-1)^{\int f \cup w_1}$ since the only group element that reverses orientation at $w_1$ is ${\bf T}$}.
       
       Banning these gauge transformations incorporates the locality of the fermion, which is crucial for obtaining physically reasonable results. For example, our exact results in Sec.~\ref{sec:examples} will show that making a transformation of either $\Gamma^{\psi \psi}_1 = -1$ or $\gamma_\psi({\bf T})=-1$ for the case $G_f=\Z_4^{{\bf T},f}$ exchanges the $\nu$ and $-\nu$ elements of the $\Z_{16}$ classification of (3+1)D SPTs in this symmetry class. If these transformations were actually gauge transformations, our path integral would indicate that the known $\Z_{16}$ classification would collapse to $\Z_2$, which is the same as the group cohomology part of the classification of \textit{bosonic} $\Z_2^{\bf T}$ SPTs in (3+1)D. That is, the presence of local fermions would be ``forgotten."

       We observe in Eq.~\eqref{eqn:gammapp1shift} that the phase shift under the vertex basis transformation $\Gamma^{\psi,\psi}_1=-1$ is expressed as the nontrivial response action $(-1)^{\int_{M_4}f\cup w_1}$. Hence, the banned transformation with $\Gamma^{\psi,\psi}_1=-1$ behaves like a global $\mathbb{Z}_2$ symmetry with an ’t Hooft anomaly, rather than a do-nothing operation. It is expected that the ’t Hooft anomaly is characterized by a  4+1D response action $(-1)^{\int a\cup f\cup w_1}$ with $a$ a background gauge field for the $\Z_2$ symmetry, which cannot be eliminated by adding a local counterterm since the anomaly is nontrivial in cohomology.
       
        The locality constraint Eq.~\eqref{eqn:Upsi1} is also important, albeit for slightly different reasons. Under a Pachner move, if Eq.~\eqref{eqn:Upsi1} is not obeyed, then the higher-form anomaly of our theory will have an additional contribution beyond those expected for a bosonic shadow. These extra contributions are not cancelled by a natural generalization of the fermion condensation procedure, and we expect that fermion condensation fails entirely. Physically, fermion condensation converts the emergent fermion into a local one, so if we insert symmetry fractionalization data which is incompatible with a local fermion, it seems natural that the resulting theory will not allow fermion condensation. It is possible that the additional contribution can be cancelled by some local $4+1$D action, but we believe that no $3+1$D generalization of the Gu-Wen path integral is able to cancel this additional contribution.

       \subsection{Anomalies of the shadow} 
       
        We now demonstrate that $Z_b$ has the correct anomalies as discussed in Section \ref{sec:anomaliesOfShadow}. In particular, we are considering two \textit{branched} triangulations of the manifold, $(M,T)$ and $(M,T')$, with background gauge fields $(A_b, f)$ and $(A_b',f')$, respectively. Furthermore, $f$ and $f'$ both define the same cohomology class $[f] \in H^3(M^4,\Z_2)$, while $A_b$ and $A_b'$ both define the same flat $G_b$ bundle over $M$. 
        As explained in Sec \ref{generalTheorySec}, we want to show that 
        \begin{equation*}
            Z_b((M,T),A_b, f) = Z_b((M,T'),A_b', f') (-1)^{\int_{M^4 \times I} \Sq^2(\tilde{f}) + \tilde{f} \cup \tilde{A}_b^*\omega_2},
        \end{equation*}
        where we pick any triangulation of $M^4 \times I$ which agrees with $T$ on $M^4 \times \{0\}$ and $T'$ on $M^4 \times \{1\}$. Furthermore, $\tilde{f}$ and $\tilde{A}_b$ are defined on $M^4 \times I$ such that they restrict to $f$, $A_b$ on $M^4 \times \{0\}$ and $f'$, $A_b'$ on $M^4 \times \{1\}$. 
        Note that this anomaly is trivial if $f=0$ throughout, which implies the topological invariance of $Z_b(M^4,A_b,f=0)$.
        
        The general strategy is to consider Pachner moves applied to the triangulation. In particular, Pachner's theorem says that any two PL-equivalent triangulations of a $d$-manifold can be connected by a set of $d$ different local moves, known as bistellar flips or ``Pachner moves'', labeled by $k=1,\dots,d$ which involving replacing $k$ $d$-simplices with $(d+2-k)$ of them. Such a move is known as a $(k,d+2-k)$-move. Our situation is complicated by the fact that $Z_b$ depends on several background quantities: the branching structure and gauge fields $A_b,f$. In Appendix \ref{appPachnerLemmas}, we show that indeed any two triangulations decorated by background gauge fields and branching structures are connected by decorated Pachner moves. This means that to prove the anomaly formula, it suffices to show that it holds for branched Pachner moves. 
        
        We demonstrate that the anomaly presented above holds explicitly for a $(3,3)$ move in Appendix \ref{ZbAnomaliesSec} and explain there how our calculation generalizes to other moves. In particular the calculation splits into two parts: one corresponding to the $\Sq^2$ anomaly and one corresponding to the $A_b^* \omega_2$ anomaly. The $\Sq^2$ part is `universal' in the sense that it does not depend on the existence of background gauge fields or symmetry fluxes. 
        The $f \cup A_b^* \omega_2 = (f_\infty A_b^*\omega_2)(f)$ part can be analyzed by a calculation using the graphical calculus of Sec.~\ref{globsym}.
        
        A special example of the anomaly, similarly described in Ref.~\cite{Gaiotto:2015zta},
        is an explicit formula for the gauge transformation $f \to f + \delta\lambda$,
        \begin{equation}
            Z_b((M,T),A_b,f + \delta \lambda) = Z_b((M,T),A_b,f) (-1)^{\int_M f \cup_{d-3} \lambda + \lambda \cup_{d-3} f + \lambda \cup_{d-3} \delta\lambda + \lambda \cup_{d-4} \lambda + \lambda \cup A_b^*\omega},
        \end{equation}
        and is related to the quadratic refinement property of the function $\sigma(f)$. This formula only works if we keep the triangulation, branching structure, and $A_b$ field fixed. We do not know of any similar explicit expressions for other kinds of changes like branching structures or $A_b$ changes with fixed triangulation, although any particular example can be in principle be evaluated by following the algorithms for implementing decorated Pachner moves in Appendix~\ref{appPachnerLemmas}. We discuss the case of $A_b$ changes in more detail in Sec.~\ref{GbSymVertexSec}.
        
        \subsection{ Equivalence to $\Z_2$ gauge theory with fermionic matter enriched with global $G_b$ symmetry} \label{equivToZ2GaugeTheory}
        
        The path integral $Z_b$ defined above, which takes the super-modular category $\mathcal{C}$ as input, is equivalent to a $\Z_2$ gauge theory with a fermionic charge, and enriched with a global $G_b$ symmetry. If we ignore the $G_b$ symmetry, this result follows from results recently proven in Ref.~\cite{johnsonfreyd2021minimal} \footnote{The work~\cite{johnsonfreyd2021minimal} shows (see the beginning of its Section 3) that, in a 2-categorical sense, every super-modular category is equivalent to the category {\bf sVec} of super-vector spaces, which corresponds to the $\{1,\psi\}$ topological order. In a physical language, this means that the bulk $(3+1)$D topological order defined by any super-modular category is in the same bulk phase as the $\{1,\psi\}$ order.}. 
        
        The above statement implies, for example, that when $M^4$ is an orientable manifold,
        \begin{align}
        \label{ZbZ2gauge}
            |Z_b(M^4, A_b = 0, f)| = \sqrt{|H^2(M^4,\Z_2)|} \delta_{[f],0}\delta_{[w_2],0},
        \end{align}
        where as we will see below, the RHS is the appropriately normalized path integral for $\Z_2$ gauge theory. Here $\delta_{[f],0} = 1$ if $[f]$ is in the trivial cohomology class, and $0$ otherwise. $\delta_{[w_2],0}$ is also defined in a similar fashion, and requires that $M^4$ be a spin manifold. Note that while $Z_b$ does not depend on a spin structure, $M^4$ must be a spin manifold (i.e. admit a spin structure) for $Z_b$ to be non-vanishing, due to the existence of a fermionic point-like particle.
        
        Furthermore, the space of ground states on any oriented closed 3-manifold $N^3$ is given by the number of non-trivial cycles on $N^3$:
        \begin{align}
        \label{ZbZ2gaugeS1}
        |Z_b(N^3 \times S^1, A_b = 0, f)| = |H_1(N^3,\Z_2)| \delta_{[f],0} \delta_{[w_2],0}. 
        \end{align}
        We can physically understand the above formula in the following way. $\Z_2$ gauge theory can be thought of in a string-net picture where strings start and end at fermionic charges. The space of states on some $N^3$ consists of all topologically distinct string-nets, of which there are $|H_1(N^3,\Z_2)|$.\footnote{We also note that Eq. \eqref{ZbZ2gaugeS1} is expected to hold for $A_b \neq 0$ as well, although we have not developed a systematic proof. The reason the dimension of the space of states can change in the presence of closed symmetry defect sheets is that the symmetry permutes the non-trivial particles so that certain Wilson lines cannot close \cite{barkeshli2019}. In $\Z_2$ gauge theory, there are no non-trivial permutations since there is only a single non-trivial particle, therefore the dimension of the space of states should not change with non-trivial $A_b$.} Furthermore, this picture implies there cannot be any states with an odd number of charges because the total number of endpoints of strings should be even; therefore $[f]$ cannot contain a fermion loop traversing the $S^1$. 
        We note that Eq.~\eqref{ZbZ2gauge} implies Eq.~\eqref{ZbZ2gaugeS1} using the Künneth formula for $H^2(N^3 \times S^1)$, which also explains the $\delta_{[f],0}$ factor in Eq.~\eqref{ZbZ2gaugeS1}.

        We note that when $A_b \neq 0$ it is possible that $Z_b(M^4,A_b,f_3) \neq 0$ even when the cohomology class of $[f_3]\neq 0$. A key example for our purposes occurs when evaluating $Z_b$ on $\mathbb{RP}^4$ with a non-trivial fermion loop. We briefly describe why this is not an issue in Sec.~\ref{z16MainText}. 
        
        In the case where the super-modular category $\mathcal{C}$ is Abelian and $A_b = 0$, we can directly prove Eq.~\eqref{ZbZ2gauge} using our state sum construction. The main reason is that in this case, the category splits as a product of a modular category and the trivial super-modular category $\{1, \psi\}$. 
        
        For general non-Abelian super-modular categories $\mathcal{C}$, we do not have an explicit proof of Eq.~\eqref{ZbZ2gauge}, however the result is expected from the general proof of equivalence between Crane-Yetter with super-modular input and $\Z_2$ gauge theory with fermionic matter as proven in Ref.~\cite{johnsonfreyd2021minimal}. There are also alternative physical arguments that can be made by passing back and forth between a Hamiltonian description of the theory. These will be reviewed below. 
        
        \subsubsection{$Z_b$ for $\{1,\psi \}$ and $\Z_2$ gauge theory}
        
        The simplest example of a super-modular category is $\mathcal{C}=\{1,\psi\}$, where $\psi$ is a fermion with $\psi\times\psi=1$. Let us consider $Z_b$ with $\mathcal{C}=\{1,\psi\}$. This special case was also studied in ~\cite{kapustinThorngren2017FermionSPT}. We will also assume that the theory does not couple to $G_b$ except through the possibility of allowing $M^4$ to be non-orientable, which we comment on further below. 
        
        In this case, the 15j symbols are completely determined by the assignment of $\psi$ to each $2$-simplex. We denote $b\in C^2(M,\mathbb{Z}_2)$ as a 2-cochain that controls the $\psi$ assignment on 2-simplices of $M^4$. Note that we have the condition $\delta b=f$ because for each 3-simplex $\braket{0123}$, the associated fusion space is only nonzero if the total number of fermion lines going into the fusion channel is $0 \text{ (mod 2)}$. This number of fermion lines is given by $f(0123) + b(012) + b(013) + b(023) + b(123) = (f + \delta b)(0123) = 0 \text{ (mod 2)}$. In particular, this implies that only trivial background $f$ with $[f] = 0 \in H^3(M,\Z_2)$ will give a nonzero $Z_b$.
        
        The 15j symbol on $\braket{01234}$ is computed as $(-1)^{(b\cup b + b \cup_1 f)(01234) = b(012)b(234) + b(034)f(0123) + b(014)f(1234)}$ which gives a factor of $(-1)$ to all crossings in the diagram. Hence 
        \begin{align}
        \begin{split}
            Z_{b}^{\{1,\psi\}}(M,f) &= \mathcal{D}_{\{1,\psi\}}^{2(N_0 - N_1)} \mathcal{D}^{-\chi}  \sum_{\substack{b\in C^2(M,\mathbb{Z}_2),\\ \delta b=f }}\exp\left\{\pi i\int_M b\cup b + b\cup_1 f\right\} \\
            &= \mathcal{D}_{\{1,\psi\}}^{2(N_0 - N_1)} \mathcal{D}^{-\chi} \frac{1}{|C^1(M,\Z_2)|}\sum_{\substack{a\in C_2(M,\mathbb{Z}_2),\\ b\in C^2(M,\mathbb{Z}_2)}}\exp\left\{\pi i\int_M (a)(\delta b+f) + b\cup b + b\cup_1 f\right\}.
            \end{split}
        \end{align}
        where we have introduced a Lagrange multiplier field $a$ to enforce the constraint $\delta b = f$ via the chain-cochain pairing $(a)(\delta b + f)$. We can check that the action is gauge invariant under $a\to a+\partial\chi$, $b\to b+\delta\lambda$ for $\chi\in C_1(M,\mathbb{Z}_2)$, $\lambda\in C^1(M,\mathbb{Z}_2)$. 
        
        Shifting $b \to b + \beta$ for $\beta \in Z^2(M,\mathbb{Z}_2)$ changes the action by $\beta \cup \beta$, which is trivial when $M^4$ is a spin manifold (i.e. admits a spin structure). Therefore when $M^4$ is a spin manifold, the shift $b \to b + \beta$ is a $\Z_2$ $2$-form symmetry. 
        We can then regard $f$ as a background gauge field of the $2$-form symmetry. We thus obtain a $\mathbb{Z}_2$ gauge theory coupled to the background gauge field $f$ of the 2-form symmetry $b\to b+\beta$. 
        
        Now we can explicitly evaluate the partition function $Z_b(M,f)$. We will evaluate the state sum in the case $f=0$ since only $[f]=0$ contributes and, by anomaly matching, $Z_b(M,f) = Z_b(M,f=0)z_c(f)$ for all $[f]=0$. Then $Z_b(M,f=0)$ is just the Crane-Yetter sum for $\{1,\psi\}$ which we will call $Z_{CY}^{\{1,\psi\}}(M)$. The above reduces to
        \begin{equation*}
        \begin{split}
            Z_{CY}^{\{1,\psi\}}(M) &= \mathcal{D}_{\{1,\psi\}}^{2(N_0 - N_1)} \mathcal{D}^{-\chi} \sum_{b \in Z^2(M,\Z_2)} (-1)^{\int b \cup b} \\
            &= \sqrt{2}^{-\chi(M)} 2^{(N_0 - N_1)} |B^2(M,\Z_2)| \sum_{[b] \in H^2(M,\Z_2)} (-1)^{\int [w_2+w_1^2] \cup [b]} \\
            &= \sqrt{2}^{-\chi(M)} 2^{(N_0 - N_1)} |B^2(M,\Z_2)| 
            \begin{cases}
                |H^2(M,\Z_2)| \text{  if } [w_2 + w_1^2] \text{ is trivial} \\
                0 \text{  if } [w_2 + w_1^2] \text{ is non-trivial}
            \end{cases}
        \end{split}
        \end{equation*}
        where the first equality uses $[b \cup b] = [\Sq^2(b)] = [(w_2 + w_1^2) \cup b]$ and the second equality is because the sum over $[b]$ will have cancelling terms $[b], [b+b']$ where $\int [w_2 + w_1^2] \cup [b'] = 1$ exists if $[w_2 + w_1^2]$ is non-trivial. In particular, this enforces that $M$ must be spin in the orientable case or $\text{pin}^-$ in the non-orientable case. 
        
        We can simplify this even further, because we claim
        \begin{equation}
        2^{(N_0 - N_1)} |B^2(M,\Z_2)| = \frac{|H^0(M,\Z_2)|}{|H^1(M,\Z_2)|}
        \end{equation}
        The proof is by considering the chain complex $1 \xrightarrow{\delta_0} (\Z_2)^{N_0} \xrightarrow{\delta_1} (\Z_2)^{N_1} \xrightarrow{\delta_2} (\Z_2)^{N_2} \xrightarrow{\delta} \to \cdots $. Note that $|B^2(M,\Z_2)| = 2^{\text{dim im}(\delta_2)}$. And by definition, $\text{dim ker}(\delta_1) - \text{dim im}(\delta_0) = \text{dim}(H^0(M,\Z_2))$ and $\text{dim ker}(\delta_2) - \text{dim im}(\delta_1) = \text{dim}(H^1(M,\Z_2))$. The rank-nullity theorem gives $\text{dim ker}(\delta_1) + \text{dim im}(\delta_1) = N_0$ and $\text{dim ker}(\delta_2) + \text{dim im}(\delta_2) = N_1$. Putting these all together gives $\text{dim}(B^2) = N_1 - N_0 - \text{dim}(H^1(M,\Z_2))+\text{dim}(H^0(M,\Z_2))$ which is equivalent to the above claim (noting that since $\Z_2$ is a field, all homology and cohomology groups are $\Z_2$ vector spaces). This gives the final result:
        \begin{equation}
           Z_b^{\{1,\psi\}}(M,f=0) = Z_{CY}^{\{1,\psi\}}(M) = \sqrt{2}^{-\chi(M)} \frac{|H^2(M,\Z_2)||H^0(M,\Z_2)|}{|H^1(M,\Z_2)|}\delta_{[w_2+w_1^2],0} = \sqrt{|H^2(M,\Z_2)|}\delta_{[w_2+w_1^2],0}
        \end{equation}
        where the last equality follows from the explicit expression of $\chi(M)$ in terms of Betti numbers.
        
        Note that the factor $\delta_{[w_2+w_1^2],0}$ implies that $M^4$ must be a spin manifold when it is orientable and a pin$^-$ manifold when it is non-orientable. The appearance of pin$^-$ arises for the following reason. The path integral $Z_{CY}^{\{1,\psi\}}(M)$ corresponds to the case where $\omega_2 = 0$. Allowing $M$ to be non-orientable implies that we are considering an anti-unitary time-reversal symmetry so that $G_b = \Z_2^{\bf T}$, but since $\omega_2 = 0$, the theory corresponds to the case $G_f = \Z_2^{\bf T} \times \Z_2^f$. As discussed in Appendix \ref{GfantiU}, defining fermions with this symmetry group requires a pin$^-$ manifold. 
        
        We consider some particular cases. Certainly $Z_{CY}^{\{1,f\}}(S^4) = 1$. For any 3-manifold $N^3$, $Z_{CY}^{\{1,f\}}(N^3 \times S^1) = |H_1(N^3,\Z_2)|\delta_{[w_2+w_1^2],0}$ using the Künneth formula for $H^\ast(N^3 \times S^1)$. This matches with the $\Z_2$ gauge theory expectation that the Hilbert space on $N^3$ is the number of distinct cycles on $N^3$. 

        \subsubsection{$Z_b$ for Abelian $\mathcal{C}$}        
        
        Let us consider the case where the super-modular category $\mathcal{C}$ is Abelian, i.e. all anyons in $\mathcal{C}$ have quantum dimension $d_a = 1$. Let us denote this as $\mathcal{C}(\mathcal{A})$, where $\mathcal{A}$ is the Abelian group arising from fusion of the Abelian anyons.  
        In this case, it is a general theorem (see, e.g., Appendix A of Ref.~\cite{mengCheng2018weakFermionSPT}) that the category splits as 
        \begin{align}
         \mathcal{C}(\mathcal{A}) = \mathcal{C}(\mathcal{A}/\Z_2^\psi) \times \mathcal{C}(\Z_2^\psi),
        \end{align}
        where $\mathcal{C}(\Z_2^\psi) = \{1,\psi\}$ and $\Z_2^\psi$ is the group $\Z_2$ generated by fusion of the fermion $\psi$ with itself. Here $\mathcal{C}(\mathcal{A}/\Z_2^\psi)$ is a unitary modular tensor category while $\mathcal{C}(\Z_2^\psi) = \{1,\psi\}$ is the trivial super-modular category.

        It follows that the Crane-Yetter path integral also factorizes:
        \begin{align}
            Z_{b;\mathcal{C}({\mathcal{A}})}(M^4, A_b = 0, f_3) 
            &= Z_{b;\mathcal{C}(\Z_2^f)}( M^d, A_b = 0, f_3) Z_{b;\mathcal{C}({\mathcal{A}}/\Z_2^f)}( M^d, A_b = 0) 
            \nonumber \\
            &= Z_{b;\mathcal{C}(\Z_2^f)}( M^d, A_b = 0, f_3)  \times e^{\frac{2 \pi i c_-}{8} \text{sgn}(M)}
            \nonumber \\
            &= \sqrt{|H^2(M^4, \Z_2)|} \delta_{[f_3], 0}\delta_{[w_2+w_1^2],0}  \times e^{\frac{2 \pi i c_-}{8} \text{sgn}(M)}
        \end{align}
        where $c_-$ is the chiral central charge of the unitary modular tensor category $\mathcal{C}(\mathcal{A}/{\Z_2^f})$ and $\text{sgn}(M)$ is the signature of the 4-manifold $M$. Above we use the general fact that the Crane-Yetter state sum gives $Z_{CY}(M^d ; \mathcal{C}) = e^{\frac{2 \pi i c_-}{8} \text{sgn}(M)}$ for any 4-manifold $M^4$ when $\mathcal{C}$ is a UMTC. Note for any oriented $N^3$ that $\text{sgn}(N^3 \times S^1) = 0$ since the signature is an oriented-bordism invariant and $N^3 \times S^1 = \partial(N^3 \times D^2)$ is null-bordant. This again gives that the number of states on $N^3$ is $Z(N^3 \times S^1, A_b = 0, f ) = |H_1(N^3,\Z_2)| \delta_{[f],0}\delta_{[w_2+w_1^2],0}$. 
        
        \subsubsection{$Z_b$ for non-Abelian $\mathcal{C}$} \label{sec:bosonZ2nonAbelianC}
        
        In the more general case where $\mathcal{C}$ is an arbitrary non-Abelian super-modular tensor category, we do not have an explicit computation that gives Eq.~\eqref{ZbZ2gauge}, although the result follows from the mathematical results of~\cite{johnsonfreyd2021minimal}. Below we provide a brief alternative explanation for why Eq.~\eqref{ZbZ2gauge} is expected. 
        
        First, we note that it is known that there is essentially a unique (3+1)D topological order (up to an anomalous one) that contains a single topologically non-trivial particle with $\Z_2$ fusion rules, which is $\Z_2$ gauge theory. There are two such theories, depending on whether the particle is a boson or a fermion. A physical account is given in~\cite{lanKongWen3DTopologicalOrder} while a mathematical account is given in~\cite{freyd3DTopologicalOrder}. 
        
        We expect on physical grounds that our path integral realizes a TQFT with a single non-trivial fermionic point-like excitation with $\Z_2$ fusion rules because it is closely related to the Crane-Yetter model whose Hamiltonian formulation is the Walker-Wang model where the input is a super-modular category. 
        
        Specifically, when $A_b, f = 0$, our path integral reduces to the Crane-Yetter path integral, which in particular gives a wave function $|\Psi(\partial M^4)\rangle$ associated to the boundary $\partial M^4$. The Walker-Wang model is an exactly solvable Hamiltonian for which the wave function $|\Psi(\partial M^4)\rangle$ is an exact ground state. Furthermore, it is expected in general that the deconfined bulk point-like excitations of the Walker-Wang model with input a premodular category correspond to the transparent anyons in the category. Our case of a super-modular category means that there is a single transparent particle, $\psi$ which in the Walker-Wang Hamiltonian formulation should correspond to there being a single species of deconfined point particle. A particular example corresponding to the super-modular category $\SO(3)_3$ \footnote{Our notation for this category differs from~\cite{fidkowski2013} and other references in the condensed matter literature, in which this category is referred to as $\SO(3)_6$. This category describes the anyon content of $\SO(3)_3$ Chern-Simons theory, which is the integer spin sector of the $\\SU(2)_6$ category. Some references, e.g.~\cite{nayak2008}, use $\SO(3)_3$ to refer to the integer spin sector of the $\SU(2)_3$ category, which is usually known as the Fibonacci anyon theory. In the present notation, Fibonacci would be written as $(G_2)_1$.} was explicitly studied in~\cite{fidkowski2013} where a Walker-Wang-like construction based on $\SO(3)_3$ was shown to have a single deconfined fermionic excitation. 
        
        Therefore, the ground state of Walker-Wang models with super-modular category as input should be described by a TQFT that is equivalent to $\Z_2$ gauge theory coupled to fermionic matter, from which we conclude that our path integral construction $Z_b(M^4, A_b = 0, f)$ is equivalent to that of $\Z_2$ gauge theory coupled to fermionic matter. 
        
        \section{ Fermionic state sum} \label{sec:fermStateSum}
        
        Since $Z_b$ has the correct higher-form anomalies to be a bosonic shadow, we may use the procedure in Sec.~\ref{sec:zpsi} to condense the fermion by writing the full partition function of our theory:
        \begin{equation*}
            Z(M,A_b, \xi_{\mathcal{G}}) = \frac{1}{\mathcal{N}} \sum_{f_3 \in H^3(M,\mathbb{Z}_2)} Z_b(M,A_b,f_3)z_c(M,\xi_{\mathcal{G}},f_3),
        \end{equation*}
        where $\mathcal{N} = \sqrt{|H^2(M^4, \Z_2)|}$ is a normalization constant that will be derived below. We emphasize that $A_b$ implicitly appears in $z_c$ via $\xi_{\mathcal{G}}$. We note that up to a normalization $\frac{1}{|B^3(M,\Z_2)|}$, we could have summed over \textit{all} closed fermion lines which would tie in more closely to the physical picture associated to fermion condensation.
        
        Although the only symmetry group which explicitly appears in the state sum is $G_b$, the structure of the group extension $G_f$ is included via the symmetry fractionalization data $U$ and $\eta$, which in general are non-trivial when their arguments include $\psi$.
        
        \subsection{ Topological invariance of $Z(M, A_b, \xi_{\mathcal{G}})$}
        
        Since the anomalies arising in $Z_b$ and $z_c$ cancel, $Z$ is actually independent of triangulation, branching structure, and changes of $A_b$ and $\xi_{\mathcal{G}_b}$ by coboundaries. More precisely, we have that
        \begin{align}
            Z( (M^4, T), A_b, \xi_{\mathcal{G}}) = Z( (M^4, T'), A_b', \xi_{\mathcal{G}}') ,
        \end{align}
        where $T$ and $T$' denote two distinct choices of triangulation, branching structure, and $\pm$ signs of simplices that are compatible with $A_b$. Furthermore, $A_b'$, $\xi_{\mathcal{G}}'$ are equivalent to $A_b$, $\xi_{\mathcal{G}}$, in the sense that one can construct a triangulated bordism $M^4 \times I$ and extend $A_b$ and $\xi_{\mathcal{G}}$ to be defined on $M^4 \times I$ and restrict to $A_b, \xi_{\mathcal{G}}$ on one boundary and $A_b', \xi_{\mathcal{G}}'$ on the other boundary. 
        
        \subsection{ Invertibility and bordism invariance of $Z(M^4,A_b,\xi_{\mathcal{G}})$}

        With an appropriate normalization, the fermionic state sum $Z(M^d, A_b, \xi_{\mathcal{G}})$ defines a path integral for a fermionic, invertible TQFT with bosonic and fermionic symmetry groups $G_b$ and $G_f$. 

        To see this, we analyze the full state sum first with background gauge field $A_b = 0$:
        \begin{align} \label{fermZ}
            Z(M^4, A_b = 0, \xi_{\mathcal{G}}) = &\frac{1}{\mathcal{N}} \sum_{[f_3]} Z_b( M^4, A_b = 0, f_3) z_c(M^d, \xi_{\mathcal{G}}, f_3 )
            \nonumber \\
            &= \frac{1}{\mathcal{N}} \sqrt{|H^2(M^4, \Z_2)|} z_c(M^4, \xi_{\mathcal{G}}, [f_3] = 0 )
        \end{align}
        The second line of follows from the identity $Z_b( M^d, A_b = 0, f_3) = \sqrt{|H^2(M^4, \Z_2)|} \delta_{[f_3],0}$ which we proved for Abelian super-modular categories and conjectured for general ones in Sec.~\ref{equivToZ2GaugeTheory}. Now we pick the normalization
        \begin{align}
            \mathcal{N} = \sqrt{|H^2(M^4, \Z_2)|} .
        \end{align}
        With this normalization,
        \begin{align}
        \label{ZAb01}
            |Z(M^4, A_b = 0, \xi_{\mathcal{G}})| = 1 .
        \end{align}
        In particular, Eq.~\eqref{ZAb01} implies that the ground state degeneracy on any 3-manifold $N^3$ is 
        $|Z(N^3 \times S^1, A_b = 0, \xi_{\mathcal{G}})| = 1$ which implies that the theory has no intrinsic topological order and therefore must be an (3+1)D FSPT. Since (3+1)D FSPTs are believed to be classified by invertible TQFTs, we have that $Z$ must describe an invertible fermionic TQFT. Furthermore, given the relation between SPT phases, invertible TQFTs, and bordism invariants \cite{freed2016,kapustin2014,Kapustin:2014dxa,yonekura2019}, we expect that $Z$ also defines a bordism invariant of manifolds with $G_b$ gauge field and $\mathcal{G}_f$ structures. 
        
        We note that from a mathematical perspective, we have not rigorously established the invertibility of $Z$, which means that $|Z(M^4, A_b, \xi_{\mathcal{G}})| = 1$ for $A_b \neq 0$ as well. A rigorous proof of invertibility could perhaps follow from explicit computations on the 4-torus $T^4$ and $S^3\times S^1$ using the results of \cite{schommerPries2018}. There, it is shown that suﬀiciently-extended TQFTs with various background structures are invertible if and only if all path integrals on $T^4 = T^3 \times S^1$ and $S^3 \times S^1$ have modulus 1. However it is not immediately obvious to us that our state sum fits into the framework of \cite{schommerPries2018} because of our use of Grassmann variables and $z_c(f)$. 
        
        Additional evidence for bordism invariance of $Z$ comes from the multiplicativity under connected-sums. We have: 
        \begin{equation}
            Z(M_1 \# M_2, A_{b,1} \# A_{b,2}, \xi_{\mathcal{G},1} \# \xi_{\mathcal{G},2}) = Z(M_1, A_{b,1},\xi_{\mathcal{G},1})Z(M_2, A_{b,2},\xi_{\mathcal{G},2})
        \end{equation}
        where $M_1 \# M_2$ is the connected-sum of the manifolds and $A_{b,1} \# A_{b,2}$, $\xi_{\mathcal{G},1} \# \xi_{\mathcal{G},2}$ are canonically-defined background gauge fields and spin structures induced from the connected-sum. See Appendix~\ref{app:multOfZunderconnSum} for more precise statements and a proof. 
        
        \subsection{ Gauging fermion parity / summing over spin structures} \label{gaugeFermionParity}
        
        The fermionic path integral $Z(M^4, A_b, \xi_{\mathcal{G}})$ can be understood in terms of condensing fermions in the shadow theory given by $Z_b(M^4, A_b, f)$. The inverse procedure to condensing fermions is gauging fermion parity, which mathematically corresponds to summing over spin structures (or $\mathcal{G}_f$ structures more generally). We thus have that 
        \begin{align}
            Z_b(M^4, A_b, f_3 = 0) = \frac{\mathcal{N}}{|H^1(M^4,\Z_2)|} \sum_{\alpha \in H^1(M^4,\Z_2)} Z(M^4, A_b, \xi_{\mathcal{G}} + \alpha). 
        \end{align}
        One can verify the above equation by using the fact that
        \begin{align}
            \frac{1}{|H^1(M^4,\Z_2)|} \sum_{\alpha \in H^1(M^4,\Z_2)} (-1)^{\int_M \alpha \cup f_3} = \delta_{0,[f_3]} ,
        \end{align}
        $z_c(M^4, A_b, f_3 = 0) = 1$, and substituting Eq.~\eqref{fermZ}. Note that $Z_b(M^4,A_b,f_3=0)$ is in its own right a topological invariant that is \textit{independent} of any background $\mathcal{G}_f$ structure $\xi_{\mathcal{G}}$.
        
        \subsection{ Stacking}
        \label{stackingSec}
        
        Given two super-modular categories $\mathcal{C}_1$ and $\mathcal{C}_2$, one can construct a third super-modular category $\mathcal{C}_{12}$ by a ``stacking construction" as follows:
        \begin{align}
            \mathcal{C}_{12} = [\mathcal{C}_{1} \boxtimes \mathcal{C}_{2}]/\{\psi_1 \psi_2 \sim 1\},
        \end{align}
        where $\boxtimes$ means we take a product of the two categories as independent copies (mathematically this is the Deligne product of braided tensor categories). The quotient by $\{\psi_1\psi_2 \sim 1\}$ means that we condense the $\psi_1 \psi_2$ anyon, which is a boson and which braids trivially with everything else in $\mathcal{C}_{1} \boxtimes \mathcal{C}_{2}$. Therefore there is a single remaining fermion $\psi_1 \sim \psi_2$, so that $\mathcal{C}_{12}$ is super-modular. Physically, this can be understood as considering a two-layer (2+1)D fermionic topological phase, where the two layers are characterized by $\mathcal{C}_1$ and $\mathcal{C}_2$ and the physical fermions $\psi_1$ and $\psi_2$ in each layer are identified to be topologically equivalent. 
        
        Note that both super-modular categories $\mathcal{C}_1$ and $\mathcal{C}_2$ should further possess a symmetry fractionalization class for the symmetry group $G_b$.
        Importantly, in order for the above condensation procedure to keep the symmetries of the system invariant, we need $\mathcal{C}_1$ and $\mathcal{C}_2$ to encode the same extension $G_f$ of $G_b$, so that
        \begin{align}
        \eta_{\psi_1}({\bf g}, {\bf h}) = \eta_{\psi_2}({\bf g}, {\bf h}) =\omega_2({\bf g,h}).
        \end{align}
        
        One can then ask how to construct $Z$ and $Z_b$ for $\mathcal{C}_{12}$ in terms of the path integrals for $\mathcal{C}_1$ and $\mathcal{C}_2$ individually. One can construct $Z_{b;\mathcal{C}_{12}}$ in terms of $Z_{b;\mathcal{C}_1}$ and $Z_{b;\mathcal{C}_2}$ by condensing $\psi_1 \psi_2$. This suggests the following identity:
        \begin{align}
        \label{ZbStack}
            Z_{b;\mathcal{C}_{12}}(M^4, A_b, f_3) = \frac{1}{\sqrt{|H^2(M^4,\Z_2)|}} \sum_{f'_3 \in H^3(M^4, \Z_2)} Z_{b;\mathcal{C}_1}(M^4, A_b, f_3 + f'_3) Z_{b;\mathcal{C}_2}(M^4, A_b, f'_3) (-1)^{\int_M (f_3 + f_3') \cup_2 f_3'}
        \end{align}
        Note that when $f_3 = 0$, the higher-form anomalies of $Z_{b;\mathcal{C}_1}$ and $Z_{b;\mathcal{C}_2}$ are exactly equal and cancel each other out. Therefore the product $Z_{b;\mathcal{C}_1} Z_{b;\mathcal{C}_2}$ is non-anomalous. The above equation is therefore precisely what one would expect, except for the additional factor of $(-1)^{\int_M f_3' \cup_2 f_3'}$, which will be discussed below.
        
        When $f_3 \neq 0$, a gauge transformation of $f_3'$ on the product $Z_{b;\mathcal{C}_1} Z_{b;\mathcal{C}_2}$ will give a (4+1)D anomaly action
        \begin{align}
        S_{4+1} &= \int_{W^{5}} [ \text{Sq}^2 (\tilde{f}_3 + \tilde{f}_3') + \text{Sq}^2(\tilde{f}_3') + (A_b^* \omega_2)(\tilde{f}_3 + \tilde{f}'_3)) + (A_b^*\omega_2)(\tilde{f}'_3) ]
        \nonumber \\
        &= \int_{W^5} [\text{Sq}^2(\tilde{f}_3) + \delta (\tilde{f}_3 \cup_2 \tilde{f}_3') + (A_b^* \omega_2)(\tilde{f}_3) ],
        \end{align}
        where recall that $\tilde{f}_3$, $\tilde{f}'_3$, and $\tilde{A}_b$ denote the extension of $f_3, f'_3, A_b$ to $W^5$. We 
        have used the identity $\text{Sq}^2 (x + x') = \text{Sq}^2(x) + \text{Sq}^2(x') + \delta (x \cup_2 x')$ for $x,x' \in Z^3(W^5,\Z_2)$. Therefore, in order for $Z_{b;\mathcal{C}_{12}}[M^4, A_b, f_3]$ to possess the expected anomalies, we need to cancel off the $\delta (\tilde{f}_3 \cup_2 \tilde{f}_3')$, which explains the additional term $f_3 \cup_2 f_3'$ in Eq.~\eqref{ZbStack}. 
        
        Equipped with Eq.~\eqref{ZbStack}, we can check by direct computation that the fermionic path integral is multiplicative under this stacking operation:
        \begin{align}
        \label{Zstack}
        Z_{\mathcal{C}_1}&(M^4,A_b,\xi_\mathcal{G}) Z_{\mathcal{C}_2}(M^4,A_b,\xi_\mathcal{G})
        \nonumber \\
        &= \bigg( \frac{1}{\mathcal{N}} \sum_{f^{(1)} \in H^3(M^4,\Z_2)} Z_{b;{\mathcal{C}_1}}(M^4,A_b,f^{(1)}) z_c(M^4,f^{(1)},\xi_{\mathcal{G}}) \bigg) \cdot \bigg( \frac{1}{\mathcal{N}} \sum_{f^{(2)} \in H^3(M^4,\Z_2)} Z_{b;\mathcal{C}_2}(M^4,A_b,f^{(2)}) z_c(M^4,f^{(2)},\xi_{\mathcal{G}}) \bigg) 
        \nonumber \\
        &=  \frac{1}{\mathcal{N}} \sum_{f \in H^3(M^4,\Z_2)} 
     \left( \frac{1}{\mathcal{N}} \sum_{f' \in H^3(M^4,\Z_2)} Z_{b;\mathcal{C}_1}(M^4,A_b,f+f') Z_{b;\mathcal{C}_2}(M^4,A_b,f') (-1)^{\int (f+f') \cup_{2} f'} \right) z_c(M^4,f,\xi_{\mathcal{G}})
     \nonumber \\
     &= \frac{1}{\mathcal{N}} \sum_{f \in H^3(M^4,\Z_2)} Z_{b;\mathcal{C}_{12}}(M^4, A_b, f) z_c(M^4,f,\xi_{\mathcal{G}})
     \nonumber \\
     &= Z_{\mathcal{C}_{12}}(M^4,A_b,\xi_\mathcal{G})
        \end{align}
        where we changed variables to set $f = f^{(1)} + f^{(2)}$ and $f' = f^{(2)}$ and used quadratic refinement, Eq.~\eqref{eq:quadRefinement_zC}, to combine two factors of $z_c$ to a single factor. 
        
        Finally, let us comment on the term $f_3' \cup_2 f_3'$ in Eq.~\eqref{ZbStack}; we do not have a completely satisfactory of the origin of this term, although~\cite{Thorngren2018bosonization} observed that an analogous term is needed in $d=2$. We can see this term is necessary in order to produce Eq.~\eqref{Zstack}. We have furthermore verified by explicit calculation that this term is necessary in the example of $M^4 = \mathbb{RP}^4$, and $\mathcal{C}_1$, $\mathcal{C}_2$ being given by the $\SO(3)_3$ and semion-fermion theories, which are discussed further in Sec.~\ref{sec:examples}. 
        This term can be recast using the Wu relation on closed manifolds $M^4$ as:
        \begin{align}
            \int_{M^4} \Sq^1(f'_3) := \int_{M^4} f_3' \cup_2 f_3' = \int_{M^4} w_1 \cup f_3' . 
        \end{align}
        Therefore, this term is only non-trivial on non-orientable manifolds. Furthermore, we can understand this term intuitively as giving a minus sign each time $\psi_1 \psi_2$ crosses the orientation-reversing wall. This is to be expected from pin$^-$ fermions, since a single fermion crossing $w_1$ should give a factor of $\pm i$; therefore two such fermions crossing together are expected to give a $-1$. However it is not clear why the fermions should be treated as pin$^-$ fermions; such a requirement may ultimately stem from the fact that the Grassmann integral that plays an important role in our constructions is naturally pin$^-$. 
        
        Note that we do not have a direct proof of Eq.~\eqref{ZbStack} from the explicit construction of our state sum. Presumably this should be possible, but we leave this for future work. Rather, with the exception of the $(-1)^{f_3' \cup_2 f_3'}$ factor, the general form of Eq.~\eqref{ZbStack} is expected on general grounds from the path integral definition of condensation of $\psi_1 \psi_2$ and from the physical expectation of the behavior of the fermionic path integral under stacking $\mathcal{C}_1$ and $\mathcal{C}_2$ as above. We leave it to future work to develop a complete proof of Eq.~\eqref{ZbStack}.

        \subsection{Modification to include explicit global $G_b$ symmetry}
        \label{GbSymVertexSec}
        
        The above construction is a perfectly legitimate state sum that allows us to define $Z$ on arbitrary $G_b$ bundles and with $\mathcal{G}_f$ structures and should produce a Walker-Wang-like Hamiltonian. However, in contrast to other related state sums defined previously \cite{bulmash2020,barkeshli2019tr,chen2013}, the path integral does not have an explicit global symmetry in the sense of being a sum over distinct terms that are equal due to a global symmetry. As a related point, we would not expect that the corresponding Walker-Wang-like Hamiltonian would explicitly possess a locally generated global $G_f$ symmetry. To remedy this, we need some additional decoration that transforms non-trivially under an action of $G_b$. 
        
        Following previous work \cite{bulmash2020,barkeshli2019tr,chen2013}, we assign to each 0-simplex $i$ an element ${\bf g}_i \in G_b$. If the particular representative for the background gauge field $A_b$ associates ${\bf h}_{ij}$ to each link, then wherever the link $ij$ appears in the state sum, we should insert ${\bf g}_i{\bf h}_{ij}\overline{\bf g}_j$ instead. The global $G_b$ action would then correspond to taking ${\bf g}_i \rightarrow {\bf g}_i \overline{\bf g}$ for any ${\bf g} \in G_b$. 
        
        The path integral is then defined by summing over all ${\bf g}_i$. This has the effect of summing over all gauge-equivalent configurations of $A_b$, which thus gives:
        \begin{equation}
            Z(M,A_b,\xi_{\mathcal{G}}) = \frac{1}{|G_b|^{N_0}} \frac{1}{\sqrt{|H^2(M,\Z_2)|}} \sum_{{\bf g}_i \in G_b, f_3 \in H^3(M,\mathbb{Z}_2)} Z_b(M,A_b,f_3,{\bf g}_i)z_c(M,A_b,f_3,\xi_{\mathcal{G}},{\bf g}_i) .
        \end{equation}
        
        We need to make consistent definitions for these ${\bf g}_i$-dependent functions $Z_b$ and $z_c$ that ensure that the full fermionic state sum is non-anomalous. In particular, the fact that the cochain-level representative of the gauge field effectively changes with each collection ${\bf g}_i$ means that all quantities involving $A_b$, including $Z_b, z_c, \xi_{\mathcal{G}}$ also depend on the ${\bf g}_i$. So all such choices should be consistent.
        
        Defining $Z_b(M,A_b,f_3,{\bf g}_i)$ is straightforward; define it to be exactly $Z_b(M,A_b',f_3)$ where $A_b'$ differs from $A_b$ by a gauge transformation given by the ${\bf g}_i$ and $Z_b(M,A_b',f)$ is defined in the absence of any group elements on the 0-simplices. Of course, $Z_b(M,A_b',f_3)$ is not quite equal to $Z_b(M,A_b,f_3)$ due to the $A_b^{\ast}\omega_2$ piece of the anomaly which gives a phase factor difference upon a bordism taking $A_b$ to $A'_b$ with all else fixed. 
        
        However, defining $z_c$ is significantly more subtle. Naively we could simply perform the above gauge transformation removing the ${\bf g}_i$ as we did for $Z_b$, but the problem is that $\xi_{\mathcal{G}}$ implicitly depends on the representative $A_b$. Hence gauge-transforming $A_b$ means we must keep careful track of $\xi_{\mathcal{G}}$ in order to ensure that every term in the sum involves the same $\mathcal{G}_f$-structure.
        
        More specifically, we define
        \begin{equation}
            z_c(M,A_b,f_3,\xi_{\mathcal{G}},{\bf g}_i) = z_c(M,A_b',f_3,\xi'_{\mathcal{G}})
        \end{equation}
        where the right-hand side uses the definition without any decoration on the 0-simplices. We must carefully define $\xi'_{\mathcal{G}}$. By the definition of $\mathcal{G}_f$-structures, we know that
        \begin{equation}
            (A_b')^{\ast}\omega_2 - (A_b)^{\ast}\omega_2 = \delta y
        \end{equation}
        for some cochain $y \in C^1(M,\Z_2)$, because the gauge fields $A_b$ and $A_b'$ are cohomologous. So we need
        \begin{equation}
            \delta\xi'_{\mathcal{G}} - \delta\xi_{\mathcal{G}} = (A'_b)^{\ast}\omega_2 - (A_b)^{\ast}\omega_2 = \delta y .
            \label{eqn:Gfstructurechange}
        \end{equation}
        To satisfy the above equation, we would like to set $\xi'_{\mathcal{G}}$ according to
        \begin{equation}
            \xi'_{\mathcal{G}}-\xi_{\mathcal{G}} = y.
        \end{equation}
        However, a priori this choice of $y$ is ambiguous. In Eq.~\eqref{eqn:Gfstructurechange}, we could have shifted $y$ by any cocycle without modifying $\delta y$, even though shifting by a non-trivial cocycle would change the $\mathcal{G}_f$-structure. To ensure that $\xi'_{\mathcal{G}}$ and $\xi_{\mathcal{G}}$ represent the same $\mathcal{G}_f$ structure, we must carefully choose $y$. We do not know an explicit formula for $y$, but it can be generated algorithmically as follows. For an elementary gauge transformation (i.e. acting on only one 0-simplex), we simply require $y$ to have support only near the vertex generating the gauge transformation. Note that the proof of connecting gauge fields via Pachner moves given in Appendix \ref{pachnerConnectBranchStruct} is local which means that $y$ can indeed by chosen to induce local changes in the $\mathcal{G}_f$ structure for local gauge transformations. This ensures that the $\mathcal{G}_f$ structures represented by the pairs $\{\xi_{\mathcal{G}}, h_{ij}\}$ and $\{\xi'_{\mathcal{G}}, {\bf g}_i{\bf h}_{ij}\overline{\bf g}_j\}$ are indeed the same. For a more general gauge transformation, we write the gauge transformation as a sequence of elementary gauge transformations ${\bf g}_i$ and sum the $y_i$ we obtain from each elementary transformation.
        
        \section{ Evaluating Examples}
        \label{sec:examples}

        \subsection{ The simplest case: $\{1,\psi\}$}
        As described in the previous Sec \ref{sec:fermStateSum}, we can explicitly carry out the condensation procedure for the $\{1,\psi\}$ theory. In this case, the state sum for a spin manifold $M^4$ gives
        \begin{align*}
            Z(M^4, A_b = 0, \xi_{\mathcal{G}})_{\{1,\psi\}} = &\frac{1}{\mathcal{N}} \sum_{[f_3]} Z_b( M^4, A_b = 0, f_3) z_c(M^d, \xi_{\mathcal{G}}, f_3 )
            \nonumber \\
            &= \frac{1}{\mathcal{N}} \frac{|H^2(M^4, \Z_2)| |H^0(M^4,\Z_2)|}{|H^1(M^4, \Z_2)|}\sqrt{2}^{-\chi(M)} \nonumber \\
            &= 1 .
        \end{align*}

        \subsection{ $\Z_{16}$ absolute anomaly for $G_f = \Z_4^{{\bf T},f}$} \label{z16MainText}
        
        With $G_f = \Z_4^{{\bf T},f}$, that is, $G_b=\Z_2^{\bf T}$ and ${\bf T}^2 = (-1)^F$, it is known that the anomaly indicator \cite{wang2017,tachikawa2016b} for a (3+1)D fermionic SPT is given by the path integral on $\mathbb{RP}^4$ \cite{barkeshli2019tr,Kobayashi2019pin}. Remarkably, we can explicitly evaluate our path integral on a (particularly simple) cellulation of $\mathbb{RP}^4$. This calculation is long and technical, so we relegate the details to Appendix~\ref{app:Z16} and sketch the calculation and results here.
        
        First, we give an explicit cellulation $T_\star$ of $\mathbb{RP}^4$. It comes from the manifold-generating procedure described in~\cite{bulmash2020}. Then, we explicitly perform the diagrammatic calculations to evaluate the state sum associated to $T_\star$ and give an analytic expression for it for any general super-modular category and symmetry fractionalization class of time-reversal symmetry. This has the interpretation of being an `anomaly indicator' for the symmetry fractionalization class. In particular, we evaluate the shadow $Z_b( (\mathbb{RP}^4, T_\star), f )$ for 
        $f = 0$ and for a certain representative $f$ of the non-trivial $[f] \neq 0 \in H^3(\mathbb{RP}^4,\Z_2) = \Z_2$. Explicitly, we find that, as suggested by the gluing argument in \cite{Kobayashi2019pin},
\begin{equation} \label{totalRP4partitionfunction}
\begin{split}
Z_b( (\mathbb{RP}^4, T_\star), f = 0) &= \frac{1}{\mathcal{D}} \sum_{x | x = {\,^{\bf T}}x} d_x \theta_x \eta^{\bf T}_x \\
Z_b( (\mathbb{RP}^4, T_\star), [f] \neq 0) &= \pm \frac{1}{\mathcal{D}} \sum_{x | x = {\,^{\bf T}}x \times \psi} d_x \theta_x \eta^{\bf T}_x
\end{split}
\end{equation}
where the $\pm$ depends on the specific representative of $[f] \neq 0 \in H^3(M,\Z_2)$. Here
        \begin{equation} \label{gaugeInvariant_T_Eta}
            \eta_a^{\bf T} := \begin{cases}
            \eta_a({\bf T},{\bf T}) & \,^{\bf T}a=a\\
            \eta_a({\bf T},{\bf T})U_{\bf T}(a,\psi;a\psi)F^{a,\psi, \psi} & \,^{\bf T}a = a \times \psi
            \end{cases} 
        \end{equation}
       is a gauge-invariant quantity that can be interpreted as the action of ${\bf T}^2$ on $a$ \cite{barkeshli2019tr,bulmashSymmFrac,aasen21ferm}. Specifically, the formula for the case $\,^{\bf T}a = a \times \psi$ defines the ``fermionic Kramers degeneracy'' \cite{bulmashSymmFrac,fidkowski2013,metlitski2014,aasen21ferm}.
       (The $F$-symbol $F^{a,\psi,\psi}$ can be canonically set to 1 and will henceforth be ignored.)

In total we find that
\begin{align}
Z(\mathbb{RP}^4, \xi_{\text{pin}^+}) &= \frac{1}{\sqrt{2}}\left(Z_b(\mathbb{RP}^4,f=0) + \sigma(f) (-1)^{\int {\xi_{\text{pin}^+}(f)}} Z_b(\mathbb{RP}^4,[f] \neq 0)\right) \\
&= \frac{1}{\sqrt{2}\mathcal{D}} \left(\sum_{x | x = {\,^{\bf T}}x} d_x \theta_x \eta^{\bf T}_x \pm i \sum_{x | x = {\,^{\bf T}}x \times \psi} d_x \theta_x \eta^{\bf T}_x\right)
\label{eqn:Z16TotalZ}
\end{align}
Here, the choice of sign in $\pm i$ depends on the choice of pin$^+$ structure $\xi_{\Pin^+(f)}$, of which there are precisely two choices. This is exactly the indicator formula that gives \cite{wang2017,tachikawa2016a,tachikawa2016b,Kobayashi2019pin}
\begin{align}
Z(\mathbb{RP}^4) = e^{2 \pi i \nu / 16} 
\end{align}
for a fixed choice of $\xi_{\text{pin}^+}$. 

At first glance, one may think that $Z_b(\mathbb{RP}^4, f)$ being nonzero for $[f] \neq 0$ is inconsistent with 
the $\Z_2$ gauge theory expectation of Eq.~\eqref{ZbZ2gauge}. However, the non-orientability implies that $A_b \neq 0$, so there is no contradiction with Eq.~\eqref{ZbZ2gauge}. Moreover, the fermion line described by $f$ intersects exactly once the $\mathbb{RP}^3$ submanifold, which represents a non-trivial homology class $0 \neq [\mathbb{RP}^3] \in H_3(\mathbb{RP}^4,\Z_2)$. Naively this would lead to a vanishing path integral because a given closed 3-dimensional slice would have an odd number of fermions, which would be expected to be impossible in $\Z_2$ gauge theory.
In fact this is not a conceptual issue because of a subtlety in how the $\mathbb{RP}^3$ is embedded into $\mathbb{RP}^4$. One can think of $\mathbb{RP}^4$ as the 4-ball $D^4$ with antipodal points on the boundary $S^3$ identified. In the gluing argument~\cite{Kobayashi2019pin} that originally suggested the expression for $Z(\mathbb{RP}^4, [f]\neq 0)$, one can see that the space of states relevant is really the boundary $S^3$ which double-covers the submanifold $\mathbb{RP}^3$. So even though the $f$ line intersects the $\mathbb{RP}^3$ only once, the state on the double-cover $S^3$ will have two fermionic charges, which is consistent with the $\Z_2$ gauge theory. 

Some representative theories that give different values of $\nu$ are well-known. In particular, the semion-fermion theory gives $\nu = \pm 2$ and is well-understood physically \cite{metlitski2014,seiberg2016gapped,wang2017}; in Appendix ~\ref{app:btcData} we provide the BTC and symmetry-fractionalization data for this example using the mathematical framework of this paper. 

In the next subsection, we study in some more detail a particular $\nu = 3$ phase that arises when $\mathcal{C}$ is taken to be $\SO(3)_3$. This example is sensitive to certain exotic smooth structures on generic triangulations and branching structures of $\Pin^+$ 4-manifolds. 

We note that given a construction for the $\nu = 3$ and $\nu = \pm 2$ phases, we can then obtain a construction for the $\nu = 1$ phase by using the stacking construction discussed in Section \ref{stackingSec}. 

        \subsection{Detecting exotic smooth structure and $\C = \SO(3)_3$} \label{sec:so33_MainText}

        Taking as input into the construction the super-modular category $\SO(3)_3$ and with a particular symmetry fractionalization class that we will specify below, we find that the anomaly indicator evaluates to $\nu = \pm 3$ with the sign depending on the pin$^+$ structure:
        \begin{align}
            Z_{\SO(3)_3}(\mathbb{RP}^4, \xi_{\text{pin}^+}) = e^{\pm 3 \times 2\pi i /16} . 
        \end{align}
        Therefore, our construction with this fractionalization data on $\SO(3)_3$ as input yields a state sum TQFT which produces the $\nu = \pm 3$ SPT of the $\Z_{16}$ classification of topological superconductors protected by a ${\bf T}^2 = (-1)^F$ symmetry \cite{kitaev2011,fidkowski2013,metlitski2014,wang2014,Kapustin:2014dxa}. 
        
        As discussed in Section \ref{sec:fermStateSum}, our state sum is expected to provide a bordism invariant. 
        Since $Z_{\SO(3)_3}(\mathbb{RP}^4, \xi_{\text{pin}^+})$ gives $\nu$ odd, it follows in particular that 
        $Z_{\SO(3)_3}( M^4, \xi_{\text{pin}^+})$ should be a $\Z_{16}$ $\Pin^+$ bordism invariant. Since smooth $\Pin^+$ bordism gives a $\Z_{16}$ group while topological $\Pin^+$ bordism gives a $\Z_8$ group, we conclude that $Z_{\SO(3)_3}( M^4, \xi_{\text{pin}^+})$ must be a smooth bordism invariant. 
        
        It is also known that the eta invariant of the pin$^+$ Dirac operator is a smooth bordism invariant, and  corresponds to the $\nu = \pm 1$ class \cite{stolz1988}. We can thus immediately use the known results for the eta invariant in distinguishing exotic smooth structure \cite{stolz1988}. 
        
        For example, it is known that there are two manifolds, $\mathbb{RP}^4$ and $\mathbb{Q}^4$, which are homeomorphic but not diffeomorphic to each other \cite{cappell1976,stolz1988,kirby_taylor_1991}. For this reason, $\mathbb{Q}^4$ is often referred to as ``fake $\mathbb{RP}^4$". Furthermore, $\mathbb{RP}^4$ and $\mathbb{Q}^4$ correspond to the $\pm 1$ and $\pm 9$ elements of $\Z_{16} = \Omega^{\text{pin}^+}_4$. Therefore, we expect that our path integral should give
        \begin{align}
            Z_{\SO(3)_3}(\mathbb{Q}^4, \xi_{\text{pin}^+}) = e^{\pm 3 \times 2\pi i 9/16}
        \end{align}

        A rigorous mathematical proof of the ability of $Z$ to distinguish exotic smooth structure may involve a rigorous proof of bordism invariance, which would directly imply the values on $\mathbb{Q}^4$ because a bordism invariant returning $\pm 1 \in \Z_{16}$ on $\mathbb{RP}^4$ and its pin structures is a generator of the $\Z_{16}$, and such a function would return $\pm 9 \in \Z_{16}$ for $\mathbb{Q}^4$. Alternatively, a detailed calculation on a triangulation of $\mathbb{Q}^4$ would be a weaker but still valid proof of distinguishing power. We do not have either of these on hand, so we leave the formal proof to future work.
        
        We note that the bosonic $\SO(3)_3$ state sum $Z_b(\mathbb{RP}^4,A_b,f=0) = \frac{1}{\sqrt{2}} ( e^{3 \cdot \frac{2 \pi i}{16}} + e^{-3 \cdot \frac{2 \pi i}{16}} )$ from summing over both $\text{pin}^+$ structures (as in Sec.~\ref{gaugeFermionParity}) is also a topological invariant that can distinguish real and fake $\mathbb{RP}^4$s. In particular, since the fake $\mathbb{RP}^4$s with pin structure are 9 times the real $\mathbb{RP}^4$ in the bordism group, we would expect that the fake manifold $\mathbb{Q}^4$ would return:
        \begin{equation*}
            Z_b(\mathbb{Q}^4,A_b,f=0) = \frac{1}{\sqrt{2}} \left( e^{9 \cdot 3 \cdot \frac{2 \pi i}{16}} + e^{-9 \cdot 3 \cdot \frac{2 \pi i}{16}}\right) \neq Z_b(\mathbb{RP}^4,A_b,f=0) .
        \end{equation*}
        Therefore the bosonic state sum should just as well have this ability to distinguish exotic smooth structure.
        
        We note that \cite{stolz1988} discusses additional examples of manifolds with exotic smooth structure that can be detected by the eta invariant, such as $\mathbb{RP}^4 \# S$, where $S$ is the connected sum of $11$ copies of $S^2 \times S^2$. 

        \subsubsection{BTC and symmetry fractionalization data for $\C = \SO(3)_3$}
        
        Here we provide the BTC and symmetry fractionalization data for $\SO(3)_3$, which is the restriction of $\SU(2)_6$ to its integer spin sectors. This BTC contains four simple objects, which we label $1,s,\tilde{s},\psi$, where the transparent fermion is $\psi$. The fusion rules are:
        \begin{align}
            \psi \times \psi &= 1\\
            \psi \times s &= \tilde{s}\\
            \psi \times \tilde{s} &= s\\
            s \times s = \tilde{s}\times \tilde{s} &= 1+s+\tilde{s}\\
            s \times \tilde{s} &= \psi + s + \tilde{s}
        \end{align}
        The quantum dimensions are $d_1=d_{\psi}=1$ and $d_s=d_{\tilde{s}}=1+\sqrt{2}$, with total quantum dimension $\mathcal{D}^2 = 8+4\sqrt{2}$. The topological spins are $\theta_1 = 1, \theta_{\psi} = -1, \theta_{s}=i, \theta_{\tilde{s}}=-i$.
        
        The $F$ and $R$-symbols can be computed using the identification of $\SO(3)_3$ as a restriction of $\SU(2)_6$. Their explicit forms are somewhat involved and unenlightening, so we relegate them to Appendix~\ref{app:SO36Data}. 
        
        Time-reversal exchanges $s \leftrightarrow \tilde{s}$. The non-trivial $U$ symbols are
        \begin{align}
            U(s,\tilde{s};\psi)=U(\tilde{s},\psi;s) = U(\psi,s;\tilde{s})=U(s,s;s)=U(\tilde{s},\tilde{s};\tilde{s})&=i\\
            U(s,\psi;\tilde{s})=U(\psi,\tilde{s};s)=U(\tilde{s},s;\psi)&=-i
        \end{align}
        and $U(a,b;c)=-i$ when two of $(a,b,c)$ are $s$ and the third is $\tilde{s}$ or vice-versa. Finally, the $\eta$ symbols are all trivial except
        \begin{equation}
            \eta_{\psi}({\bf T},{\bf T})=-1
        \end{equation}
        One can check exhaustively by computer that these data satisfy all of the consistency conditions.
        
        At this point, it is straightforward to plug this BTC and fractionalization data into Eq.~\eqref{eqn:Z16TotalZ} to find 
        \begin{equation}
            Z_{\SO(3)_3}(\mathbb{RP}^4, \xi_{\text{pin}^+})= \exp\left(\pm 3 \times \frac{2\pi i}{16}\right)
        \end{equation}

\section{ Discussion}\label{sec:Discussion}

In this paper, we have shown that given symmetry fractionalization data of a super-modular category, we can define a topologically invariant path integral on triangulated manifolds with flat $G_b$ bundles and which depends on a generalized spin structure (a $\mathcal{G}_f$ structure). We have explained why this construction should give a fermion SPT, due to the fact that it condenses an emergent fermion of a bosonic shadow that is expected in general to contain a single non-trivial fermionic excitation. As a special case, we have reproduced the $\Z_{16}$ anomaly indicator for topological superconductors in class DIII, protected by a time-reversal symmetry ${\bf T}$ with ${\bf T}^2 = (-1)^F$. We have also argued that the state sum with input $\SO(3)_3$ can distinguish some exotic smooth structures, like the real $\mathbb{RP}^4$ and its homeomorphic but not-diffeomorphic cousin $\mathbb{Q}^4$. We will now conclude with some discussion and open questions about our construction, possible generalizations therein, and possible applications to geometry. 

First, it would be good to have a fully rigorous argument about why our state sum represents an invertible spin TQFT and a $\mathcal{G}_f$-bordism-invariant. While our argument for topological invariance is indeed rigorous, we do not yet have the tools to rigorously prove its bordism invariance. One line of proof may involve theories of vector spaces and skein-modules associated to 3-manifolds \cite{Roberts1995}, as in previous works. 

We also expect that the fermionic nature of our state sum leads to \textit{super-vector-space} structures associated to 3-manifolds which would be closely related to anti-symmetry of fermionic wave-functions (see also~\cite{bhardwajGaiottoKapustin2017}). In particular, a manifold $M$ with boundary where fermion-lines $f$ end at $\partial M$ should be associated with \textit{dangling} Grassmann variables in $z_c(f)$. The anti-symmetry of Grassmann variables should be related to anti-symmetry of fermions. 

Another useful direction would be to rephrase our sum in terms of handlebody decompositions and Kirby calculus on 4-manifolds, like~\cite{walker2006,barkeshli2019tr,barenzStateSumTQFT}. Such a handlebody viewpoint of the state sum could probably involve some scheme to generalize Kirby calculus to encode spin structures and gauge fields (see also~\cite{barenzGCrossedStateSum}), and thus some way to encode the `winding' viewpoint of the induced spin structures encoded in $z_c(f)$. In a similar vein, it would be good to rigorously establish the $\Z_2$ gauge theory aspects of our state sum, which remains to be done rigorously with $A_b \neq 0$ and when $\mathcal{C}$ is a non-Abelian category. 

We are also curious whether there is a way to derive our diagrams from a different, perhaps more geometrical perspective. For example, all pairs of crossings in the diagram for the generalized 15j symbol either have a $\cup$-product structure $(a_{012},a_{234})$, $(a_{012},\{{\bf g}_{23},{\bf g}_{24}\})$,$(\{a_{013},a_{023},a_{123}\},{\bf g}_{34})$,$(f_{0123},{\bf g}_{34})$ or a $\cup_1$-product structure $(a_{034},f_{0123})$,$(a_{014},f_{1234})$. We believe this is not an accident, and that maybe the $\cup_i$ products' geometric interpretation (c.f. Appendix~\ref{prelimSecB}) can be used to motivate a handlebody picture. See also~\cite{douglas2018} for related pictures of 15j symbols associated to higher categories.

Next, while the $\mathbb{RP}^4$ calculation gives a full set of indicators for ${\bf T}^2 = (-1)^F$ phases, it would be useful to have a general algorithm to generate a full set of anomaly indicators for FSPT phases of general symmetry groups. Given the construction of this paper, the remaining issue is an algorithm to determine generating manifolds and $G_b$ bundles for the relevant $\mathcal{G}_f$-structure decorated bordism groups. In~\cite{bulmash2020}, an algorithm was given that can generate anomaly indicators to fully determine a group cocycle in $\H^4(G,\U)$ for any finite group $G$. A similar algorithm for bordism groups with $\mathcal{G}_f$ structure would be useful. 

We note that our state sum is valid for any flat $G_b$ bundle, including the case where $G_b$ is continuous. In the group cohomology case, it is known that flat $G_b$ bundles for discrete subgroups of $G_b$ are sufficient to fully characterize a $\mathcal{H}^4(G_b, \U)$ cocycle \cite{thorngren,contGNote}. It is an interesting question whether flat $G_b$ bundles are also sufficient to fully characterize fermionic SPTs as well, i.e. to fully characterize an element of the appropriate $\mathcal{G}_f$ bordism group. 

Another open question in the case of continuous $G_b$ is if our state sum could be generalized to curved bundles. This would also be useful for obtaining a full set of anomaly indicators for the case of continuous $G_b$. For example, anomaly indicators for SPTs with $\U$ symmetry often are studied in the literature on bundles with non-zero curvature.\footnote{The most famous case of this is the fermionic topological insulator with topological action given by the $\theta$ term $\propto \theta \int F \wedge F$, where $\theta = \pi$ and $F$ is the field strength of a $\U$ bundle.\cite{qi2008b}}

From a physical standpoint, there are many interesting open questions and directions. For example, topological path integrals in terms of exact state sums are often associated to commuting projector Hamiltonians and tensor network descriptions. It would be interesting to work this out explicitly and see what subtleties arise. For example, recently schemes for bosonizing fermionic Hilbert spaces and operator algebras have appeared and have been shown to have explicit spin structure dependence (e.g.~\cite{Chen2019bosonization} and references therein); it would be nice to see how this fermionization interplays with our shadow theory $Z_b$. 

The Grassmann integral $\sigma(f)$ that we discussed in this paper was naturally associated with pin$^-$ structures (see Sec.~\ref{inducedPinStruct}). One can ask if there is an analog of $\sigma(f)$ with a definition more closely related to pin$^+$ structures. For example, in Appendix D of~\cite{tata2020}, pin$^+$ structures were defined combinatorially and there seemed to be an analog of a quadratic function on \textit{directed} loops $Z_1(M^\vee,\Z)$.

As we mentioned above, it is known~\cite{witten2016rmp} that the eta-invariant of pin$^+$ Dirac operators provides the $\nu = \pm 1 \in \Z_{16}$ element in the Pontryagin dual of the 4-dimensional pin$^+$ bordism group. Furthermore, as discussed in Section \ref{z16MainText}, our state sum with the appropriate super-modular category and with symmetry ${\bf T}^2 = (-1)^F$ also gives a $\nu = \pm 1$ element. As such, our state sum carries the same topological information as the pin$^+$ Dirac operators. A simpler example along these lines is the $d=2$ eta-invariant for pin$^-$ Dirac operators given by the Arf-Brown-Kervaire=$ABK$ invariant. This has a state sum expression~\cite{Kobayashi2019pin,Gaiotto:2015zta} similar to ours, $ABK(M,\xi) \propto \sum_{f} Z_b(f) z_c(f)$, except where the bosonic shadow $Z_b$ is a trivial normalization factor.
This raises the question of whether there is a more direct way of understanding the relationship between our state sum and Dirac operators, perhaps by developing a combinatorial theory of them (see~\cite{Kenyon2003AnIT,cimasoniReshetikhin2007} for statistical mechanics interpretations in $d=2$).

We note that the idea of fermion condensation leading to spectral and index-theoretic information about Dirac operators has also appeared in previous works; for example, \cite{friedanWindey1984,alvarezGaume1983} have phrased heat-kernel derivations of index theorems in terms of supersymmetric path integrals that (roughly) sum over fermion worldlines. 
Given our state sum's apparent sensitivity to smooth structure and this possible relation to Dirac operators, it would be natural to wonder whether there is any relation to other non-trivial invariants of 4-manifolds, such as Seiberg-Witten invariants where spin$^c$ Dirac operators play a central role. 
Finally, it would be mathematically useful to see if generalizations of these ideas in higher dimensions can be used to distinguish other types of exotic smooth structure, such as the exotic spheres.

        \begin{acknowledgments}

        We thank Michael Freedman, Parsa Bonderson, and Corey Jones for discussions. In particular, MB is grateful to P. Bonderson for discussions on fermionic topological phases of matter and fermionic symmetry fractionalization and for sharing unpublished results. 

        This work is supported by NSF CAREER (DMR- 1753240) and JQI- PFC-UMD. 
        The work of R.~K. is supported by Japan Society for the Promotion of Science (JSPS) through Grant No.~19J20801. 
  
        \end{acknowledgments}
        
        \appendix 
                
        \section{ Topological preliminaries} \label{topologicalPreliminaries}
        \label{prelimSec}
        
        Here we will provide some mathematical background necessary to understand various technical details and geometric content of this paper, and particularly for derivations in Appendix \ref{sec:windingDefOfSigma}, \ref{appPachnerLemmas},  \ref{ZbAnomaliesSec}, \ref{app:Z16}. We start by reviewing chains/cochains on a triangulation and Poincaré duality. Then we discuss how branching structures connect geometric to algebraic constructions by inducing geometrical structures like $\cup_i$ products, Stiefel-Whitney classes, (s)pin structures. 
        
        Throughout this appendix we will refer to $M^d$ as a $d$-manifold equipped with a triangulation, as opposed to the notation $(M^d,T)$ we used in the main text. 
        
        \subsection{ Triangulations and cellulations of manifolds}
        
\begin{figure}[h!]
            \centering
            \includegraphics[width=0.25\linewidth]{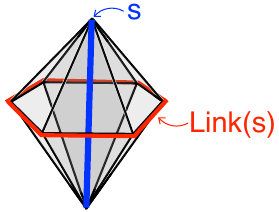}
            \caption{$\mathrm{Link}(s)$ for a 1-simplex $s$ in $d=3$. In general $\mathrm{Link}(s_{d-k})$ of a $(d-k)$-simplex is a $(k-1)$-sphere.}
            \label{linkOfSimplexEx}
        \end{figure}

         Let $M^d$ be a triangulated $d$-manifold. We denote a $k$-simplex of $M$ as $\braket{r_0 \cdots r_k}$, with $r_i$ labeling vertices. We will often abuse language in discussing these $k$-simplices, as it could be possible that multiple $r_i$ are identified on a single $k$-simplex. Most precisely, we will be working in the setting of PL cellulations of $d$-dimensional manifolds, which means that for every $k$-simplex $\braket{r_0 \cdots r_k}$, $\mathrm{Link}(\braket{r_0 \cdots r_k})$ is homeomorphic to a $(d-k-1)$-sphere. Here $\mathrm{Link}(\braket{r_0 \cdots r_k})$ is the collection of $(d-k-1)$-simplices $\braket{y_0 \cdots y_{d-k-1}}$ such that the vertices $\{r_0 \cdots r_k,y_0 \cdots y_{d-k-1}\}$ form a $d$-simplex. This collection of simplices can be visualized as `linking' with the original $k$-simplex (see Fig.~\ref{linkOfSimplexEx} for an example). This is the setting for which Pachner's theorem about equivalent PL cellulations of manifolds applies. A reader interested in PL-manifold theory might find~\cite{rourke1982introduction} helpful, although we will not make much use of it here.
         
         We note that often a triangulation is defined such that all subsimplices consist of distinct vertices and for which any two $k$-simplices can only share at most one $(k-1)$-simplex. This is too restrictive for our purposes and we instead use PL-cellulations so we can deal with more compact cellulations. In any case, passing to a barycentric subdivision (see Sec.~\ref{pachnerConnectBranchStruct} for a definition) of a PL-cellulation will give a triangulation in the standard sense and one can pass to the barycentric subdivision via Pachner moves (again see Sec.~\ref{pachnerConnectBranchStruct}). Since our state sum is invariant under Pachner moves, this abuse of language will not be an issue for us. 
         
         Often we will refer to $k$-simplices as $\braket{0 \cdots k}$ for ease.
        
        \subsubsection{Chains and cochains}
        
        To set notation, here we briefly review the concepts of $\Z_2$-valued chains and cochains.
       
        For $0 \le k \le d$, the set of `$k$-chains' on $M$ is denoted $C_k(M,\Z_2)$. Each $C_k(M,\Z_2)$ is a $\Z_2$ vector space with one basis element per $k$-simplex, spanned by basis vectors $\{\ket{r_0 \cdots r_k}\}$. 
        So chains $c \in C_k(M,\Z_2)$ are in bijection with subsets of the $k$-simplices of the triangulation. We write
        \begin{equation}
        c = \sum_{\braket{0 \cdots k} \in M} c_{\braket{0 \cdots k}} \ket{0 \cdots k}, \;\;\; c_{\braket{0 \cdots k}} \in \{0,1\}.
        \end{equation}
        The sum of two chains $c+d$ returns a sum over $k$-simplices that are part of exactly one (not both) of $c$ or $d$. 
        
        The `boundary' operator $\partial$ from $C_k(M,\Z_2) \to C_{k-1}(M,\Z_2)$ is a linear operator determined by its action on the basis elements $\ket{0 \cdots k}$, which is given as
        \begin{equation}
        \partial \ket{0 \cdots k} = \sum_{i=0}^k \ket{0 \cdots \hat{i} \cdots k}
        \end{equation}
        where $\hat{i}$ means skipping over $i$ in the vertex collection. This corresponds exactly to the $(k-1)$-simplices on the boundary of $\braket{0 \cdots k}$. So $\partial c$ consists of the $(k-1)$-simplices that are contained in an \textit{odd} number of $k$-simplices of $c$, and that set can be identified with the `mod 2' boundary of the $k$-simplices of $c$. 
        We say that $c$ is a `closed' chain, or a `cycle', if $\partial c \equiv 0$ is the zero vector. The set of closed $k$-chains is denoted $Z_k(M,\Z_2)$. We say that $c$ is a `boundary' if $c = \partial b$ for some $b \in C_{k+1}(M,\Z_2)$, and denote $B_k(M,\Z_2)$ as the set of boundaries. Every boundary is closed, so that $\partial \partial c = 0$ for any chain. 
        
        Now, we describe cochains on $M$, denoted by $C^k(M,\Z_2)$. These are all linear functionals from $C_k(M,\Z_2) \to \Z_2$. In particular, we can write a cochain $\alpha \in C^k(M,\Z_2)$ in terms of the dual basis vectors $\bra{0 \cdots k}$, so
        \begin{equation}
        \alpha = \sum_{\braket{0 \cdots k} \in M} \alpha_{\braket{0 \cdots k}} \bra{0 \cdots k}, \;\;\; \alpha_{\braket{0 \cdots k}} \in \{0,1\}
        \end{equation}
        and equivalently $\alpha$ can be thought of as a collection of $k$-simplices.
        
        The `coboundary' operation $\delta$, which is a linear function $\delta : C^k(M,\Z_2) \to C^{k+1}(M,\Z_2)$, is defined as the adjoint of $\partial$. Specifically,
        \begin{equation}
        \delta \alpha =  \sum_{\braket{\tilde{r}_0 \cdots \tilde{r}_{k+1}} \in M} \alpha(\partial \ket{\tilde{r}_0 \cdots \tilde{r}_{k+1}}) \bra{\tilde{r}_0 \cdots \tilde{r}_{k+1}}
        \end{equation}
        which can be interpreted as the set of $(k+1)$-simplices whose boundaries give an odd number of $k$-simplices of $\alpha$. We say $\alpha$ is `closed', or a `cocycle', if $\delta \alpha = 0$ and that $\alpha$ is a `coboundary' if $\alpha = \delta \beta$ for some $\beta \in C^{k-1}(M,\Z_2)$. The set of closed and coboundary $k$-cochains are denoted $Z^k(M,\Z_2)$ and $B^k(M,\Z_2)$ respectively.
        
        The (mod 2) `homology' groups of $M$ are $H_k(M,\Z_2) := Z_k(M,\Z_2) / B_k(M,\Z_2)$. The (mod 2) `cohomology' groups are $H^k(M,\Z_2) := Z^k(M,\Z_2) / B^k(M,\Z_2)$.
        
        \subsubsection{Poincaré Duality and the dual cellulation} \label{app:poincareDuality}
         
\begin{figure}[h!]
            \centering
            \includegraphics[width=0.25\linewidth]{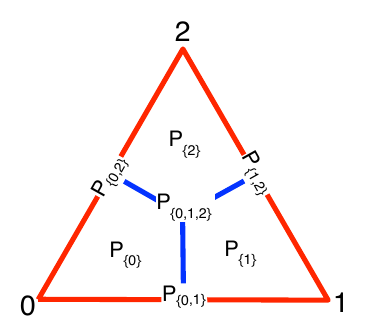}
            \caption{Dual cellulation (blue) of a triangulation (red) in $d=2$}
            \label{dualCellulation}
        \end{figure}

        We denote by $M^\vee$ the dual cellulation of $M$. The dual of a triangulation will in general not be a triangulation, but a cell complex (see, e.g.,~\cite{Hatcher:book}). Every $k$-simplex $\braket{r_0 \cdots r_k}$ is dual to a $(d-k)$-cell $P_{\braket{r_0 \cdots r_k}}$, as shown for the $d=2$ case in Fig.~\ref{dualCellulation}. A precise statement of this construction is not important for our purposes, but can be done by gluing together pieces from the barycentric subdivision (definition in Appendix~\ref{defOfBarycentricSubdivision}) of each $d$-simplex. 
        
        We can also consider chains of the dual cellulation as formal linear combinations of dual $k$-cells, which give the elements of a vector space $C_k(M^\vee,\Z_2)$ . An element $c \in C_k(M^\vee,\Z_2)$ can be written as
        \begin{equation}
            c = \sum_{P_{0 \cdots {d-k}}} c_{P_{0 \cdots {d-k}}} \ket{P_{0 \cdots {d-k}}}
        \end{equation}
        
        There is also a boundary operator $\partial$ acting on $M^\vee$ which gives the boundaries of the dual cells. The boundary of a dual $k$-cell $P_{0 \cdots {d-k}}$ can be expressed in terms of the cochain $\bra{0 \cdots d-k} \in C^{d-k}(M,\Z_2)$ of the original triangulation. In particular we have
        \begin{equation}
        \partial \ket{P_{0 \cdots {d-k}}} = \sum_{\braket{\tilde{r}_0 \cdots \tilde{r}_{d-k+1}}} \big( (\delta \bra{0 \cdots {d-k}})\ket{\tilde{r}_0 \cdots \tilde{r}_{d-k+1}} \big) \ket{P_{\tilde{r}_0 \cdots \tilde{r}_{d-k+1}}} .
        \end{equation}
        That is, the boundary of a dual $k$-cell consists of all the $(k-1)$-cells whose dual $(d-k+1)$-simplices' boundaries contain the $k$-cell's corresponding dual $(d-k)$ simplex. Using the boundary operator, we can define closed dual chains $Z_k(M^\vee,\Z_2)$ and boundary dual chains $B_k(M^\vee,\Z_2)$ as usual. 
        
        $k$-cochains on $M^\vee$, denoted by $C^k(M^\vee,\Z_2)$, are linear functionals $C_k(M^\vee,\Z_2) \to \Z_2$. There's an analogous formula for coboundaries,
        \begin{equation}
        \delta \bra{P_{0 \cdots {d-k}}} = \sum_{\braket{\tilde{r}_0 \cdots \tilde{r}_{d-k+1}}} \big( \bra{0 \cdots {d-k}} \partial \ket{\tilde{r}_0 \cdots \tilde{r}_{d-k+1}} \big) \bra{P_{\tilde{r}_0 \cdots \tilde{r}_{d-k+1}}}
        \end{equation}
        
        Note that this means we can interchangeably refer to chains on $M$ and cochains on $M^\vee$ and vice-versa, since the boundary, coboundary operators on $C_k(M,\Z_2),C^k(M,\Z_2)$ give the same action as the coboundary, boundary operators on $C^{d-k}(M^\vee,\Z_2),C_{d-k}(M^\vee,\Z_2)$. This is the \textit{chain-level} statement of Poincaré duality. And also we will have that $Z_k(M,\Z_2) \cong Z^{d-k}(M^\vee,\Z_2)$ and $Z^k(M,\Z_2) \cong Z_{d-k}(M^\vee,\Z_2)$, and the same statements for the boundaries.
        
\begin{figure}[h!]
            \centering
            \includegraphics[width=0.8\linewidth]{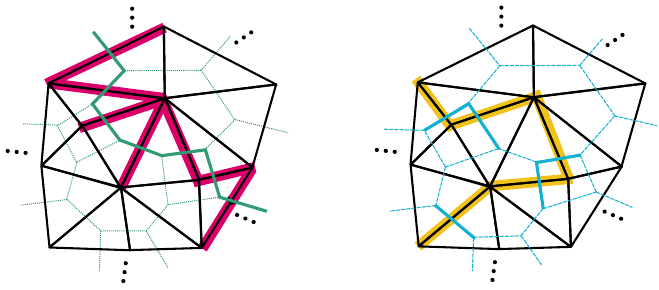}
            \caption{(Left) A closed cochain in $C^1(M,\Z_2)$ (red) dual to a closed cochain in $C_1(M^\vee,\Z_2)$ (green). (Right) A closed chain in $C_1(M,\Z_2)$ (yellow) dual to a closed cochain on $C^1(M^\vee,\Z_2)$ (blue). Black lines are 1-simplices of $M$, and green/blue lines are 1-cells of $M^{\vee}$.}
            \label{chainsCochainsDuality}
        \end{figure}

        See Fig.~\ref{chainsCochainsDuality} for examples of closed chains and cochains on a two-dimensional triangulation and its dual.
        
        \subsection{ Branching structures, $\cup_i$ products, and induced geometrical structures}
        \label{prelimSecB}
        
        A branching structure on a triangulation is a local ordering of vertices, which can be specified by an arrow on each 1-simplex $\braket{ij}$, such there are no closed loops on any 2-simplices. One can show this defines a total ordering of vertices on \textit{every} $k$-simplex $\braket{r_0 \cdots r_k}$. 
        
        Below we will review how the branching structure gives definitions of cochain-level product operations and underlying geometric structures associated with them. 
        In particular, the branching structure canonically defines a frame of vector fields in the triangulation. These vector fields can be used to give a geometric account of the cochain-level formulas of the $\cup$ and $\cup_i$ products. Next, via so-called obstruction theory, we review how the frame of vector fields define the Stiefel-Whitney classes $w_1$ and $w_2$ and how they are related to the theory of (s)pin structures. Therefore the branching structure gives a bridge between various algebraic formulas and the concrete geometry that underlies the path integral. 
        
        For a discussion of spin structures similar to one used here, the reader may find the description in textbook~\cite{scorpan2005wild} useful and related. We also note that~\cite{Thorngrenthesis} discusses many of these ideas and was an instrumental resource in formulating~\cite{tata2020}, which much of this subsection is summarizing.
        
        \subsubsection{$\cup_i$ products}
        
        The $\cup$ product is a function
        \begin{equation*}
            - \cup -: C^k(M,\Z_2) \times C^\ell(M,\Z_2) \to C^{k + \ell}(M,\Z_2)
        \end{equation*}
        whose explicit formula for $\alpha \in C^k(M,\Z_2)$ and $\beta \in C^\ell(M,\Z_2)$ is
        \begin{equation}
            (\alpha \cup \beta)(0 \cdots {k+\ell}) = \alpha(0 \cdots k) \beta(k \cdots k+\ell)
        \end{equation}
        Note that the branching structure is needed in order to give the vertices $r_0 \to \cdots \to r_{k+\ell}$ an ordering to unambiguously define the formula. There is the important Leibniz rule
        \begin{equation}
            \delta(\alpha \cup \beta) = (\delta \alpha) \cup \beta + \alpha \cup (\delta \beta)
        \end{equation}
        that is true on the cochain level. 
        
        The $\cup$ product can be interpreted geometrically in the Poincaré dual picture. It is well known that, for closed $\alpha,\beta$, the $\cup$ product induces a cohomology operation $H^k(M,\Z_2) \times H^\ell(M,\Z_2) \to H^{k+\ell}(M,\Z_2)$ that can be interpreted as the intersection product of the $(d-k)$-manifold $\alpha^\vee$ and the $(d-\ell)$-manifold $\beta^\vee$ dual to $\alpha,\beta$. In particular, the dual of $\alpha \cup \beta$ is homologous to the intersection $\alpha^\vee \cap \beta^\vee$. On the cochain level, this means that for any cohomologous pairs of closed $\alpha, \alpha + \delta A$ and $\beta, \beta + \delta B$, we will have $(\alpha + \delta A) \cup (\beta+ \delta B) = \alpha \cup \beta + \delta(A \cup \delta B + \alpha \cup B + A \cup \beta)$, which means that the cup product induces a product on cohomology classes. Even though the cochain level formulas all depend on a branching structure, on the cohomology level everything is independent of branching structure.
        
\begin{figure}[h!]
            \centering
            \includegraphics[width=0.27\linewidth]{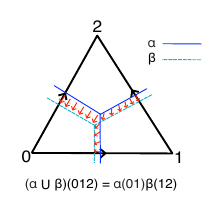}
            \caption{The cup product $(\alpha \cup \beta) = \alpha(01)\beta(12)$ is interpreted as shifting the cells dual to $\beta$ by a vector field (red arrows). The only potential intersection between dual 1-cells is between $\alpha(01)$ and $\beta(12)$.}
            \label{cupProductIntersection2D}
        \end{figure}

        While this intersection property is well-known at the level of homology and cohomology, there is a way to directly visualize it on the cochain level. In particular, given cochains $\alpha \in C^k(M,\Z_2)$ and $\beta \in C^\ell(M,\Z_2)$ and their dual chains $\alpha^\vee \in C_{d-k}(M^\vee,\Z_2)$ and $\beta^\vee \in C_{d-\ell}(M^\vee,\Z_2)$, the dual chain $(\alpha \cup \beta)^\vee$  can be thought of as the intersection of $\alpha^\vee$ with a \textit{shifted} version $\beta^\vee_\text{shifted}$ of $\beta^\vee$, where the shifting vector is determined by the branching structure  \footnote{Technically, $\alpha \cup \beta$ is dual to some limit of $\alpha^\vee \cap \beta^\vee_\text{shifted}$ as the shifting goes to zero. Furthermore $\alpha^k \cup \beta^\ell$ only picks out the pairs of cells who stay at the full-dimension of $(d-k-\ell)$ and throws away `lower-dimensional' cells that survive the limit \cite{tata2020}.}. We illustrate this in Fig.~\ref{cupProductIntersection2D} for the simple case of a 2-manifold with $\alpha,\beta \in C^1(M,\Z_2)$. 
        The vector field used to define the shifting is discussed in the subsequent subsection.

        The $\cup_i$ products for $i > 0$ are product operations
        \begin{equation*}
            - \cup_i -: C^k(M,\Z_2) \times C^\ell(M,\Z_2) \to C^{k + \ell - i}(M,\Z_2)
        \end{equation*}
        whose explicit formula can be written
        \begin{equation}
            (\alpha \cup_i \beta)(0 \cdots k+\ell-i) = \sum_{0 \le j_0 < \cdots < j_i \le k+\ell-i} \alpha(0 \to j_0, j_1 \to j_2, \dots) \beta(j_0 \to j_1, j_2 \to j_3, \dots)
        \end{equation}
        where we use $\{0,1,2,\dots\}$ to refer to the vertices of the $(k+\ell-i)$ simplex, the notation $m \to n$ refers to all vertices $\{m, m+1, \dots, n-1, n\}$, and we restrict our attention to those combinations $\{j_0,\dots,j_i\}$ such that the sets $\{0 \to j_0, j_1 \to j_2,\dots\}$ and $\{j_0 \to j_1, j_2 \to j_3,\dots\}$ consist of exactly $(k+1)$ and $(\ell + 1)$ elements respectively. The $\cup_i$ products also have a generalized Leibniz rule
        \begin{equation}
            \delta(\alpha \cup_i \beta) = (\delta \alpha) \cup_i \beta + \alpha \cup_i (\delta \beta) + \alpha \cup_{i-1} \beta + \beta \cup_{i-1} \alpha
        \end{equation}
        that inductively relate $\cup_i$ products to $\cup_{i-1}$ products. From this formula one can check examples that $\alpha \cup_i \beta$ for $i>0$ are generally not even closed for closed $\alpha,\beta$. Therefore there cannot be a cohomology level interpretation of $\alpha \cup_i \beta$, for general $\alpha,\beta$. Nevertheless, the operations $\alpha \mapsto \alpha \cup_i \alpha$ are indeed cohomology operations that map cohomology classes to cohomology classes, and define the so-called `Steenrod operations' $\Sq$. These are written as
        \begin{equation}
        \begin{split}
            &\Sq^{k-i}: Z^k(M^d,\Z_2) \to Z^{2k-i}(M^d,\Z_2), \text{  with} \\
            &\Sq^{k-i} \alpha := \alpha \cup_i \alpha
        \end{split}
        \end{equation}
        because the generalized Leibniz rule above gives $(\alpha + \delta A) \cup_i (\alpha + \delta A) = \alpha \cup_i \alpha + \delta(\alpha\cup_i A+A\cup_i\alpha+A\cup_{i-1}A+A\cup_i\delta A)$. Additionally, changing the branching structure leaves the cohomology class of the Steenrod operations invariant.
        
        The Steenrod operations similarly have an interpretation~\cite{thom1950} in terms of Poincaré dual submanifolds (see also~\cite{SelfintersectionsofimmersionsandSteenrodoperations}). Concretely, we can \textit{thicken} $\alpha^\vee$ along some $(k-p)$ generic vector fields and shift it along a $(k-p+1)^{th}$ vector field to produce some $\alpha^\vee_\text{thickened,shifted}$. Then Sq$^p(\alpha)$ is Poincaré dual to the intersection $\alpha^\vee \cap \alpha^\vee_\text{thickened,shifted}$. Abstractly, this means that Sq$^p(\alpha)$ can be interpreted in terms of pushforwards of Stiefel-Whitney classes of the normal bundles of the embedded $\alpha^\vee \subset M$; this translates to the concrete statement above via `obstuction-theoretic' definitions of Stiefel-Whitney classes. 
        
        While globally the above statement has been known since the 1950's at the level of cohomology, there is a way to see this locally simplex-by-simplex, directly on the cochain-level. In particular, the cells of $\alpha \cup_i \beta$ can be interpreted in terms of some canonical frame of vector fields defined in the neighborhood of the dual 1-skeleton. In particular, $\alpha \cup_i \beta$ is given by first \textit{thickening} $\beta^\vee$ in $i$ directions and shifting it along an $(i+1)^{th}$ direction to produce $\beta^\vee_\text{thickened,shifted}$. Then, $\alpha \cup_i \beta$ on a $(k+\ell-i)$ simplex is obtained by counting the number of intersections of $\alpha^{\vee} \cap \beta^\vee_\text{thickened,shifted}$. A brief description of the frame of vector fields used to define the shifting is given next.
        
        \subsubsection{A frame of vector fields from a branching structure}
        
        We now briefly review the explicit vector fields used to define the $\cup$ and $\cup_i$ products, and which we later use to define the first and second Stiefel-Whitney classes, spin structures, and induced spin structures on loops. We refer the reader to \cite{tata2020} for a more thorough treatment. 
        
        Specifically, we use the branching structure to define $d$ vector fields, ${\tilde{v}}_1,\cdots, {\tilde{v}}_d$, which are defined in the neighborhood of the dual $1$-skeleton.  
        
        First, we recall that a $d$-simplex $\Delta_d$ can be defined as the subset $\Delta_d \subset \R^{d+1}$ defined in coordinates, 
        \begin{equation*}
            \Delta_d = \{(x_0 \cdots x_{d}) | x_0 + \cdots x_{d} = 1, \text{ and each } x_i \ge 0\}
        \end{equation*}
        Next, we define a collection of vectors $\vec{b}_1,\dots,\vec{b}_d \in \R^{d+1}$, 
        \begin{equation}
            \vec{b}_j = \left(1^j, \frac{1}{2^j}, \frac{1}{3^j},\dots, \frac{1}{(d+1)^j}\right). 
        \end{equation}
        The fact that $\vec{b}_0 = (1 \cdots 1)$ and $\vec{b}_j$ together form a Vandermonde matrix leads to some nice algebra that give the formulas for $\cup_i$. 
        
        At the center $C$ of each $d$ simplex, we define ${\tilde{v}}_j$ such that
         \begin{equation}
            {\tilde{v}}_j (C) = \vec{b}_j - \vec{b}_j \cdot \vec{b}_0  
            \text{  for } j=1,\dots,d . 
        \end{equation}
        Thus ${\tilde{v}}_j$ in the center corresponds to $\vec{b}_j$ with the direction along $\vec{b}_0$ projected out. Next, we define ${\tilde{v}}_j$ along the dual 1-skeleton by gradually projecting away the direction normal to the $(d-1)$ simplex as it is approached along the dual $1$-skeleton. The precise manner in which this projection gradually occurs is arbitrary and unimportant for our purposes. Once ${\tilde{v}}_j$ is defined along the dual $1$-skeleton, it can then be smoothly extended to a neighborhood of the dual $1$-skeleton. The Vandermonde structure of the ${\tilde{v}}_j$ will show that any $(d-1)$ of these ${\tilde{v}}_j$ are independent after projecting away that coordinate normal to the $(d-1)$ simplex being approached.
        
        We note that one can continue and use the vector defined at the center of each $(d-1)$-simplex to extend the vector field to be defined along the dual $1$-skeleton of the $(d-1)$-simplex. If we inductively continue this process until we approach the $1$-simplices, the fact that the coordinates of each $\vec{b}_i$ are decreasing leads to the fact that ${\tilde{v}}_i$ degenerate \textit{opposite} to the branching structure. This inductive procedure is expected to allow us to define a frame everywhere in the triangulation, although the precise technical construction would involve detailed constructions in PL-manifold theory beyond the scope of this paper. 
        
        The vector fields ${\tilde{v}}_j$ along the dual 1-skeleton can be used to give a geometric interpretation of higher cup products $\alpha \cup_i \beta$ when $\beta$ is a $(d-1)$-cochain. Namely, we define $\beta^\vee_\text{thickened,shifted}$ 
        for a collection of $1$-cells $\beta^\vee$ by thickening $\beta^\vee$ along $i$ directions ${\tilde{v}}_1,\cdots, {\tilde{v}}_i$ and shifting it along ${\tilde{v}}_{i+1}$. 
        
        When $\beta$ is an $\ell$-cochain, we can define the thickening and shifting in principle by using ${\tilde{v}}_i$ extended to the dual $\ell$-skeleton, although a prescription with $\ell>1$ for such ${\tilde{v}}_i$ has not yet been given rigorously in the scope of PL-manifold theory. (See~\cite{HalperinToledo,Goldstein} for earlier related but more rigorous constructions along these lines.) Ref.~\cite{tata2020} gave a prescription to define $\beta^\vee_\text{thickened,shifted}$ by utilizing instead the constant vector field defined by $\vec{b}_j$ for the thickening and shifting directions, which gives a geometric interpretation of higher cup products evaluated on an individual $d$-simplex. 
        
        \subsubsection{$\pm$ assignments of simplices, $w_1$, and vector fields on the dual 1-skeleton} \label{w1PmAssignments_AndVectorFields}
 
        Given a triangulation, we can assign an arbitrary orientation $\pm$ to each $d$-simplex. As we will explain, this choice, together with the branching structure, determines a closed chain $w_1 \in Z_{d-1}(M,\Z_2)$ consisting of the $(d-1)$-simplices for which the labeling of the neighboring $d$-simplices is inconsistent. This chain $w_1$ defines a 1-cochain on the dual cellulation, which with some abuse of notation we also refer to as $w_1$, and which gives a representative of the first Stiefel-Whitney class.

\begin{figure}[h!]
            \centering
            \begin{minipage}{0.6\textwidth}
                \centering
                \includegraphics[width=\linewidth]{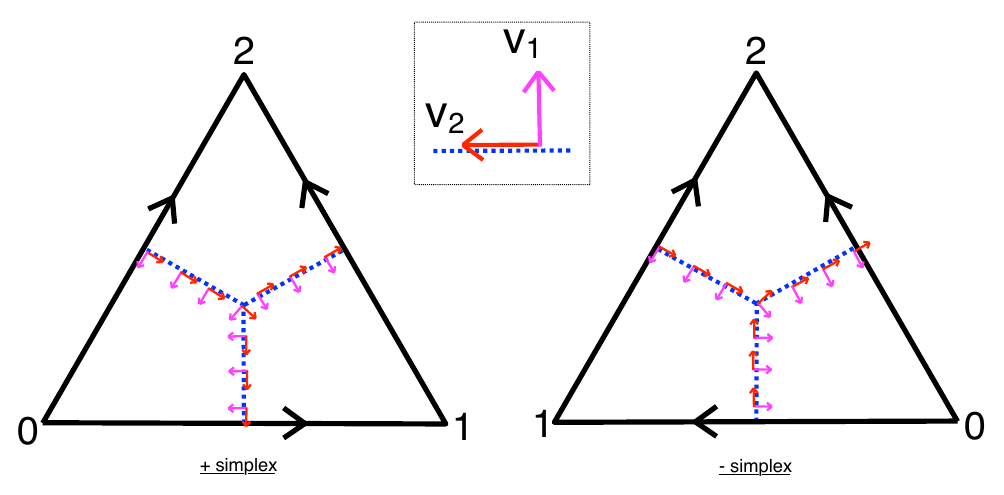}
                \end{minipage} \quad \quad \quad
            \begin{minipage}{0.6\textwidth}
                 \centering
                 \includegraphics[width=\linewidth]{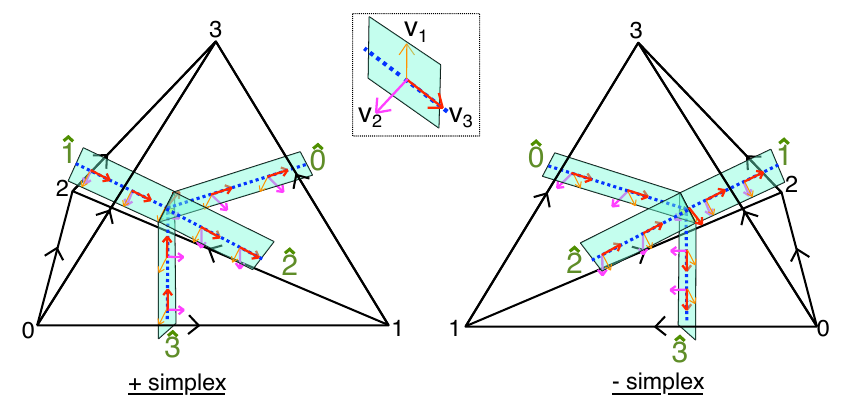}
            \end{minipage}
            \caption{ Illustration of vector field frames with respect to $\pm$ assignments of simplices. Note that for each $d=2,3$ $v_d$ will typically be parallel to the dual 1-skeleton except near the barycenter. (Top) Vector fields in $d=2$. (Bottom) $v_1$ is the `thickening' direction that determines the framing of curves on the dual skeleton. For each $i=0,\dots,3$, $\hat{i}$ refers to the 1-cell opposite to vertex $i$. For ease of drawing, the $v_2$ drawn here is \textit{not} the one described in the main text, but is instead related to $v_1,v_3$ by the right-hand rule. The $v_2$ described in the main text will be \textit{almost} parallel to $v_3$, but still lie on the same side of the thickened sheet as the one drawn.}
            \label{fig:vectorFields}
        \end{figure}

        Let us define a collection of vector fields ${v}_i$, for $i = 1,\cdots, d$, such that
        \begin{align}
            {v}_i = {\tilde{v}}_i, \;\; i = 1,\cdots, d-1 ,
        \end{align}
        where ${\tilde{v}}_i$ were defined in the previous subsection. We remarked there that these ${v}_i = {\tilde{v}}_i$ with $i \le d-1$ are all linearly independent near the dual 1-skeleton even after projecting away the components normal to the face.
      The $d$th vector field ${v}_d$ is chosen as follows. 
      Away from the center of the $d$-simplices, ${v}_d$ is chosen to be parallel to the dual 1-skeleton, with the direction chosen so that the determinant of the frame in the local coordinates is positive on a $+$ simplex and negative on a $-$ simplex. Near the centers of the $d$-simplices, ${v}_d$ is chosen to smoothly interpolate from its values away from the center. See Fig.~\ref{fig:vectorFields} for a depiction of these vector fields in $d = 2$ and $3$. In particular, note that on a $\pm$ simplex along the dual 1-skeleton edge $P_{\braket{0 \cdots \hat{i} \cdots d}}$, ${v}_d$ points away from the center of the $d$-simplex if $\pm (-1)^i = +1$ and points towards the center if $\pm (-1)^i = -1$. 
        
        \begin{figure}[h!]
            \centering
            \includegraphics[width=0.6\linewidth]{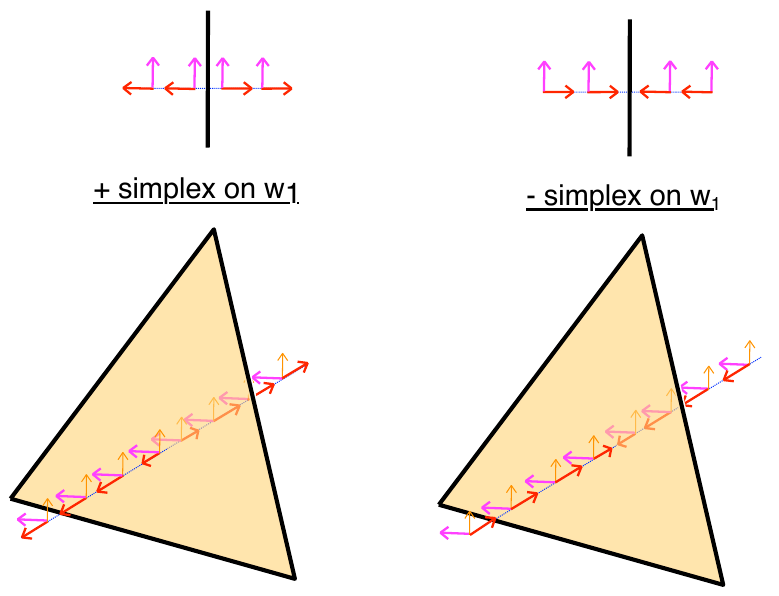}
            \caption{The dual of $w_1$ is the collection of all $(d-1)$-simplices for which the assignments of vectors $v_d$ along the dual 1-skeleton are inconsistent crossing it. In particular, this is equivalent to the places where the frame of vectors $\{v_1 \dots v_d\}$ become degenerate. (Top) In $d = 2$. (Bottom) In $d = 3$.}
            \label{fig:pmSimplicesCrossingW1}
        \end{figure}

        The direction of ${v}_d$ along $P_{\braket{0 \cdots \hat{i} \cdots d}}$ determines the induced orientation of the boundary $(d-1)$-simplex from the perspective of the $d$-simplex: the induced orientation is, say, positive (negative) if ${v}_d$ points towards (away from) the center of the $d$-simplex, i.e. away from (towards) the $(d-1)$-simplex ${\braket{0 \cdots \hat{i} \cdots d}}$. 
        
        If the manifold is orientable, then there is always a choice of $\pm$ assignments for the $d$-simplices such that the induced orientations on each $(d-1)$-simplex would be opposites as induced from both its $d$-simplex neighbors. However if the induced orientations are the same due to an inconsistent $\pm$ assignment, which necessarily occurs for non-orientable manifolds, this implies that the frame of vector fields must become degenerate crossing the $(d-1)$-simplex. Since the first Stiefel-Whitney class $w_1$ is the obstruction to defining a frame of independent vectors, we can identify such $(d-1)$-simplices with the dual of $w_1$. This leads us to define the chain $w_1 \in Z_{d-1}(M,\Z_2)$ from the branching structure and $\pm$ assignments as all such $(d-1)$-simplices with the same, `inconsistent', induced orientations. In addition, we can label the $(d-1)$-simplices of $w_1$ as $\pm$ depending on their shared induced orientations, positive or negative, from the neighboring $d$-simplices. In fact, this assignment can show that the dual of $w_1$ is orientable with a consistent labeling of $\pm$ given by the shared induced orientation. See Fig.~\ref{fig:pmSimplicesCrossingW1} for an illustration.
        
        For an orientable manifold, if we require that the chain representative $w_1 = 0$, then we recover the more familiar situation where the branching structure directly determines the $\pm$ assignments of the $d$-simplices, up to an overall global choice.
        
        We can package the $\pm$ assignment all into a single equation as follows. Consider the \textit{canonical} dual of $w_1$, which we name $w_1^{\text{canonical}}$, which is formed by the procedure above if \textit{all} simplices are chosen as $+$. Then $w_1^{\text{canonical}}$ on a $(d-1)$-simplex $\braket{r_0 \cdots r_{d-1}}$ can be found in terms of the two $d$-simplices $\{r_0 \cdots r_{d-1} A\},\{r_0 \cdots r_{d-1} B\}$ that it is a part of, where $A$ and $B$ denote two additional vertices. In particular, let $\ell(A), \ell(B)$ be the coordinate in $\{0,\cdots,d\}$ in which $A,B$ respectively appear in their $d$-simplices. Then since the induced orientation on $\hat{i}$ on a $+$ simplex is $-(-1)^i$, we have
        \begin{equation}
            w_1^{\text{canonical}}(\braket{r_0 \cdots r_{d-1}}) = -(-1)^{\ell(A) + \ell(B)}
        \end{equation}
       Flipping the assignment to $-$ on some region changes matching$\leftrightarrow$mismatching on the boundary of said region. Thus we can obtain a $\pm$ assignment consistent with a global orientation by choosing the $-$ region to satisfy
        \begin{equation}
            \partial(- \text{ region}) = w_1^{\text{canonical}}
        \end{equation}
        which only has a solution if and only if $w_1$ is a coboundary, i.e. if $w_1$ is trivial. If a solution exists, then there are exactly $2^{\text{\# of connected components of }M}=|Z_0(M,\Z_2)|$ solutions.
        
        \subsubsection{Obstruction Theory, $w_2$, and spin structures on a triangulated manifold} \label{app:SpinStructReview}
        
        Now we are in a position to define the second Stiefel-Whitney class $w_2$ and spin structures. To do this, we need the definition of $w_2(M)$ in obstruction theory. Given $M^d$, we pick any collection of $(d-1)$ `generic' vector fields. Generic means that these fields will be linearly independent everywhere except for a closed codimension-2 submanifold for which the zeros in the determinants defining the linear dependency are linear zeroes in local coordinates. This codimension-2 submanifold is Poincaré dual to $w_2(M)$. Another way to define $w_2(M)$ is to first trivialize the tangent bundle $TM$ on the 1-skeleton (which is possible if and only if $M$ is orientable). On every 2-cell, $c$, this trivialization can be extended with some number $n(c)$ generic singular points where the the vector fields are not independent. Then, $w_2(M)$ is defined as the cocycle whose values on 2-cells are $w_2(c)=(-1)^{n(c)}$, so is the (mod 2) obstruction to extending the trivialization to the entire 2-skeleton.
        
        A spin structure $\xi$ can be thought of as a `fix' of the vector fields so that all singularities become eliminated (mod 2). This can be implemented by choosing some codimension-1 submanifold whose boundary is $w_2$ and adding a $360^\circ$ `twist' to the vector fields along this submanifold. Equivalently, it is some collection of edges such that twisting the vector fields on those edges allows the trivialization to extend (mod 2) along the 2-skeleton. These are both equivalent to specifying an $\xi \in C^1(M,\Z_2)$ with $\delta \xi = w_2$. 
        
\begin{figure}[h!]
    \centering
    \includegraphics[width=\linewidth]{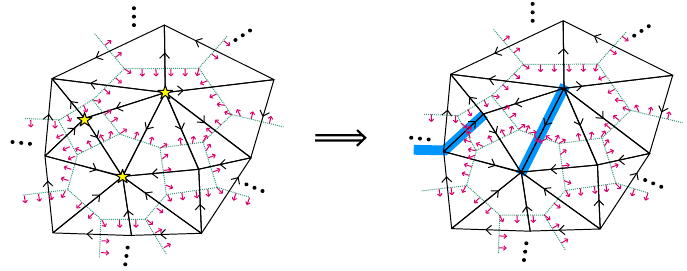}
    \caption{(Left) Vector fields on the dual 1-skeleton; one is drawn and the other is given by the right-hand-rule. Continuing the vector fields into the 2-skeleton leads to $(d-2)$-simplices (in this case vertices, starred) with odd-index singularities that are dual to $w_2$. (Right) A spin structure $\xi$ is dual to a collection of edges (thick blue ones here) whose boundary is dual to $w_2$. The vector fields get `twisted' by $360^\circ$ at every crossing with $\xi$.}
    \label{fig:spinStruct2D}
\end{figure}

\begin{figure}[h!]
    \centering
    \begin{minipage}{0.45\textwidth}
        \centering
        \includegraphics[width=\linewidth]{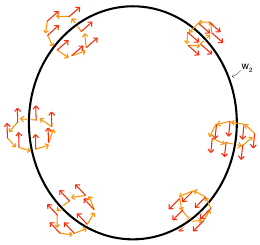}
    \end{minipage} \quad \quad
    \begin{minipage}{0.45\textwidth}
         \centering
         \includegraphics[width=\linewidth]{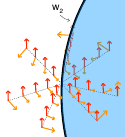}
    \end{minipage}
    \caption{In $d=3$, $w_2$ is dual to a one-dimensional submanifold where a frame of vector fields has an odd-index singularity. A spin structure is dual to a two-dimensional submanifold (blue region) whose boundary is $w_2$ and where the vector fields gain a twist by $360^\circ$ to turn the singularities into even index ones.} 
    \label{fig:spinStructs3D}
\end{figure}

        In our case of a triangulated manifold, the vector fields ${v}_1,\cdots, {v}_d$ defined in the previous sections give a trivialization of $TM$ along the \textit{dual} 1-skeleton. We can therefore use these vector fields to define a representative of $w_2$ and choice of $\xi$. See Figures \ref{fig:spinStruct2D} and \ref{fig:spinStructs3D} for an illustration of the vector fields associated to a triangulation and how a spin structure acts to `correct' odd-index singularities in the cases $d = 2$ and $d = 3$.
      
        Given a branched triangulation, there are known explicit chain-level formulas for duals of general Stiefel-Whitney classes \cite{Goldstein}. We will see in Appendix \ref{sec:windingDefOfSigma} that the Grassmann integral can be used to give an alternate formula for $w_2$ in terms of the winding of the vector fields ${v}_i$ around a $(d-2)$-simplex. 
        
        Finally, we state a nice canonical formula for $w_2$ that we use in Section \ref{app:w2AndPinPlusRep} and that was recently discovered in~\cite{Chen2019bosonization}, which expresses the dual of $w_2$ on a branched triangulation in terms of higher cup products.
        In particular, given a $(d-2)$-simplex $\braket{r_0 \cdots r_k}$, $\mathrm{Link}(r_0 \cdots r_k)$ comprises of a collection of $k$ $(d-1)$-simplices that form a circle  $\Delta_{d-1}^1 \to \Delta_{d-1}^2 \to \cdots \to \Delta_{d-1}^{k} \to \Delta_{d-1}^{k+1} \equiv \Delta_{d-1}^1$. The formula for $w_2$ is 
        \begin{equation} \label{YuAnFormula}
        (-1)^{w_2(\Delta_{d-2})} = - \prod_{i=1}^k (-1)^{\int_M \boldsymbol{\Delta}_{d-1}^i \cup_{d-2} \boldsymbol{\Delta}_{d-1}^{i+1}}
        \end{equation}
        where each $\boldsymbol{\Delta}_{d-1}^i$ is the cochain that's an indicator on $\Delta_{d-1}^i$. This formula turns out to be equivalent to the winding definition reviewed in Appendix \ref{sec:windingDefOfSigma}, which is shown in Appendix \ref{windingFormulaSameAsYuAns}.

        \subsubsection{Induced spin structures on framed curves} \label{sec:inducedSpinStructOnCurve}
        
        Let $M$ be an orientable manifold. Another way to think about spin structures in arbitrary dimensions is how they define induced spin structures on \textit{ framed } loops.
        
        Given a loop embedded in a manifold $M$, we define a `framing' on it to be a collection of $(d-2)$ vectors of $TM$ independent of the tangent of the loop. Since $M$ is orientable this is equivalent to specifying a trivialization of the normal bundle of the loop in $M$, since the global orientation will specify the $d^{th}$ vector. We will call the trivialization of $\restr{TM}{\text{loop}}$ the `tangent framing' of the loop because one of the vectors is always tangent to the loop's direction.
        
        Given a framing of the loop, the induced spin structure on the loop is defined as follows. As we reviewed in the previous section, a spin structure on $M$ is equivalent to a frame of vector fields that become degenerate only at some even-index singularities localized on a codimension-2 submanifold of $M$. Refer to this framing as the `background framing' of $M$; in the previous section, this framing was given by the $d$ vector fields ${v}_1,\cdots, {v}_d$. WLOG the curve lies on the 1-skeleton (which for us is the \textit{dual} 1-skeleton) of $M$ where $TM$ is trivial, so we can refer to these framings as literally a $d \times d$ matrix with positive determinant for every point on the loop. Also WLOG, these framings will be homotopic to maps $[0,1] \to \SO(d)$ since the manifold of positive-determinant matrices $GL^+(d)$ deformation retracts onto $\SO(d)$ by the orthogonalization procedure.
        
        Call the background framing $F_\text{bckd}: [0,1] \to \SO(d)$ and the tangent framing $F_\text{tang}: [0,1] \to \SO(d)$. Note that $F_\text{bckd}(0) = F_\text{bckd}(1)$ and $F_\text{tang}(0) = F_\text{tang}(1)$. The relative path of these framings is the function $(F_\text{bckd}^{-1} F_\text{tang})(t) : [0,1] \to \SO(d)$. Note that if we lift this map to $\tilde{F}: [0,1] \to \Spin(d)$, we will have that $\tilde{F}(0) = \pm \tilde{F}(1)$ since $\Spin(d)$ is a double-cover of $\SO(d)$.
        
        Now we can define the induced spin structure in terms of this $\pm$ sign of $\tilde{F}(0) = \pm \tilde{F}(1)$. If $\tilde{F}(0) = -\tilde{F}(1)$, then the induced spin structure on the loop is the bounding/anti-periodic one. And if $\tilde{F}(0) = +\tilde{F}(1)$, then the induced spin structure is the non-bounding/periodic one.
        
\begin{figure}[h!]
    \centering
    \includegraphics[width=0.6\linewidth]{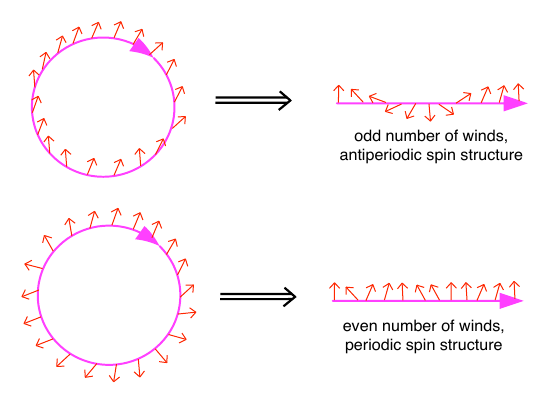}
    \caption{Induced spin structures on a curve in $d=2$ can be visualized by the number of rotations the background framing makes with respect to the curve. (Top) An odd number of winds makes antiperiodic boundary conditions. (Bottom) An even number of winds makes periodic boundary conditions.}
    \label{fig:inducedSpinStruct2D}
\end{figure}

        In $d = 2$, this induced spin structure can be thought of in terms of the amount of times one of the vectors of the background frame `winds' around the tangent of the loop. In particular, if it winds around an even number of times then the induced spin structure will be the periodic one. If it winds around an odd number of times then the induced spin structure will be the anti-periodic one. In particular, loops that are homologically trivial will always have an induced anti-periodic boundary condition. See Fig.~\ref{fig:inducedSpinStruct2D}.
        
\begin{figure}[h!]
    \centering
    \includegraphics[width=0.8\linewidth]{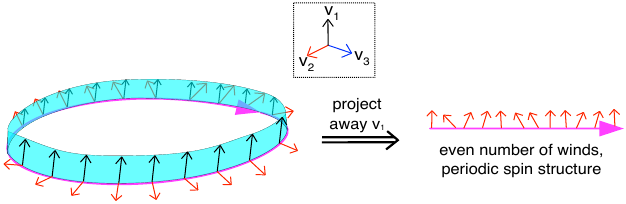}
    \caption{A framed curve with `shared framing' (blue sheet spanned by $v_1$). The background frame consists of the vectors $v_1, v_2, v_3$ as labeled in the box; $v_3$ is not shown but is chosen to make the orientation positive. The tangent frame consists of the shared $v_1$, the tangent of the pink curve, and some unshown vector such that orientation is positive. The induced spin structure can be computed by projecting away the shared framing and measuring the winding of one of the background frame's projected vectors ($v_2$ here) with respect to the curve's tangent in the projected two-dimensional picture.
    }
    \label{fig:inducedSpinStruct3D}
\end{figure}

        In higher dimensions, a very similar picture of the winding of vector fields can be used. In particular, note that the definition of induced spin structure is invariant under continuous deformations of either of the background or tangent frames. And because $\pi_1(\SO(d)) = \Z_2$, where the two elements can be generated from paths of rotations in a single two-dimensional plane, the relative framing $(F_\text{bckd}^{-1} F_\text{tang})(t)$ can be deformed to a rotation in a two-dimensional plane. This means we can deform the tangent and background frames to have the functional form:
        \begin{equation*}
        (F_\text{bckd}^{-1} F_\text{tang})(t) = 
            \begin{pmatrix}
            1 &        &   & \\
              & \ddots &   & \\
              &        & 1 & \\
              &        &   & F_{\text{rel } 2 \times 2}(t)
            \end{pmatrix}
        \end{equation*}
        where $F_{\text{rel } 2 \times 2}(t)$ is some $2 \times 2$ matrix that tells us how the remaining two vectors in the framing wind with respect to each other. In particular, the first $(d-2)$ vectors of this deformed frame can be thought of as a `shared framing' which is shared by both the tangent and background frames. And, the induced spin structure can be visualized by measuring the number of $2\pi$ rotations the last two vectors of the background frame make with respect to the loop after \textit{projecting away} the shared framing. See Fig.~\ref{fig:inducedSpinStruct3D} for an example.
        
        This point of view of a shared framing will become useful to us when reviewing the geometric interpretation of the Gu-Wen/Gaiotto-Kapustin Grassmann integral $\sigma(f)$ later. In particular, the $(d-2)$ vectors $v_1,\dots,v_{d-2}$ introduced earlier from the branching structure play the role of this shared framing.

        \subsubsection{$\det(TM)$, $w_1^2$, and pin$^-$ structures} 
        \label{pinMinusStructs}
        
        In the previous sections we discussed spin structures and $w_2$ for orientable manifolds. Here we will generalize the discussion to non-orientable manifolds, $w_1^2$, and spin-like structures on them. The main difficulty in describing a spin structure on a non-orientable manifold is the fact that $TM$ is non-orientable. In other words, it is impossible to trivialize $TM$ on the 1-skeleton, meaning that the constructions above do not directly apply.
        
        However we can consider the orientation bundle $\det(TM)$. The bundle $TM \oplus \det(TM)$ is always orientable since $w_1(TM \oplus \det(TM)) = w_1(TM) + w_1(\det(TM)) = w_1(\det(TM)) + w_1(\det(TM)) = 0$: the first step is the Whitney sum formula and the second step uses $w_1(TM) = w_1(\det(TM))$. So it is always possible to trivialize $TM \oplus \det(TM)$ along a 1-skeleton.
        
        We will see that defining a spin structure on $TM \oplus \det(TM)$ defines a $\text{pin}^-$ structure on $M$. The reason that $TM \oplus \det(TM)$ is related to the $\Pin^-$ groups is that $\Pin^-(d)$ can be thought of as a subgroup of $\Spin(d+1)$. In particular, $\Pin^-(d) \subset \Spin(d+1)$ consists of all elements whose projections onto $\SO(d+1)$ restrict to $O(d)$ on the first $d$ coordinates. This implies that $\text{pin}^-$ structures can be viewed in terms of $TM \oplus \det(TM)$, since this $\det(TM)$ direction plays the role of this $(d+1)^{th}$ direction that parameterizes orientation-reversal. See~\cite{kirby_taylor_1991} for a more detailed explanation.
        
\begin{figure}[h!]
    \centering
    \includegraphics[width=\linewidth]{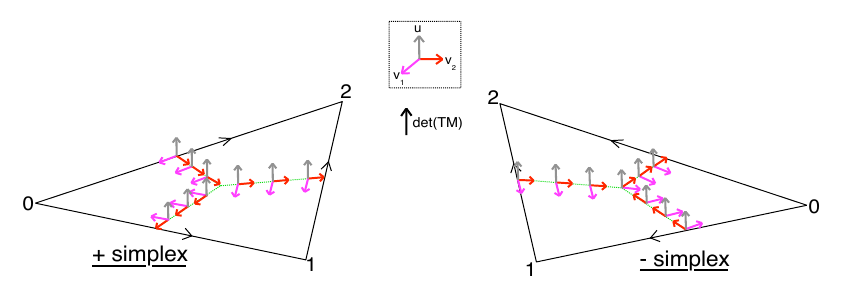}
    \caption{The vector fields ${u},{v}_1,{v}_2$ to trivialize $TM \oplus \det(TM)$ in $d=2$, on the interior of $\pm$ 2-simplices. $\det(TM)$ can be thought of as a direction `transverse' to the manifold and the vector field ${u}$ locally keeps track of the orientations. The relevant vector fields' colors match those in the box and illustrate what directions their sections point along the dual 1-skeleton.}
    \label{fig:vectorFields2D_detTM}
\end{figure}

        Let us start by describing this trivialization for a 2-manifold. We can visualize this $\det(TM)$ piece as an `extra dimension' sticking out transverse to the surface. So overall we will have that $TM \oplus \det(TM)$ is trivialized by three vector fields ${u},{v}_1,{v}_2$. Given an assignment $\pm$ of the 2-simplices, the fields ${v}_1,{v}_2$ are the same inside the 2-simplices as in the orientable case. For example see Fig.~\ref{fig:vectorFields2D_detTM}.
        
\begin{figure}[h!]
    \centering
    \includegraphics[width=0.8\linewidth]{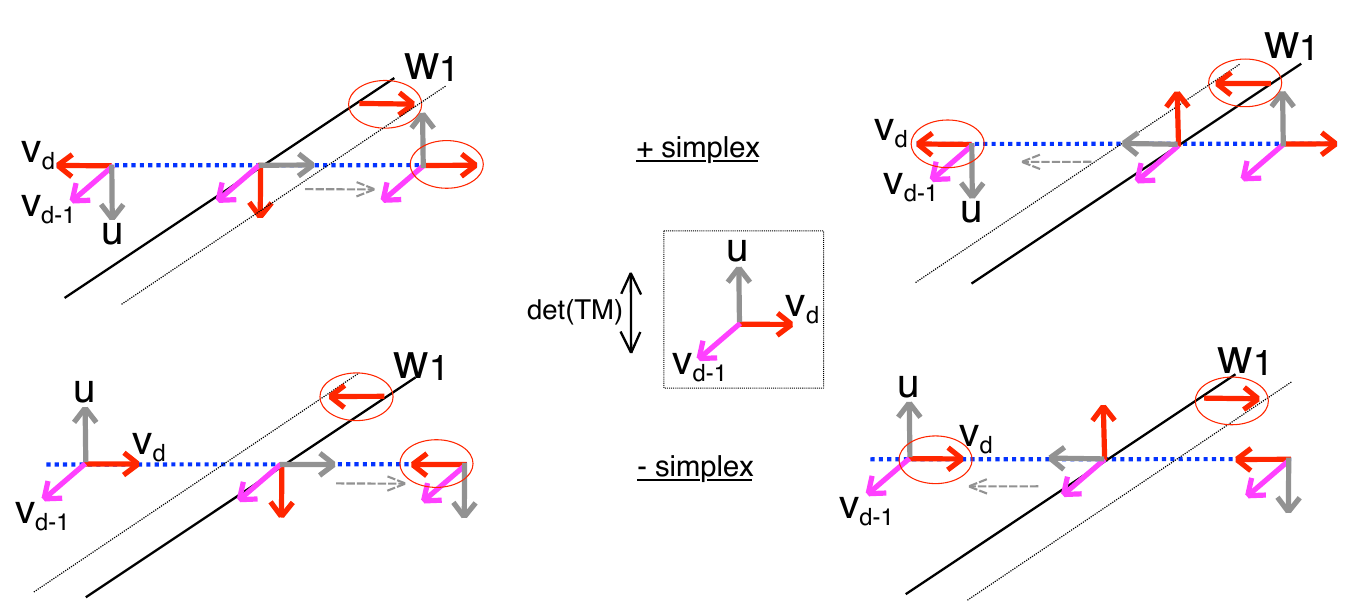}
    \caption{Choices of how to rotate ${u}$ and ${v}_d$ into each other across the dual of $w_1$ with respect to the different possibilities of $\pm$ $(d-1)$-simplices on $w_1$. These different choices correspond to a perturbation of the $w_1$ wall. The relevant vector fields' colors match those in the box and illustrate what directions their sections point along the dual 1-skeleton.}
    \label{fig:vectsCrossingW1}
\end{figure}

        In general dimensions, the same idea applies, where we have the same vector fields ${v}_1,\dots,{v}_d$ spanning the $TM$ directions within the $d$-simplices and the extra ${u}$ vector field in the $\det(TM)$ direction. These constructions are uniquely defined away from the orientation-reversing wall. But since our manifold is non-orientable we have to modify the vectors along the wall dual to $w_1$ since the vector fields $\{{v}_1, \dots,{v}_d\}$ by themselves degenerate near $w_1$. Given this extra $\det(TM)$ direction, we can in fact trivialize the bundle if we allow the $\{{v}_1,\dots,{v}_d\}$ to rotate into the $\det(TM)$ direction with respect to the $u$ vector. In particular we choose a scheme for which only ${v}_d$ and ${u}$ rotate into each other (recall that ${v}_1,\cdots, {v}_{d-1}$ always remain independent). In fact there are two ways to arrange this and roughly correspond to choosing to rotate them into each other by an angle of either $+180^\circ$ or $-180^\circ$. See Fig.~\ref{fig:vectsCrossingW1}. 
        
\begin{figure}[h!]
    \centering
    \includegraphics[width=0.6\linewidth]{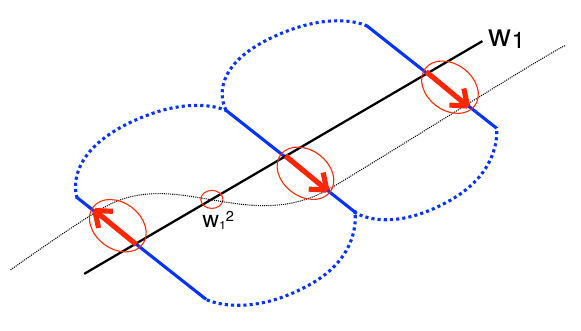}
    \caption{Representative of dual of $w_1^2$ from a perturbation of the orientation-reversing wall, as defined in Fig.~\ref{fig:vectsCrossingW1}.}
    \label{fig:w1SquaredRep}
\end{figure}

        Also in Fig.~\ref{fig:vectsCrossingW1}, we can see that the particular choice of how ${v}_d$ and ${u}$ rotate into each other determines a canonical perturbation of the $w_1$ wall as follows and is explained pictorially in the figure. The vector ${u}$ rotates into being parallel to the dual 1-skeleton towards one of the two $d$-simplices that share this $(d-1)$-simplex. The perturbing direction is the direction that ${v}_d$ points on the side specified by ${u}$. This perturbation can be used to define a `self-intersection' of $w_1$. This self-intersection will be some closed collection of $(d-2)$-simplices that form a cycle in $Z_{d-2}(M,\Z_2)$ dual to $w_1^2$. See Fig.~\ref{fig:w1SquaredRep}. This particular convention for defining $w_1^2$ is chosen to match the representative of $w_1^2$ used in the Grassmann integral $\sigma(f)$ later on. It also gives the winding matrices in Eq.~\eqref{windingGrassmannExpr2} a somewhat aesthetic form.
        
        Now we are in a position to define a pin$^-$ structure on a manifold. Geometrically, a pin$^-$ structure can be thought of as a trivialization of $TM \oplus \det(TM)$ on the 1-skeleton of a manifold that extends (mod 2) to a trivialization on the 2-skeleton. This is essentially the same as a spin structure except we replace $TM$ with $TM \oplus \det(TM)$. The obstruction to doing this is $w_2(TM \oplus \det(TM)) = w_2(TM) + w_1(TM)w_1(\det(TM)) = w_2(TM) + w_1(TM)^2 = w_2 + w_1^2$. For us, $w_2$ is the same canonical representative as in Eq.~\eqref{YuAnFormula} and $w_1^2$ is the same representative based on the perturbation of the dual of $w_1$. So a spin structure is represented by a cochain $\xi$ dual to a collection of $(d-1)$-simplices along which we twist the background vector fields to fix all singularities to be even-index.
        
        A $\text{pin}^+$ structure can also be defined similarly, but is instead a trivialization of $TM \oplus 3 \cdot \det(TM)$ on the 1-skeleton that extends (mod 2) to the 2-skeleton. This is because the group $\Pin^+$ embeds into $\Spin(d+3)$, such that projections onto $\SO(d+3)$ restrict to $O(d)$ in the first $d$ coordinates. We do not further consider this in detail because the discussion and geometric constructions of $\text{pin}^-$ structures are more closely related to the definition of Grassmann integral $\sigma(f)$. We do not know if there is a meaningful Grassmann integral that is more closely related to pin$^+$ structures. 
        
        \subsubsection{Induced pin$^-$ structures on framed curves} \label{inducedPinStruct}
        Similarly to spin structures, pin$^-$ structures give a way to induce spin/pin structures on embedded loops. The idea is exactly the same in the sense that a framing of $(d-2)$ vectors of $TM$ along an embedded loop induces a particular trivialization of $TM \oplus \det(TM)$ along the loop which is a `tangent framing' of the loop $F_\text{tang}$. And similarly the background framing $F_\text{bckd}$ is the trivialization of $TM \oplus \det(TM)$ given by the vector fields ${{v}_1,\dots,{v}_d,{u}}$ along the dual $1$-skeleton. In this case, $F_\text{tang}$ and $F_\text{bckd}$ are instead $(d+1) \times (d+1)$ matrices.
        
        Just as before, the relative framing is a function $(F_\text{bckd}^{-1} F_\text{tang})(t): [0,1] \to \SO(d+1)$ which lifts to a path $\tilde{F}(t): [0,1] \to \Spin(d+1)$. A subtlety is that in this non-orientable case, we may not have $(F_\text{bckd}^{-1} F_\text{tang})(0) = (F_\text{bckd}^{-1} F_\text{tang})(1)$ since curves passing through the orientation-reversing wall will have their local orientation be the opposite at the end of the path with respect to what it was in the beginning.
        
\begin{figure}[h!]
    \centering
    \includegraphics[width=0.85\linewidth]{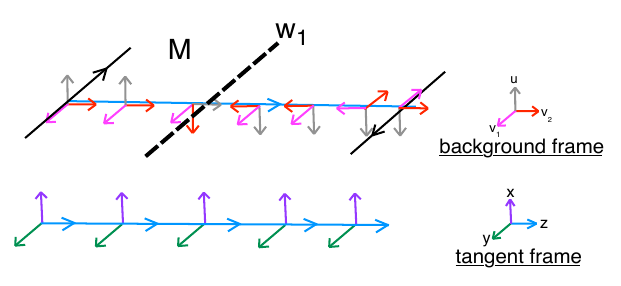}
    \caption{The tangent and background framings for a loop in $d=2$ that crosses the orientation-reversing wall dual to $w_1$ exactly once. The two blue lines going to the right should be thought of as on top of each other but visually displaced to make it easier to see the vector fields. A neighborhood of the loop in this manifold looks like a Möbius strip; note the depiction is that the two black lines identified via opposite arrows. The vector $u$ is in the $\det(TM)$ direction `transverse' to the 2-manifold $M$ except when it crosses $w_1$. Note that the vector $u$ switches sign relative to the loop's framing from the beginning to the end. The net rotation in $\SO(3)$ between the frames $\{x,y,z\},\{u,v_1,v_2\}$ is the diagonal matrix $\text{diag}(-1,-1,1)$ which lifts to $\pm i Z \in \Spin(3) = \SU(2)$.}
    \label{fig:windingAcrossW1_2D}
\end{figure}

        We illustrate this in $d = 2$ in Fig.~\ref{fig:windingAcrossW1_2D} for a path that crosses an orientation-reversing loop. Note that in that case 
        \begin{equation*}
        (F_\text{bckd}^{-1} F_\text{tang})(0) = \begin{pmatrix} -1 & 0 & 0 \\ 0 & -1 & 0 \\ 0 & 0 & 1 \end{pmatrix} (F_\text{bckd}^{-1} F_\text{tang})(1)
        \end{equation*}
        in the coordinate basis given by $\{{u},{v}_1,{v}_2\}$ for the background frame and $\{x,y,z\}$ for the tangent frame. Indeed, in general for orientation-reversing loops, the relative framing gets changed by $\begin{pmatrix} -1 & 0 & 0 \\ 0 & -1 & 0 \\ 0 & 0 & 1 \end{pmatrix}$ across the loop, while for orientation-preserving loops $(F_\text{bckd}^{-1} F_\text{tang})(0) = (F_\text{bckd}^{-1} F_\text{tang})(1)$. 
        
        Continuing with the $d = 2$ example, we can consider the lifts to $\Spin(3) = \SU(2)$ that these paths in $\SO(3)$ induce. In particular, the lifting to $\SU(2)$ will be (fixing WLOG $\tilde{F}(0) = \mathbbm{1}$)
        \begin{equation}
        \tilde{F}(1) =
        \begin{cases}
        \pm \mathbbm{1} \text{  if orientation-preserving} \\
        \pm i Z \text{  if orientation-reversing}
        \end{cases}
        \end{equation}
        where $Z = \begin{pmatrix} 1 & 0 \\ 0 & -1 \end{pmatrix}$. In particular, this lets us define a function on $(d-1)$-cochains $f_L$ dual to a loop $L$
        \begin{equation}
        \sigma(f_L) = -
        \begin{pmatrix} 1 & 0 \end{pmatrix} \tilde{F}(1) \begin{pmatrix} 1 \\ 0 \end{pmatrix} =
        \begin{cases}
        \pm 1 \text{  if orientation-preserving} \\
        \pm i \text{  if orientation-reversing}
        \end{cases}
        \end{equation}
        that encodes this lift. Note the minus sign in front of the matrix element. More precisely, each orientation-preserving loop is endowed with an induced spin structure, which corresponds to $+1$ for $\sigma(f_L)$ if it is anti-periodic and $-1$ if it is periodic.
        Orientation-reversing loops are endowed with an induced pin$^-$ structure on their Möbius bundle labeled by $\pm i$. 
        
        The function $\sigma(f_L)$ defined above generalizes beyond $d = 2$ to higher dimensions after homotoping the first $(d-2)$ vectors of the frame to match, as we discussed previously in the orientable case.  Also, even though we technically phrased everything in terms of a given $\text{pin}^-$ structure on the manifold where all singularities are even-index, the same procedure also gives a quantity when there are odd-index singularities, although isotopy invariance of moving the curve across these singularities is lost.
        
        In this more general context, the function $\sigma(f_L)$ is an important part of the winding number definition of the Grassmann integral discussed in Section \ref{windingGrassDefSec} and Appendix \ref{sec:windingDefOfSigma}.

\section{ Explicitly computing winding number definition of Grassmann integral, $\sigma(f)$} 
\label{sec:windingDefOfSigma}

In this appendix, we fill in the details sketched in Sec.~\ref{windingGrassDefSec} and expand on the discussion of Appendix \ref{sec:inducedSpinStructOnCurve},\ref{inducedPinStruct} for how to explicitly compute the winding definition of the Grassmann integral $\sigma(f)$. We further show how the geometric winding number based definition of Section \ref{windingGrassDefSec} is equivalent to the algebraic definition given in Section \ref{grassIntegral} based on an integration over Grassmann variables. 

As discussed in the main text, for a $(d-1)$-cocycle $f_{d-1}$, we can (using an appropriate trivalent decomposition) decompose its Poincar\'e dual chain into a set of non-intersecting loops on the dual 1-skeleton. Using Eq.~\eqref{eqn:GrassmannAsProductOfLoops}, in order to compute $\sigma(f_{d-1})$ we need only explain how to compute $\sigma$ on a single such loop $L$.
        
    \subsection{ Orientable Manifolds}
    
        We begin with the simpler case of $M$ orientable. In Appendix \ref{sec:inducedSpinStructOnCurve}, we explained how a frame of $d$ ``background'' vector fields along a loop defines an induced spin structure on that loop, or alternatively a winding of a tangent framing of the loop with respect to the background framing. In that context, the background vector fields arose from a spin structure, but as we showed in Appendix \ref{w1PmAssignments_AndVectorFields}, even in the absence of a spin structure, the branching structure of the triangulation of $M$ together with the $\pm$ assignments of $d$-simplices defines a nonsingular background framing along the dual 1-skeleton consisting of the vector fields ${v}_1,\dots,{v}_d$. The discussion in Appendix \ref{sec:inducedSpinStructOnCurve} then applies directly. Given a $(d-1)$-cocycle $f_L$ dual to a single loop $L$, the background vector fields, together with the sign assignments $\epsilon(\Delta_d)$, determine a tangent framing $F_\text{tang}$ of $L$.
        Technically there are two choices for the tangent framing depending on which direction we pick the tangent vector along the curve, but this choice will not change the winding.
        We can write the background and tangent framings as maps $F_\text{bckd},F_\text{tang}: [0,1] \rightarrow \SO(d)$, lift the relative path $F^{-1}_\text{bckd}F_\text{tang}$ to a map $\tilde{F} : [0,1] \rightarrow \Spin(d)$, and use the sign $\tilde{F}(0) = \pm \tilde{F}(1)$ to define the $\Z_2$ winding. 
        
        The above prescription is still somewhat abstract; we presently explain how to calculate this winding number locally. First, pick a direction for $L$. Then on each simplex $\Delta_d = \braket{0 \cdots d}$ (here the branching structure is assumed to be $0 \to 1 \to \cdots \to d$), $L$ will traverse between two dual 1-cells, going from from $\hat{i} \to \hat{j}$, where $\hat{i}$ is the 1-cell dual to the sub-simplex $\braket{0 \cdots \hat{i} \cdots d}$ (which is also opposite to the vertex $i$). The winding along this leg of the curve depends on $i,j$ as well as the sign assignment $\epsilon(\Delta_d)$ to the $d$-simplex, and will be 
        \begin{align} \label{windingsInHigherDimensions}
        \text{wind}(\hat{i} \to \hat{j}) = 
        \begin{cases}
        0 \text{ if } i \not\equiv j \text{ (mod 2)} \\
        \epsilon(\Delta_d) \pi \text{ if } i \equiv j \text{ (mod 2) and } i < j \\
        -\epsilon(\Delta_d) \pi \text{ if } i \equiv j \text{ (mod 2) and } i > j
        \end{cases}.
        \end{align}
        
\begin{figure}[h!]
            \centering
            \includegraphics[width=0.6\linewidth]{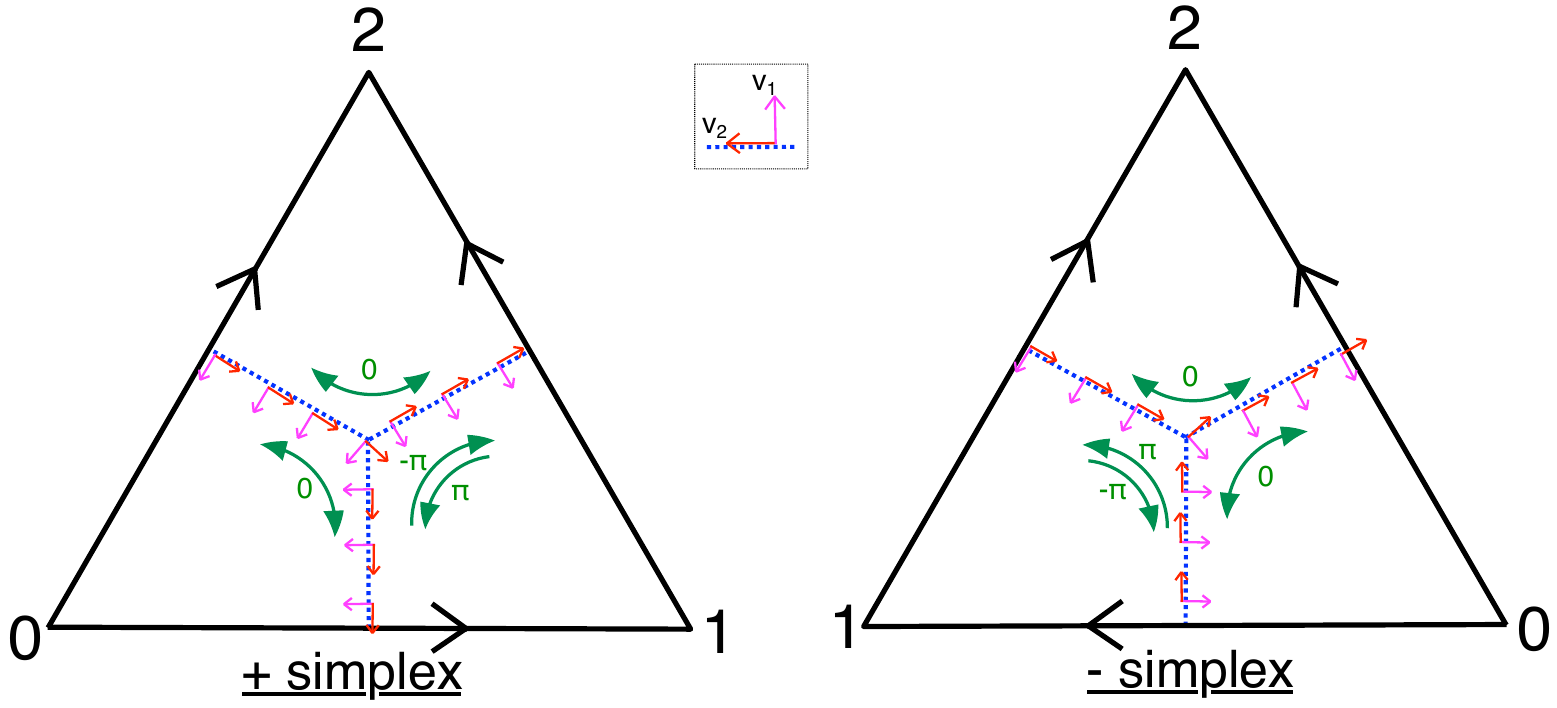}
            \caption{In $d=2$, frame of background vector fields (red, pink) on the dual $1$-cells of a $2$-simplex. The tangent framing of a fermion loop traveling along a green arrow winds by the given angle relative to this background framing.}
            \label{winding_vectorFields2D}
        \end{figure}

\begin{figure}[h!]
            \centering
            \includegraphics[width=0.6\linewidth]{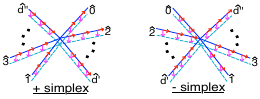}
            \caption{Projection to $2$D of frame of background vector fields (red and pink) on the dual $1$-cells of a $d$-simplex after projecting out the $(d-2)$ vectors shared by the background and tangent framings. The solid lines are the dual 1-skeleton and the dashed lines are the shift of the dual 1-skeleton along the pink vector field. As described in the main text, the pairs of solid and dashed lines that intersect correspond to the $\cup_{d-2}$ pairing. The tangent framing of a fermion loop traversing the interior of this $d$-simplex experiences the windings in Eq.~\eqref{windingsInHigherDimensions}. Here, $d'$ ($d''$) is the largest even (odd) number less than or equal to $d$. The clockwise order of the dual $1$-cells and the directions of the vector fields are determined by the triangulation, branching structure, and orientation.}
            \label{windingsGeneralDimensions}
        \end{figure}

        These windings are obtained in two dimensions by examining Fig.~\ref{winding_vectorFields2D}; the background vector fields are shown on the dual 1-skeleton, and a tangent framing can be drawn in for any particular curve passing through the 2-simplex that is drawn. The winding can then be obtained graphically. In higher dimensions, similar logic can be used to obtain Eq.~\eqref{windingsInHigherDimensions} after projecting away the $(d-2)$ vectors shared by the background and tangent framings. The projected background vector fields are shown in Fig.~\ref{windingsGeneralDimensions}, and the winding of the tangent vector field relative to this background vector field can be computed by inspection. We emphasize that the tangent framing is determined by the loop $L$, the $d$-simplex orientations $\epsilon(\Delta_d)$, and, for $d>2$, the background framing.
        
        Then the total winding is just the product of windings along each segment of the loop $L$, and $\sigma(f_L)$ is defined from that winding:
        \begin{equation}
            \sigma(f_L) = -\prod_{(\hat{i} \rightarrow \hat{j}) \in L} e^{ \frac{i}{2} \times \text{wind}(\hat{i} \rightarrow \hat{j})}.
        \end{equation}
        As we saw above, the partial windings are either 0 or $\pi$, and since there must always be an even number of $\pi$ rotations, $\sigma(f_L) = \pm 1$. The minus sign in front is present in order to match Eq.~\eqref{sigmaLbc} and is needed to correctly reproduce quadratic refinement Eq.~\eqref{quadRefinementEq}.
        
        As a technical comment, the order in which the dual 1-cells appear and the relative directions of the vector fields in Fig.~\ref{windingsGeneralDimensions} is not arbitrary and follows from the fact that the background vector fields, which are completely defined by the triangulation, branching structure, and orientation, were defined to give a geometric meaning to the higher cup product~\cite{tata2020}.
        This choice of background vector fields ensures that the higher cup product properly appears in the quadratic refinement property Eq.~\eqref{quadRefinementEq}. The pink vector field in Fig.~\ref{windingsGeneralDimensions} defines a shift of the dual 1-cells, and in this projected picture, the product $\alpha(\hat{i})\beta(\hat{j})$ appears in the formula for $\alpha \cup_{d-2} \beta$ if and only if the shifted $\hat{j}$ 1-cell intersects the unshifted 1-cell $\hat{i}$. One can check that Fig.~\ref{windingsGeneralDimensions} correctly reproduces the formula for the $\cup_{d-2}$ product:
        \begin{equation}
        (\alpha \cup_{d-2} \beta)(0 \dots d) = \sum_{\substack{i < j \text{ both odd, OR} \\ i > j \text{ both even}}} \alpha(\hat{i}) \beta(\hat{j})
        \label{cupDminus2Formula}
        \end{equation}

        As an aside, we note that the trivalent resolution used throughout was chosen in an \textit{ad hoc} way to reproduce quadratic refinement formulas and equivalence to the Grassmann integral definitions. However, the cyclic order that appears in Fig.~\ref{windingsGeneralDimensions} looks quite similar to the trivalent resolution Fig.~\ref{trivalentResolutionsAndWindings}. We hope that the appearance of this ordering in Fig.~\ref{windingsGeneralDimensions} may lead to a more first-principles explanation for why certain trivalent resolutions work, although we have not figured out a precise connection.

        \subsubsection{Winding formula for $w_2$ is equivalent to Eq.~\eqref{YuAnFormula}} \label{windingFormulaSameAsYuAns}
        
        As a cross-check, we show in the case of orientable manifolds that inputting the winding formula into Eq.~\eqref{sigmaElLoop} gives the same answer for $w_2$ as Eq.~\eqref{YuAnFormula}. More precisely, consider the loop of $(d-1)$-simplices $\Delta_{d-1}^1 \to \cdots \to \Delta_{d-1}^k \to \Delta_{d-1}^1$ comprising the link of 2-simplex $\braket{012}$. The winding formula expresses $\sigma$ in terms the partial windings $\text{wind}(\Delta_{d-1}^\ell \to \Delta_{d-1}^{\ell+1})$ as in Eq.~\eqref{windingsInHigherDimensions}. On the other hand, Eq.~\eqref{YuAnFormula} expresses $w_2$ in terms of the indicator cochains $\boldsymbol{\Delta}_{d-1}^\ell$ on each $(d-1)$-simplex. The equality we want to show is
        \begin{equation}
        \begin{split}
            \sigma(\delta {\lambda_{\braket{012}}}) &= -(-1)^{\frac{1}{2\pi} \sum_{\ell=1}^k \text{wind}(\Delta_{d-1}^\ell \to \Delta_{d-1}^{\ell+1})} \\
            &= -(-1)^{\sum_{\ell=1}^k \int {\boldsymbol{\Delta}}_{d-1}^{\ell} \cup_{d-2} {\boldsymbol{\Delta}}_{d-1}^{\ell+1}}
        \end{split}
        \end{equation}
        where we set $\Delta_{d-1}^{k+1} = \Delta_{d-1}^{1}$. Thus we just need to show that $\sum_{\ell=1}^k \int \boldsymbol{\Delta}_{d-1}^{\ell} \cup_{d-2} \boldsymbol{\Delta}_{d-1}^{\ell+1}$ is an alternate way to express the total winding $\frac{1}{2\pi} \sum_{\ell=1}^k \text{wind}(\Delta_{d-1}^\ell \to \Delta_{d-1}^{\ell+1})$, modulo $2$. 
        
        To do this, we can redefine the partial windings as follows. In the following pictures, we will say that the black arrows on the bottom are the path along $\mathrm{Link}(\braket{012})$ and always points to the right. The red arrows on top are the background vector field $v_d$ which we can use to define the relative windings. The windings of $\pm \pi$ fall into four cases of counterclockwise / clockwise and $v_d$ starting in the same direction / opposite direction as the path. We will denote $\{\curvearrowleft, \curvearrowbotright\} / \{\curvearrowright, \curvearrowbotleft\}$ as a counterclockwise / clockwise rotations of the red $v_d$, depending on the relative starting orientatations. Below, we list the various possible rotations and give the cases in which they occur for paths $\hat{i} \to \hat{j}$ on a $\pm$ simplex:
        
        \begin{center}
          \begin{tabular}{ccccc}
            &$\begin{matrix} {\color{red} \longrightarrow} \\ \longrightarrow \end{matrix}$
            $\curvearrowleft$
            &$\begin{matrix} {\color{red} \longleftarrow} \\ \longrightarrow \end{matrix}$
            &= $+\pi$ rotation
            &$\begin{cases}
               \hat{i} \to \hat{j} \text{ both odd on } + \text{ simplex, and } i < j \\ 
               \hat{i} \to \hat{j} \text{ both even on } - \text{ simplex, and } i > j
            \end{cases}$
            \\
            \hline
            &$\begin{matrix} {\color{red} \longrightarrow} \\ \longrightarrow \end{matrix}$
            $\curvearrowbotleft$
            &$\begin{matrix} {\color{red} \longleftarrow} \\ \longrightarrow \end{matrix}$
            &= $-\pi$ rotation
            &$\begin{cases}
               \hat{i} \to \hat{j} \text{ both odd on } + \text{ simplex, and } i > j \\ 
               \hat{i} \to \hat{j} \text{ both even on } - \text{ simplex, and } i < j
            \end{cases}$
            \\
            \hline
            &$\begin{matrix} {\color{red} \longleftarrow} \\ \longrightarrow \end{matrix}$
            $\curvearrowbotright$
            &$\begin{matrix} {\color{red} \longrightarrow} \\ \longrightarrow \end{matrix}$
            &= $+\pi$ rotation
            &$\begin{cases}
               \hat{i} \to \hat{j} \text{ both even on } + \text{ simplex, and } i < j \\ 
               \hat{i} \to \hat{j} \text{ both odd on } - \text{ simplex, and } i > j
            \end{cases}$
            \\
            \hline
            &$\begin{matrix} {\color{red} \longleftarrow} \\ \longrightarrow \end{matrix}$
            $\curvearrowright$
            &$\begin{matrix} {\color{red} \longrightarrow} \\ \longrightarrow \end{matrix}$
            &= $-\pi$ rotation
            &$\begin{cases}
               \hat{i} \to \hat{j} \text{ both even on } + \text{ simplex, and } i > j \\ 
               \hat{i} \to \hat{j} \text{ both odd on } - \text{ simplex, and } i < j
            \end{cases}$
          \end{tabular}
        \end{center}
        
        However, note that these we can modify these values of these \textit{partial} windings in such a way that the \textit{total} windings stay the same. Any angle $\theta$ for which 
        \begin{equation*}
        \theta = \text{wind}(\curvearrowleft) = -\text{wind}(\curvearrowright) \quad\text{ and }\quad 2\pi - \theta = \text{wind}(\curvearrowbotright) = - \text{wind}(\curvearrowbotleft)
        \end{equation*}
        will give the same total windings. The first conditions that 
        \begin{equation*}
        \text{wind}(\curvearrowleft) = -\text{wind}(\curvearrowright) \quad\text{ and }\quad \text{wind}(\curvearrowbotright) = - \text{wind}(\curvearrowbotleft)
        \end{equation*}
        enforces that windings that can be deformed to the identity have zero total winding. The conditions \begin{equation*}
        \text{wind}(\curvearrowleft)+\text{wind}(\curvearrowbotright) = -(\text{wind}(\curvearrowbotleft)+\text{wind}(\curvearrowright)) = 2\pi
        \end{equation*} 
        are to ensure that a full $2\pi$ rotation indeed has winding $2\pi$. The above cases that we used to define the winding correspond to $\theta = \pi$. However, we just as well could choose $\theta = 2\pi$ to get alternate expressions in terms of the modified ``rotations'':
        
        \begin{center}
        \begin{equation} \label{modifiedRotations}
          \begin{tabular}{ccccc}
            &$\begin{matrix} {\color{red} \longrightarrow} \\ \longrightarrow \end{matrix}$
            $\curvearrowleft$
            &$\begin{matrix} {\color{red} \longleftarrow} \\ \longrightarrow \end{matrix}$
            &= $+2\pi$ ``rotation''
            &$\begin{cases}
               \hat{i} \to \hat{j} \text{ both odd on } + \text{ simplex, and } i < j \\ 
               \hat{i} \to \hat{j} \text{ both even on } - \text{ simplex, and } i > j
            \end{cases}$
            \\
            \hline
            &$\begin{matrix} {\color{red} \longrightarrow} \\ \longrightarrow \end{matrix}$
            $\curvearrowbotleft$
            &$\begin{matrix} {\color{red} \longleftarrow} \\ \longrightarrow \end{matrix}$
            &= $0$ ``rotation''
            &$\begin{cases}
               \hat{i} \to \hat{j} \text{ both odd on } + \text{ simplex, and } i > j \\ 
               \hat{i} \to \hat{j} \text{ both even on } - \text{ simplex, and } i < j
            \end{cases}$
            \\
            \hline
            &$\begin{matrix} {\color{red} \longleftarrow} \\ \longrightarrow \end{matrix}$
            $\curvearrowbotright$
            &$\begin{matrix} {\color{red} \longrightarrow} \\ \longrightarrow \end{matrix}$
            &= $0$ ``rotation''
            &$\begin{cases}
               \hat{i} \to \hat{j} \text{ both even on } + \text{ simplex, and } i < j \\ 
               \hat{i} \to \hat{j} \text{ both odd on } - \text{ simplex, and } i > j
            \end{cases}$
            \\
            \hline
            &$\begin{matrix} {\color{red} \longleftarrow} \\ \longrightarrow \end{matrix}$
            $\curvearrowright$
            &$\begin{matrix} {\color{red} \longrightarrow} \\ \longrightarrow \end{matrix}$
            &= $-2\pi$ ``rotation''
            &$\begin{cases}
               \hat{i} \to \hat{j} \text{ both even on } + \text{ simplex, and } i > j \\ 
               \hat{i} \to \hat{j} \text{ both odd on } - \text{ simplex, and } i < j
            \end{cases}$
          \end{tabular}
        \end{equation}
        \end{center}
        
        From here, we can directly compare to the formula in terms of higher cup products. For a path $\hat{i} \to \hat{j}$ inside a $d$-simplex, the corresponding contribution $\int \boldsymbol{\Delta}_{d-1}^{\ell} \cup_{d-2} \boldsymbol{\Delta}_{d-1}^{\ell+1}$ will be:
        \begin{equation*}
            \int \boldsymbol{\Delta}_{d-1}^{\ell} \cup_{d-2} \boldsymbol{\Delta}_{d-1}^{\ell+1} = 
            \begin{cases}
                1 \text{ if } i > j \text{ both even, OR if } i < j \text{ both odd} \\
                0 \text{ otherwise}
            \end{cases}
        \end{equation*}
        using the equation Eq.~\eqref{cupDminus2Formula}. Since either of the $\pm 2\pi$ rotations contribute $(-1)$ to the winding expression for $w_2$, comparing the $\cup_{d-2}$ expression above to the table of modified windings Eq.~\eqref{modifiedRotations}, we can see that there is a $\pm 2\pi$ rotation if and only if $\int \boldsymbol{\Delta}_{d-1}^{\ell} \cup_{d-2} \boldsymbol{\Delta}_{d-1}^{\ell+1} = 1$. 
        
        This proves the equivalence of the two formulas for $w_2$.

    \subsection{ Non-orientable Manifolds} \label{sigmaNonorientableWindingDefn}
    
        On a non-orientable manifold, the above procedure needs to be modified because any assignment $\epsilon(\Delta_d)$ of local orientations to $d$-simplices produces inconsistent induced orientations on the $(d-1)$-simplices. Equivalently, it is impossible to define a nondegenerate frame of vector fields everywhere along the 1-skeleton.
        The solution, as discussed in Sec.~\ref{pinMinusStructs}, is to instead define a frame of vectors on the bundle $TM \oplus \det(TM)$ instead of $TM$; the former is always orientable.
        
        Such a ``background'' framing $F_{\text{bckd}}$ on the dual 1-skeleton can be constructed from the branching structure, assignments $\epsilon(\Delta_d)$ of $\pm$ signs to $d$-simplices, and some additional (arbitrary, for present purposes) choices of how the framing behaves across the orientation-reversing wall. The first two pieces of data determine a cochain representative of $w_1$, and the rest of this process determines a perturbation of the orientation-reversing wall and thus a cochain representative of $w_1^2$.
        This construction is explained in Appendices~\ref{w1PmAssignments_AndVectorFields} and~\ref{pinMinusStructs}. As discussed in Appendix~\ref{inducedPinStruct}, the background framing and the assignments $\epsilon(\Delta_d)$ determine\footnote{Strictly speaking we also need an arbitrary choice of a vector in the $\det(TM)$ direction at one point on the curve, but one can check that the windings are independent of this choice.} a ``tangent" framing $F_{\text{tang}}$ of each fermion loop $L$ and use it to define induced (s)pin structures on the curve.

        We need to describe how to actually compute the windings and $\sigma$ on a $(d-1)$-cocycle $f_L$ dual to a single loop $L$. In the orientable case, we computed the number of times the tangent framing of a loop winds around the background framing. In the presence of an orientation-reversing wall, we need something like a  ``half-winding"; following the naming scheme of~\cite{kirby_taylor_1991}, we will call this $\text{(\# of right-half-twists)(L)}$.
        The reason for this name is that~\cite{kirby_taylor_1991} considers a different scheme for background framings in $d=2$ where the background framing always shares one vector with the tangent of the curve and all rotations are about this axis. In their scheme, the total rotation is always quantized as a multiple of $\pm \pi$. For orientation-reversing loops it will be in total $\pm \pi$ (mod $2\pi$), i.e. a half-twist, because the vectors in the $\det(TM)$ direction switch sides relative to each other around the loop; a similar behavior of the vectors in the $\det(TM)$ direction occurs in our scheme when loops cross the orientation-reversing wall, as in Fig.~\ref{fig:windingAcrossW1_2D}. In fact, the relative frames in our scheme can be deformed to one more like~\cite{kirby_taylor_1991} where the relative framing rotates by $\pm \pi$ about the $z$ axis, although this is not trivial to see by inspection. For orientation-preserving loops, the total rotation is always an even multiple of $\pm \pi$, i.e., a full twist. We want $\sigma=\pm 1$ for orientation-preserving loops, i.e., an even number of half-twists, so we demand $\sigma = \pm i$ for orientation-reversing loops, i.e., an odd number of half-twists.
        
        Returning to our scheme, similarly to the orientable case, there is a contribution to the windings that can be computed from how the loop traverses between $\hat{i} \to \hat{j}$ within simplices, although in the non-orientable case, it is important that this traversal direction matches the direction of the tangent vector. There is an additional contribution to the winding when the loop crosses $w_1$. We thus divide the loop $L$ into $N$ legs $k \in 1,\dots,N$, where each leg consists of either traveling between the $(d-1)$-simplices $\hat{i} \to \hat{j}$ \textit{within} a $d$-simplex or traveling \textit{between} two $d$-simplices along a single dual $1$-cell. The loop can only cross $w_1$ in the latter case. Associate to the $k$th leg of the loop a $4 \times 4$ matrix $W_k$ as follows. If the $k$th leg is traversing $\hat{i} \to \hat{j}$ within the $d$-simplex $\Delta_d$, then
        \begin{align} \label{windingGrassmannExpr1}
            W_k = \epsilon(\Delta_d)
            \begin{cases}
            \quad\quad\mathbbm{1}_4 &\text{ if } i \not\equiv j \text{ (mod 2)} \\
            \begin{pmatrix} i X & 0 \\ 0 & -i X \end{pmatrix} &\text{ if } i \equiv j \text{ (mod 2) and } i < j \\
            \begin{pmatrix} -i X & 0 \\ 0 & i X \end{pmatrix} &\text{ if } i \equiv j \text{ (mod 2) and } i > j 
            \end{cases}
        \end{align}
        
        If instead the $k$th leg travels between two $d$-simplices through a $(d-1)$-simplex $\Delta_{d-1}$ (i.e. along the 1-cell dual to $\Delta_{d-1}$), then if $w_1(\Delta_{d-1}) = 0$, assign $W_k = \mathbbm{1}_4$. If $w_1(\Delta_{d-1}) = 1$, then both $d$-simplices neighboring $\Delta_{d-1}$ induce the same orientation $\pm$ on $\Delta_{d-1}$; given that induced orientation, define
        \begin{align} \label{windingGrassmannExpr2}
            W_k = \pm
            \begin{cases}
            \begin{pmatrix} 0 & -iY \\ iY & 0 \end{pmatrix}   &\text{ if crossing in the perturbing direction} \\
            \begin{pmatrix} 0 & iY \\ -iY & 0 \end{pmatrix}   &\text{ if crossing opposite to the perturbing direction}
            \end{cases}
        \end{align}
        
        Here, $X={\begin{pmatrix} 0 & 1 \\ 1 & 0 \end{pmatrix}}, Y={\begin{pmatrix} 0 & -i \\ i & 0 \end{pmatrix}},Z={\begin{pmatrix} 1 & 0 \\ 0 & -1 \end{pmatrix}}$ are the Pauli matrices. The winding of the loop is related to the product of these winding matrices $W_1 \cdots W_N$ by
        
        \begin{align}
            i^\text{\# of right-half-twists} = \begin{pmatrix} 1 & 0 & 1 & 0 \end{pmatrix} W_1 \cdots W_N \begin{pmatrix} 1 \\ 0 \\ 0 \\ 0 \end{pmatrix}.
        \end{align}

        We can now express $\sigma(f_L)$ as
        \begin{align}
            \sigma(f_L) = -i^\text{\# of right-half-twists}.
        \end{align}
        It will be the case that $\sigma(f_L)$ will always be in $\{\pm 1, \pm i\}$, and will be $\pm i$ if and only if the loop is orientation-reversing (i.e. if $\int_{w_1} f = 1$).
        For a general cocycle $f$, we can write
        \begin{align}
            \sigma(f) = (-1)^\text{\# of loops} \prod_\text{loops} i^\text{(\# of right-half-twists)(loop)}
        \end{align}
        
        Now, we give a brief description of how the winding matrices are derived, deferring details to~\cite{tata2020}. These winding matrices can be derived by inspecting Figs.~\ref{fig:vectsCrossingW1}, \ref{windingsGeneralDimensions}. First, this block form  of zeros and $\SU(2)$ elements is related to the two possible configurations of the tangent framing along the segment; these two possibilities correspond to whether the `orientation vector' of the tangent framing in the $\det(TM)$ direction starts out in the same direction as the $u$ of the background framing or starts out opposite. This $2 \times 2$ block structure of $0$'s and $\SU(2)$ matrices is a way to compactly encode these choices. As for why the matrices are in $\SU(2)$, recall that the lift of the relative framing $(F_\text{bckd}^{-1} F_\text{tang})(t)$ involves three vector fields ${u},{v}_{d-1},{v}_d$ of the background framing and three vectors of the tangent framing, thus forming a path in $\SO(3)$ that lifts to a path in $\SU(2)$. To derive the relevant winding matrices, one has to consider how these partial windings from the figure lift from elements of $\SO(3)$ to elements of $\SU(2)$. For example, all winding matrices inside a $d$-simplex involve no rotation or a $\pm \pi$ rotation about the common $u=x$ axis (see Fig.~\ref{fig:windingAcrossW1_2D} for axis labels) which lift to $\pm i X$ rotations. This process does not reverse orientation locally, so the block structure should be diagonal. Also, one gets opposite rotations if the $\det(TM)$-direction vectors are in the same direction vs. opposite directions, accounting for the sign difference between the blocks. Whereas, the winding matrices across $w_1$ involve $\pm i Y$ corresponding to $\pm \pi$ rotations about the $v_{d-1}=y$ axis. This process reverses orientation and thus gives the off-diagonal block matrices; one can convince themselves via the figures that these lifts to $\SU(2)$ are opposites if the $\det(TM)$ vectors start out in the same vs. opposite directions. 
        
        The final winding will always be a rotation about the $z$-axis and thus have an eigenvector $\begin{pmatrix} 1 \\ 0 \end{pmatrix}$. We assume, without loss of generality, that the loop starts with the vectors aligned in the $\det(TM)$ direction, which is why we apply the winding matrices to $\begin{pmatrix} 1 \\ 0 \\ 0 \\ 0 \end{pmatrix}$. After traversing the loop, either the vectors in the $\det(TM)$ direction are aligned or anti-aligned; we sum over both possibilities, only one of which will give a nonzero contribution. This can be accomplished by taking an inner product with $\begin{pmatrix} 1 & 0 & 1 & 0 \end{pmatrix}$ which sums over the eigenvalues of the top-left and bottom-left blocks of the product $W_1 \cdots W_N$.
        
        Again, it turns out that $\sigma(f)$ satisfies the main properties Eqs.~\eqref{sigmaElLoop} and \eqref{quadRefinementEq}, where the representative of $w_1^2$ is the one as specified above and $w_2$ is the `canonical' representative that can be expressed by Eq.~\eqref{YuAnFormula}.
        
\subsection{ Equivalence of winding and Grassmann definitions}
\label{grassEquivsec}
        
Now, we demonstrate that the winding and Grassmann definitions are indeed equivalent to each other, i.e., $\sigma(f) = \sigma^\text{gr}(f)$ for all $\delta f = 0$. We start with the orientable case before proceeding to the non-orientable case.

We claim that it is actually sufficient to show $\sigma(f_L) = \sigma^\text{gr}(f_L)$, where $f_L$ is dual to a single closed loop $L$, which stems from our matching choice of trivalent resolution and Grassmann variable orderings. Proving this will be the contents of Secs.~\ref{grassmannEqualsWindingOrientable},\ref{grassmannEqualsWindingNonorientable}. Presently we will show $\sigma(f) = \sigma^\text{gr}(f)$ for \textit{all} $f$ \textit{assuming} they match on single loops.

First, it is clear that if $f$ is nonzero on only two $(d-1)$-simplices per every $d$-simplex, then even without the trivalent resolution $f$ decomposes into distinct loops that never meet in a common $d$-simplex. In this case, $f = f_\text{loop 1} + \cdots + f_\text{loop n}$ and it is simple to see that $\sigma(f) = \sigma(f_\text{loop 1}) \cdots \sigma(f_\text{loop n}) = \sigma^\text{gr}(f_\text{loop 1}) \cdots \sigma^\text{gr}(f_\text{loop n}) = \sigma^\text{gr}(f)$.

Now, consider the case where on some $(d-1)$-simplex ${\Delta_{d}}$, $f$ is nonzero on more than two $(d-1)$-simplices, with $\{\hat{i}_1 \cdots \hat{i}_{2k}\}$ listed in the order of the trivalent resolution and Grassmann variables. Then the trivalent resolution gives $f$ to a decomposition into distinct loops for whom the edges are paired up locally as $\{\hat{i}_1, \hat{i}_2\},\dots,\{\hat{i}_{2k-1} \hat{i}_{2k}\}$. In addition, the $u({\Delta_{d}}) = \vartheta_{\hat{i}_1}^{f(\hat{i}_1)} \cdots \vartheta_{\hat{i}_{2k}}^{f(\hat{i}_{2k})}$ means that we can split $u({\Delta_{d}})$ into a product of $k$ terms $u({\Delta_{d}}) = (\vartheta_{\hat{i}_1}^{f(\hat{i}_1)} \vartheta_{\hat{i}_2}^{f(\hat{i}_2)}) \cdots (\vartheta_{\hat{i}_{2k-1}}^{f(\hat{i}_{2k-1})} \vartheta_{\hat{i}_{2k}}^{f(\hat{i}_{2k})})$. 

Using this trivalent resolution we can organize $f = f_{L_1} + \cdots + f_{L_n}$. The Grassmann formulation gives $\sigma^\text{gr}(f) = \sigma^\text{gr}(f_{L_1}) \cdots \sigma^\text{gr}(f_{L_n})$ because of the freedom to reorganize the pairs of Grassmann variables from each trivalent resolution. And the trivalent resolution order and Eq.~\eqref{cupDminus2Formula} mean that 
\begin{equation*}
\sigma(f) = \sigma(f_{L_1}) \cdots \sigma(f_{L_n}) \prod_{1 \le \ell_1 < \ell_2 \le n} (-1)^{\int(f_{L_{\ell_1}} \cup_{d-2} f_{L_{\ell_2}})}
\end{equation*}
simplifies to $\sigma(f) = \sigma(f_{L_1}) \cdots \sigma(f_{L_n})$ since each $(-1)^{\int(f_{L_{\ell_1}} \cup_{d-2} f_{L_{\ell_2}})}$ will return $1$. 

This shows that $\sigma(f) = \sigma^\text{gr}(f)$ always assuming each $\sigma(f_L) = \sigma^\text{gr}(f_L)$. Now we prove that each $\sigma(f_L) = \sigma^\text{gr}(f_L)$.

\subsubsection{Orientable case} \label{grassmannEqualsWindingOrientable}
In the case of an orientable manifold, we now verify that $\sigma(f_L) = \sigma^\text{gr}(f_L)$ for $f_L$ dual to a single closed loop $L$ on the dual 1-skeleton. 

The argument is inductive. In the winding definition, every segment of the loop that introduces a nonzero winding involves a $\pm \pi$ rotation of the vector $v_d$ along the dual 1-skeleton relative to the tangent of the loop. We thus choose to induct on the number of times the direction of the vector $v_d$ along the dual 1-skeleton switches direction ${\color{red} \leftarrow} \Rightarrow {\color{red} \rightarrow} \Rightarrow {\color{red} \leftarrow}$ along the course of the loop, where the tangent to the loop is in the $\rightarrow$ direction. This number of switches is always even. The idea is that each pair of direction switches can be `removed' from the loop independently from each other at the cost of introducing a sign of $\pm 1$ to $\sigma$. In the winding definition, we will get a sign of $+1$ or $-1$ depending on whether the direction switches came from a $0$ or $\pm 2\pi$ rotation. In the Grassmann definition, there will be some sign associated to reordering the Grassmann variables in the segment between those direction switches. These signs can be computed for each segment independently; the winding is additive and removing a segment does not affect the rest of the loop, while the reorderings of Grassmann variables on different segments are independent. If the winding and Grassmann definitions give the same sign for each pair of direction switches, and if the base case of no direction switches match, then this implies that both definitions of $\sigma$ agree on the loop.

Before proceeding, we recall from Appendix~\ref{w1PmAssignments_AndVectorFields} that within a $+$ $d$-simplex $\Delta_d$ while traversing along the dual 1-cell $\hat{i}$ away from the barycenter of $\Delta_d$ (i.e. near the boundary of $\Delta_d$), $v_d$ points towards the barycenter of $\Delta_d$ if $i$ is odd and away if $i$ is even, and vice-versa for $-$ $d$-simplex.
In parallel, recall from Section~\ref{grassIntegral} that (away from $w_1$) on a $+$ $d$-simplex, a Grassmann variable associated to $\hat{i}$ is labeled $\theta$/black if $i$ is even and is labeled $\overline{\theta}$/white if $i$ is odd, and vice-versa on a $-$ $d$-simplex. When we say that we are considering labelings of $(d-1)$-simplices, sign assignments, and Grassmann variables to the $d$-simplices which are compatible with a particular set of $v_d$, we mean that the labelings and sign assignments obey these rules.

\begin{figure}[h!]
    \centering
    \includegraphics[width=0.8\linewidth]{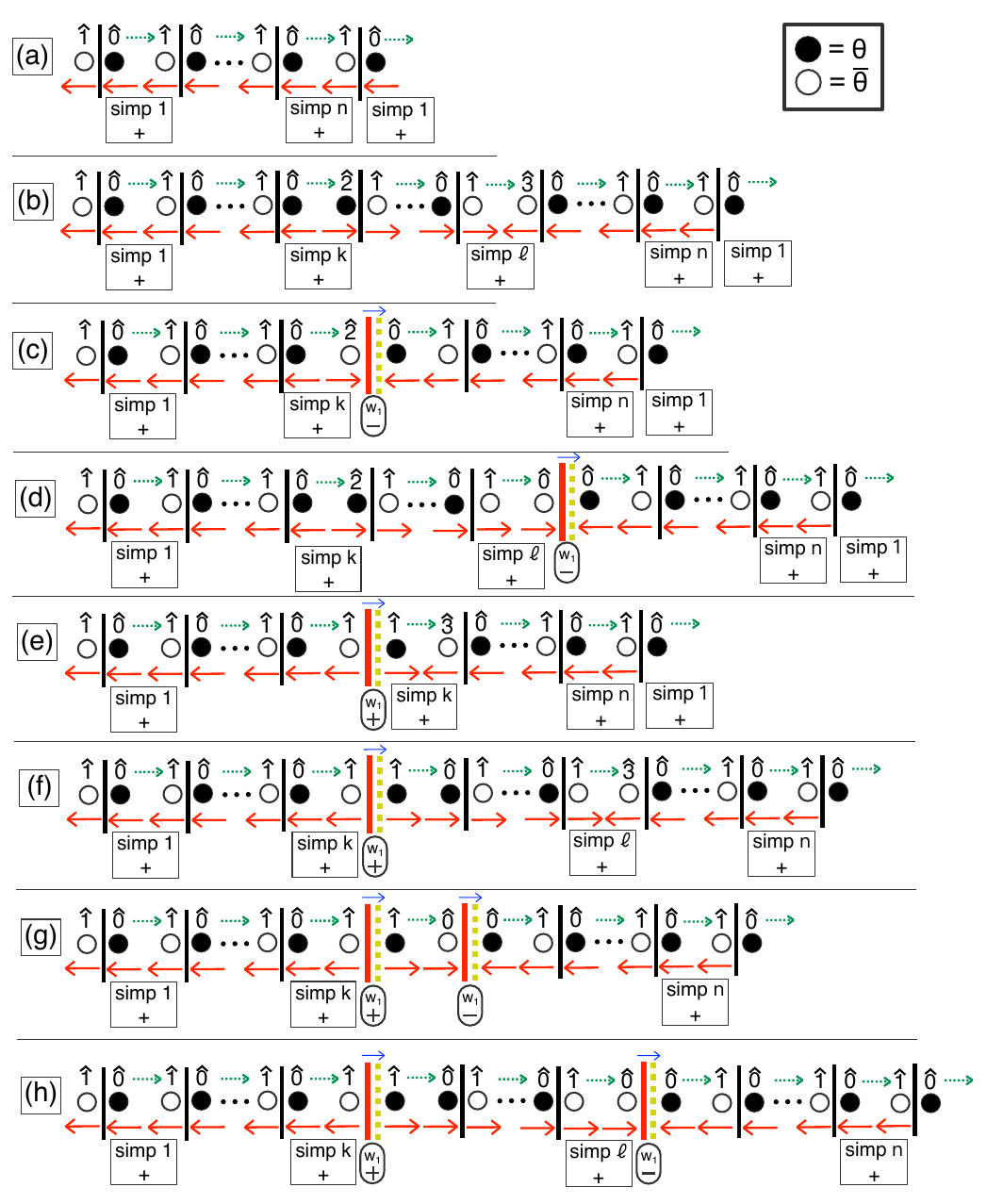}
    \caption{
    Examples of assignments of Grassmann variables going from $d$-simplices (labeled `simp') $1 \to n$ and corresponding $\hat{i} \to \hat{j}$ assignments for loops traversing $n$ $d$-simplices (where $\hat{i},\hat{j}$ refer to the $(d-1)$-simplex with vertices ordered as $i,j \in \{0,\cdots,d\}$ missing.). All $d$-simplices are $+$ oriented; see the main text for $-$ orientations. We consider traversing the loops from left to right. The red arrows on the bottom represent the directions of $v_d$ along the dual 1-skeleton that correspond to the choices of $\hat{i} \to \hat{j}$.  (a,b) give representative cases where there are no crossings with $w_1$. (c,d,e,f,g,h) give representative cases when there are crossings with $w_1$. Here, $w_1$ is located at the solid red and dashed yellow line. The solid red line represents $w_1$ and the dashed yellow line is the perturbation of $w_1$. The small blue arrow on top of $w_1$ gives a direction of perturbation of $w_1$ in the direction of $\theta \to \overline{\theta}$, which is used to define $w_1^2$ as described in the main text. The $(d-1)$-simplices representing $w_1$ are labeled as $\pm$ as described in Fig.~\ref{fig:vectsCrossingW1}.
    (a) The case when there are no direction switches of $v_d$. (b) The case when there are exactly two direction switches of $v_d$. (c), and (e) are edge cases of (d) and (f) respectively where a direction switch within a $d$-simplex involves a $(d-1)$-simplex on the orientation-reversing wall. (g) is an edge case of (h) where both crossings of $w_1$ occur on the boundary of the same $d$-simplex.
    }
    \label{fig:grassmannWinding}
\end{figure}

First, let us verify that the functions equal each other when there are no direction switches of $v_{d}$ along the dual 1-skeleton. For an example path see Fig.~\ref{fig:grassmannWinding}(a). Even though the figure shows some specific $\hat{i} \to \hat{j}$ and $\pm$ assignments on simplices, one can readily verify that the same computation follows through with different data compatible with the directions of $v_d$. The winding definition trivially gives $\sigma(f_L) = -1$, since there is no winding. Now consider the Grassmann integral definition. Labeling the $(d-1)$-simplices along the loop as $e_1, \dots, e_n$, we have 
\begin{equation*}
\sigma^\text{gr}(f_L) = \int \prod_{\Delta_{d-1}} d\theta_{\Delta_{d-1}} d\overline{\theta}_{\Delta_{d-1}} (\theta_{e_1} \overline{\theta}_{e_2})(\theta_{e_2} \overline{\theta}_{e_3}) \cdots (\theta_{e_{n-1}} \overline{\theta}_{e_n}) (\theta_{e_n} \overline{\theta}_{e_1}) = -1
\end{equation*}
So we have verified $\sigma = \sigma^\text{gr}$ in the base case.

Now let us consider through the example Fig.~\ref{fig:grassmannWinding}(b) a case where there are two direction switches of $v_d$. The winding definition for the specific choices in that figure produces
\begin{equation*}
\sigma(f_L) = - 
\begin{pmatrix}
1 & 0 & 1 & 0
\end{pmatrix}
\begin{pmatrix}
i X & 0 \\
0 & -i X
\end{pmatrix}
\begin{pmatrix}
i X & 0 \\
0 & -i X
\end{pmatrix}
\begin{pmatrix}
1 \\ 0 \\ 0 \\ 0
\end{pmatrix} = +1
\end{equation*}
since the winding matrices for $\hat{0} \to \hat{2}$ on $d$-simplex $k$ and $\hat{1} \to \hat{3}$ on $d$-simplex $\ell$ are both $\begin{pmatrix} i X & 0 \\ 0 & -i X \end{pmatrix}$ on $+$-oriented simplices. Similarly, one can check by writing out the Grassmann integral in this case that $\sigma^\text{gr}(f_L) = +1$.

Now we claim that this example is independent of the choices made in Fig.~\ref{fig:grassmannWinding}(b), specifically of the choices of $(d-1)$-simplex labelings $\hat{i}$ and $d$-simplex orientations $\pm$. First, recall that the winding matrices and the Grassmann variable orderings for $\hat{i} \to \hat{j}$ both only depend on whether $i = j \text{ (mod 2)}$, whether $i < j$, and the sign $\pm$ of the simplex; and the $i < j$ vs $i > j$ only matter if indeed $i = j \text{ (mod 2)}$. So, the results for $\hat{0} \to \hat{2}$ and $\hat{1} \to \hat{3}$ etc imply the same results for general $i < j$ both even and $i < j$ both odd. Next, note that every switch from $i < j$ both even (resp. odd) to $i > j$ both even (resp. odd) introduces a minus sign in the corresponding winding matrix (see Eq.~\eqref{windingGrassmannExpr1}), which changes $\sigma$ by a $-1$ and permutes $\vartheta_{\hat{i}}$ and $\vartheta_{\hat{j}}$ in the Grassmann integral, introducing one minus sign in $\sigma^\text{gr}$. So all such cases of $\hat{i} \to \hat{j}$ will have $\sigma$ and $\sigma^\text{gr}$ match. Since the example we gave consists of all $+$ simplices, we have now verified all cases compatible with the $v_d$ directions when all simplices are $+$.

Now we want to check, for these $v_d$ directions, the compatible cases when some simplices are $-$.
In particular, changing $i < j$ (resp. $i > j$) both even on a $+$ simplex to $i > j$ (resp. $i < j$) both odd on a $-$ simplex, will be compatible with the $v_d$ directions. By inspection, this leaves the winding matrices and Grassmann orderings unaffected. Likewise, changing $i < j$ (resp. $i > j$) both even on a $+$ simplex to $i < j$ (resp. $i > j$) both odd on a $-$ simplex will introduce a $-1$ to both $\sigma$ and $\sigma^\text{gr}$. 

The changes to Fig.~\ref{fig:grassmannWinding}(b) that we have considered above generate all possible $(d-1)$-simplex labelings and $d$-simplex orientations that are compatible with these $v_d$, and the winding and Grassmann definitions agree on all of them. This proves that $\sigma(f_L) = \sigma^\text{gr}(f_L)$ for loops which do not intersect an orientation-reversing wall.

\subsubsection{Non-Orientable case} \label{grassmannEqualsWindingNonorientable}

    We will use similar methods to prove that $\sigma(f_L) = \sigma^\text{gr}(f_L)$ for loops that cross $w_1$, and thus that $\sigma = \sigma^\text{gr}$. As before we will prove the statement for the case when the number of direction switches of $v_d$ is two, and this implies the general equivalence by induction. The main difference from the orientable case is that $v_d$ always switches direction when the loop crosses $w_1$; when this occurs, $\sigma$ picks up a factor of $\pm i$.
    The inductive step now follows from the short calculation that for any of the non-trivial winding matrices
    \begin{equation*}
    W_{1,\cdots,2n} \in \left\{ \pm \begin{pmatrix} i X & 0 \\ 0 & -iX \end{pmatrix} , \pm \begin{pmatrix} 0 & i Y \\ -i Y & 0 \end{pmatrix} \right\}
    \end{equation*}
    we have that 
    
    \begin{equation*}
        \begin{pmatrix} 1 & 0 & 1 & 0 \end{pmatrix} W_1 \cdots W_{2n} \begin{pmatrix} 1 \\ 0 \\ 0 \\ 0 \end{pmatrix} = \left( \begin{pmatrix} 1 & 0 & 1 & 0 \end{pmatrix} W_1 W_2 \begin{pmatrix} 1 \\ 0 \\ 0 \\ 0 \end{pmatrix} \right) \left( \begin{pmatrix} 1 & 0 & 1 & 0 \end{pmatrix} W_3 \cdots W_{2n} \begin{pmatrix} 1 \\ 0 \\ 0 \\ 0 \end{pmatrix} \right).
    \end{equation*}
    
    As before, we will pick some representative examples of paths with given arrow switches and locations of $w_1$ along the loop, and it will be easiest to illustrate in the cases where all the simplices are $+$ oriented. The representative examples we will check are drawn in Fig.~\ref{fig:grassmannWinding}(c-h). When considering a loop with two direction switches, we need to consider two classes of cases; either the loop crosses $w_1$ once or twice. In the latter case, since there are only two direction switches, the crossings must be on $(d-1)$-simplices with opposite induced orientation, as can be examined from Fig.~\ref{fig:grassmannWinding}(g,h).
    
    For the cases $(c),(d)$ in Fig.~\ref{fig:grassmannWinding}, there is exactly one crossing of $w_1$ at a $-$ oriented $(d-1)$ simplex; (c) is an edge case of (d) where the direction switch within a $d$-simplex involves a simplex on the orientation-reversing wall. First, we compute the winding definition of $\sigma(f_L) = - \begin{pmatrix} 1 & 0 & 1 & 0 \end{pmatrix} \begin{pmatrix} i X & 0 \\ 0 & -i X \end{pmatrix} \begin{pmatrix} 0 & i Y \\ -i Y & 0 \end{pmatrix} \begin{pmatrix} 1 \\ 0 \\ 0 \\ 0 \end{pmatrix} = +i$. The first matrix comes from the $\hat{0} \to \hat{2}$ on simplex $k$ and the second matrix comes from crossing $w_1$ at a $-$ simplex in the perturbing direction.
    Next we will compute $\sigma^\text{gr}(f_L)$. The crossing of $w_1$ is associated to a factor of $-i$. And, both of the cases of $(c),(d)$ can be seen to have a Grassmann integral part $\int \prod_{\Delta_{d-1}} d\theta_{\Delta_{d-1}} d\overline{\theta}_{\Delta_{d-1}} \prod_{\Delta_{d}} u({\Delta_{d}}) = -1$. So in total, $\sigma^\text{gr}(f_L) = +i$. So the definitions match for this example.
    
    The cases $(e),(f)$ in Fig.~\ref{fig:grassmannWinding} are the same as (c),(d) but the crossing of $w_1$ occurs at a $+$ oriented $(d-1)$ simplex. Then $\sigma(f_L) = - \begin{pmatrix} 1 & 0 & 1 & 0 \end{pmatrix} \begin{pmatrix} i X & 0 \\ 0 & -i X \end{pmatrix} \begin{pmatrix} 0 & -i Y \\ i Y & 0 \end{pmatrix} \begin{pmatrix} 1 \\ 0 \\ 0 \\ 0 \end{pmatrix} = -i$. In the Grassmann definition, the crossing with $w_1$ is associated to a factor of $+i$. And again, both the cases of $(e),(f)$ can be seen to have a Grassmann integral part $\int \prod_{\Delta_{d-1}} d\theta_{\Delta_{d-1}} d\overline{\theta}_{\Delta_{d-1}} \prod_{\Delta_{d}} u({\Delta_{d}}) = -1$. So in total, $\sigma^\text{gr}(f_L) = -i$. Again the definitions match for this example.
    
    In the cases $(g),(h)$ in Fig.~\ref{fig:grassmannWinding}, there are two crossings of $w_1$, one at a $+$ oriented $(d-1)$-simplex and one at a $-$ oriented one; (g) is an edge case of (h) where the crossings both occur on the boundary of the same $d$-simplex. The winding definition gives $\sigma(f_L) = - \begin{pmatrix} 1 & 0 & 1 & 0 \end{pmatrix} \begin{pmatrix} 0 & -i Y \\ i Y & 0 \end{pmatrix} \begin{pmatrix} 0 & i Y \\ -i Y & 0 \end{pmatrix} \begin{pmatrix} 1 \\ 0 \\ 0 \\ 0 \end{pmatrix} = +1$. In the Grassmann definition, the crossings of $w_1$ at a $+$ oriented $(d-1)$-simplex and at a $-$ oriented $(d-1)$-simplex give signs of $(+i) \cdot (-i)=1$. The Grassmann part can be shown to be $\int \prod_{\Delta_{d-1}} d\theta_{\Delta_{d-1}} d\overline{\theta}_{\Delta_{d-1}} \prod_{\Delta_{d}} u({\Delta_{d-1}}) = +1$, so $\sigma^\text{gr}(f_L)=+1$. Again, the definitions agree.
    
    Above we have only checked the equivalence explicitly for some representative examples. But these examples are sufficient to show the equivalence of $\sigma=\sigma^\text{gr}$ for the same reasons as the orientable case, that $\sigma(f_L)$ and $\sigma^\text{gr}(f_L)$ get changed by the same sign for any $\pm$ assignments of the $d$ simplices and $\hat{i} \to \hat{j}$ assignments compatible with the $v_d$ directions along the path. The only additional thing we have to worry about are how the perturbing directions of $w_1$ affect $\sigma$ and $\sigma^\text{gr}$. Changing the perturbing direction will change the winding matrix across $w_1$ by a sign, and it will switch the order of two Grassmann variables which adds a sign to $\sigma^\text{gr}(f_L)$. This implies that $\sigma = \sigma^\text{gr}$ in all the representative cases, and therefore for all cases.

\section{ Spin, pin, and $\mathcal{G}_f$ structures $\xi_{\mathcal{G}}$}
\label{twistedSpinSec}
        
        The purpose of this appendix is to derive the formula that defines a $\mathcal{G}_f$ structure $\xi_{\mathcal{G}}$ as a trivialization of the 2-cocycle $w_2 + w_1^2 + A_b^*\omega_2$, which is used in the definition of our path integral. Along the way, we provide some detailed explanations of spin and pin structures that may be useful for physicists, and which we have not seen in other treatments of the subject. 
        
        \subsection{ Spin structure review: algebraic definition}
        
        In order to define a quantum field theory path integral that describes a fermionic system with $G_f = \Z_2^f$ internal symmetry, it is well-known that a choice of spin structure $\xi$ is required. Mathematically, a spin structure $\xi$ is a trivialization of the 2nd Stiefel-Whitney class, $\delta \xi = w_2$, and two distinct spin structures on the space-time manifold $M^d$ are related to each other by $H^1(M^d, \Z_2)$. The need for a spin structure arises for the following reason. A relativistic quantum field theory, of which a TQFT can be thought of as a special case, possesses on flat space an $\SO(d)$ symmetry. However fermions, described in relativistic field theory by spinors, transform according to the double cover of $\SO(d)$, which is $\Spin(d)$. 
        To define the field theory on curved space-time, one needs to consider a $\SO(d)$ bundle, which can be thought of as the tangent bundle $TM$ of an orientable manifold $M$. The transition functions $\phi_{ij} \in \SO(d)$ between overlapping patches $U_i$ and $U_j$ must be lifted to $\tilde{\phi}_{ij} \in \Spin(d)$. The lift $\tilde{\phi}$ is determined by $\phi_{ij}$ together with a choice of sign $(-1)^{\xi_{ij}} = \pm 1$, for $\xi_{ij} = 0,1$ (mod 2), which we can thus refer to as the spin structure. The combination $\xi_{ij} + \xi_{jk} + \xi_{ki}$ is precisely the definition of the second Stiefel-Whitney class $w_2(M^d)$ in \v{C}ech cohomology. We thus find that the spin structure $\xi$ must satisfy $\delta \xi = w_2$. 
        
        It is useful to understand this in a bit more detail. First, note that we can define a cohomology class $[\overline{w}_2]$ as the non-trivial element in $\H^2(\SO(n), \Z_2) = H^2(B\SO(n), \Z_2) = \Z_2$. In particular, the class $[\overline{w}_2]$ specifies a non-trivial $\Z_2$ extension of $\SO(n)$, which is $\Spin(n)$. A general $\SO(d)$ bundle on $M^d$ can be understood as a map $\phi : M^d \rightarrow B\SO(d)$. Therefore, using the pullback $\phi^*$, one obtains a characteristic class $[w_2(M)] = \phi^* [\overline{w}_2] \in H^2(M^d, \Z_2)$. 
        
        Now consider our transition functions $\tilde{\phi}_{ij} \in \Spin(d)$, which are lifts of $\phi_{ij} \in \SO(d)$. We can in general write $\tilde{\phi}_{ij} = (\phi_{ij}, \xi_{ij})$, where $\xi_{ij} = 0,1$ (mod 2) determines the choice of lift. Now consider
        $\tilde{\phi}_{ij} \tilde{\phi}_{jk} \tilde{\phi}_{ik}^{-1}$. We have
        \begin{align}
            \tilde{\phi}_{ij} \tilde{\phi}_{jk} = (\phi_{ij} \phi_{jk}, \xi_{ij} + \xi_{jk} + \overline{w}_2(\phi_{ij}, \phi_{jk}))
        \end{align}
        Next, note that $\tilde{\phi}_{ik}^{-1}$, defined as the right inverse such that $\tilde{\phi}_{ik} \tilde{\phi}_{ik}^{-1} = 1$, is given by 
        \begin{align}
        \tilde{\phi}_{ik}^{-1} = (\phi_{ik}^{-1}, \xi_{ik} +  \overline{w}_2(\phi_{ik}, \phi_{ik}^{-1})). 
        \end{align}
        Therefore, we find
        \begin{align}
            \tilde{\phi}_{ij} \tilde{\phi}_{jk}\tilde{\phi}_{ik}^{-1} = (1, \xi_{ij} + \xi_{jk} + \xi_{ik} +  \overline{w}_2( \phi_{ij}, \phi_{jk})) . 
        \end{align}
        In order to have a well-defined $\Spin(d)$ bundle, we require
        $\tilde{\phi}_{ij} \tilde{\phi}_{jk}\tilde{\phi}_{ik}^{-1} = 1 \in \Spin(d)$, which leads us to
        \begin{align}
            \xi_{ij} + \xi_{jk} + \xi_{ik} = \overline{w}_2( \phi_{ij}, \phi_{jk})
        \end{align}
        In coordinate-free notation, this is precisely the \v{C}ech cohomology version of the equation
        \begin{align}
        \delta \xi = \phi^* \overline{w}_2 = w_2(M). 
        \end{align}
        
        It is useful to note that if a spin structure $\xi$ is not specified, we can canonically pick the lift $\phi_{ij} \rightarrow \tilde{\phi}_{ij} = (\phi_{ij},0)$. In this case, we can see that $w_2$ effectively acts as a source of fermion parity flux: a fermion traversing the three overlapping regions $U_i$, $U_j$, $U_k$ will pick up a minus sign according to $\overline{w}_2(\phi_{ij},\phi_{jk})$. This sign is effectively canceled by the specification of a spin structure $\xi$ that trivializes $w_2$. An alternate viewpoint of this is also explained in Appendix \ref{app:SpinStructReview}.
        
        When $G_f$ is a non-trivial central extension of $G_b$ by $\Z_2^f$, characterized by $[\omega_2] \in \H^2(G_b, \Z_2)$, then we need to consider a generalization which leads us to the notion of a $\mathcal{G}_f$ structure, $\xi_{\mathcal{G}}$. This can be understood as follows. 
        
        \subsection{ $\mathcal{G}_f$ structure derivation for unitary symmetries}
        
        In the case where $G_b$ only contains unitary symmetries, then we have the bosonic space-time symmetry group $\mathcal{G}_b = \SO(d) \times G_b$. The fermionic space-time symmetry group is $\mathcal{G}_f = (\Spin(d) \times G_f)/\Z_2$, where here the $\Z_2$ equivalence identifies the $-1$ element of $\Spin(d)$ with the fermion parity $\Z_2^f \subset G_f$. 
        
        Here, $\mathcal{G}_f$ can be understood as a $\Z_2$ extension of $\mathcal{G}_b$, specified by a class $[\overline{w}_2(\mathcal{G}_b)] \in \H^2( \mathcal{G}_b, \Z_2) = H^2( B\mathcal{G}_b, \Z_2)$.
        Then, one can define a lift of a $\mathcal{G}_b$ bundle to the double cover $\mathcal{G}_f$ with a \v{C}ech 1-cochain $\xi \in C^1(M^d,\Z_2)$. Letting $\phi_{ij} \in \mathcal{G}_b$ be the transition functions and $\tilde{\phi}_{ij} = (\phi_{ij}, \xi_{ij})$ be the lift, with $\xi_{ij} \in \Z_2$, we require $\tilde{\phi}_{ij} \tilde{\phi}_{jk} \tilde{\phi}_{ik}^{-1} = 1 \in \mathcal{G}_f$. If we specify the $\mathcal{G}_b$ bundle in terms of a map $\mathcal{A}_b : M^d \rightarrow B\mathcal{G}_b$, then this is equivalent to 
        \begin{align}
            \delta \xi = \mathcal{A}_b^* \overline{w}_2(\mathcal{G}_b)  . 
        \end{align}
        Next, note that, via the K\"{u}nneth decomposition, we have $\H^2(\mathcal{G}_b, \Z_2) = \H^2(\SO(d), \Z_2) \oplus \H^2(G_b, \Z_2)$. Therefore,
        \begin{align}
            \overline{w}_2(\mathcal{G}_b) = \overline{w}_2(\SO(d)) + \omega_2. 
        \end{align}
        It then follows that we have
        \begin{align}
        \label{etaGen}
            \delta \xi = w_2(M) + A_b^* \omega_2 ,
        \end{align}
        where $A_b: M \rightarrow BG_b$ is the projection of $\mathcal{A}_b: M \rightarrow B \mathcal{G}_b = B\SO(d) \times BG_b$ onto $BG_b$. 
        
        We can go through the above calculation in somewhat more detail as follows. Let us denote an element of $\mathcal{G}_b$ as $(\alpha, g)$, where $\alpha \in \SO(d)$ and ${\bf g} \in G_b$. Let $\tilde{\alpha} = (\sigma, \alpha)$ be an element in $\Spin(d)$, where $\sigma \in \{0,1\}$ specifies the lift of $\alpha \in \SO(d)$ to $\tilde{\alpha} \in \Spin(d)$, and $\tilde{{\bf g}} = ({\bf g},\mu) \in G_f$, where $\mu \in \{0,1\}$ specifies the lift of ${\bf g} \in G_b$ to $\tilde{{\bf g}} \in G_f$. Finally we denote an element of $\mathcal{G}_f$ as $(\sigma, \alpha, {\bf g}, \mu)$. The equivalence relation under $\Z_2$ implies $(\sigma, \alpha, {\bf g}, \mu) \sim (\sigma + \mu, \alpha, {\bf g}, 0)$. Let us now consider a transition function $\phi_{ij} \in \mathcal{G}_b$, and its lift $\tilde{\phi}_{ij} \in \mathcal{G}_f$. We thus have
        \begin{align}
            \tilde{\phi}_{ij} = (\sigma_{ij}, \alpha_{ij}, {\bf g}_{ij}, \mu_{ij}) \sim (\sigma_{ij} + \mu_{ij}, \alpha_{ij}, {\bf g}_{ij},0) = (\xi_{ij}, \alpha_{ij}, {\bf g}_{ij}, 0),
        \end{align}
        where we defined $\xi_{ij} = \sigma_{ij} + \mu_{ij}$. 
        
        Now, to have a well-defined $\mathcal{G}_f$ bundle, we require
        \begin{align}
        \tilde{\phi}_{ij} \tilde{\phi}_{jk} \tilde{\phi}_{ik}^{-1} = 1
        \end{align}
        First, note that 
        \begin{align}
            \tilde{\phi}_{ij} \tilde{\phi}_{jk} = (\xi_{ij} + \xi_{jk} + \bar{w}_2(\alpha_{ij}, \alpha_{jk}), \alpha_{ij} \alpha_{jk}, {\bf g}_{ij} {\bf g}_{jk}, \omega_2({\bf g}_{ij}, {\bf g}_{jk})). 
        \end{align}
        Next, consider
        \begin{align}
            \tilde{\phi}_{ik} \tilde{\phi}_{ik}^{-1} &= 1
            \nonumber \\
            \tilde{\phi}_{ik} \tilde{\phi}_{ki} &= (\xi_{ik} + \xi_{ki} + \bar{w}_2(\alpha_{ik}, \alpha_{ki}), \alpha_{ik} \alpha_{ki}, {\bf g}_{ik} {\bf g}_{ki}, \omega_2({\bf g}_{ik}, {\bf g}_{ki}) ) 
            \nonumber \\
            &= (\bar{w}_2(\alpha_{ik}, \alpha_{ki}), 1, \omega_2({\bf g}_{ik}, {\bf g}_{ki})) 
            \nonumber \\
            &= (\bar{w}_2(\alpha_{ik}, \alpha_{ki}) + \omega_2({\bf g}_{ik}, {\bf g}_{ki}), 1, 0) 
        \end{align}
        Here we use the fact that ${\bf g}_{ik} = {\bf g}_{ki}^{-1}$, $\alpha_{ik} = \alpha_{ki}^{-1}$, and $\xi_{ik} = \xi_{ki}$.
        Therefore, we see that the right inverse of $\tilde{\phi}_{ik}$ is:
        \begin{align}
            \tilde{\phi}_{ik}^{-1} = \tilde{\phi}_{ki} \times (\bar{w}_2(\alpha_{ik}, \alpha_{ki}) + \omega_2({\bf g}_{ik}, {\bf g}_{ki}), 1, 0) . 
        \end{align}
        Therefore, 
        \begin{align}
            \tilde{\phi}_{ij} \tilde{\phi}_{jk} \tilde{\phi}_{ik}^{-1} = & 
            (\xi_{ij} + \xi_{jk} + \xi_{ki} + \bar{w}_2(\alpha_{ij}, \alpha_{jk})+  \bar{w}_2(\alpha_{ik}, \alpha_{ki}), 1, 1, \omega_2({\bf g}_{ij}, {\bf g}_{jk}) + \omega_2({\bf g}_{ik}, {\bf g}_{ki})) 
            \nonumber \\
            &\times (\bar{w}_2(\alpha_{ik}, \alpha_{ki}) + \omega_2({\bf g}_{ik}, {\bf g}_{ki}), 1, 0)
            \nonumber \\
            = &(\xi_{ij} + \xi_{jk} + \xi_{ki} + \bar{w}_2(\alpha_{ij}, \alpha_{jk}) + \omega_2({\bf g}_{ij}, {\bf g}_{jk}), 1,1, 0) ,
        \end{align}
        where we have used that $\alpha_{ij} \alpha_{jk} \alpha_{ki} = 1$ and similarly for ${\bf g}_{ij}$. 
        
        We see that the $\mathcal{G}_f$ bundle is determined from the $\mathcal{G}_b$ bundle by the choice of lift characterized by $\xi_{ij}$, and to have a well-defined $\mathcal{G}_f$ bundle we need
        \begin{align}
            \delta \xi [ijk] + \bar{w}_2(\alpha_{ij}, \alpha_{jk}) + \omega_2({\bf g}_{ij}, {\bf g}_{jk}) = 0 . 
        \end{align}
        Expressing this in coordinate-free notation then gives Eq.~\eqref{etaGen}. 
        
        As an example, let us consider the case of $G_f = \U^f$ (i.e. the group $\U$ where the $\pi$ rotation equals fermion parity). In this case $\xi_{\mathcal{G}}$ should be a $\Spin^c$ connection, which requires
        \begin{align}
            \delta \xi_{\Spin^c} = w_2 + c \text{ mod 2} ,
        \end{align}
        where $c$ is the first Chern class of the $\U$ bundle specified by $A_b$. To check this against our explicit formula above, we need to see that $A_b^*\omega_2$ is the mod $2$ reduction of the first Chern class.
        For $\U^f$, we have $\omega_2(a,b) = (-1)^{a + b - [a+b]}$, where the $\U$ group elements are $e^{2\pi i a}$, $e^{2\pi i b}$ and $[a] \equiv a \text{ mod } 1$. We thus have $\omega_2(A_b[01], A_b[12]) = e^{i \pi (A_b[01] + A_b[12] - [A_b[01] + A_b[12]])}$. For a flat $\U$ gauge field, we have
        $A_b[01] + A_b[12] -A_b[13] \in \Z$. Therefore we can write $A_b[01] + A_b[12] = n + A_b[13]$, and $[ A_b[01] + A_b[12] ] = A_b[13]$. Therefore 
        \begin{align}
        \omega_2(A_b[01], A_b[12]) = e^{i \pi (A_b[01] + A_b[12] - A_b[13])} = (-1)^{dA},
        \end{align}
        which is precisely the mod $2$ reduction of the first Chern class. 
        
        \subsection{ $\mathcal{G}_f$ structure derivation for anti-unitary symmetries}
        \label{GfantiU}
        
        The generalization to the case where $G_b$ contains anti-unitary symmetries is given by
        \begin{align}
        \delta \xi_{\mathcal{G}} = w_2 + w_1^2 + A_b^* \omega_2. 
        \end{align}
        
        Let us go through the derivation of this in the case where $G_b$ is a split extension of $\Z_2^{\bf T}$ by the unitary subgroup $G_b^u$: $G_b = G_b^u \times \Z_2^{\bf T}$ or $G_b = G_b^u \rtimes \Z_2^{\bf T}$.
        In this case, the two cases above imply that 
        \begin{align}
        \mathcal{G}_b =
        \begin{cases}
        O(d) \ltimes G_b^u = \Z_2^{\bf T} \ltimes [G_b^u \times \SO(d)] = G_b \ltimes \SO(d)
        \\
        O(d) \times G_b^u 
        \end{cases}
        \end{align}
        
        Next, we need to lift the $\mathcal{G}_b$ bundle to its double cover $\mathcal{G}_f$. Here, there is an important subtlety. We are interested in the Euclidean space-time symmetry group for the fermions. This requires us to do a Wick rotation, which changes ${\bf T}^2 = 1$ to ${\bf T}^2 = (-1)^F$ in $\mathcal{G}_f$ and vice versa \cite{Kapustin:2014dxa,witten2016rmp}.
        Therefore, we consider a different extension $[\omega_2^E] \in \mathcal{H}^2(G_b, \Z_2)$, given by 
        \begin{align}
            \omega_2^E = \omega_2 + \epsilon_2.
        \end{align}
        Here $\epsilon_2 = s^{\ast}\overline{\epsilon}_2$, where $[\overline{\epsilon}_2]$ is the non-trivial class in $\H^2(\Z_2^{\bf T}, \Z_2)=\Z_2$ and $s:G_b \rightarrow \Z_2^{\bf T}$ is non-trivial on anti-unitary symmetries. If we let $\phi: M^d \rightarrow B\Z_2$ specify the orientation bundle on $M^d$, then $\phi^*\epsilon_2 = w_1^2$. 
        
        Then, $\mathcal{G}_f$ should be considered to be a $\Z_2$ extension of $\mathcal{G}_b$ specified by \footnote{In the semi-direct product case, one can use the Lyndon-Hochschild-Serre spectral sequence together with some general arguments to derive Eq.~\eqref{Gbcoho}. }
        \begin{align}
        \label{Gbcoho}
        \overline{w}_2(\mathcal{G}_b) \in \H^2(\mathcal{G}_b, \Z_2) = \H^2(G_b,\Z_2) \oplus \H^2(\SO(d), \Z_2).
        \end{align}
        We thus have
        \begin{align}
            \overline{w}_2(\mathcal{G}_b) = \overline{w}_2(\SO(d)) + \omega_2^E
        \end{align}
        Then, to specify the lift of the $\mathcal{G}_b$ bundle to a $\mathcal{G}_f$ bundle, we need a choice of $\xi \in \Z_2$, such that 
        \begin{align}
            \delta \xi &= \mathcal{A}_b^* \overline{w}_2(\mathcal{G}_b) = w_2 + A_b^* ( \omega_2 + \epsilon_2) 
            \nonumber \\
            &= w_2 + w_1^2 + A_b^* \omega_2 
            \label{generalEta}
        \end{align}

       As an example, consider $G_b = \Z_2^{\bf T}$ with trivial $\omega_2$, so that $G_f = \Z_2^{\bf T} \times \Z_2^f$. In this case Eq.~\eqref{generalEta} gives $\delta \xi_{\mathcal{G}} = w_2 + w_1^2$, which is what we expect for $\Pin^-$ structures. Next consider $G_b = \Z_2^{\bf T}$ with $G_f = \Z_4^{ {\bf T}, f}$. In this case $\omega_2({\bf T}, {\bf T}) = 1$. Then we identify $A_b = w_1$, so $\omega_2(A_b[01], A_b[12]) = w_{01}w_{12}$, so $A_b^*\omega_2 = w_1^2$. It then follows that Eq.~\eqref{generalEta} reduces to $\delta \xi_{\mathcal{G}} = w_2$, which is what we expect for a $\Pin^+$ structure. 

        As another example, consider $G_b = \U \rtimes \Z_2^{\bf T}$ and $G_f = [\U^f \rtimes \Z_4^{{\bf T},f}]/\Z_2$. Here we have $\omega_2((z,t), (z',t') ) = z + \,^{t'}z' - [z + \,^{t'}z'] + tt'$, where $z \in [0,1)$ parameterizes the $\U$ part, $t = 0,1$ parameterizes the $\Z_2^{\bf T}$ part, and $[a] = a \text{ mod } 1$. Note that $\,^t z = -z$ if $t = 1$ to account for the non-trivial action of $\Z_2^{\bf T}$ on $\U$. The gauge field $A_b[ij] = (A_b^u[ij], w_1[ij])$ where $A_b^u$ is the unitary part and $w_1$ sets the anti-unitary part to coincide with the orientation bundle. In this case, Eq.~\eqref{generalEta} reduces to $\delta \xi_{\mathcal{G}} = w_2 + (w_1^2 + \tilde{F} \mod 2) = w_2 + w_1^2 + \tilde{F} \mod 2$. Here $\tilde{F}[012] = A_b^u[01] + w_1[01]A_b^u[12] - A_b^u[02]$ is an element of the local cohomology $H^2(M, \Z)$ which is twisted by $w_1$. 

        The case where $G_b = \U \times \Z_2^{\bf T}$ and $G_f = \U^f \rtimes \Z_2^{\bf T}$ is similar except $\omega_2((z,t), (z',t') ) = z + \,^{t'}z' - [z + \,^{t'}z']$, so that $\delta \xi_{\mathcal{G}} = w_2 + \tilde{F} \mod 2$, which defines a $\Pin_-^{\tilde{c}}$ structure. 

        Let us consider the case $G_b = G_b^u \times \Z_2^{\bf T}$ and $G_f = G_b^u \times \Z_2^{\bf T} \times \Z_2^f$. In this case, $\omega_2$ is trivial, so we expect $\delta \xi_{\mathcal{G}} = w_2 + w_1^2$, which gives a $\Pin^-$ structure. If we instead consider $G_f = G_b^u \times \Z_4^{{\bf T}, f}$, then $A_b^* \omega_2$ reduces to $w_1^2$, and we get $\delta \xi_{\mathcal{G}} = w_2$, which is a $\Pin^+$ structure.

\begin{table*}[t]
	\centering
	\begin{tabular} {l |l |l |l}
      $G_b$ & $\mathcal{G}_b$ & $G_f$ & $\mathcal{G}_f$ \\
      \hline
      && $G_b^u \times \Z_2^{\bf T} \times \Z_2^f$ &$\Pin^- \times G_b^u$ \\
   $G_b^u \times \Z_2^{\bf T}$  & $O(d) \times G_b^u$ & $G_b^u \times \Z_4^{{\bf T},f}$  & $\Pin^+ \times G_b^u$ \\
      && $G_f^u \times \Z_2^{\bf T}$ &  $[\Pin^- \times G_f^u]/\Z_2$ \\
      && $[G_f^u \times \Z_4^{ {\bf T}, f}]/\Z_2$ &  $[\Pin^+ \times G_f^u]/\Z_2$ \\
          \hline
      && $G_b^u \rtimes \Z_2^{\bf T} \times \Z_2^f$ & $\Pin^- \ltimes G_b^u$ \\
   $G_b^u \rtimes \Z_2^{\bf T}$ & $O(d) \ltimes G_b^u $  & $G_b^u \rtimes \Z_4^{{\bf T},f}$ & $\Pin^+ \ltimes G_b^u$ \\
      && $G_f^u \rtimes \Z_2^{\bf T}$ & $[\Pin^- \ltimes G_f^u]/\Z_2$ \\
      && $[G_f^u \rtimes \Z_4^{ {\bf T}, f}]/\Z_2$ & $[\Pin^+ \ltimes G_f^u]/\Z_2$ \\
          \hline
	\end{tabular}
        \caption{\label{GtildeList} Some possible cases for $G_f$ and $\mathcal{G}_f$ for $G_b = G_b^u \times \Z_2^{\bf T}$ and $G_b = G_b^u \rtimes \Z_2^{\bf T}$.}
\end{table*}

We can also consider the case of the symmetry groups associated with the ``relativistic'' 10-fold way, as shown in Table \ref{10foldway}, in order to connect our results to periodic table of free fermion topological insulators and superconductors.

\begin{table*}[t]
	\centering
	\begin{tabular} {l |l |l||l}
          Cartan & $G_b$ & $G_f$ & $\mathcal{G}_f$  \\
          \hline
          A & $\U$ & $\U^f$  & $\Spin^c$ \\
          AI & $\U \rtimes \Z_2^{\bf T}$ & $\U^f \rtimes \Z_2^{\bf T}$ & $[\Pin^- \ltimes \U]/\Z_2$\\
          AII & $\U \rtimes \Z_2^{\bf T}$ & $[\U^f \rtimes \Z_4^{ {\bf T},f}]/\Z_2$ & $[\Pin^+ \ltimes \U]/\Z_2$ \\
          AIII & $\U \times \Z_2^{\bf T}$ &  $\U^f \times \Z_2^{\bf T}$ & $\Pin^c$ \\
          D & 1 & $\Z_2^f$ & $\Spin$ \\
          DIII & $\Z_2^{\bf T}$ & $\Z_4^{ {\bf T}, f}$ & $\Pin^+$ \\
          BDI & $\Z_2^{\bf T}$ & $\Z_2^{\bf T} \times \Z_2^f$ & $\Pin^-$ \\
          C & $\SO(3)$ & $\SU(2)^f = [\SU(2) \times \Z_2^f]/\Z_2$ & $[\Spin\times \SU(2)]/\Z_2$ \\
          CI & $\SO(3) \times \Z_2^{\bf T}$ & $(\SU(2)^f \times \Z_4^{{\bf T},f})/\Z_2$ & $[\Pin^+ \times \SU(2)]/\Z_2$ \\
          CII & $\SO(3) \times \Z_2^{\bf T}$ & $\SU(2)^f \times \Z_2^{\bf T}$ & $[\Pin^- \times \SU(2)]/\Z_2$ \\          
          \hline
	\end{tabular}
        \caption{\label{10foldway} Relativistic 10-fold way and the relevant $\mathcal{G}_f$ groups. Note that $\Pin^c = [\Pin^+ \times \U]/\Z_2 \simeq[\Pin^- \times \U]/\Z_2$. $\U^f$ and $\SU(2)^f$ denote the $\U$ and $\SU(2)$ groups with the $-1$ element identified with fermion parity. A similar table was also given in Ref.~\cite{guo2018}}
\end{table*}
        
\section{ The $f_\infty$ map and turning cochains into chains} \label{appendixFInfty}
Here we describe the $f_\infty$ map of~\cite{Thorngrenthesis} that shows up in several places in our constructions. 

\subsection{ Definition of $f_\infty$}
\label{app:FInfty}

On a manifold $M^d$ equipped with a branched triangulation, there is a map
\begin{align}
f_\infty : Z^k(M^d,\Z_2) \to Z_{d-k}(M^d,\Z_2)
\end{align}
that turns a $\Z_2$ cocycle on the triangulation of $M^d$ to a cycle on the triangulation of $M^d$. Note that this is distinct from the usual cochain-level Poincaré duality, which maps $k$-cocycles on the triangulation to $(d-k)$-cycles on the dual cellulation. One can thus think of $f_\infty$ as implementing a somewhat different cochain-level Poincaré duality. Since $\Z_2$ cocycles are dual to submanifolds living on the dual cellulation of $M$, one can equivalently think of this map as taking submanifolds living on the dual cellulation to submanifolds living on the original triangulation of a manifold. This map readily generalizes to more general coefficient groups and non-closed cochains, but we will not need it for our purposes. 

Recall that chains $C_{d-k}(M,\Z_2)$ uniquely represent the most general kind of linear function on the cochains $C^{d-k}(M,\Z_2)$. In particular, any linear functional $C^{d-k}(M,\Z_2) \to \Z_2$ can be represented as $\alpha \mapsto \int_c \alpha$ for some choice of chain $c \in C_{d-k}(M,\Z_2)$. The main idea of the $f_\infty$ map is that the cup product pairing $C^{d-k}(M,\Z_2) \times C^{k}(M,\Z_2): (\alpha, \beta) \mapsto \int_M \alpha \cup \beta$ for some fixed $\beta$ is also a linear functional on $\alpha$. This observation leads us to the definition of $f_\infty \beta$. Linearity tells us that $\beta$ can be mapped to a \textit{unique} chain, which we will call $f_\infty \beta$, for which $\int \alpha \cup \beta = \int_{f_\infty \beta} \alpha$ for all $\alpha \in C^{d-k}(M,\Z_2)$. The fact that (away from $\partial M$) $f_\infty \beta$ is a \textit{closed} submanifold for $\beta$ closed is because $\int \delta\lambda \cup \beta = 0$ for any $\lambda \in C^{d-k-1}(M,\Z_2)$, and a chain $c$ representing a nonclosed collection of $(d-k)$-simplices will always have a $\lambda$ for which $\int_c \delta \lambda = \int_{\partial c}\lambda \neq 0$.

\begin{figure}[h!]
  \centering
  \includegraphics[width=\linewidth]{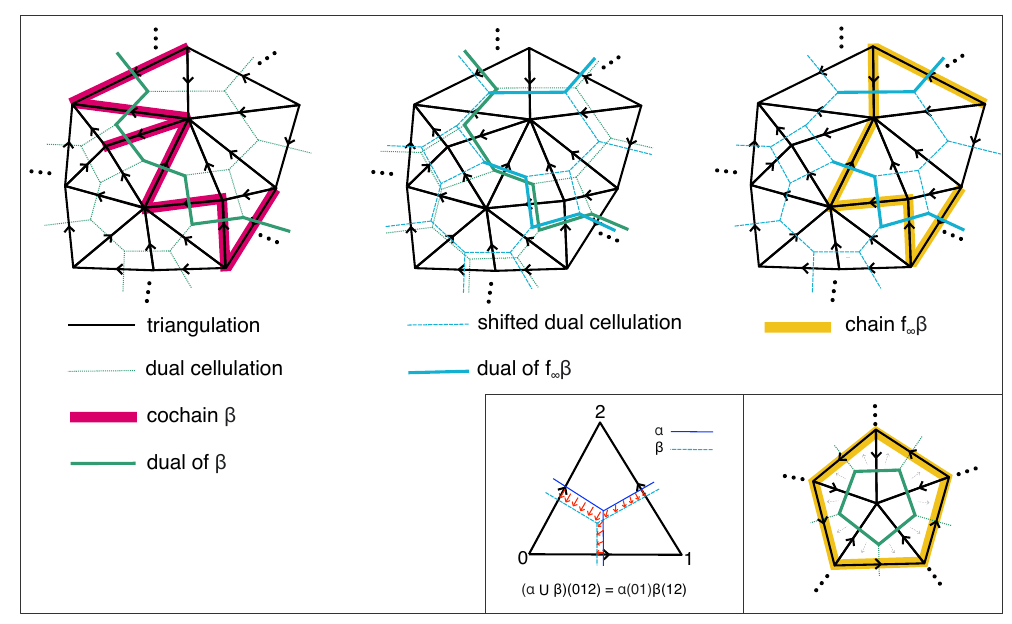}
  \caption{Illustration of $f_\infty \beta$ in $d=2$. The triangle on the bottom right illustrates a way to think of $(\alpha \cup \beta)(012)=\alpha(01)\beta(12)$ as an intersection of the dual of $\alpha$ with a shifted version of $\beta$. Note that the only intersection of $\alpha$ with $\beta$ occurs with $\alpha(01)$ and $\beta(12)$. To compute $f_\infty \beta$ for $\beta \in Z^1(M^2,\Z_2)$, we collect all the dual 1-simplices for which their shifted version intersects the dual of $\beta$. Then, the dual of this dual collection becomes $f_\infty \beta$. In the bottom-right panel, a simpler example illustrates how $f_\infty$ can be visualized as flowing the dual of a cochain opposite to the branching structure.}
  \label{f_infty_illustration}
\end{figure}

To actually compute $f_\infty \beta$, one needs to find all of the $(d-k)$-simplices $S$ for which the indicator cochain $\alpha_S$ satisfies $\int \alpha_S \cup \beta = 1$. We illustrate a $d=2$ example for this in Figure~\ref{f_infty_illustration}. Note that the end result of the chain can roughly be visualized as `flowing' the dual of $\beta$ opposite to the branching structure until it reaches the original triangulation. This perspective is expanded on in~\cite{Thorngrenthesis} in reference to a so-called `Morse Flow' associated to the branching structure, and we reference it in our discussions about how the geometry is encoded in the diagrammatics.

\subsection{ Use of $f_\infty$ to define orientation-reversing walls} \label{usingFInftyOrientationWall}

\begin{figure}[h!]
 \centering
 \includegraphics[width=0.28\linewidth]{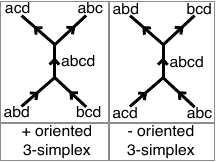}
 \caption{Induced orientations on a 3-simplex and how they appear in a 15j symbol.}
 \label{inducedOrientation3Simplex}
\end{figure}

As discussed in Sec.~\ref{sec:geometricChoices}, the $G_b$ gauge field determines an element $A_b^{\ast}s \in Z^1(M,\Z_2)$, where $s:G_b \rightarrow \Z_2$ is non-trivial on anti-unitary group elements. The interpretation of $A_b^{\ast}s$ is that it is nonzero on 1-simplices across which the local orientation is reversed. The object that plays a crucial role in our constructions of both $z_c$ and $Z_b$ is the orientation-reversing wall $w_1 \in Z^1(M^{\vee},\Z_2) \sim Z_{d-1}(M,\Z_2)$. The $f_{\infty}$ map is a natural way to obtain a cochain representative of $w_1$ from a cochain representative of $A_b$ via
\begin{equation}
w_1 := f_\infty A_b^*s.    
\label{eqn:w1FInfty}
\end{equation}
As mentioned in Sec.~\ref{subsubsec:shadowChoices}, using this representative is important to ensure that the bosonic shadow is independent of various choices. It also plays a role in specifying the anomaly of $Z_b$, as described in Appendix~\ref{app:abOmegaAnomalyCalc}. In particular, to ensure that the shadow is independent of our trivalent resolution of the five objects that meet at a 3-simplex, we need the $F$- and $R$-moves relating different trivalent resolutions to cancel. This cancellation requires consistency with the way the diagrams are drawn in both 4-simplices' 15j symbols, which is related to the orientations induced on the 3-simplex from the 4-simplices as in Fig.~\ref{inducedOrientation3Simplex}.

\begin{figure}[h!]
  \centering
  \includegraphics[width=\linewidth]{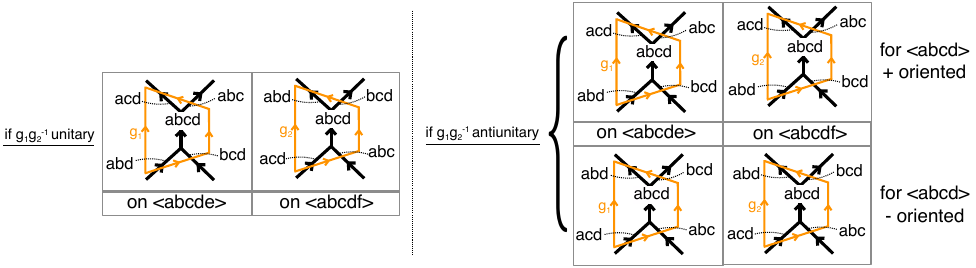}
  \caption{Induced orientations for a 3-simplex $\braket{abcd}$ (ordered $a \to b \to c \to d$) compared to the group elements ${\bf g}_1, {\bf g}_2$ surrounding them in 15j symbols of the 4-simplices $\braket{a b c d e},\braket{a b c d f}$ (unordered, where $e$,$f$ can be in \textit{any} position relative to $a \to b \to c \to d$). (Left) In the case ${\bf g}_1 {\bf g}_2^{-1}$ is unitary, $\braket{abcd}$ has opposite induced orientations on $\braket{a b c d e},\braket{a b c d f}$. (Right) In the case ${\bf g}_1 {\bf g}_2^{-1}$ is anti-unitary, $\braket{abcd}$ has the same induced orientation on $\braket{a b c d e},\braket{a b c d f}$, either both $+$ or both $-$ as in Fig.~\ref{inducedOrientation3Simplex}. The shared induced orientation means that $w_1(\braket{abcd}) \neq 0 \in \Z_2$, so that $\braket{abcd}$ is labeled as a $\pm$ $(d-1)$-simplex on the orientation-reversing wall.}
  \label{inducedOrientationUnitarity}
\end{figure}

We claim that the above representative of $w_1$ causes the following fact to be true: a 3-simplex lies on the orientation-reversing wall of a closed manifold $M$ (i.e., it has the same induced orientation from both 4-simplices which contain it), if and only if a domain wall for an anti-unitary group element surrounds that 3-simplex in exactly one 15j symbol. That is, the 3-simplices would then appear in 15j symbols in the way shown in Fig.~\ref{inducedOrientationUnitarity}. By inspection, this fact means that $F$- and $R$-moves do indeed cancel and the shadow is independent of trivalent resolution.

To see this, first examine the 15j symbols. Suppose that the 3-simplex $\braket{abcd}$, with vertices ordered $a \to b \to c \to d$ according to the branching structure, appears in the two 4-simplices $\braket{abcde}$ and $\braket{abcdf}$, where this notation does \textit{not} specify how $e$ or $f$ are ordered in the branching structure. Then in the 15j symbol for $\braket{abcde}$, $\braket{abcd}$ appears surrounded by a ${\bf g}_1$ domain wall, where 
\begin{equation}
    {\bf g}_1 = 
    \begin{cases}
        {\bf g}_{ed} & \text{branching structure orders }a \to b \to c \to d \to e \\
        {\bf 1} & \text{else}
    \end{cases}.
\end{equation}
Likewise, in the 15j symbol for $\braket{abcdf}$, $\braket{abcd}$ appears surrounded by a ${\bf g}_2$ domain wall where 
\begin{equation}
    {\bf g}_2 = \begin{cases}
    {\bf g}_{fd} & \text{branching structure orders }a \to b \to c \to d \to f \\
    {\bf 1} & \text{else}
    \end{cases}.
\end{equation}
Note that $s({\bf g}_1 {\bf g}_2^{-1})$ is the non-trivial element of $\Z_2$ if and only if exactly one of ${\bf g}_1$ or ${\bf g}_2$ is anti-unitary. Thus a domain wall for an anti-unitary group element surrounds $\braket{abcd}$ in exactly one 15j symbol if and only if $s({\bf g}_1 {\bf g}_2^{-1})$ is the non-trivial element of $\Z_2$. 

Although they are motivated by the 15j symbols, the group elements ${\bf g}_{1,2}$ are well-defined independent of the diagrams. With that definition, one may directly compute for a general branching structure that
\begin{equation}
f_\infty A_b^*s (\braket{abcd}) = s({\bf g}_1 {\bf g}_2^{-1}).
\end{equation}
By definition $w_1(\braket{abcd})$ is nonzero if and only if $\braket{abcd}$ is on the orientation-reversing wall; with our choice of representative of $w_1$ given by Eq.~\eqref{eqn:w1FInfty}, $\braket{abcd}$ is on the orientation-reversing wall if and only if $s({\bf g}_1 {\bf g}_2^{-1})$ is the non-trivial element of $\Z_2$. As shown in the previous paragraph, these are also exactly the 3-simplices which are surrounded by a domain wall for an anti-unitary group element in exactly one 15j symbol; this proves our assertion.

\subsection{ Perturbation of orientation-reversing wall and $w_1^2$ in relation to $f_\infty$} \label{fInftyAndPerturbationOf_w1}

Now that we have explained how the gauge field determines a representative orientation-reversing wall $w_1 \in Z_3(M,\Z_2)$ and how this representative of $w_1$ is encoded in the diagrammatics, we explain the same for a representative of $w_1^2 \in Z_2(M,\Z_2)$. Given Eq.~\eqref{eqn:w1FInfty}, there is a natural representative
\begin{equation}
    w_1^2 = f_{\infty} \left(A_b^\ast s \cup A_b^\ast s \right).
\end{equation}
Recall that $w_1^2$ is the intersection of $w_1$ with a perturbed version of itself. Note that there are two distinct perturbations of $w_1$ compatible with $w_1^2$: any given one and its reversal. It is important to distinguish these and fix one, since the Grassmann integral depends not only on a cochain representative of $w_1^2$, but also the particular perturbation. We claim that $f_{\infty}$ actually encodes such a perturbation. Specifically, take a particular 3-simplex $\braket{abcd}$, which appears in two 4-simplices. Suppose that in the 15j symbols, it is surrounded by group elements ${\bf g}_1,{\bf g}_2$ as in Fig.~\ref{inducedOrientationUnitarity}. If $\braket{abcd}$ is part of the orientation-reversing wall, then exactly one of the ${\bf g}_1,{\bf g}_2$ will have an anti-unitary action. We claim that $f_{\infty}$ encodes a perturbation of $w_1$ into the 4-simplex whose surrounding bubble is anti-unitary.

This follows from the interpretation in~\cite{Thorngrenthesis} of $f_\infty$ corresponding to a `Morse flow' of the dual of the cochain to the output chain (see Fig.~\ref{f_infty_illustration}).
We can imagine flowing the dual of $A_b^* s \in Z^1(M,\Z_2)$ along some vector fields until it reaches the orientation-reversing wall $f_\infty A_b^* s$. The Morse flow works as follows. For each edge $e$ for which $A_b^*s(e) \neq 0$, break its dual 3-cell into pieces, each of which is contained in a single 4-simplex. For a given piece contained in, say, 4-simplex $\braket{01234}$ (where the vertices are given in order of the branching structure), there are two possibilities. If $e=\braket{34}$ for this 4-simplex, then the piece flows within $\braket{01234}$ to the 3-simplex $\braket{0123}$. Otherwise, the piece degenerates under the flow. This connects to the diagrammatics because as shown in Sec.~\ref{usingFInftyOrientationWall}, a 15j symbol for $\braket{01234}$ contains an anti-unitary domain wall loop if and only if ${\bf g}_{34}$ is anti-unitary, i.e. if $A_b^*s({\braket{34}}) \neq 0$. Hence, the piece of the 3-cell which has a nondegenerate flow is, at any intermediate point in the flow, always contained inside the 4-simplex whose 15j symbol contains an anti-unitary domain wall loop. Therefore, any 3-simplex in $f_{\infty} A_b^* s$ originates from the flow of portions of 3-cells contained in 4-simplices whose 15j symbol contains an anti-unitary domain wall loop 
\footnote{If both ${\bf g}_1, {\bf g}_2$ are anti-unitary, then the 3-simplex $\braket{0123}$ is approached from nondegenerate cells from both sides. This can be thought of as the domain wall being `folded' on itself on that 3-simplex. This does not correspond to being part of the orientation-reversing wall because crossing that 3-simplex gives two orientation-reversals, or equivalently $f_\infty A_b^*s(\braket{0123}) = 1 + 1 = 0 \text{ (mod 2)}$. }.

\begin{figure}[h!]
  \centering
  \includegraphics[width=\linewidth]{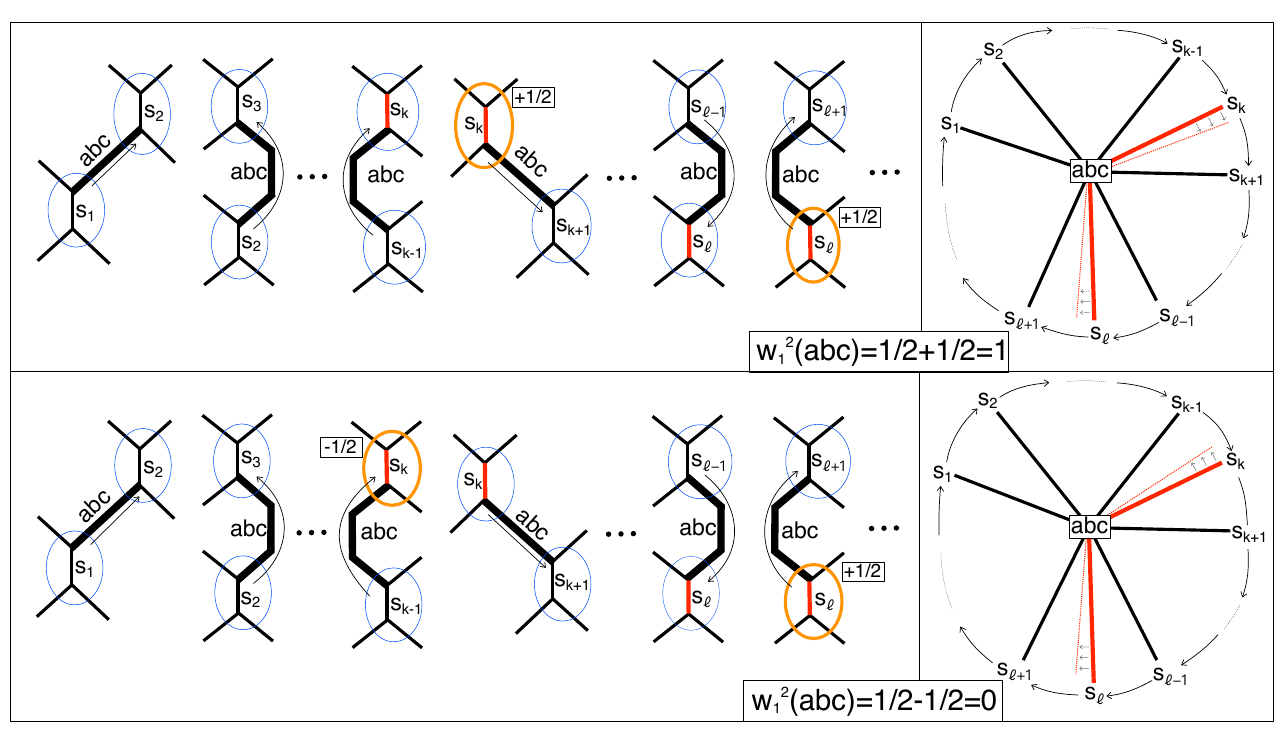}
  \caption{The link of a 2-simplex $\braket{abc}$ can be seen from the 15j symbols, consisting of simplices $s_1 \to s_2 \to \cdots$ that contain $\braket{abc}$. In the top left and bottom left portions of the figure, we drawn part of each 15j symbol that contains $\braket{abc}$.
  The one total solid orange line around each $s_k, s_\ell$ mean that $w_1(s_k) = w_1(s_\ell) = 1$ and that $w_1$ is perturbed into the 4-simplex consisting of that 15j symbol. The thick red line on $s_k$ and $s_{\ell}$ means $w_1(s_k) = w_1(s_{\ell}) = 1$. 
  One can compute $w_1^2(\braket{abc})$ by adding up $\pm 1/2$ for every crossing of the arrows with an anti-unitary (thick-orange) circle and $0$ for every unitary (thin-blue) circle.  (Top) Crossing $w_1$ in the perturbing direction twice giving $w_1^2(\braket{abc}) = 1$. (Bottom) Crossing in the perturbing direction once and in the opposite direction once gives $w_1^2(\braket{abc}) = 0$.}
  \label{linkOf2SimplexAndw1squared}
\end{figure}

For completeness, we give an algorithmic way to see if a given 2-simplex $\braket{abc}$ is part of $w_1^2 \in Z_2(M,\Z_2) = f_\infty \left(A_b^\ast s \cup A_b^\ast s \right)$ using the diagrammatics. We first collect the loop of 3-simplices that contain it, which forms $\text{Link}(\braket{abc})$. Recall these are the 3-simplices for which $\delta \alpha_{\braket{abc}}$ acting on them is nonzero, where $\alpha_{\braket{abc}}$ is the indicator cochain.  Then draw out all the 15j symbols that contain $\braket{abc}$. For all the $\braket{a'b'c'd'} \ni \braket{abc}$ in $w_1$, one can draw a thin blue circle around $\braket{a'b'c'd'}$ in all the 15j symbols where the group element surrounding it is unitary. Likewise draw a thick orange circle around it in a 15j symbol if the group element is anti-unitary. Then, as in Fig.~\ref{linkOf2SimplexAndw1squared} we draw a directed loop along $\text{Link}(\braket{abc})$ by going between the two 3-simplices containing $\braket{abc}$ in each 15j symbol in sequence, drawing an arrow within each 15j symbol. (Note that we have two choices for the direction of the loop, of which we arbitrarily pick one.) We sum up a quantity going around the loop that adds up to $w_1^2(\braket{abc}) \text{ (mod 2)}$ as follows. For a particular 15j symbol, add $0$ or $+1/2$ if the tail of the arrow is adjacent to a thin-blue or thick-orange circle respectively, and add $0$ or $-1/2$ if the head of the arrow is adjacent to a thin-blue or thick-orange circle. Since $\delta \alpha_{\braket{abc}}$ is cohomologically trivial, this loop of 3-simplices crosses $w_1$ an even number of times and will therefore pick up an even number of $\pm 1/2$ signs, giving us something that adds up to $w_1^2(\braket{abc}) \text{ (mod 2)}$. Conceptually, since each 3-simplex on $w_1$ has been perturbed towards the anti-unitary group element, the above algorithm associates a $+1/2$ to a crossing of Link$(\braket{abc})$ in the direction of the perturbation and a $-1/2$ when the crossing is opposite the perturbation. The only way that the loop can cross $w_1$ twice in the same direction relative to the perturbation is if the perturbed $w_1$ crosses the unperturbed version on $\braket{abc}$, that is, if $w_1^2(\braket{abc})=1$.

As an example, we can compute $w_1^2$ on the triangulation of $\mathbb{RP}^4$ as in Appendix \ref{app:Z16} with 15j symbols drawn out in Fig.~\ref{rp4_Step1}. The orientation-reversing wall consists of the 3-simplices labeled $\{{\color{OliveGreen} 17, 20}\}$, and the only 2-simplices $\braket{abc}$ which could possibly have $w_1^2(abc)$ non-trivial are labeled $\{{\color{OliveGreen} 5,7,16,19}\}$. Following the procedure in Fig.~\ref{linkOf2SimplexAndw1squared} gives:
\begin{equation}
\begin{split}
w_1^2({\color{OliveGreen} 7}) &= w_1^2({\color{OliveGreen} 16}) = 0 \\
w_1^2({\color{OliveGreen} 5}) &= w_1^2({\color{OliveGreen} 19}) = 1
\end{split}
\end{equation}

        \section{ Lemmas about Pachner moves with branching structure and background gauge fields} \label{appPachnerLemmas}
        
\subsection{ Introduction}
Pachner's theorem states that any two triangulations that define PL-equivalent manifolds are connected by a series of combintorial moves, called ``Pachner moves''. In our state sum, it would be useful to have more general results than this because the triangulations we deal with are decorated with more general objects, like branching structures and gauge fields. 

In this appendix, we first extend Pachner's result to include branching structures, in particular showing that any two \textit{branched} triangulations of a PL-manifold can be connected by Pachner moves which respect branching structures. Then, we add background flat gauge fields, showing that any two gauge fields on the triangulation that are gauge equivalent can be connected by Pachner moves. This is proved for general groups in the case of 1-form gauge fields and for Abelian higher-form gauge fields.

Note that in this appendix, we will often use the symbols $v_{\cdots},w_{\cdots}$ to refer to vertices. These symbols will not have anything to do with the vector fields $v$ and Stiefel-Whitney classes $w$ from earlier.

\subsection{ Pachner-connectedness of branching structures} \label{pachnerConnectBranchStruct}
In this section, we will show that any two branching structures are connected by Pachner moves. The strategy is to connect an arbitrary branched triangulation to a canonical ``inwards'' branching structure on the barycentric subdivision of the triangulation. This proves the theorem since any two branching structures can be connected using this common refinement.

First, we describe in some detail what we mean by branched Pachner moves and the barycentric subdivision. Then we prove that the barycentric subdivision can be obtained from branched Pachner moves. We start the proof in $d=2$ and $d=3$, which will then prepare us for the proof in arbitrary dimensions.

\subsubsection{Preliminaries}
\paragraph{Branched Pachner moves}

Recall that given a triangulation of a $d$-dimensional manifold and $k \in \{1 \cdots d+1\}$, a $(k,d+2-k)$ Pachner move is a transformation that can be thought of as a transformation of the triangulation consisting of the following steps:

\begin{itemize}
    \item Attach a $(d+1)$-simplex to the manifold along $k$ adjacent simplices.
    \item Remove the original $k$ simplices and replace them with the other `leftover' $(d+2-k)$ simplices
\end{itemize}

\begin{figure}[h!]
  \centering
  \includegraphics[width=0.75\linewidth]{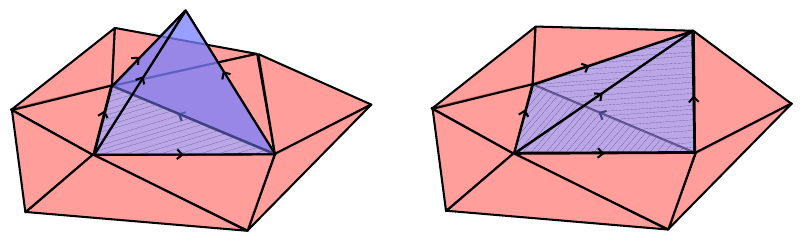}
  \caption{Pachner moves in $d=2$ as attaching a $3$-simplex to a $2$-manifold and discarding the attaching region. The red part represents the original $2$-manifold, the blue part are the `new' $2$-simplices of the $3$-simplex, and the hatched regions are the shared triangles of the 3-simplex and the manifold, which get discarded. The final triangulation is the red part plus blue part. (Left) A branched (1,3) Pachner move. (Right) A branched (2,2) move.}
  \label{Pachnercobordism}
\end{figure}

This can be thought of as an elementary bordism of the manifold with itself. See for example Fig.~\ref{Pachnercobordism}. 

\begin{figure}[h!]
  \centering
  \includegraphics[width=\linewidth]{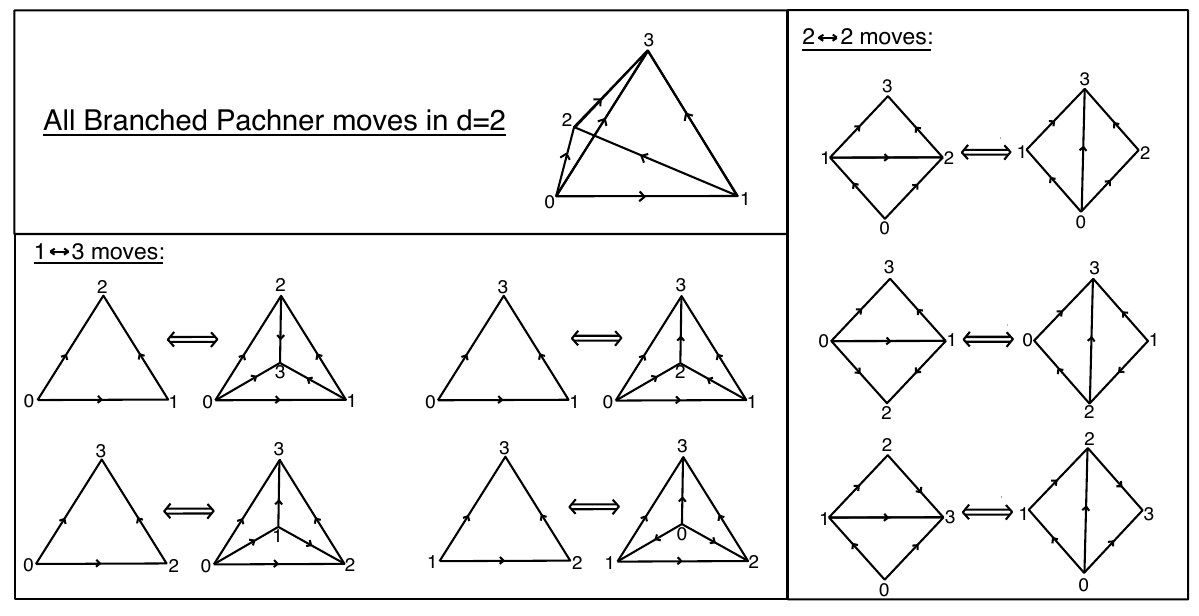}
  \caption{All possible branched Pachner moves for 2-manifold triangulations. The vertex labels and the $3$-simplex corresponding to them are in the top-left cell.}
  \label{2D_Pachner}
\end{figure}

\begin{figure}[h!]
  \centering
  \includegraphics[width=\linewidth]{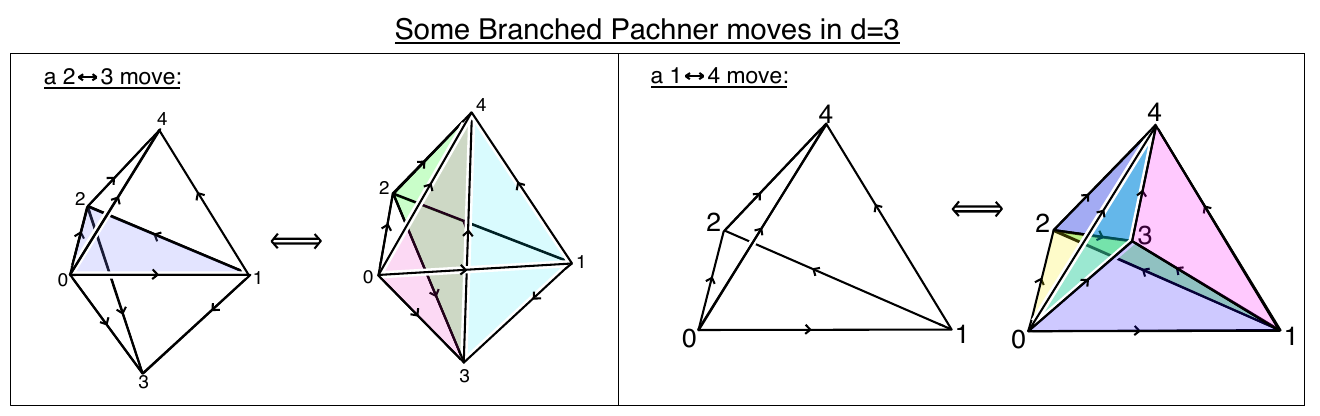}
  \caption{Some examples of branched $d=3$ Pachner moves, which involve a 4-simplex $\braket{01234}$. All $d=3$ Pachner moves will be $\{(1,4),(2,3),(3,2),(4,1)\}$-moves, so will look like the ones drawn, except with different vertex orders and branching structures. The differently-colored triangles illustrate boundaries between distinct 3-simplices. (Left) A (2,3) move $\{\braket{0123},\braket{0124}\} \to \{\braket{1234},\braket{0234},\braket{0134}\}$. (Right) A (1,4) move $\{\braket{0124}\} \to \{\braket{1234},\braket{0234},\braket{0134},\braket{0123}\}$}
  \label{3D_Pachner}
\end{figure}

A branched Pachner move is entirely analogous, except the original triangulation and the attaching $(d+1)$-simplex are endowed with branching structures, and the branching structure of the manifold must agree with that of the attaching $(d+1)$-simplex. All branched Pachner moves in $d=2$ are given in Fig.~\ref{2D_Pachner}, and some $d=3$ ones are given in Fig.~\ref{3D_Pachner}.

\paragraph{Barycentric subdivision} \label{defOfBarycentricSubdivision}

Given a triangulation of a manifold, there is a well-known canonical refinement of the triangulation called the barycentric subdivision. A property of this subdivision is that for each $n \in \{0 \cdots d\}$, every $n$-simplex of the triangulation gets split into $(n+1)!$ different $n$-simplices. We give a sketch of its definition as follows. 

First, for $n\in \{0 \cdots d\}$ define $S_n$ as the $n$-simplices of the original triangulation. We can define the vertex set $\Tilde{S}_0$ of the barycentric subdivision as having one vertex for every simplex of the original: 
\begin{equation}
\Tilde{S}_0 := \bigsqcup_{m=0}^d \{ \braket{m_{a}} | a \in S_m \}
\end{equation}
Each element of $\Tilde{S}_0$ represents the barycenter of a simplex in $S_m$. To an $m$-simplex $a$, we label an element of $\Tilde{S}_0$ as $\braket{m_{a}}$, so the barycenters of the original $m$-simplices all have a label specifying the dimension of their simplex.

The $n$-simplices $\Tilde{S}_n$ of a barycentric subdivision are labelled by ascending sequences of simplices of length-$(n+1)$, i.e. sequences of simplices $\{a_0 \subsetneq \cdots \subsetneq a_n\}$ with dimensions $\{k_0 < \cdots < k_n\}$, where $a \subsetneq b$ means that simplex $a$ is a subsimplex of $b$ not equal to $b$. We can write this as:
\begin{equation}
\Tilde{S}_n = \bigsqcup_{0 \le k_0 < \cdots < k_n \le d} \{\braket{(k_0)_{a_0} \cdots (k_n)_{a_n}} | a_0 \subsetneq \cdots \subsetneq a_n \}
\end{equation}
where each $(k_i)_{a_i}$ is a vertex in $\Tilde{S}_0$ coming from a $k_i$-simplex $a_i \in S_{k_i}$. 

\begin{figure}[h!]
  \centering
  \includegraphics[width=\linewidth]{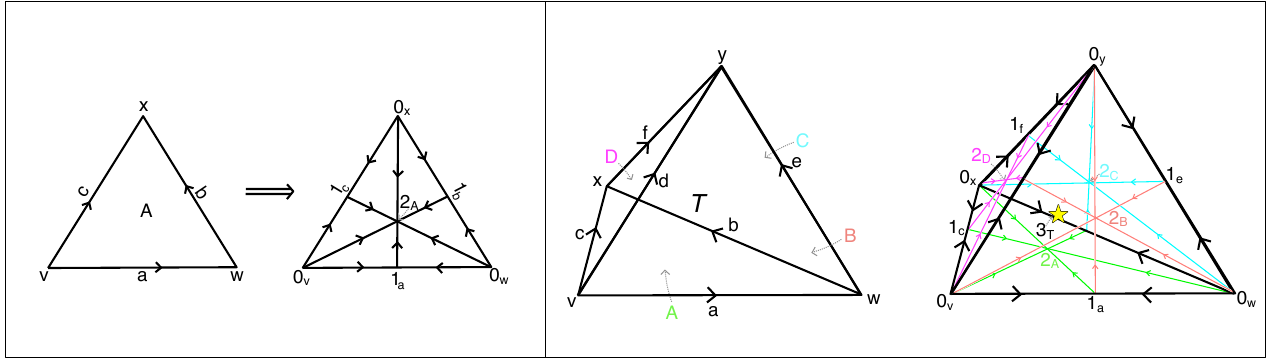}
  \caption{A barycentric subdivision of a simplex in $d=2$/$d=3$ and its canonical branching structure. (Left) $d=2$, vertices are $\{v,w,x\}$, edges are $\{a,b,c\}$ and the face is $\{A\}$. (Right) $d=3$, vertices are $\{v,w,x,y\}$, edges are $\{a,b,c,d,e,f\}$, 2-faces are $\{A,B,C,D\}$, and the 3-simplex is $\{T\}$. For visual clarity, we do not draw the edges going towards the barycenter vertex $3_T$ (shown as a yellow star). The reader should imagine an edge going from each of the shown vertices directed towards $3_T$. }
  \label{BaryCentricSubdivision}
\end{figure}

Every top-dimensional $d$-simplex of the barycentric subdivision is given a canonical vertex ordering coming from the inclusion ordering, $(k_0)_{a_0} \to \cdots \to (k_d)_{a_d}$. it is clear that these vertex orderings are consistent and define a branching structure on the subdivision. See Fig.~\ref{BaryCentricSubdivision} for illustration of the barycentric subdivision and the induced branching structures in $d=2$ and $d=3$. 

\subsubsection{Proof of branched barycentric subdivision from Pachner moves}

The idea of our proof borrows heavily from~\cite{traversingThreeManifoldTriangulationsAndSpines}, which illustrates, without branching structures, how to obtain a barycentric subdivision from Pachner moves in $d=2$ and $d=3$. We essentially fill in arrows to their $d=2$ and $d=3$ pictures to prove the desired statement in $d=2$/$d=3$. The structure of their construction (and decorating branching structures) readily generalize to arbitrary dimensions. Since the dimension we care about is $d=4$, we will not be able to prove the statement pictorially and will need to translate the construction into symbols.

We distinguish the original simplices of dimensions between $0 \to 4$ as follows. 0-simplices are lower-case roman $\{v,w,x,...\}$ in the upper-alphabet. 1-simplices are lower-case roman $\{e,f,...\}$ near the lower alphabet. 2-simplices are upper-case Roman $\{A,B,C,...\}$ near the lower-alphabet (although in the $d=3$ proof we find it convenient to deviate from this convention). 3-simplices are upper-case roman $\{S,T,...\}$ near the upper-alphabet. 4-simplices are lower-case Greek $\{\alpha,\beta\}$ near the lower-alphabet.

\paragraph{Proof in $d=2$}

First we will reproduce their pictorial $d=2$ argument making sure to carefully account for the branching structure. Then we will translate the $d=2$ proof into symbols in a way that will make it apparent how to generalize to arbitrary dimensions.

\begin{figure}[h!]
  \centering
  \includegraphics[width=\linewidth]{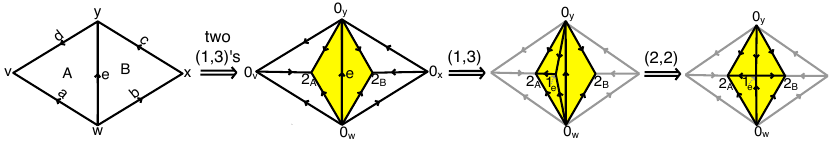}
  \caption{Obtaining the branched barycentric subdivision of a $d=2$ triangulation using branched Pachner moves. We illustrate it with a specific initial choice of branching structure, but the same argument applies for any such choice. The last two moves are restricted to the yellow regions surrounding the original edges, and can they can be done independently for each such region. }
  \label{PachnerBarycentric2D}
\end{figure}

The proof in $d=2$ is summarized in Fig.~\ref{PachnerBarycentric2D}. First, rename every vertex `$v$' as $0_v$. Then for every 2-simplex `$A$', we perform a $(1,3)$-move which creates a new vertex $2_A$ such that all new edges point towards $2_A$. Then, the link of any of the original edges `$e$' will consist of exactly two triangles. For each of these $e$, another $(1,3)$-move on one of these triangles creates an additional vertex $1_e$, and another $(2,2)$ move completes the subdivision. 

Symbolically, we can express this in the following multistep process. The branching structure is accounted for through the ordering of the vertices of each simplex.

\begin{enumerate}[start=0,label=\text{Step} {\arabic*}:,leftmargin=1.25cm]
    \item For all the vertices of the original triangulation, change notation $v \to 0_v$.
    \item For every 2-simplex $A=\braket{0_v 0_w 0_x}$, perform a $(1,3)$-move creating a new vertex $2_A$: 
        \begin{align*} \braket{0_v 0_w 0_x} \to \{\braket{0_v 0_w 2_A},\braket{0_v 0_x 2_A},\braket{0_w 0_x 2_A}\} \end{align*}
    \item This step has two sub-parts.
    \begin{enumerate}[leftmargin=0cm,label=({\alph*})]
        \item After Step 1, for each of the \textit{original} 1-simplices $e=\braket{0_w 0_y}$, there are exactly two 2-simplices $\braket{0_w 0_y 2_A},\braket{0_w 0_y 2_B}$ adjacent to $e$. Do a $(1,3)$-move on one of these, creating a new vertex $1_e$:
        \begin{align*} \braket{0_w 0_y 2_A} \to \{\braket{0_w 1_e 2_A},\braket{0_y 1_e 2_A},\braket{0_w 0_y 1_e}\} \end{align*}
        \item Do the $(2,2)$-move:
        \begin{align*} \{\braket{0_w 0_y 1_e},\braket{0_w 0_y 2_A}\} \to \{\braket{0_w 1_e 2_A},\braket{0_w 1_e 2_A}\} \end{align*}
    \end{enumerate}
\end{enumerate}

At the end of this process every $2$-simplex will be of the form $(0_v 1_e 2_f)$ for each possible $v \subsetneq e \subsetneq f$ of the original triangulation, so we have produced a barycentric subdivision with the correct branching structure.

\paragraph{Proof in $d=3$}
The argument in $d=3$ will proceed similarly, but will be a bit more work. We will omit drawing some of the pictures in $d=3$, which is already done nicely in~\cite{traversingThreeManifoldTriangulationsAndSpines} (and which we recommend the reader look at while reading our symbolic argument). Instead we translate their pictorial argument into a symbolic argument parallel to the $d=2$ case, allowing us to track the branching structure using the vertex orderings. We note  the top row of the bottom panel of Fig.~\ref{Pachner2FormGaugeField3D} depicts a dualized version of the process although in the context of a 2-form gauge field.

\begin{enumerate}[start=0,label=\text{Step} {\arabic*}:,leftmargin=1.25cm]
    \item For all the vertices of the original triangulation, change notation $v \to 0_v$.
    \item For every 3-simplex $T=\braket{0_v 0_w 0_x 0_y}$, perform a $(1,4)$-move creating a new vertex $3_T$: 
        \begin{align*} \braket{0_v 0_w 0_x 0_y} \to \{\braket{0_v 0_w 0_x 3_T},\braket{0_v 0_w 0_y 3_T},\braket{0_v 0_x 0_y 3_T},\braket{0_w 0_x 0_y 3_T}\} \end{align*}
    \item This step again has two sub-parts.
    \begin{enumerate}[leftmargin=0cm,label=({\alph*})]
        \item After Step 1, for each of the \textit{original} 2-simplices $A=\braket{0_v 0_w 0_x}$, there are exactly two 3-simplices $\braket{0_v 0_w 0_x 3_S},\braket{0_v 0_w 0_x 3_T}$ adjacent to $e$. Do a $(1,4)$-move on one of these, creating a new vertex $2_A$:
        \begin{align*} \braket{0_v 0_w 0_x 3_T} \to \{\braket{0_v 0_w 2_A 3_T},\braket{0_v 0_x 2_A 3_T},\braket{0_w 0_x 2_A 3_T},\braket{0_v 0_w 0_x 2_A}\} \end{align*}
        \item Do the $(2,3)$-move:
        \begin{align*} \{\braket{0_v 0_w 0_x 2_A},\braket{0_v 0_w 0_x 3_S}\} \to \{\braket{0_v 0_w 2_A 3_S},\braket{0_v 0_x 2_A 3_S},\braket{0_w 0_x 2_A 3_S}\} \end{align*}
    \end{enumerate}
    At this point every $3$-simplex in the triangulation will be of the form $\braket{0_v 0_w 2_A 3_T}$ for the original edges $e=\braket{0_v 0_w}$. And, we will have one of these 3-simplices for every sequence $e \subsetneq A \subsetneq T$.
    \item The goal of this step is, for each $e=\braket{0_v 0_w}$, to transform the set of the $\{\braket{0_v 0_w 2_A 3_T}|e \subsetneq A \subsetneq T\}$ into the set $\{\braket{0_x 1_e 2_A 3_T}|x \subsetneq e \subsetneq A \subsetneq T\}$, which would complete the branched barycentric subdivision. We first note that the link of the edge $e=\braket{0_v 0_w}$ consists of 1-simplices $\{\braket{2_T 3_A}|e \subsetneq A \subsetneq T\}$. These 1-simplices taken together will be homeomorphic to a circle, and will in fact be the barycentric subdivision of the link of $e$ in the original triangulation.
    \begin{enumerate}[leftmargin=0cm,label=({\alph*})]
        \item There are an even number, call it $2 m$, of edges and $2 m$ vertices in the link of $e$. Of these vertices, $m$ of them will be of the form $3_T$ for a 3-simplex $T$, and $m$ of them will be of the form $2_{(S T)}$ where $(ST)$ is the 2-simplex adjacent to 3-simplices $S,T$. We label the 3-simplices as $T_1,\cdots,T_m$, and the edges in the link of $e$ will be:
       \begin{align*} \braket{2_{(T_1 T_2)} 3_{T_1}},\braket{2_{(T_1 T_2)} 3_{T_2}},\braket{2_{(T_2 T_3)} 3_{T_2}},\braket{2_{(T_2 T_3)} 3_{T_3}},\cdots,\braket{2_{(T_m T_1)} 3_{T_m}},\braket{2_{(T_m T_1)} 3_{T_1}} \end{align*}
       \item For the first $3$-simplex $\braket{0_v 0_w 2_{(T_1 T_2)} 3_{T_1}}$ perform a $(1,4)$-move, creating a new vertex $1_e$:
       \begin{align*} \braket{0_v 0_w 2_{(T_1 T_2)} 3_{T_1}} \to \{\braket{0_v 1_e 2_{(T_1 T_2)} 3_{T_1}},\braket{0_w 1_e 2_{(T_1 T_2)} 3_{T_1}},\braket{0_v 0_w 1_{e} 3_{T_1}},\braket{0_v 0_w 1_{e} 2_{(T_1 T_2)}}\} \end{align*}
       Then, do a $(2,3)$-move replacing $\braket{0_v 0_w 1_{e} 2_{(T_1 T_2)}},\braket{0_v 0_w 2_{(T_1 T_2)} 3_{T_2}}$:
       \begin{align*} \{\braket{0_v 0_w 1_{e} 2_{(T_1 T_2)}},\braket{0_v 0_w 2_{(T_1 T_2)} 3_{T_2}}\} \to \{\braket{0_v 0_w 1_e 3_{T_2}},\braket{0_w 1_e 2_{(T_1 T_2)} 3_{T_2}},\braket{0_v 1_e 2_{(T_1 T_2)} 3_{T_2}}\} \end{align*}
       \item Do two more $(2,3)$-moves. First, replace:
       \begin{align*} \{\braket{0_v 0_w 1_{e} 3_{T_2}},\braket{0_v 0_w 2_{(T_2 T_3)} 3_{T_2}}\} \to \{\braket{0_v 0_w 1_e 2_{(T_2 T_3)}},\braket{0_v 1_e 2_{(T_2 T_3)} 3_{T_2}},\braket{0_2 1_e 2_{(T_2 T_3)} 3_{T_2}}\} \end{align*}
       Then, replace:
       \begin{align*} \{\braket{0_v 0_w 1_{e} 2_{(T_2 T_3)}},\braket{0_v 0_w 2_{(T_2 T_3)} 3_{T_3}}\} \to \{\braket{0_v 0_w 1_e 3_{T_3}},\braket{0_v 1_e 2_{(T_2 T_3)} 3_{T_3}},\braket{0_w 1_e 2_{(T_2 T_3)} 3_{T_3}}\} \end{align*}
       \item Inductively repeat this process until the replacement 
       \begin{align*} \{\braket{0_v 0_w 1_{e} 2_{(T_m T_1)}},\braket{0_v 0_w 2_{(T_m T_1)} 3_{T_m}}\} \to \{\braket{0_v 0_w 1_e 2_{(T_m T_1)}},\braket{0_v 1_e 2_{(T_m T_1)} 3_{T_m}},\braket{0_v 1_e 2_{(T_m T_1)} 3_{T_m}}\} \end{align*}
       Now, there are three $3$-simplices $\{\braket{0_v 0_w 1_e 2_{(T_m T_1)}},\braket{0_v 0_w 1_e 3_{T_1}},\braket{0_v 0_w 2_{(T_m T_1)} 3_{T_1}}\}$ on which we perform $(3,2)$-move:
       \begin{align*} \{\braket{0_v 0_w 1_e 2_{(T_m T_1)}},\braket{0_v 0_w 1_e 3_{T_1}},\braket{0_v 0_w 2_{(T_m T_1)} 3_{T_1}}\} \to \{\braket{0_v 1_e 2_{(T_m T_1)} 3_{T_1}},\braket{0_w 1_e 2_{(T_m T_1)} 3_{T_1}}\}\end{align*}
    \end{enumerate}
    At this point we have done all the desired replacements, and have completed the barycentric subdivision. And, the branching structure matches the canonical one.
\end{enumerate}

\paragraph{Proof in $d=4$ and higher} \label{4DbarycentricProof}

The proof in $d=4$ is again quite similar. The Steps 0-3 of the $d=3$ proof have a direct analog in $d=4$. Except instead, Steps 1-3 will begin with $(1,5)$-moves:
\begin{align*} 
&\braket{0_v 0_w 0_x 0_y 0_z} \to \{\braket{0_v 0_w 0_x 0_y 4_\alpha},\braket{0_v 0_w 0_x 0_z 4_\alpha},\braket{0_v 0_w 0_y 0_z 4_\alpha},\braket{0_v 0_x 0_y 0_z 4_\alpha},\braket{0_w 0_x 0_y 0_z 4_\alpha}\} \\
&\dots \text{etc.}
\end{align*}

where $4_\alpha$ is a new vertex created for an original 4-simplex $\alpha$. And in general of the $(a,5-a)$-moves done for $a=1,2,3$ will be replaced with $(a,6-a)$ moves. 

At the end of Step 3, we will have that all of the 4-simplices in the triangulation will be of the form:
\begin{align*}
\{\braket{0_v 0_w 2_A 3_T 4_\alpha} | e=\braket{vw} \text{ was originally a 1-simplex and } e \subsetneq A \subsetneq T \subsetneq \alpha\}
\end{align*}

The goal of Step 4 would be similar, in trying to use Pachner moves to replace the above set (with fixed 1-simplex $e$) with the set $\{\braket{0_x 1_e 2_A 3_T 4_\alpha}|x \subsetneq e \subsetneq A \subsetneq T \subsetneq \alpha\}$. After Step 3, we will have a similar statement that the link of $e$ is the set $\{\braket{2_T 3_A 4_\alpha}|e \subsetneq A \subsetneq T\}$. But instead, $\mathrm{Link}(e)$ will be homoemorphic to a 2-sphere, and will be the barycentric subdivision of the link of $e$ in the original triangulation.

The strategy to do these replacements is basically the same as in the $d=3$ case (or Step 3 in general) where we instead tried to do the replacements with on a circle instead of on a sphere. The only complication is that it was more straightforward to do on a circle, since there was a natural cyclic order to do the replacements. But the same argument still applies.

\begin{figure}[h!]
  \centering
  \includegraphics[width=\linewidth]{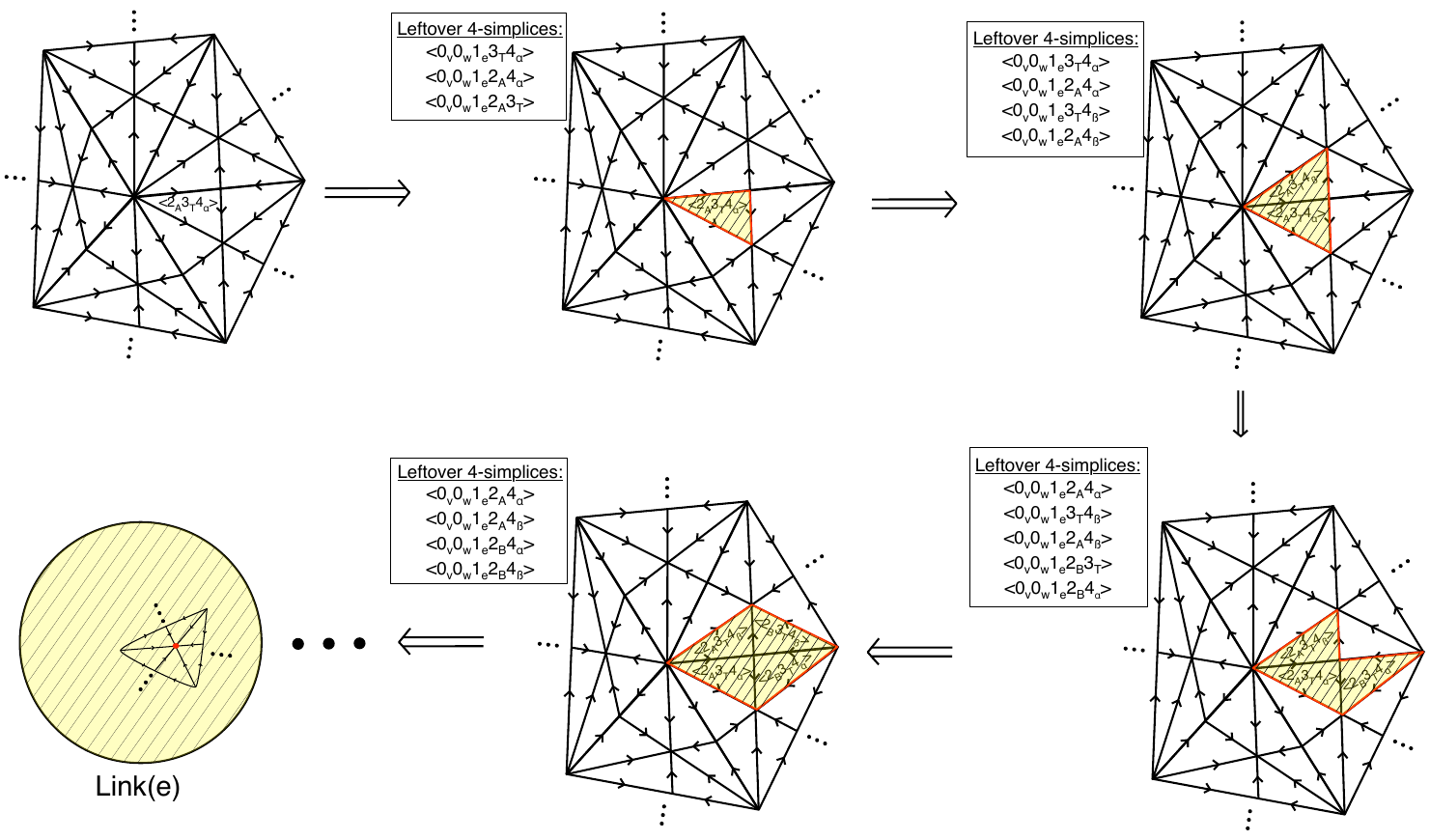}
  \caption[]{Step 4 of barycentric subdivision via Pachner moves in $d=4$. (Top-left) $\mathrm{Link}(e)$ for a particular 1-simplex $e=\braket{v, w}$ of the original triangulation after the first three steps of subdivision. $\mathrm{Link}(e)$ is homeomorphic to a 2-sphere and has been barycentrically subdivided in the other steps. (Other diagrams) Each large arrow corresponds to a subdivision process involving a particular 2-simplex in $\mathrm{Link}(e)$. After a 2-simplex's corresponding 4-simplex is subdivided, we shade it yellow and hatch it, so $\mathcal{R}_{\text{used}}$ (see main text) corresponds to the shaded region. Red edges are in correspondence with ``leftover 4-simplices", which are described in the main text and listed at the beginning of each step. For each such subdivision process, a leftover simplex is changed for each edge. If the edge was black before subdivision, a leftover simplex is produced, while if the edge was red before subdivision, the corresponding leftover simplex is removed. The leftover simplices are in correspondence with the boundary of the yellow hatched region, i.e., with $\partial \mathcal{R}_{\text{used}}$. At the end, the entirety of $\mathrm{Link}(e) \cong S^{2}$ will be part of $\mathcal{R}_{\text{used}}$ which completes the barycentric subdivision.}
  \label{PachnerBarycentric4D}
\end{figure}

We illustrate $\mathrm{Link}(e)$ and more generally the argument for Step 4 in Fig.~\ref{PachnerBarycentric4D}. The idea is as follows. For some $\braket{2_A 3_T 4_\alpha} \in \mathrm{Link}(e)$, perform a $(1,5)$-move on $\braket{0_v 0_w 2_A 3_T 4_\alpha}$:

\begin{align*}
\braket{0_v 0_w 2_A 3_T 4_\alpha} \to \{\braket{0_v 1_e 2_A 3_T 4_\alpha},\braket{0_w 1_e 2_A 3_T 4_\alpha},\braket{0_v 0_w 1_e 3_T 4_\alpha},\braket{0_v 0_w 1_e 2_A 4_\alpha},\braket{0_v 0_w 1_e 2_A 3_T}\}
\end{align*}

Note that the first two of these simplices after the subdivision $\{\braket{0_v 1_e 2_A 3_T 4_\alpha},\braket{0_w 1_e 2_A 3_T 4_\alpha}\}$ are of the final form we want, so we can keep them and forget about them. The other three $\{\braket{0_v 0_w 1_e 3_T 4_\alpha},\braket{0_v 0_w 1_e 2_A 4_\alpha},\braket{0_v 0_w 1_e 2_A 3_T}\}$ are \textit{not} in the final form we seek, so we will need to keep subdividing them away. We will call these other three simplices `\textit{leftover simplices}' that we will need to deal with.

Note that the leftover simplices are of the form $\{\braket{0_v 0_w 1_e \eta \, \xi} | \braket{\eta \, \xi} \in \partial\braket{2_A 3_T 4_\alpha}\}$ and correspond directly to the 1-simplices on the boundary of the 2-simplex $\braket{2_A 3_T 4_\alpha}$. 

The procedure will be to iterate this subdivision process by subdividing 4-simplices corresponding to neighboring $2$-simplices in $\mathrm{Link}(e)$. In general, we will keep track of the set $\mathcal{R}_{\text{used}}$ of 2-simplices $\braket{2_A 3_T 4_\alpha}$ in $\mathrm{Link}(e)$ for which $\braket{0_v 0_w 2_A 3_T 4_\alpha}$ has been subdivided. This will turn out to be the same as the set of 2-simplices in $\mathrm{Link}(e)$ for which both $\braket{0_v 1_e 2_A 3_T 4_\alpha},\braket{0_w 1_e 2_A 3_T 4_\alpha}$ have been created. And, it will turn out that the `leftover simplices' will always be of the form $\braket{0_v 0_w 1_e \eta \, \xi}$ for $\braket{\eta \, \xi} \in \partial{\mathcal{R}_{\text{used}}}$, which are the 1-simplices $\braket{\eta \, \xi}$ on the boundary of the set $\mathcal{R}_{\text{used}}$ of `used' 2-simplices. 

In general, a $(2,4)$-move will correspond to adding a simplex to the set $\mathcal{R}_{\text{used}}$ so that the size of $\partial \mathcal{R}_{\text{used}}$ increases by one. A $(3,3)$-move will cause $\#(\partial \mathcal{R}_{\text{used}})$ to decrease by one. And a $(4,2)$-move will cause $\#(\partial \mathcal{R}_{\text{used}})$ to decrease by three. And, every `leftover' 4-simplex that is removed from $\mathcal{R}_{\text{used}}$ at a step will be an input to the Pachner move. See, for example Fig.~\ref{PachnerBarycentric4D}. In that Figure, the Pachner moves corresponding to each of the arrows is:

\begin{enumerate}
    \item $\braket{0_v 0_w 2_A 3_T 4_\alpha} \to \{\braket{0_v 1_e 2_A 3_T 4_\alpha},\braket{0_w 1_e 2_A 3_T 4_\alpha},\braket{0_v 0_w 1_e 3_T 4_\alpha},\braket{0_v 0_w 1_e 2_A 4_\alpha},\braket{0_v 0_w 1_e 2_A 3_T}\}$ 
    \item $\{\braket{0_v 0_w 1_e 2_A 3_T},\braket{0_v 0_w 2_A 3_T 4_\beta}\} \to \{\braket{0_v 1_e 2_A 3_T 4_\beta},\braket{0_w 1_e 2_A 3_T 4_\beta},\braket{0_v 0_w 1_e 3_T 4_\beta},\braket{0_v 0_w 1_e 2_A 4_\beta}\}$ 
    \item $\{\braket{0_v 0_w 1_e 3_T 4_\alpha},\braket{0_v 0_w 2_B 3_T 4_\alpha}\} \to \{\braket{0_v 1_e 2_B 3_T 4_\alpha},\braket{0_w 1_e 2_B 3_T 4_\alpha},\braket{0_v 0_w 1_e 2_B 3_T},\braket{0_v 0_w 1_e 2_A 4_\alpha}\}$ 
    \item $\{\braket{0_v 0_w 1_e 2_B 3_T},\braket{0_v 0_w 1_e 3_T 4_\alpha},\braket{0_v 0_w 2_B 3_T 4_\beta}\} \to \{\braket{0_v 1_e 2_B 3_T 4_\beta},\braket{0_w 1_e 2_B 3_T 4_\beta},\braket{0_v 0_w 1_e 2_B 4_\beta}\}$
\end{enumerate}

At the end of this whole process, we indeed complete the barycentric subdivision with the correct branching structure!

We note that essentially the same process holds in higher dimensions. These Steps 0-4 will go through exactly the same in higher dimensions. Except in dimension $d$, we will need a series of Steps 0-$d$. A step $k$ will work in the exact same way, except we are dealing with the link of a $(d-k+1)$-simplex which is a barycentrically subdivided $(k-2)$-sphere. This process of subdividing the $d$-simplices corresponding to the $(k-2)$ simplices in the link, one simplex at a time, goes through in the exact same manner. (In fact, we could have used this picture in $d=2$ and $d=3$ as well.)

\subsection{ Pachner connectedness of equivalent flat gauge connections}
Now we consider triangulations decorated with flat gauge connections. Consider a triple $(M^d,T,\alpha)$, where $T$ is a triangulation of $d$-manifold $M^d$ and $\alpha$ is some flat gauge connection. We show that any gauge equivalent gauge fields on $(M^d,T)$ can be connected by Pachner moves, in the case of 1-form gauge fields for general groups and higher-form gauge fields for Abelian groups.

First, we explain the preliminaries of defining gauge fields on triangulations. Then we prove the connected of equivalent gauge fields.

\subsubsection{Preliminaries}
\paragraph{Gauge fields on a triangulation}

Let us review how to formulate gauge fields and higher-form gauge fields on a triangulation. Here, we will be assuming all gauge fields are \textit{flat}. Throughout this section, we will work with a pair $(M,T)$ of a manifold $M$ and triangulation $T$. 

A flat 1-form gauge field with gauge group $G$ on $M$ is an assignment of group elements ${\bf g}_{ij} \in G$ on all of the directed 1-simplices $\braket{ij}$ satisfying certain consistency constraints. The first constraint is that ${\bf g}_{ij} = {\bf g}_{ji}^{-1}$, that the group elements are inverse in opposite directions. Second, for every 2-simplex $\braket{i j k}$, there is a `flatness' condition ${\bf g}_{ij}{\bf g}_{jk}={\bf g}_{ik}$. This definition does not depend on whether $G$ is Abelian or non-Abelian. Note that if $G$ is Abelian, we can express the gauge field ${\bf g}_{ij}$ as a cochain ${\bf g} \in C^1(M,G)$. The flatness condition reduces to $\delta {\bf g} = 1$ (using the multiplicative notation that $\delta {\bf g} (\braket{ijk}) = {\bf g}_{ij}{\bf g}_{jk}{\bf g}_{ik}^{-1}$). For Abelian $G$ this means that ${\bf g}$ would be a closed cocycle, that ${\bf g} \in Z^1(M,G)$. Two 1-form gauge fields ${\bf g,g}'$ are \textit{gauge equivalent} if there exists a function ${\bf h}_i$ on vertices of the triangulation such that ${\bf g}'_{ij} = {\bf h}_i {\bf g}_{ij} {\bf h}_j^{-1}$. For Abelian gauge fields, this equivalence condition is the same as transforming the cocycle $a \to a + \delta \lambda$ where $\lambda \in C^0(M,A)$ is some function on vertices. So, equivalence classes of Abelian gauge fields are given by elements of $H^1(M,A)$. For non-Abelian gauge fields, the more general statement is that gauge fields are in correspondence with $\mathrm{Hom}(\pi_1(M), G) / G$ \footnote{$\mathrm{Hom}(\pi_1(M),G)$ are homomorphisms $\pi_1(M) \xrightarrow{\phi} G$, and ${\cdots}/G$ refers to identification under the equivalence under conjugation by $\phi \sim {\bf g} \phi {\bf g}^{-1}$ for all ${\bf g} \in G$.}.

Higher-form gauge fields for Abelian groups $A$ are defined similarly. A $k$-form gauge field with gauge group $A$ can be identified with an element $a \in Z^{k}(M,A)$. So, there would be group elements on each $k$-simplex with analog of the `flatness' condition being $\delta a = 0$ (using the additive notation for $\delta a$). Two higher-form gauge fields $a,a'$ are equivalent if $a' = a + \delta \lambda$ where $\lambda \in C^{k-1}(M,A)$ is some $(k-1)$-cochain. This means that equivalence classes of higher-form $A$-gauge fields are in bijection with $H^{k}(M,A)$. 

\paragraph{Pachner moves with background gauge fields}
First, we describe Pachner moves in the presence of background gauge fields. The idea is that since a Pachner move changes the simplices in the triangulations, we can assign gauge fields to the new simplices that are consistent with the flatness conditions and the fields on the other unchanged simplices. Another way to think about a gauge field on series of Pachner moves on $M$ is that it is a gauge field on the bordism geometry that gives a triangulation of $M \times I$. 

\begin{figure}[h!]
  \centering
  \includegraphics[width=\linewidth]{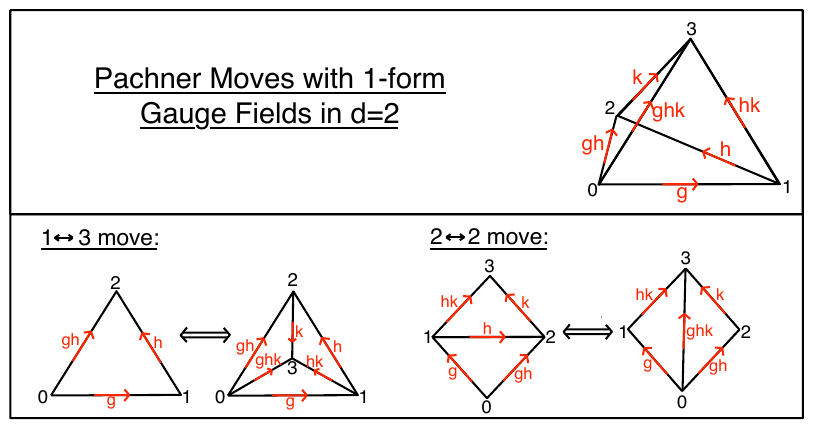}
  \caption{General Pachner move in $d=2$ with background gauge field. The 3-simplex with the gauge field group elements is on top, and the gauge fields in the presence of a Pachner move is on the bottom.}
  \label{PachnerGaugeField2D}
\end{figure}

\begin{figure}[h!]
  \centering
  \includegraphics[width=\linewidth]{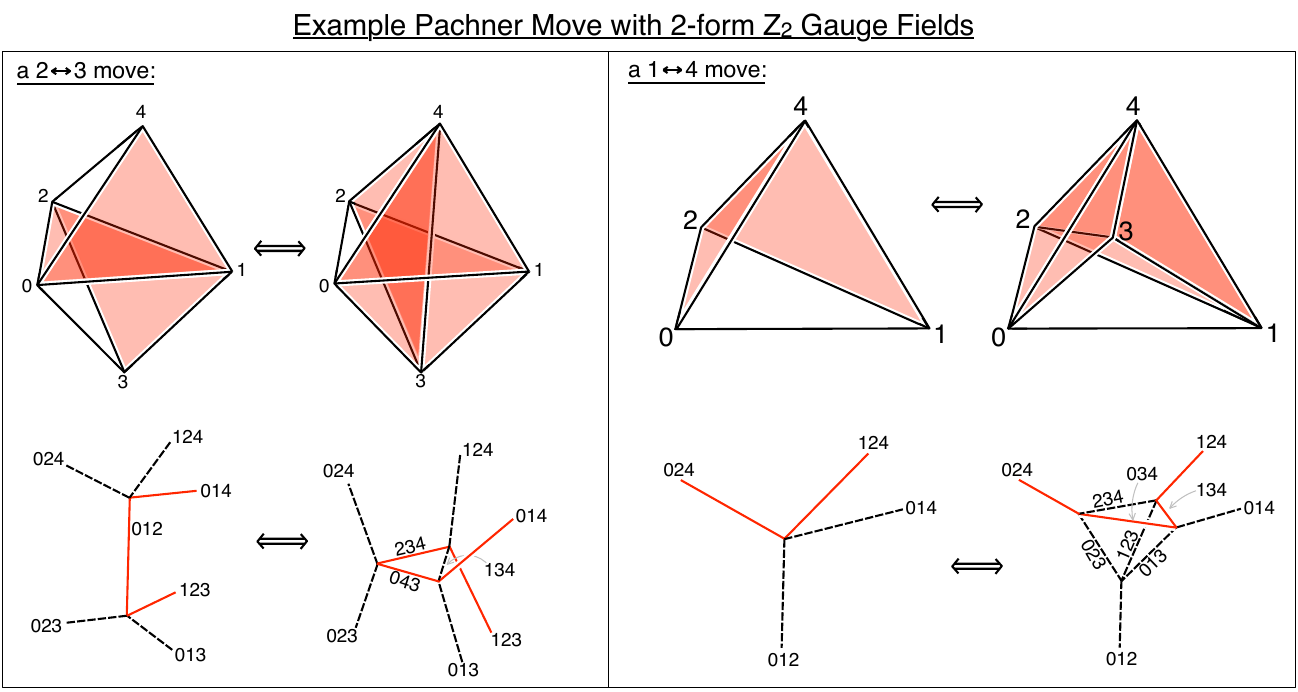}
  \caption{Example Pachner moves in $d=3$ with background 2-form $\Z_2$ gauge field, $a$. A 2-simplex $\braket{ijk}$ with $a(ijk)$ non-trivial is shaded in red. The Poincaré dual pictures are below, where $a$ is dual to a closed set of lines.}
  \label{Pachner2FormGaugeField3D}
\end{figure}

We give some examples of Pachner moves in the presence of a background gauge field in Fig.~\ref{PachnerGaugeField2D} for 1-form fields and Fig.~\ref{Pachner2FormGaugeField3D} for higher-form ones. Note that for a 1-form gauge field, the only situation in which there are several possibilities for the final gauge field after the Pachner move is the $(1,d+1)$ move in which a new vertex is created. This is because if all the vertices involved are in the triangulation both before and after the move, then the flatness condition ${\bf g}_{ij}{\bf g}_{jk}={\bf g}_{ik}$ fixes all the links' group elements. But if a new vertex $v$ is created inside a simplex $\braket{0 \cdots d}$, then there are different choices of gauge field are parameterized by any choice $g_{0 v} \in G$. However, note as in Fig.~\ref{Pachner2FormGaugeField3D} that for higher-form gauge fields, there are usually more possibilities.

Note that the branching structure plays no role in these gauge transformations. As such, for the remainder of this section, any simplex $\braket{0 \cdots n}$ we write down will be \textit{unordered}. 

\subsubsection{Gauge transformations via Pachner moves for 1-form gauge fields} \label{zeroFormProof}
First, we explain the procedure to do a gauge transformation for 1-form gauge fields. The method and picture for the proof will be quite similar to the last step of the construction of the barycentric subdivision in Section \ref{4DbarycentricProof}.

First, let's phrase more precisely what we want to achieve. Given a 1-form gauge field ${\bf g}_{ij}$, we want to implement the gauge transformations ${\bf g}_{ij} \to {\bf h}_i {\bf g}_{ij} {\bf h}_j^{-1}$ for functions ${\bf h}_i$ on vertices $\{i\}$. All such transformations can be implemented by elementary transformations such that ${\bf h}_i = {\bf h} \neq {\bf 1}$ on exactly a single vertex $i$. This means that for any given vertex $i$ and some ${\bf h} \in G$, we want to change ${\bf g}_{ik} \to {\bf h} {\bf g}_{ik}$ for each neighboring vertex $k$.

\begin{figure}[h!]
  \centering
  \includegraphics[width=\linewidth]{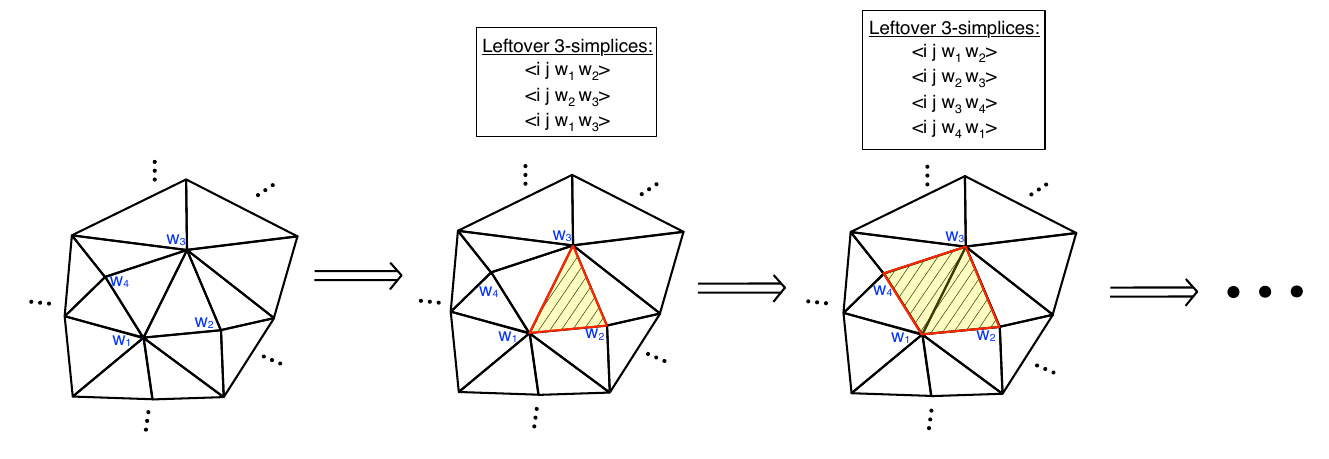}
  \caption{Pachner moves that will implement a gauge transformation of a 1-form gauge field on a vertex $i$ in $d=3$ as in Section \ref{zeroFormProof}. The $w_{\cdots}$ are in $\mathrm{Link}(i)$ (note $i$ is not shown), which is homeomorphic to a 2-sphere. Each 3-simplex $\braket{i w_{\alpha} w_{\beta} w_{\gamma}}$ is in correspondence with a 2-simplex $\braket{w_\alpha w_\beta w_\gamma} \in \mathrm{Link}(i)$. We perform a set of Pachner moves for each 3-simplex, one at a time, and shade the corresponding 2-simplex after completing those moves. The first move creates a vertex $j$, and the last move removes the original $i$; suitable decorations of gauge fields (described in the main text) effectively implement the gauge transformation. Each set of Pachner moves creates and removes ``leftover'' 3-simplices in correspondence with the (depicted red) edges of the 2-simplex such that the set of leftover 3-simplices correspond with the boundary of the yellow shaded region.
  }
  \label{0FormGaugeFieldPachner}
\end{figure}

To implement this process, we consider the link of the vertex $i$, $\mathrm{Link}(i)$, which consists of the $(d-1)$-simplices $\braket{v_0 \cdots v_{d-1}}$ for which $i \not\in \{v_0 \cdots v_{d-1}\}$ and that $\braket{i \, v_0 \cdots v_{d-1}}$ is a simplex in the triangulation. In general, we will have that $\mathrm{Link}(i)$ is homeomorphic to a $(d-1)$-sphere.

For some arbitrary choice of $\braket{v_0 \cdots v_{d-1}} \in \mathrm{Link}(i)$, we will implement the $(1,d+1)$-move creating a new vertex $j$
\begin{align*}
\braket{i \, v_0 \cdots v_{d-1}} \to \{\braket{j \, v_0 \cdots v_{d-1}}, \braket{i \, j \, v_1 \cdots v_{d-1}}, \dots, \braket{i \, j \, v_0 \cdots \hat{v}_k \cdots v_{d-1}}, \dots \braket{i \, j \, v_0 \cdots v_{d-2}}\}
\end{align*}
where $\braket{i \, j \, v_0 \cdots \hat{v}_k \cdots v_{d-1}}$ is a $(d-1)$-simplex consisting of all of vertices $\{i, j, v_0, \cdots, v_{d-1}\}$ \textit{except} for $v_k$. Along with this move, we will choose the gauge field ${\bf g}'$ at the end of this move so that ${\bf g}_{j i} = {\bf h}$ which will enforce that ${\bf g}_{j v_k} = {\bf g}_{j i} {\bf g}_{i v_k} = {\bf h} {\bf g}_{i v_k}$. 

From here, the strategy will be similar to Section \ref{4DbarycentricProof}. However, the end goal will be to replace the vertex $i$ with $j$ in all connections involving vertex $i$ and at the end of the day remove $i$ with a $(d+1,1)$-move. If we are able to accomplish this without doing any more $(1,d+1)$-moves (i.e. without creating any new vertices), then the fact that the gauge fields are uniquely determined at the end of each move will mean that the final gauge fields ${\bf g}_{j v_k}^\text{final}$ will all be ${\bf g}_{j v_k}^\text{final} = {\bf h} {\bf g}_{i v_k}$. In other words, we will have accomplished the desired gauge transformation.

Note that at the end of the first move above, there is one simplex $\braket{j \, v_0 \cdots v_{d-1}}$ that is of the final form we want, and the rest $\braket{i \, j \, v_0 \cdots \hat{v}_k \cdots v_{d-1}}$ are `leftover' $d$-simplices that we want to eliminate via Pachner moves. 

To complete the process, we will iteratively perform Pachner moves, one for each $(d-1)$-simplex in $\mathrm{Link}(i)$. At each step, define $\mathcal{R}_{\text{used}}$ to be the set of $(d-1)$-simplices $\{\braket{w_0 \cdots w_{d-1}}\}$ for which we have performed a Pachner move involving $\braket{i \, w_0 \cdots w_{d-1}}$. Next, choose a $(d-1)$-simplex $\braket{w_0 \cdots w_{d-1}}$ such that the next $\mathcal{R}_{\text{used}}' = \mathcal{R}_{\text{used}} \sqcup \{\braket{w_0 \cdots w_{d-1}}\} $ is connected. Perform a $(k+1,d+1-k)$ Pachner move, where $k$ is the number of $(d-2)$-simplices that the boundary $\partial\braket{w_0 \cdots w_{d-1}}$ shares with $\partial\mathcal{R}_{\text{used}}$. The $(k+1)$ ``input" $d$-simplices consist of $\braket{i w_0 \cdots w_{d-1}}$ and the ``leftover simplices" corresponding to the shared $(d-2)$-simplices. The outputs of the Pachner move are the $(d+1-k)$ ``leftover simplices" corresponding to the unshared $(d-2)$-simplices of $\partial \{\braket{w_0 \cdots w_{d-1}}\}$. This Pachner move produces some new leftover simplices so that the new set of all leftover simplices corresponds to $\partial \mathcal{R}_{\text{used}}'$. Repeat this process until $\mathcal{R}_{\text{used}} = \mathrm{Link}(i)$; the last step will be a $(d+1,1)$-move that entirely eliminates the vertex $i$. 

We illustrate the steps of Pachner moves in $d=3$ in Fig.~\ref{0FormGaugeFieldPachner}, where the link of $i$ is some triangulation of the sphere $S^2$. Note the similarity of the process to Fig.~\ref{PachnerBarycentric4D}. The first two moves written out are:

\begin{enumerate}
    \item $\braket{i \, w_1 w_2 w_3} \to \{\braket{j \, w_1 w_2 w_3},\braket{i \, j \, w_2 w_3},\braket{i \, w_1 \, j \, w_3},\braket{i \, w_1 w_2 \, j}\}$ 
    \item $\{\braket{i \, w_1 w_3 w_4},\braket{i \, j \, w_1 w_3}\} \to \{\braket{j \, w_1 w_3 w_4},\braket{i \, j \, w_1 w_4},\braket{i \, w_3 \, j \, w_4}\}$
\end{enumerate}

\subsubsection{Higher-form gauge transformations via Pachner moves}

Now, we will explain how a similar process applies to implement gauge transformations of higher-form gauge fields. Given an $n$-form gauge field $\alpha \in Z^n(M,A)$, we want to implement the transformation $\alpha \to \alpha + \delta\lambda$ for $\lambda \in C^{n-1}(M,A)$. If we can implement the transformation for the basic gauge transformations for which $\lambda$ is nonzero on a single $(n-1)$-simplex, then any general gauge transformation can be implemented. 

\begin{figure}[h!]
  \centering
  \includegraphics[width=\linewidth]{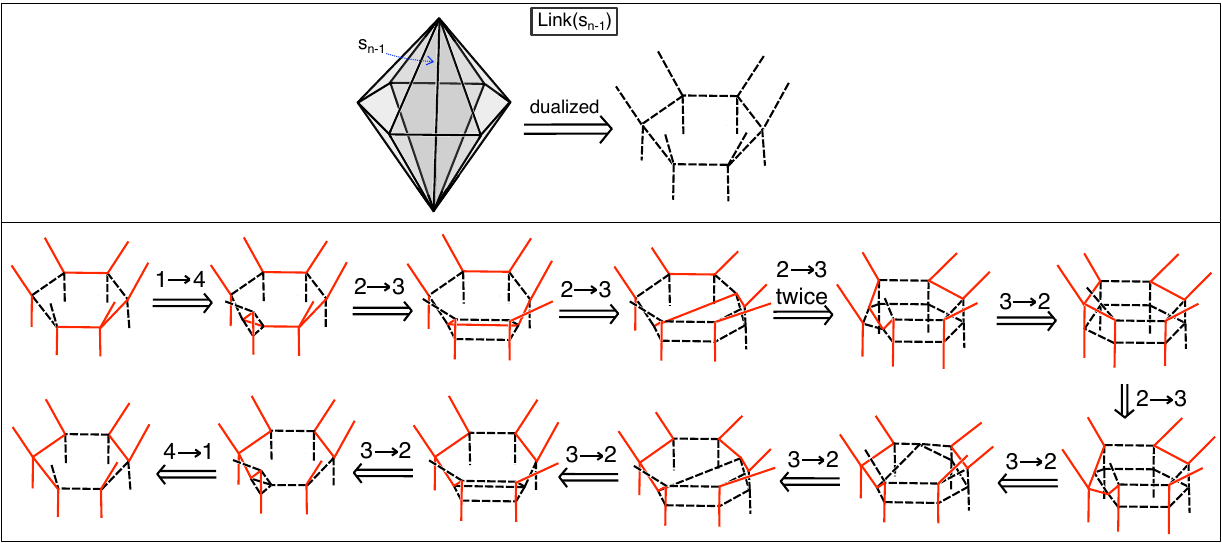}
  \caption{Implementing $\alpha \to \alpha + \delta\lambda$ for 2-form $\Z_2$ gauge fields in $d=3$. (Top) The original link of an $(n-1)$-simplex and its dualized version. No gauge field are specified in this part. (Bottom) Decoration with gauge fields and (as viewed from the dual persepective) Pachner moves implementing $\alpha \to \alpha + \delta \lambda$ for $\lambda(s_{n-1}) = 1$ and $\lambda(t)=0$ for $t \neq s_{n-1}$. Note that throughout the top line of moves, all dual edges on newly created edges on `bottom hexagon' are always zero. And, at the end of the top line that the dual edges on the `top hexagon' are equal to the `orignal hexagon' except gauge transformed. This is illustrating the $\alpha(t) = 0$ for $n$-simplices $t = \braket{v \, j \, i_0 \cdots \hat{i}_k \cdots i_{n-1}}, k \neq 0$ and how at the end of the first half of moves $\alpha(v \, j \, i_1 \cdots i_{n-1}) = a + \alpha^\text{original}(v i_0 \cdots i_{n-1})$ as described in the main text. The bottom line of moves brings us back to the original triangulation and shows that $\alpha \to \alpha + \delta \lambda$ has indeed been implemented.}
  \label{exampleHigherFormPachner}
\end{figure}

The set-up will be similar in that we will implement the transformation for $\lambda = a$ on $(n-1)$-simplex $s_{n-1} = \braket{i_0 \cdots i_{n-1}}$ by considering $\mathrm{Link}(s_{n-1})$, which will be some triangulation of the sphere $S^{d-n}$. First we will describe the series of Pachner moves and then describe how to decorate them with gauge fields. We recommend the reader to look at the Fig.~\ref{exampleHigherFormPachner} for an illustration of the moves and the 2-form $\Z_2$ gauge fields in $d=3$ while reading the rest of the argument.

The procedure will be similar to the 1-form case and the barycentric subdivision. First pick some $\braket{w_0 \cdots w_{d-n}} \in \mathrm{Link}(s_n)$ and do a $(1,d+1)$-move creating a new vertex $j$.

\begin{equation} \label{firstMoveHigherForm}
\begin{split}
\braket{i_0 \cdots i_n \, w_0 \cdots w_{d-n}} \to &\{\braket{j \, i_1 \cdots i_{n-1} \, w_0 \cdots w_{d-n}},\braket{i_0 \, j \, i_2 \cdots i_{n-1} \, w_0 \cdots w_{d-n}}, \dots, \braket{i_0 \, i_1 \cdots i_{n-2} \, j \, w_0 \cdots w_{d-n}}, \\ 
& \braket{i_0 \cdots i_{n-1} \, j \, w_1 \cdots w_{d-n}}, \braket{i_0 \cdots i_{n-1} \, w_0 \, j \, w_2 \cdots w_{d-n}}, \dots, \braket{i_0 \cdots i_{n-1} \, w_0 w_1 \cdots w_{d-n-1} j}\}
\end{split}
\end{equation}

And proceeding in the same fashion as for the 1-form case, we can keep doing moves one at a time and collect all of the `leftover' simplices that involve both $j$ and the \textit{entire} simplex $s_n$; leftover simplices  contain all of $\{j, i_0, \dots, i_{n-1}\}$ and are to be gotten rid of one at a time. After each step, there will be a `used' region $\mathcal{R}_{\text{used}} \subset \mathrm{Link}(s_n)$ that corresponds to those $(d-n)$-simplices $\braket{v_0 \cdots v_{d-n}}$ for which the $d$-simplices
\begin{equation*}
\{\braket{j \, i_1 \cdots i_{n-1} \, v_0 \cdots v_{d-n}},\braket{i_0 \, j \, i_2 \cdots i_{n-1} \, v_0 \cdots v_{d-n}}, \dots, \braket{i_0 \, i_1 \cdots i_{n-2} \, j \, v_0 \cdots v_{d-n}}\}
\end{equation*}
have all been created. And, each step will leave `leftover' $d$-simplices that are in bijection with $\partial\mathcal{R}_{\text{used}}$. So for some $\braket{x_0 \cdots x_{d-n-1}} \in \partial\mathcal{R}_{\text{used}}$, the $d$-simplex
\begin{equation*}
\braket{j i_0 \cdots i_{n-1} \, x_0 \cdots x_{d-n-1}}
\end{equation*}
is a `leftover' simplex which still exists and which we will eliminate. After every step, we do a Pachner move on a simplex neighboring the `used' region $\mathcal{R}_{\text{used}}$ and eventually, $\mathcal{R}_{\text{used}}$ will consist of the entire $\mathrm{Link}(s_{n-1})$. We can collect all of the terms at the end of this process and they will all be of the form:
\begin{align*}
\braket{j \, i_0 \, \cdots \hat{i}_k \cdots i_{n-1} \, v_0 \cdots v_{d-n}}
\end{align*}
where $\hat{i}_k$ is the vertex that is not included in the simplex and $\braket{v_0 \cdots v_{d-n}}$ is any $(d-n)$-simplex in the link. Note that the number of $d$-simplices in the link at this point is $n$ times the number of $d$-simplices originally, so our triangulation is \textit{not} the same as it was originally. To remedy this, we will reverse the order of the Pachner moves to bring us back to the original triangulation. The hope is that at the end of this process, if we are careful about which group elements we place on the links, then the change $\alpha \to \alpha + \delta\lambda$ can be implemented, for $\lambda(s_{n-1}) = a \in A$ and $\lambda = 0$ on all other $(n-1)$-simplices. 

To do this, we first specify the group elements on $n$-simplices after the first move (\ref{firstMoveHigherForm}) in the sequence. Consider a move for which the $n$-simplex $\braket{j \, i_0 \cdots i_{n-1}}$ is given value $\alpha(\braket{j \, i_0 \cdots i_{n-1}})=a$ and for which $\alpha(t) = 0$ for every $n$-simplex $t$ of the form $t = \braket{v \, j \, i_0 \cdots \hat{i}_k \cdots i_{n-1}}, k \neq 0$. Note that this condition together with $\alpha(\braket{j \, i_0 \cdots i_{n-1}})=a$ and the values of the unchanged $n$-simplices completely specify each $\alpha(\braket{v j \, i_1 \cdots i_{n-1}})$ as 
\begin{equation*}
    \alpha(\braket{v j \, i_1 \cdots i_{n-1}}) =  \alpha^{\text{original}}(\braket{v i_0 \, i_1 \cdots i_{n-1}}) + a
\end{equation*}
after this first move because of the condition $\delta \alpha = 0$. (The values on other $n$-simplices can be fixed arbitrarily and do not matter.)

Then, we proceed by doing Pachner moves as described above. In general, for each vertex $v \in \mathrm{Link}(s_{n-1})$ and each $n$-simplex $t = \braket{v \, j \, i_0 \cdots \hat{i}_k \cdots i_{n-1}}, k \neq 0$, we will require that $\alpha(t) = 0$. Note that as long as $\partial\mathcal{R}_{\text{used}}$ is nonempty $\braket{j \, i_0 \cdots i_{n-1}}$ will still exist in the simplicial complex with $\alpha(\braket{j \, i_0 \cdots i_{n-1}}) = a$, and it only disappears at the last step when $\partial \mathcal{R}_{\text{used}}$ vanishes and there are no `leftover' simplices. This together with the values of $\alpha$ on the other $n$-simplices will again specify $\alpha(\braket{v j \, i_1 \cdots i_{n-1}}) =  \alpha^{\text{original}}(\braket{v i_0 \cdots i_{n-1}}) + a$ since $\delta \alpha = 0$. 

Now consider the end of this first half of the steps when the $d$-simplices are all of the form $\braket{j \, i_0 \, \cdots \hat{i}_k \cdots i_{n-1} \, v_0 \cdots v_{d-n}}$ for $k \in \{0 \cdots n-1\}$ and $\braket{v_0 \cdots v_{d-n}} \in \mathrm{Link}(s_{n-1})$. We will have that for each \textit{vertex} $v \in \mathrm{Link}(s_{n-1})$, $\alpha(v j \, i_1 \cdots i_{n-1})$ will be $a + \alpha^\text{original}(v i_0 \cdots i_{n-1})$, where $\alpha^\text{original}(v i_0 \cdots i_{n-1})$ was the value of the cochain at the beginning of this process. 

Now the second half of the steps consists in doing the Pachner moves backwards to the original triangulation. Since $\braket{j i_0 \cdots i_{n-1}}$ was removed at the last Pachner move, we will be free to reassign $\alpha(\braket{j i_0 \cdots i_{n-1}}) = 0$ this time around when it reappears. Continuing this backwards process with this value of $\alpha(\braket{j i_0 \cdots i_{n-1}}) = 0$ together with $\delta\alpha = 0$ will completely specify the cochain everywhere. And at the end, we will have that for each vertex $v \in \mathrm{Link}(s_{n-1})$, $\alpha(v i_0 \cdots i_{n-1}) = a + \alpha^\text{original}(v i_0 \cdots i_{n-1})$. This is the change $\alpha \to \alpha + \delta\lambda$ we seek and we are done. 

\section{Effect of vertex-basis transformation $\Gamma^{\psi \psi}_1$} \label{app:GammaPsiPsi_1}

Here, we examine the effect of vertex-basis transformations $\Gamma^{a b}_c$ (see Eq.~\eqref{eq:vertexBasisTransformation}) on the bosonic shadow $Z_b(M,A_b,f)$ for closed $M$. As explained in Secs.~\ref{subsubsec:shadowChoices},\ref{subsubsec:fermionLocality}, the restriction of only considering $\Gamma^{\psi \psi}_1 = +1$ leaves the path integral invariant using the same proof as in~\cite{bulmash2020}. In this section we will review the proof and examine how the path integral changes under more general $\Gamma^{\psi \psi}_1 \neq 1$. 

First, we will see that the presence of an anti-unitary symmetry action constrains $\Gamma^{\psi,\psi}_1$. Gauge transformations must preserve the constraint Eq.~\eqref{eqn:Upsi1}, and therefore must leave $U_{\bf g}(\psi,\psi;1)$ invariant. Using the transformation rules Eq.~\eqref{eq:U_change_vertex_basis} for $U$ under a vertex basis transformation, we find $U_{\bf g}$ is invariant provided
\begin{equation}
    \Gamma^{\psi \psi}_1 [(\Gamma^{\psi \psi}_1)^{-1}]^{s({\bf g})}=1.
\end{equation}
If ${\bf g}$ has a unitary action (i.e. if $s({\bf g}) = +1$), then the above equation is automatically satisfied for any $\Gamma^{\psi \psi}_1 \in \U$. However, if ${\bf g}$ is anti-unitary (i.e. $s(\bf g) = *$), then the above constraint gives $(\Gamma^{\psi \psi}_1)^2 = +1$, which is only true if 
\begin{equation}
 \Gamma^{\psi \psi}_1 = \pm 1.
\end{equation}
The only nontrivial transformation allowed in the presence of anti-unitary symmetries is thus $\Gamma^{\psi \psi}_1 = -1$.

Now let the amplitude after a vertex basis gauge transformation be $Z_b'(M,A_b,f)$. As above, if there are \textit{no} anti-unitary symmetry actions in the category, we may consider $\Gamma^{\psi \psi}_1$ to be any phase. In this case, we will show that
\begin{equation}
    Z_b'(M,A_b,f) = Z_b(M,A_b,f),
\end{equation}
i.e. that the amplitude is fully invariant under any $\Gamma^{\psi \psi}_1$ transformation. However, the case of an anti-unitary action will give the change
\begin{equation}
    Z_b'(M,A_b,f) = Z_b(M,A_b,f) \cdot (-1)^{\int w_1 \cup f}, \text{ if } \Gamma^{\psi \psi}_1 = -1
\end{equation}
for the only nontrivial choice of $\Gamma^{\psi \psi}_1$.

\begin{figure}[h!]
  \centering
  \includegraphics[width=\linewidth]{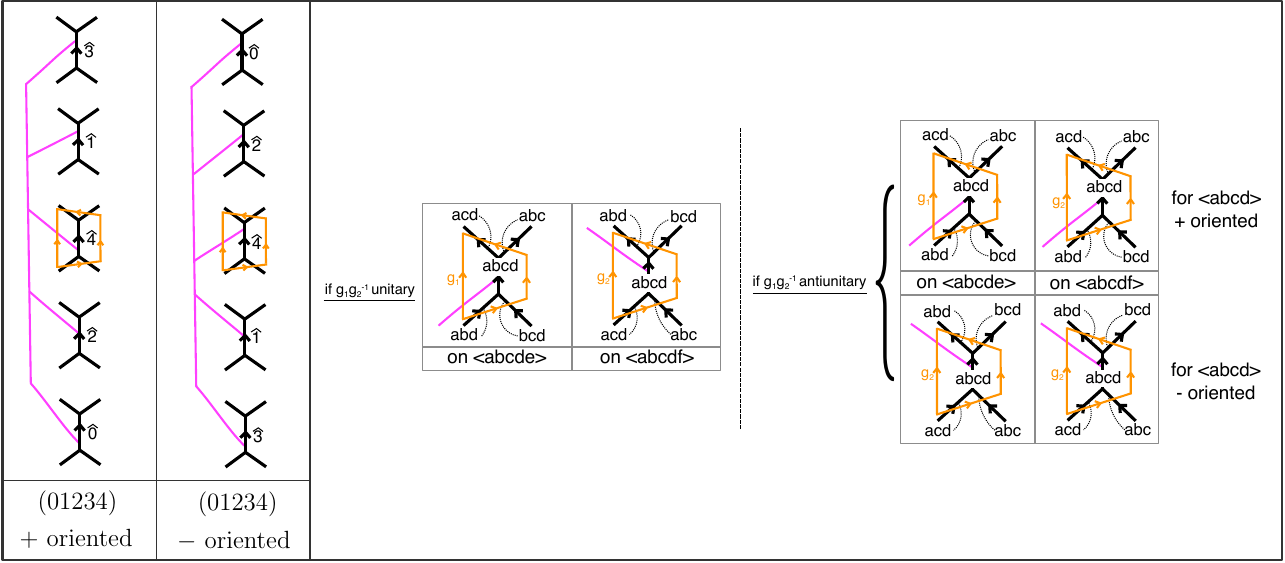}
  \caption{Relevant parts of the diagrammatics needed to analyze the effects of vertex-basis transformations $\Gamma^{a b}_c$. (Left) The orientations of the background fermion lines emanating from the 3-simplices, where each $\hat{i}$ represents the 3-simplex missing $i \in \{0,1,2,3,4\}$. On a $+$-oriented [$-$-oriented] 4-simplex, the lines emanating from $\hat{i}$ for $i$ even go up [down] and those from $\hat{i}$ for $i$ odd go down [up]. (Right) The relative orientations of the 3-simplices on its two adjacent 4-simplices. The left-side shows how it looks away from $w_1$ and the right-side shows how they look when the 3-simplex is part of $w_1$. See Fig.~\ref{inducedOrientationUnitarity} and the surrounding text for more discussion regarding induced orientation on the 3-simplices.}
  \label{fig:gammaPsiPsi_diagrams}
\end{figure}

Now, we explain the argument for the above statements. In the process, we will review how the path integral is totally invariant under all transformations $\Gamma^{a b}_c$ that are \textit{not} $\Gamma^{\psi \psi}_1$. The relevant parts of the diagrammatics to consider are given in Fig.~\ref{fig:gammaPsiPsi_diagrams}, which show the directions that the background fermion lines exit from a 3-simplex and how the diagrams look with respect to relative induced orientations on 3-simplices.

There are two potential classes of fusion vertices that we need to consider. First are those on a 3-simplex $\Delta_3$ that involve the anyons $b_{\Delta_3},b_{\Delta_3} \times \psi^{f(\Delta_3)}$. There are three such fusion vertices for each 3-simplex in each diagram. For example, this class of fusion vertices for $\braket{0123}$ would be $\{(012 , 023 \leftrightarrow 0123), (0123 , \psi^{f(0123)} \leftrightarrow 0123 \times \psi^{f(0123)}),  (0123 \times \psi^{f(0123)} \leftrightarrow 013 , 123\}$. Since $M$ is closed, $\Delta_3$ occurs in exactly two 4-simplices. If $\Delta_3$ is not part of $w_1$, then each fusion vertex in this class appears twice, but with opposite orientation. Under a vertex basis transformation, one appearance contributes $\Gamma^{a b}_c$ and the other contributes $(\Gamma^{a b}_c)^{-1}$, which cancel out in the product over 4-simplices. The same argument also applies if $\Delta_3$ \textit{is} part of $w_1$, since even though the fusion vertices appear with the same induced orientation, one factor will be complex conjugated by being in a region of the diagram with an anti-unitary twist. 

The second class of fusion vertices involve the junctions where the five potential background fermion lines on a 4-simplex fuse into each other. Here, the only relevant basis transformations are $\Gamma^{\psi \psi}_1$ since 
we only consider gauges such that all transformations $\Gamma^{\psi 1}_\psi,\Gamma^{1 \psi}_\psi$ are $1$\footnote{If we declare that, in our diagrammatics, all fusion vertices only appear on 3-simplices, our path integral is actually invariant under basis transformations $\Gamma^{a,1}_a$ and $\Gamma^{1,a}_a$ for $a \neq \psi$, but is not gauge-invariant for $a=\psi$ (nor is there a simple expression for the transformation in this case). Vertices with $a \neq \psi$ only appear on 3-simplices, so the reason for invariance is the same as the reason for invariance under generic $\Gamma^{a,b}_c$.}. First, we will use the trivalent resolution in the diagrams to split the background fermion lines into a set of distinct loops which enter and exit a 4-simplex through two different 3-simplices. This splits the loop up in the same way as it would the trivalent resolution of the dual 1-skeleton, as in Fig.~\ref{trivalentResolutionsAndWindings}. Without loss of generality, we will consider the case that $f$ is dual to a single loop (note the final result given by the multiplicative factor $(-1)^{\int w_1 \cup f}$ is linear in $f$). Since the factors can be considered independently on different loops, the proof for a single loop implies the general case.

\begin{figure}[h!]
  \centering
  \includegraphics[width=\linewidth]{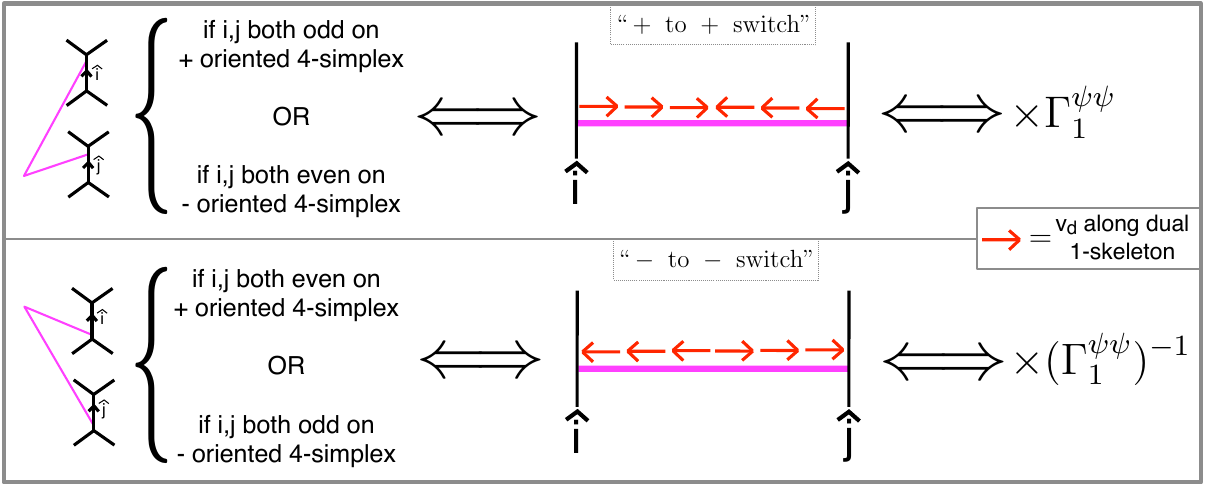}
  \caption{Factors of $\Gamma^{\psi \psi}_1$ that multiply the amplitude $Z_b$ and how they relate to the parity of the 4-simplex and the 3-simplices in question. All other cases of $\hat{i},\hat{j}$ give a trivial factor of $\Gamma^{\psi 1}_\psi = 1$ or $\Gamma^{1 \psi}_\psi = 1$. The top case of the amplitude being multiplied by $\Gamma^{\psi \psi}_1$ corresponds to the case where the vector $v_d$ along the dual 1-skeleton (see Sec.~\ref{prelimSecB}) points as ${\color{red} \rightarrow \leftarrow}$ between $\hat{i}$ and $\hat{j}$. The bottom case of a factor $(\Gamma^{\psi \psi}_1)^{-1}$ corresponds to $v_d$ pointing as ${\color{red} \leftarrow \rightarrow}$ along this leg. As such, the only times when such factors appear correspond to these `direction switches' of $v_d$ along the loop. The top/bottom cases can be referred to as a ``$+ \text{ to } +$ switch'' / ``$- \text{ to } -$ switch'' because they interpolate between two 3-simplices both with induced $+$-orientation / both $-$-orientation (see Fig.~\ref{inducedOrientation3Simplex}).}
  \label{fig:gammaPsiPsi_fermionFactors}
\end{figure}

In Fig.~\ref{fig:gammaPsiPsi_fermionFactors}, the different cases of when the amplitude gets multiplied by $\Gamma^{\psi \psi}_1$ or $(\Gamma^{\psi \psi}_1)^{-1}$ are enumerated. It is convenient to think in terms of the vector $v_d$ along the dual 1-skeleton, as explained in the text below Fig.~\ref{fig:gammaPsiPsi_fermionFactors}.

\begin{figure}[h!]
  \centering
  \includegraphics[width=0.8\linewidth]{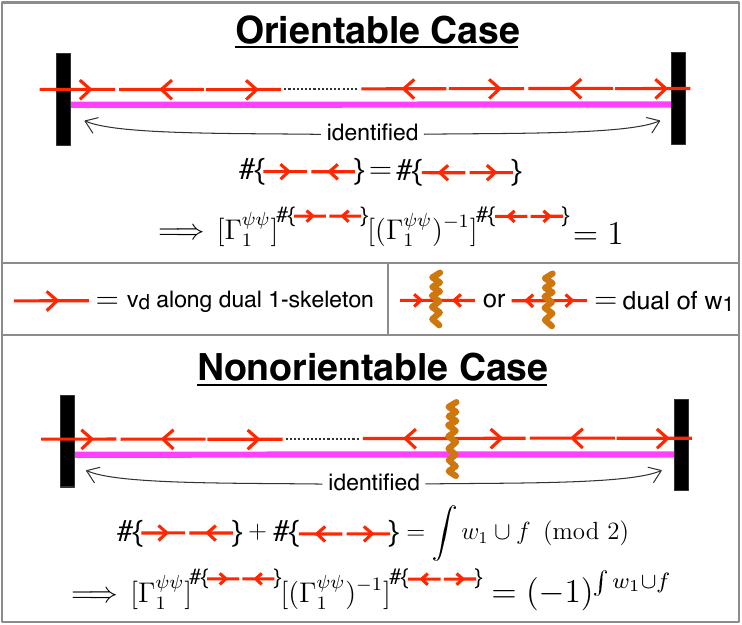}
  \caption{Pictorial argument for the changes in the amplitude under $\Gamma^{\psi \psi}_1$ in the orientable and non-orientable cases. The pink line is a single fermion loop considered with ends at the black rectangles being identified, and the red arrows are the $v_d$ vector field along the loop. $\#\{{\color{red} \leftarrow \rightarrow}\},\#\{{\color{red} \rightarrow \leftarrow}\}$ refer to the number of direction switches of $v_d$ that occur \textit{inside} a 4-simplex; in the non-orientable case this means they \textit{do not} count those coming from crossing $w_1$.}
  \label{fig:gammaPsiPsi_proof}
\end{figure}

From here, we can compute the contribution of the amplitude change from this loop. The logic is summarized pictorially in Fig.~\ref{fig:gammaPsiPsi_proof}, and is explained below.

Let us first consider the case with no anti-unitary symmetries. Denote $\#\{{\color{red} \rightarrow \leftarrow}\}$ as the number of $+ \text{ to } +$ switches and $\#\{{\color{red} \leftarrow \rightarrow}\}$ as the number of $- \text{ to } -$ switches of $v_d$ in the loop (see Fig.~\ref{fig:gammaPsiPsi_fermionFactors}). The amplitude change from the loop is
\begin{equation*}
    [\Gamma^{\psi \psi}_1]^{\#\{{\color{red} \rightarrow \leftarrow}\}} \cdot [(\Gamma^{\psi \psi}_1)^{-1}]^{\#\{{\color{red} \leftarrow \rightarrow}\}} = [\Gamma^{\psi \psi}_1]^{\#\{{\color{red} \rightarrow \leftarrow}\} - \#\{{\color{red} \leftarrow \rightarrow}\}}.
\end{equation*}
Note that since there are no anti-unitary symmetries, the manifold must be orientable. All of the direction switches that occur inside a loop are inside a 4-simplex and accounted for above. This means that $\#\{{\color{red} \rightarrow \leftarrow}\} = \#\{{\color{red} \leftarrow \rightarrow}\}$. So, the change in amplitude is  $[\Gamma^{\psi \psi}_1]^{\#\{{\color{red} \rightarrow \leftarrow}\} - \#\{{\color{red} \leftarrow \rightarrow}\}} = 1$, which is invariant as we wanted.

In the anti-unitary case, the above argument needs modification due to the presence of $w_1$. In the orientable case, the $v_d$ vector always points in the same direction going across a 3-simplex. In the non-orientable case, the $v_d$ vector changes directions going across $w_1$, like in Fig.~\ref{fig:vectsCrossingW1}. However, only direction switches \textit{inside} a 4-simplex contribute to the amplitude change: direction switches across $w_1$ are not accounted for in the diagrammatics and don't contribute any factors. As such, we will define $\#\{{\color{red} \leftarrow \rightarrow}\},\#\{{\color{red} \rightarrow \leftarrow}\}$ to be the number of such direction switches occurring \textit{only inside} a 4-simplex. Since the total number of direction switches (i.e. including those coming from $w_1$) is even, we have the equality $\#\{{\color{red} \leftarrow \rightarrow}\}+ \#\{{\color{red} \rightarrow \leftarrow}\} = \int {w_1 \cup f} \text{ (mod 2)}$. And if we consider the only possible nontrivial $\Gamma^{\psi \psi}_1 = -1$ in the antiunitary case, we find
\begin{equation}
    Z_b' = Z_b \cdot (-1)^{\#\{{\color{red} \leftarrow \rightarrow}\}+ \#\{{\color{red} \rightarrow \leftarrow}\}} = Z_b \cdot (-1)^{\int w_1 \cup f}
\end{equation}
which is what we set out to show.

\section{ $\Sq^2 + A_b^{\ast} \omega_2$ Anomaly of $Z_b(M,f)$}
\label{ZbAnomaliesSec}

Here, we demonstrate how the $\Sq^2 + A_b^{\ast} \omega_2$ anomaly of the bosonic shadow can be derived. We derive this by considering a Pachner move $M \to M'$ in which $M$ and $M'$ differ by gluing on a $5$-simplex to $M$ on some collection of $4$-simplices and removing the original $4$-simplices, analogously to Fig.~\ref{2D_Pachner} in two dimensions higher. 

The Pachner move $M \to M'$ can thought of as a $5$-manifold $W$ with boundary $\partial W = M \bigsqcup \overline{M'}$ and consists of a single $5$-simplex $\braket{012345}$. In addition, after decorating the $3$-form field $f$ dual fermion lines and the $G_b$ gauge field $A_b$ onto these Pachner moves, the fields $f$ and $A_b$ can be thought to be extended onto this $5$-manifold $W$.

The invariance property we want to show is:
\begin{equation}
Z_b\left(M,\restr{f}{M},\restr{A_b}{M}\right) = (-1)^{\int_{\braket{012345}} \Sq^2(f) + f \cup A_b^{\ast} \omega_2} Z_b\left(M',\restr{f}{M'},\restr{A_b}{M'}\right)
\end{equation}
We can alternatively express the sign difference as:
\begin{equation*}
(-1)^{\int_{\braket{012345}} \Sq^2(f) + f \cup A_b^{\ast} \omega_2} = (-1)^{\int_{\braket{012345}} ( \Sq^2 + f_\infty A_b^{\ast} \omega_2 )(f)}
\end{equation*}

First we will describe how the $\Sq^2(f)$ part of the anomaly arises in the Pachner calculation. This part is `universal' in the sense that it would occur even in the absence of symmetry domain walls. 

Then we will describe how the $U$ and $\eta$ symbols coming from the symmetry domain walls in the Pachner move gives a contribution of $f \cup A_b^{\ast} \omega_2$ to the $Z_b$ anomaly.

Technically, we need to show this for all possible Pachner moves with all possible decorated branching structures and fermion and gauge field decorations. We will only describe these calculations in detail for the case of the 3-3 Pachner move for which we change the simplices from $\braket{12345},\braket{01345},\braket{01235}$ in $M$ to $\braket{01234},\braket{01245},\braket{02345}$ in $M'$. However the arguments will indeed translate to the other cases. We will explain how at the end of each sub-part of the calculation.

\subsection{The $\Sq^2$ part}
Here we describe the $\Sq^2$ part of the anomaly. The present calculation is valid when $A_b=0$, implying there is no orientation-reversing wall in the vicinity. We will see in the next subsection that the calculation for when there \textit{are} orientation-reversing walls and nonzero $A_b$ can in fact mapped onto this case.

Since there is no orientation-reversing wall, the move $\braket{12345},\braket{01345},\braket{01235}$ to $\braket{01234},\braket{01245},\braket{02345}$ can be checked to take three 4-simplices of the same orientation (say all $+$'s) to three other $+$ 4-simplices. It will eliminate the 2-simplex $\braket{024}$ and the 3-simplices $\braket{0124},\braket{0234},\braket{0245}$ on $M$ and create the new 2-simplex $\braket{135}$ and the 3-simplices $\braket{0135},\braket{1235},\braket{1345}$ on $M'$. This means that the equation we have to verify is roughly
\begin{equation*}
\text{``}
\sum_{024,0124,0234,0245} Z_b^+(02345)Z_b^+(01245)Z_b^+(01234) = \sum_{135,0135,1235,1345} Z_b^+(12345)Z_b^+(01345)Z_b^+(01235)
\text{''}
\end{equation*}
where the quotation marks mean we have not accounted for quantum dimension factors. The normalizing quantum-dimension factors for the amplitudes in $Z_b(M)$ in front of each product of diagram evaluations on 4-simplices are: 
\begin{equation*}
\mathcal{D}^{2(N_0 - N_1) - \chi(M)}
\frac{\prod_{\text{all 2-simplices } \Delta_2} d_{\Delta_2}}{\prod_{\text{all 3-simplices } \Delta_3} d_{\Delta_3}}
\prod_{\text{4-simplices } \Delta_4} \sqrt{
\frac{\prod_{\text{3-simplices } \Delta_3 \subset \Delta_4} d_{\Delta_3}}{\prod_{\text{2-simplices } \Delta_2 \subset \Delta_4} d_{\Delta_2}}
}
\end{equation*}
where the product over 4-simplices involves $\mathcal{N}_{\Delta_4}$ (see Eq.~\eqref{eq:normalizationFactor_15j}). Define the above as $\mathcal{N}_\text{q-dims}$. Note that every 3-simplex is present in exactly two 4-simplices. This means that the second product over 4-simplices gives a factor of $d_{\Delta_3}$ for every 3-simplex. So the two products over 3-simplices cancel out. The right-hand product over 2-simplices gives a factor of $(\frac{1}{\sqrt{d_{\Delta_2}}})^{|\text{Link}(\Delta_2)|}$, since $|\text{Link}(\Delta_2)|$ is the number of $3$-simplices that $\Delta_2$ is a part of, which is also the number of $4$-simplices $\Delta_2$ is a part of (recalling that the link of a $(d-2)$-simplex is always a circle). This gives
\begin{equation} \label{quantumDimContribution}
\mathcal{N}_\text{q-dims} =
\mathcal{D}^{2(N_0 - N_1) - \chi(M)}
\prod_{\text{all 2-simplices } \Delta_2} \sqrt{d_{\Delta_2}}^{2-|\text{Link}(\Delta_2)|}
\end{equation}

For the 3-3 move $\braket{01234},\braket{01245},\braket{02345}$ to $\braket{12345},\braket{01345},\braket{01235}$, $\text{Link}(\braket{024})$ consists of only the 4-simplices $\braket{01234},\braket{01245},\braket{02345}$ which are in the original $M$ and that $\text{Link}(\braket{135})$ consists of only the 4-simplices $\braket{01235},\braket{01345},\braket{12345}$ on the transformed $M'$. This is because those 4-simplices all border their respective 2-simplices and form a loop, so they must be the entire link. 

In addition, one can check (by counting the number of 3-simplices containing them before and after) that the sizes of the links of the 2-simplices $\{\braket{045},\braket{245},\braket{025},\braket{014},\braket{124},\braket{012},\braket{034},\braket{023},\braket{234}\}$ will all decrease by one whereas those of $\{\braket{123},\braket{125},\braket{235},\braket{013},\braket{035},\braket{015},\braket{134},\braket{145},\braket{345}\}$ increase by one. 

Finally, note that the number of 2-,3-,4-simplices in the triangulation stays the same before and after the Pachner move. So $N_0-N_1$ stays the same before and after since $N_0 - N_1 + N_2 - N_3 + N_4 = \chi(M)$ is a topological invariant throughout the process.

All together means that the precise equation we need to check is
\begin{equation} \label{sq2Anomaly_3_3_pachner_eqn}
\begin{split}
LHS' &:= \frac{1}{\sqrt{d_{045}d_{245}d_{025}d_{014}d_{124}d_{012}d_{034}d_{023}d_{234}}} \sum_{024,0124,0234,0245} \frac{1}{\sqrt{d_{024}}}\frac{Z_b^+(01234)Z_b^+(01245)Z_b^+(02345)}{\mathcal{N}_{01234}\mathcal{N}_{01245}\mathcal{N}_{02345}} \\
&=
\frac{(-1)^{f(0345)f(0123)+f(0145)f(1234)+f(0125)f(2345)}}{\sqrt{d_{123}d_{125}d_{235}d_{013}d_{035}d_{015}d_{134}d_{145}d_{345}}} \sum_{135,0135,1235,1345} \frac{1}{\sqrt{d_{135}}}\frac{Z_b^+(01235)Z_b^+(01345)Z_b^+(12345)}{\mathcal{N}_{01235}\mathcal{N}_{01345}\mathcal{N}_{12345}}\\
& =: RHS' \cdot (-1)^{\Sq^2(f)(012345)}
\end{split}
\end{equation}
where each $\frac{Z_{\text{4-simplex}}}{\mathcal{N}_{\text{4-simplex}}}$ is an unnormalized 15j symbol and all the quantum dimension factors are there to compensate for the deletion of $024$, the addition of $135$, and the changes in sizes of other 2-simplex links as stated above. The factor $(-1)^{f(0345)f(0123)+f(0145)f(1234)+f(0125)f(2345)} = (-1)^{(f \cup_1 f)(012345)} = (-1)^{\Sq^2(f)(012345)}$ is the anomaly factor that we wanted to find in this subsection. 

\begin{figure}[h!]
    \centering
    \begin{minipage}{0.98\textwidth}
        \centering
        \includegraphics[width=\linewidth]{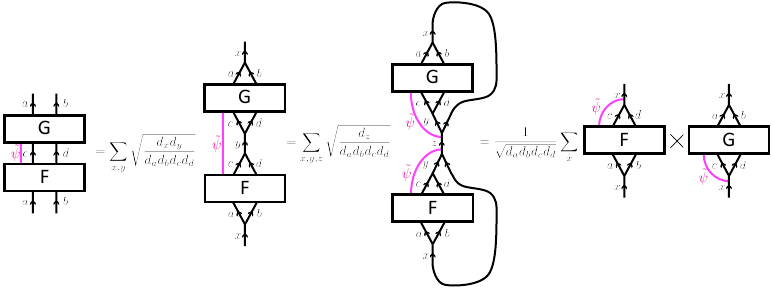}
        \end{minipage} \quad \quad \quad
    \begin{minipage}{0.98\textwidth}
         \centering
         \includegraphics[width=\linewidth]{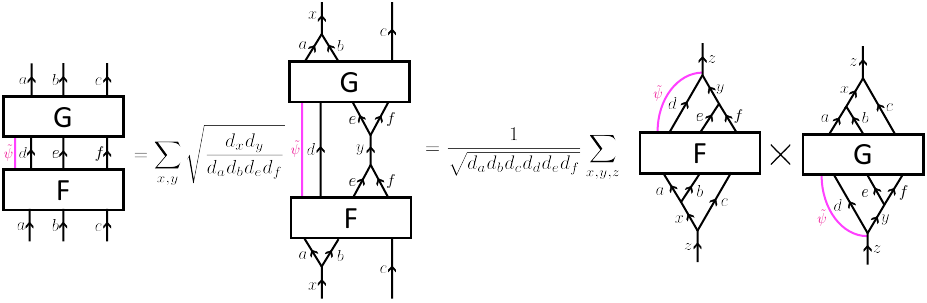}
    \end{minipage}
    \caption{(Top) Merging Lemma I and (bottom) Merging Lemma II with diagrammatic proof. The second equality of Merging Lemma II uses Merging Lemma I. Here, we take $\tilde{\psi}$ to be either the identity anyon or the fermion $\psi$. Each step uses resolutions of the identity as in Eq.~\eqref{resOfIdentity}. In addition, at one point the fermion lines can be `peeled off' using the same resolution of identity as well as the freedom to `bend' the fermion lines discussed in Sec.~\ref{sec:superModularCategories}.}
    \label{fig:mergeLemmas}
\end{figure}

\begin{figure}[h!]
  \centering
  \includegraphics[width=\linewidth]{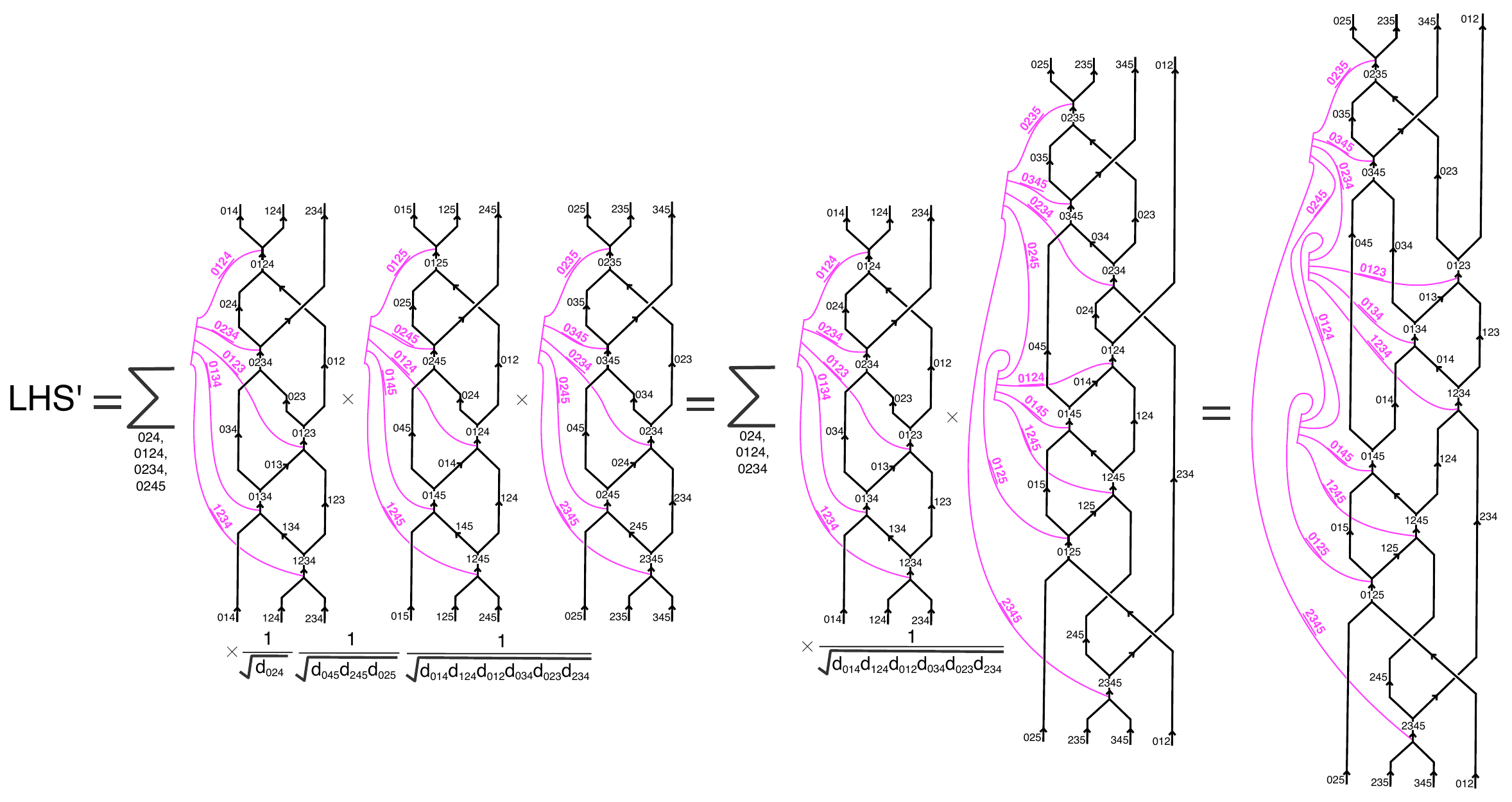}
  \caption{Diagrammatic simplification of $LHS'$ of Eq.~\eqref{sq2Anomaly_3_3_pachner_eqn}. The first step uses Merging Lemma I for $x = 0245$ and the second uses Merging Lemma II for $x=0124,y=0234,z=024$}
  \label{pachner_3_3_LHS}
\end{figure}

\begin{figure}[h!]
  \centering
  \includegraphics[width=\linewidth]{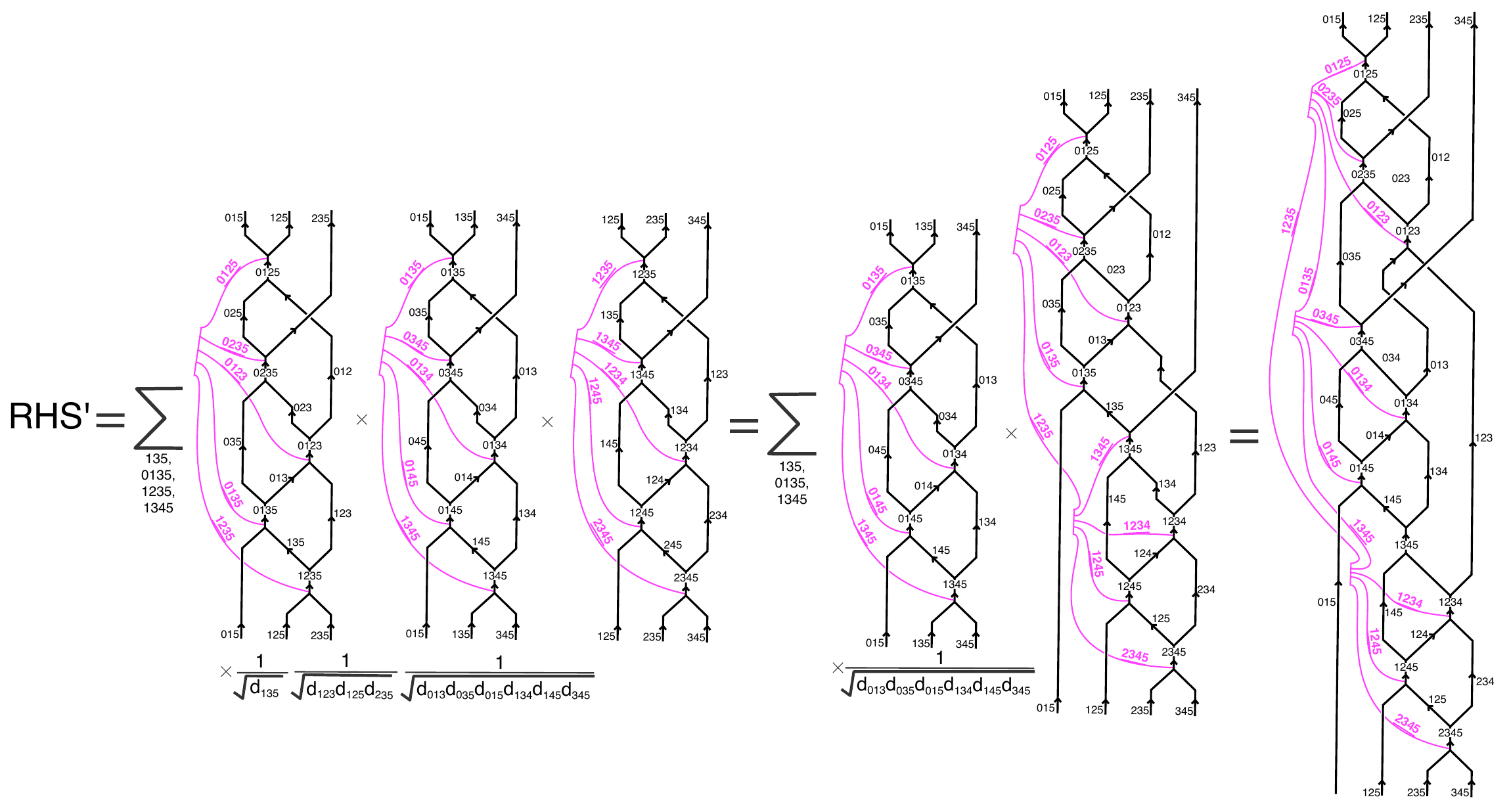}
  \caption{Diagrammatic simplification of $RHS'$ of Eq.~\eqref{sq2Anomaly_3_3_pachner_eqn}. The first step uses Merging Lemma I for $x = 1235$ and the second uses Merging Lemma II for $x=1345,y=0135,z=135$}   \label{pachner_3_3_RHS}
\end{figure}

\begin{figure}[h!]
  \centering
  \includegraphics[width=0.7\linewidth]{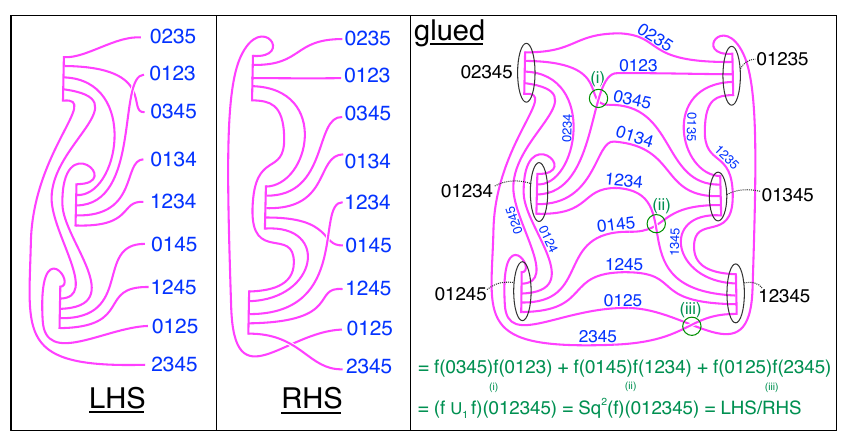}
  \caption{Comparison of the fermion lines of Figs.~\ref{pachner_3_3_LHS},\ref{pachner_3_3_RHS} shows that $LHS'/RHS' = (-1)^{\Sq^2(f)(012345)}$, which is the $\Sq^2$ anomaly. Note that one has to arrange the fermion lines in matching orders to do this comparison, which entails rearranging the fusion channels in this diagram to go in the order shown here.}
  \label{pachner_3_3_fermionSq2}
\end{figure}

To verify this, we will first need the ``merging lemmas" in Fig.~\ref{fig:mergeLemmas} that allow us to convert products of diagrams into single diagrams, slightly modifying the ones in~\cite{cuiGcrossed,bulmash2020} to account for the fermion lines. Using the merging lemmas, Eq.~\eqref{sq2Anomaly_3_3_pachner_eqn} follows from the diagrammatic calculation in Figs.~\ref{pachner_3_3_LHS},\ref{pachner_3_3_RHS},\ref{pachner_3_3_fermionSq2}.

This calculation generalizes to more general Pachner moves, and we leave it to the reader to verify this explicitly. The idea is the same for each possible move. First, one has to check that the quantum dimension factors all match up as before. Then one can use the merging lemmas to verify the $\Sq^2$ as before. In all cases, the final comparison of fermion lines will look like Fig.~\ref{pachner_3_3_fermionSq2}. This is because to produce that figure, we \textit{reflected} the fermion lines fusing from the $RHS$, which means that their orientation in that fused figure is the same as how it would be for a $-$ simplex. For general moves, the fusion of lines from $\braket{01234},\braket{01245},\braket{02345}$ will appear as they do in a $+$ 4-simplex and those from $\braket{12345},\braket{01345},\braket{01235}$ will appear as they do in a $-$ 4-simplex.

\subsection{The $A_b^{\ast} \omega_2$ part} \label{app:abOmegaAnomalyCalc}
We will find that the $f \cup A_b^{\ast} \omega_2$ part of the anomaly arises by considering how the $U$ and $\eta$ symbols coming from the diagrams all differ. We again consider the same 3-3 Pachner move. We denote ${\bf g}_{ij} = A_b(ij)$ in this subsection.

We now have to deal with how potential orientation-reversing walls coming from the gauge fields interplay with the orientations of the 4-simplices in the move. The relative orientations of a 3-simplex can be related to the gauge fields surrounding each of the two 4-simplices a 3-simplex via the $f_\infty$ map as in Sec.~\ref{usingFInftyOrientationWall}.

Examining the orientations of the simplices (for which one can look at Figs.~\ref{pachner_3_3_LHS},\ref{pachner_3_3_RHS} along with Fig.~\ref{fig:15jSymbols}) shows that for all the orientations of simplices to be consistent in the 3-3 Pachner move $\braket{01234},\braket{01245},\braket{02345}$ to $\braket{12345},\braket{01345},\braket{01235}$, we will need the orientations of $\braket{01245},\braket{02345},\braket{12345},\braket{01345},\braket{01235}$ to be the same orientation (say $+$) and that the orientation of $\braket{01234}$ should be $s({\bf g}_{45}) = s(45)$.

Contract all of the domain walls in the 15j symbols, collect all the $U$ and $\eta$ symbols from the 15j symbols on the $LHS$, $\braket{01234},\braket{01245},\braket{02345}$, and define their product as
\begin{equation}
\begin{split}
\text{symm}_\text{LHS} = \bigg(&\frac{1}{\eta_{012}(23,34)^{s(24)}} \Big( \frac{U_{34}(013,123 ; 0123 \times f_{0123})}{U_{34}(023, {^{32}}012 ; 0123 ) U_{34}(f_{0123},0123 ; 0123 \times f_{0123})} \Big)^{s(34)} \bigg)^{s(45)} \\
\times &\frac{1}{\eta_{012}(24,45)^{s(25)}} \Big( \frac{U_{45}(014,124 ; 0124 \times f_{0124})}{U_{45}(024, {^{42}}012 ; 0124 ) U_{45}(f_{0124},0124 ; 0124 \times f_{0124})} \Big)^{s(45)} \\
\times &\frac{1}{\eta_{023}(34,45)^{s(35)}} \Big( \frac{U_{45}(024,234 ; 0234 \times f_{0234})}{U_{45}(034, {^{43}}023 ; 0234 ) U_{45}(f_{0234},0234 ; 0234 \times f_{0234})} \Big)^{s(45)}
\end{split}
\end{equation}
where the exponent $s(45)$ on the first row corresponds to the the orientation of $\braket{01234}$. And similarly, we define the corresponding product on the $RHS$, $\braket{01235},\braket{01345},\braket{12345}$, as
\begin{equation}
\begin{split}
\text{symm}_\text{RHS} = &\frac{1}{\eta_{012}(23,35)^{s(25)}} \Big( \frac{U_{35}(013,123 ; 0123 \times f_{0123})}{U_{35}(023, {^{32}}012 ; 0123 ) U_{35}(f_{0123},0123 ; 0123 \times f_{0123})} \Big)^{s(35)} \\
\times &\frac{1}{\eta_{013}(34,45)^{s(35)}} \Big( \frac{U_{45}(014,134 ; 0134 \times f_{0134})}{U_{45}(034, {^{43}}013 ; 0134 ) U_{45}(f_{0134},0134 ; 0134 \times f_{0134})} \Big)^{s(45)} \\
\times &\frac{1}{\eta_{123}(34,45)^{s(35)}} \Big( \frac{U_{45}(124,234 ; 1234 \times f_{1234})}{U_{45}(134, {^{43}}123 ; 1234 ) U_{45}(f_{1234},1234 ; 1234 \times f_{1234})} \Big)^{s(45)}
\end{split}
\end{equation}

\begin{figure}[h!]
  \centering
  \includegraphics[width=0.7\linewidth]{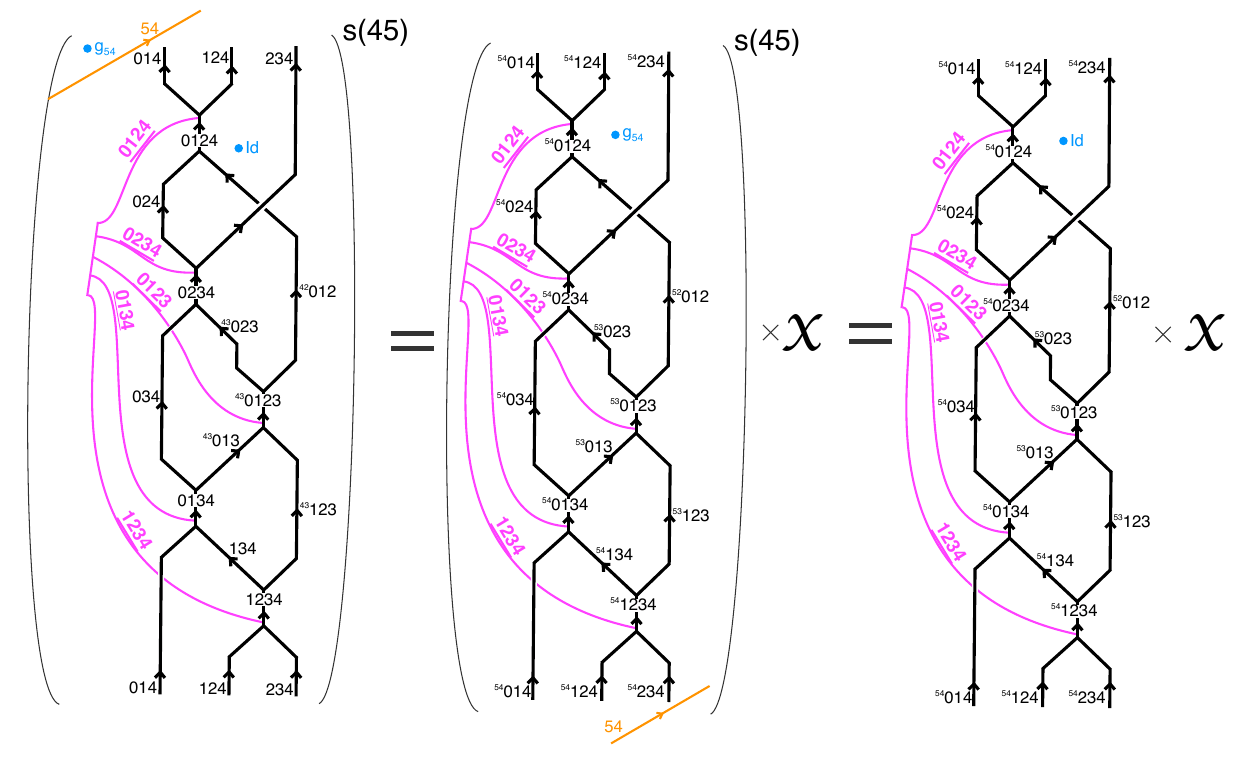}
  \caption{Sweeping a domain wall across the $\braket{01234}$ to make the anyon labels match the rest of those in the Pachner calculation gives us an extra factor $x$ of $U$ and $\eta$ symbols. }
  \label{pachner_3_3_sweepingDomainWall}
\end{figure}

The hope is that after contracting the domain walls, we will be able to apply the argument of the previous subsection to the resulting diagrams. However, following~\cite{bulmash2020}, there are two problems. First is that the diagram on $\braket{01234}$ is twisted by $s(45)$ relative to the calculation in the previous subsection. Second, looking at the resulting diagrams shows that the anyon lines on $\braket{01234}$ differ from the lines on the rest of the diagrams up to group multiplication by ${\bf g}_{45}$. To fix these issues, we ``sweep'' a ${\bf g}_{45}$ domain wall across the entire $\braket{01234}$ diagram, as in Fig.~\ref{pachner_3_3_sweepingDomainWall}, which has two effects. First, the domain wall sweep gives us an extra factor of 
\begin{equation}
\begin{split}
x = &\frac{U_{54}({^{54}}014,{^{54}}124;{^{54}}0124 \times f_{0124})}
{U_{54}({^{54}}024,{^{52}}012;{^{54}}0124) U_{54}(f_{0124},{^{54}}0124 \times f_{0124};{^{54}}0124)} \\
\times &\frac{U_{54}({^{54}}024,{^{54}}234;{^{54}}0234 \times f_{0234}) U_{54}(f_{0234},{^{54}}0234 \times f_{0234};{^{54}}0234)}{U_{54}({^{54}}034,{^{53}}023;{^{54}}0234)} \\
\times &\frac{U_{54}({^{53}}023,{^{52}}012;{^{53}}0123) U_{54}(f_{0123},{^{53}}0123; {^{53}}0123 \times f_{0123})}{U_{54}({^{53}}013,{^{53}}123;{^{53}}0123 \times f_{0123})} \\
\times &\frac{U_{54}({^{54}}034,{^{53}}013;{^{54}}0134) U_{54}(f_{0134},{^{54}}0134; {^{54}}0134 \times f_{0134})}{U_{54}({^{53}}014,{^{54}}134;{^{54}}0134 \times f_{0134})} \\
\times &\frac{U_{54}({^{54}}134,{^{53}}123;{^{54}}1234) U_{54}(f_{1234},{^{54}}1234; {^{54}}1234 \times f_{1234})}{U_{54}({^{54}}124,{^{54}}234;{^{54}}1234 \times f_{1234})} 
\end{split}
\end{equation}
that will multiply $\text{symm}_\text{LHS}$. This formula assumes that the locality constraint Eq.~\ref{eqn:Upsi1} is obeyed, or else additional factors of $U$ may appear. Then, note that after the sweep the diagram lies in a region of group element ${\bf g}_{45}$. This means by the graphical calculus rules in Fig.~\ref{fig:graphicalCalculus_defs} that the diagram evaluation in this region should be raised to the power of $s(45)$ as compared to when the diagram lies in a region with the identity group element. So in effect, this twist of $s(45)$ from the ${\bf g}_{45}$ region cancels out the twist $s(45)$ from the orientation compatibility, leaving a diagram in the identity region \textit{without} any total twist by $s(45)$. 

Now, we need to evaluate this big product of $U$ and $\eta$ in $\frac{\text{symm}_\text{LHS} \cdot x}{\text{symm}_\text{LHS}}$. The result is that this large product simplifies to
\begin{equation}
\frac{\text{symm}_\text{LHS} \cdot x}{\text{symm}_\text{RHS}} = \eta_{f_{0123}}({\bf g}_{34},{\bf g}_{45})
\end{equation}
which is exactly the evaluation of $f \cup A_b^{\ast} \omega_2 = (f_\infty A_b^\ast \omega_2)(f)$ on the 5-simplex $\braket{012345}$.

The first step is to organize all of the $U$ symbols coming from the same fusion vertices. Then, insert several copies of unity in the form $U_{{\bf 1}}(a,b;c) = 1$ to put the combination of $U$ symbols from each fusion vertex into the form $\left( \frac{U_{\bf g}(a,b;c)U_{\bf h}({^{\bar{\bf g}}}a,{^{\bar{\bf g}}}b;{^{\bar{\bf g}}}c)}{U_{\bf gh}(a,b;c)} \right)^{\pm s({\bf k})}$ where sometimes ${\bf gh} = {\bf 1}$. The relation $s(12)s(23)=s(13)$ is important. Then, repeatedly use the relation Eq.~\eqref{UUoverU_equals_EtaOverEtaEta} to obtain
\begin{equation*}
\begin{split}
\frac{\text{symm}_\text{LHS} \cdot x}{\text{symm}_\text{RHS}} = 
&\left(
\frac{\eta_{013}(35,54) \eta_{123}(35,54)}{\eta_{0123 \times f_{0123}}(35,54)}
\frac{\eta_{0123}(35,54)}{\eta_{023}(35,54) \eta_{{^{32}}012}(35,54)}
\frac{\eta_{0123 \times f_{0123}}(35,54)}{\eta_{0123}(35,54) \eta_{f_{0123}}(35,54)}
\right)^{s(35)} \\
\times &\left(
\frac{\eta_{0124 \times f_{0124}}(45,54)}{\eta_{014}(45,54) \eta_{124}(45,54)}
\frac{\eta_{024}(45,54) \eta_{{^{42}}012}(45,54)}{\eta_{0124}(45,54)}
\frac{\eta_{f_{0124}}(45,54) \eta_{0124}(45,54)}{\eta_{0124 \times f_{0124}}(45,54)}
\right)^{s(45)} \\
\times &\left(
\frac{\eta_{014}(45,54) \eta_{134}(45,54)}{\eta_{0134 \times f_{0134}}(45,54)}
\frac{\eta_{0134}(45,54)}{\eta_{{^{43}}013}(45,54) \eta_{034}(45,54)}
\frac{\eta_{0134 \times f_{0134}}(45,54)}{\eta_{f_{0134}}(45,54) \eta_{0134}(45,54)}
\right)^{s(45)} \\
\times &\left(
\frac{\eta_{0234 \times f_{0234}}(45,54)}{\eta_{024}(45,54) \eta_{234}(45,54)}
\frac{\eta_{034}(45,54) \eta_{{^{43}}023}(45,54)}{\eta_{0234}(45,54)}
\frac{\eta_{0234}(45,54)}{\eta_{f_{0234}}(45,54) \eta_{0234 \times f_{0234}}(45,54)}
\right)^{s(45)} \\
\times &\left(
\frac{\eta_{124}(45,54) \eta_{234}(45,54)}{\eta_{1234 \times f_{1234}}(45,54)}
\frac{\eta_{1234}(45,54)}{\eta_{134}(45,54) \eta_{{^{43}}123}(45,54)}
\frac{\eta_{1234 \times f_{1234}}(45,54)}{\eta_{1234}(45,54)\eta_{f_{1234}}(45,54)}
\right)^{s(45)} \\
\times &\left(
\frac{\eta_{012}(23,35)^{s(25)}\eta_{013}(34,45)^{s(35)}\eta_{123}(34,45)^{s(35)}}{\eta_{012}(23,34)^{s(24)}\eta_{012}(24,45)^{s(25)}\eta_{023}(34,45)^{s(35)}}
\right),
\end{split}
\end{equation*}
where the first five rows were obtained using Eq.~\eqref{UUoverU_equals_EtaOverEtaEta} and the last row is the remaining $\eta$ symbols from the original terms.

The next step is is to cancel many of the above $\eta$ symbols. In addition, we insert unity a few times in the form of
\begin{equation*}
\frac{1 \cdot 1}{1 \cdot 1} \cdot (1)^{s(25)} = \frac{\eta_{023}(34,{\bf 1})^{s(35)}\eta_{012}(24,{\bf 1})^{s(25)}}{\eta_{013}(34,{\bf 1})^{s(35)} \eta_{123}(34,{\bf 1})^{s(35)}} \left(\frac{\eta_{012}(25,54)}{\eta_{012}(25,54)}\right)^{s(25)}
\end{equation*}
to obtain
\begin{equation*}
\begin{split}
\frac{\text{symm}_\text{LHS} \cdot x}{\text{symm}_\text{RHS}} = 
&\left(
\frac{\eta_{013}(35,54)\eta_{013}(34,45)}{\eta_{{^{43}}013}(45,54)^{s(34)}\eta_{013}(34,{\bf 1})}
\right)^{s(35)} \left(
\frac{\eta_{123}(35,54)\eta_{123}(34,45)}{\eta_{{^{43}}123}(45,54)^{s(34)}\eta_{123}(34,{\bf 1})}
\right)^{s(35)} \\
\times &\left(
\frac{\eta_{{^{43}}023}(45,54)^{s(34)}\eta_{023}(34,{\bf 1})}{\eta_{023}(35,54)\eta_{023}(34,45)}
\right)^{s(35)} \\
\times &\left(
\frac{\eta_{{^{42}}012}(45,54)^{s(24)}\eta_{012}(24,{\bf 1})}{\eta_{012}(24,45)\eta_{012}(25,54)}
\frac{\eta_{012}(23,35)\eta_{012}(25,54)}{\eta_{{^{32}}012}(35,54)^{s(23)}\eta_{023}(23,34)}
\right)^{s(25)} \\
\times &\frac{1}{\eta_{f_{0123}}(35,54)^{s(35)}}
\left(\frac{\eta_{f_{0124}}(45,54)\eta_{f_{1234}}(45,54)}{\eta_{f_{0234}}(45,54)\eta_{f_{0134}}(45,54)}\right)^{s(45)}
\end{split}
\end{equation*}

The first three rows of the above vanish due to the identity Eq.~\eqref{EtaEta_equals_EtaEta}. In the final row, all the terms $\eta_{f_\text{3-simplex}}$ are all $\pm 1$ so we may as well ignore the twists $s({\bf g})$. We can also use the fact that $\delta f = 0$ implying $f_{0124} \times f_{1234} \times f_{0234} \times f_{0134} = f_{0123}$. These mean we can write the above as 
\begin{equation*}
\begin{split}
\frac{\text{symm}_\text{LHS} \cdot x}{\text{symm}_\text{RHS}} = \frac{\eta_{f_{0123}}(45,54)}{\eta_{f_{0123}}(35,54)} = \eta_{f_{0123}}(34,45)\eta_{f_{0123}}(34,{\bf 1}) = \eta_{f_{0123}}(34,45)
\end{split}
\end{equation*}
where the second equality again uses Eq.~\eqref{EtaEta_equals_EtaEta}. This is exactly what we wanted.

For the other Pachner moves, this calculation is identical, except which terms go into the $LHS$ and which go into the $RHS$ get switched around. However, they get switched around in a way that keeps the overall expression $\frac{\text{symm}_\text{LHS} \cdot x}{\text{symm}_\text{RHS}}$ the same because the relative orientations $\pm$ of the 4-simplices will add powers of $\pm 1$ which will make the $LHS$ and $RHS$ ratio be the same.  

\section{ Multiplicativity of $Z$ under connected-sums} \label{app:multOfZunderconnSum}

It is a quick calculation to show that $Z$ is multiplicative under taking connected-sums, as follows. Let $M, M'$ be manifolds equipped with background fields $A_{b},A_{b}'$, $f,f'$ and with twisted spin structures $\xi_{\mathcal{G}},\xi_{\mathcal{G}}'$. Furthermore, suppose WLOG that the triangulations of $M,M'$ have been subdivided sufficiently finely so that we can choose attaching spheres on $M,M'$ without any background gauge fields or spin structures living on them. This entails choosing a subdivision so that there is at least one 4-simplex on $M$ and one on $M'$ for which all $A_b = f = \xi = 0$, which can always be done for flat gauge fields $A_b$, $f$. Then we can form a manifold $M \# M'$ with background gauge fields $A_{b} \# A_{b}'$, $f \# f'$ and spin structure $\xi_{\mathcal{G}} \# \xi_{\mathcal{G}}'$. To show that $Z$ is multiplicative, we first use the fact that each bosonic shadow piece $Z_b$ is multiplicative, i.e. that:
\begin{equation}
    Z_b(M \# M', A_{b} \# A_{b}', f \# f') = Z_b(M, A_{b}, f) Z_b(M', A_{b}', f') 
\end{equation}
This follows from the similar well-known fact that the Crane-Yetter sum (even for \textit{non-modular} $\mathcal{C}$) is multiplicative under connected-sums\footnote{One can use the skein module construction of the Crane-Yetter TQFT and the gluing formula \cite{walker2006} to show on general grounds that the Crane-Yetter path integral $Z_{CY}(M) = Z_b(M, A_b = 0, f = 0)$ satisfies $Z_{CY}(M \# M') = Z_{CY}(M) Z_{CY}(M')/Z_{CY}(S^4)$ and picking the normalization where $Z_{CY}(S^4) = 1$. A more direct way to do this with finite triangulations is by identifying a 4-simplex from each triangulated $M,M'$ and identifying all the labels associated with that $4$-simplex before summing over all labels.} together with locality so that one can focus on the region where the background structures vanish to use the Crane-Yetter result.

Since the connected-sum can be chosen on 4-simplices with no representatives of $f,f'$ or $\xi_{\mathcal{G}}, \xi_{\mathcal{G}}'$, it also follows that
\begin{equation}
    z_c(M \# M', f \# f', \xi_{\mathcal{G}} \# \xi_{\mathcal{G}}') = z_c(M, f, \xi_{\mathcal{G}}) z_c(M', f', \xi_{\mathcal{G}}') 
\end{equation}
since the loop decompositions of $f,f'$ are disjoint and unaffected by the connected-sum. We can also use for $*=0 \cdots 4$ the canonical identifications $H_{*}(M,\Z_2) \oplus H_{*}(M',\Z_2) \equiv H_{*}(M \# M',\Z_2)$ with respect to the natural inclusion maps $H_{*}(M,\Z_2),H_{*}(M',\Z_2) \xhookrightarrow{} H_{*}(M \# M',\Z_2)$. Note these are equivalent to $H^{4-*}(M,\Z_2) \oplus H^{4-*}(M',\Z_2) \equiv H^{4-*}(M \# M',\Z_2)$. From here, multiplicativity under connected-sum is proved as:
\begin{equation}
\begin{split}
    Z(M & \# M', A_{b} \# A_{b}', \xi_{\mathcal{G}} \# \xi_{\mathcal{G}}') \\
    =& \frac{1}{\sqrt{|H^2(M \# M', \Z_2)|}} \sum_{f = f \# f' \in H^3(M \# M',\Z_2)} Z_b(M \# M', A_{b} \# A_{b}', f \# f') z_c(M \# M', f \# f', \xi_{\mathcal{G}} \# \xi_{\mathcal{G}}') \\
    =& \left( \frac{1}{\sqrt{|H^2(M, \Z_2)|}} \sum_{f \in H^3(M,\Z_2)} Z_b(M, A_{b}, f) z_c(M, f, \xi_{\mathcal{G}}) \right) \\
    &\quad \cdot \left( \frac{1}{\sqrt{|H^2(M', \Z_2)|}} \sum_{f' \in H^3(M',\Z_2)} Z_b(M', A_{b}', f') z_c(M', f', \xi_{\mathcal{G}}') \right) \\
    =& Z(M, A_{b},\xi_{\mathcal{G}})Z(M', A_{b}',\xi_{\mathcal{G}}')
\end{split}
\end{equation}

\section{ Explicit data for semion-fermion theory and $\SO(3)_3$ with time-reversal symmetry ${\bf T}^2 = (-1)^F$} \label{app:btcData}

In this appendix, we state the relevant BTC and symmetry fractionalization data for the semion-fermion and $\SO(3)_3$ theories, which are the input categories that produce the $\nu=2,\nu=3$ phases in the $\Z_{16}$ classification for $G_f=\Z_4^{{\bf T},f}$. 

   \subsection{ Semion-fermion} \label{app:semionFermionData}
   
        The anyon content of the semion-fermion theory can be labeled $1,s,\tilde{s},\psi$. Although this label set matches that of $\SO(3)_3$, semion-fermion is an Abelian theory. It is a direct product of $(1,\psi)$ and the semion theory $\U_2$. The fusion rules are generated by the fusion rules:
        \begin{align}
            s \times s = \tilde{s} \times \tilde{s} = \psi \times \psi &= 1\\
            s \times \psi &= \tilde{s}
        \end{align}
        The non-trivial $F$-symbols are $F^{abc}=-1$ when $(a,b,c)$ is any combination of only $s$ and $\tilde{s}$. The $R$-symbols, with the anyons ordered $(1,s,\psi, \tilde{s})$, are
        \begin{equation}
            R^{ab} = \begin{pmatrix}
            1 & 1 & 1 & 1\\
            1 & i & 1 & i\\
            1 & 1 & -1 & -1\\
            1 & i & -1 & -i
            \end{pmatrix}
        \end{equation}
        Time-reversal must exchange $s$ and $\tilde{s}$.
        
        Gauge-fixing $U_{\bf T}(s,s;1)=U_{\bf T}(\psi,\psi;1)=U_{\bf T}(s,\psi;\tilde{s})=1$, we find that, with the same anyon ordering as before,
        \begin{equation}
            U_{\bf T}(a,b;a\times b) = \begin{pmatrix}
            1 & 1 & 1 & 1\\ 
            1 & 1   & 1 & 1\\
            1 & -1 & 1 & -1\\
            1 &-1 & 1 & -1
            \end{pmatrix}
        \end{equation}
        Solving the consistency equations for $\eta$ in this gauge, we find
        $\eta_{\psi}({\bf T},{\bf T}) = \pm 1$ and $\eta_{s,\tilde{s}}({\bf T},{\bf T}) = \pm i$, subject to the constraint $\eta_s \eta_{\tilde{s}}/\eta_{\psi}=-1$. (We drop the explicit ${\bf T}$ from now on - there is only one non-trivial $\eta$ per anyon.)
        
        Inputting the $\eta_{\psi}=-1$, $\eta_s = \pm i$, $\eta_{\tilde{s}}=\mp i$ case into our fermionic state sum, we can evaluate Eq.~\eqref{eqn:Z16TotalZ} to find that
        \begin{equation}
            Z(\mathbb{RP}^4) = \exp\left(\mp 2 \times \frac{2\pi i}{16}\right) 
        \end{equation}

   \subsection{ $\SO(3)_3$} \label{app:SO36Data}
        We list for completeness the explicit $F$ and $R$ symbols of $\SO(3)_3$. The fractionalization data of $U,\eta$ symbols is given in Sec.~\ref{sec:so33_MainText}.

        In the usual labeling of anyons of $\SU(2)_6$ where anyons are labeled by integers from $0$ to $6$, we identify $s \sim 2$, $\tilde{s} \sim 4$, and $\psi \sim 6$. With these identifications, the $F$- and $R$-symbols can be obtained by standard formulas in, e.g.~\cite{ardonne2010}.  
        
        Let $q=e^{\pi i/4}$. The $R$-symbols are simple to write:
        \begin{equation}
            R^{a, b}_c = (-1)^{(a+b-c)/2}q^{\frac{1}{8}\left(c(c+2)-a(a+2)-b(b+2)\right)}
        \end{equation}
        Note that by halving the labels (i.e., instead label $s \sim 1$, $\tilde{s} \sim 2$, $\psi \sim 3$) we could eliminate some factors of 2, but we retain these factors of 2 for consistency with $\SU(2)_6$ labelings.
        
        The $F$-symbols require a set of auxiliary functions
        \begin{align}
            \lfloor n \rfloor &= \sum_{m=1}^n q^{(n+1)/2-m}\\
            \lfloor n \rfloor! &= \lfloor n \rfloor \lfloor n-1 \rfloor \cdots \lfloor 1 \rfloor\\
            \Delta(a,b,c) &= \sqrt{\frac{\lfloor (a+b-c)/2\rfloor!\lfloor (a-b+c)/2\rfloor!\lfloor (-a+b+c)/2\rfloor!}{\lfloor (a+b+c+2)/2\rfloor!}}
        \end{align}
        for $n\geq 1$, and with $\Delta$ defined when $a,b,c$ satisfy the triangle inequality. We also define $\lfloor 0\rfloor!=1$.
        
        With this definition, we can define the $F$-symbols by the following rather complicated formula:
        \begin{align}
            &F^{abc}_{def} = (-1)^{(a+b+c+d)/2}\Delta(a,b,e)\Delta(c,d,e)\Delta(b,c,f)\Delta(a,d,f) \sqrt{\lfloor e+1 \rfloor \lfloor f+1 \rfloor} \times\nonumber \\
            &\times \sum_{n}' \frac{(-1)^{n/2}\lfloor (n+2)/2\rfloor!}{\lfloor(a+b+c+d-n)/2\rfloor!\lfloor(a+c+e+f-n)/2\rfloor!\lfloor b+d+e+f-n\rfloor!} \times \nonumber \\
            &\times \frac{1}{\lfloor (n-a-b-e)/2\rfloor ! \lfloor (n-c-d-e)/2\rfloor ! \lfloor (n-b-c-f)/2\rfloor ! \lfloor (n-a-d-f)/2\rfloor !}
        \end{align}
        
        These formulae match the gauge choices of~\cite{fidkowski2013}.

\section{ $\Z_{16}$ anomaly indicator for ${\bf T}^2 = (-1)^F$ symmetry}
\label{app:Z16}

Here we perform an explicit calculation using our state sum for the case of time-reversal symmetry squaring to $(-1)^F$. 

This corresponds to the total symmetry group $G_f=\Z_4^{{\bf T},f}$ which corresponds to the group extension $\Z_2^f \to \Z_4^{{\bf T},f} \to \Z_2^{\bf T}$ described by the cocycle $\omega_2$ with $\omega_2({\bf T},{\bf T})=-1$ and all others $+1$. This means that $A_b^{\ast} \omega_2 = w_1^2$ in cohomology and given by $A_b^*s \cup A_b^*s \in Z^2(M,\Z_2)$ on the cochain level with $A_b^*s \in Z^1(M,\Z_2)$ the anti-unitary flux representing $[w_1]$, with $w_1 = f_\infty A_b^* s \in H^1(M^\vee,\Z_2^{\bf T})$. 

The (dual of the) $\mathcal{G}_f$-structure is a cycle $\xi \in Z_{d-1}(M,\Z_2)$ that trivializes $w_2 + w_1^2 + f_\infty A_b^{\ast} \omega_2 = w_2 \in Z_{d-2}(M,\Z_2)$. Note the $f_\infty$ map will push $A_b^{\ast} \omega_2 = A_b^*s \cup A_b^*s \in Z^2(M,\Z_2)$ to the same cycle representative of $w_1^2 \in Z_{d-2}(M,\Z_2)$ since the representative of $[w_1^2]$ coming from $A_b^{\ast} \omega_2$ is the same as $A_b^*s \cup A_b^*s$.

As discussed in the main text, the outline of the calculation is as follows. First, we give an explicit cellulation that we call $T_{\star}$ of $M=\mathbb{RP}^4$ and determine a set of cochain representatives for $w_2, A_b$, and $w_1$. Next, we perform the diagrammatic calculations to explicitly evaluate the state sum associated to the triangulation and give an analytic expression for it for any general super-modular category and symmetry fractionalization class of time-reversal symmetry. This has the interpretation of being an `anomaly-indicator' for the symmetry fractionalization class. In particular, we will evaluate the shadow $Z_b((\mathbb{RP}^4 ,T_{\star}),A_b,f)$ for two choices $f = 0 \in Z^3((\mathbb{RP}^4 ,T_{\star}),A_b,\Z_2)$ and $f$ with $[f] \neq 0 \in H^3(\mathbb{RP}^4 ,\Z_2)$. The bosonic shadow results given in Eq.~\eqref{totalRP4partitionfunction} are reproduced below,
\begin{equation}
\begin{split}
Z_b((\mathbb{RP}^4 ,T_{\star}),A_b,f=0) &= \frac{1}{\mathcal{D}} \sum_{x | x = {^{\bf T}}x} d_x \theta_x \eta^{\bf T}_x \\
Z_b((\mathbb{RP}^4 ,T_{\star}),A_b,[f]\neq0) &= \pm \frac{1}{\mathcal{D}} \sum_{x | x = {^{\bf T}}x \times \psi} d_x \theta_x \eta^{\bf T}_x
\end{split}
\end{equation}
where the $\pm$ depends on the specific representative of $[f] \neq 0 \in H^3(\mathbb{RP}^4 ,\Z_2)$. The FSPT partition function will be the result 
\begin{equation}
Z((\mathbb{RP}^4 ,T_{\star}),A_b,\xi_{\text{pin}^+})=\frac{1}{\sqrt{2}\mathcal{D}} \left( \sum_{x | x = {\,^{\bf T}}x} d_x \theta_x \eta^{\bf T}_x + i \sum_{x | x = {\,^{\bf T}}x \times f} d_x \theta_x \eta^{\bf T}_x \right)
\end{equation}
Here, the fact that the last term has a $+i$ indicates our choice of pin$^+$ structure and is independent of the specific representatives of $f$. The other pin$^+$ structure would have a $-i$ there. This is exactly the $\Z_{16}$ indicator formula giving $e^{2 \pi i \nu / 16}$. 

\subsection{ Compact cellulation of $\mathbb{RP}^4$}
Now, we list the cellulation $T_{\star}$ of $\mathbb{RP}^4$. This cellulation comes from the gluing procedure of generating manifolds from~\cite{bulmash2020}. In particular we use a cellulation generated from $\Z_2^{\bf T}$ in that paper. One technical note is that the gluing procedure in~\cite{bulmash2020} actually produces two plausible ways of gluing things together. One of them is not a manifold because it corresponds to a degenerate gluing in the sense that links of 2-simplices are not circles. We choose to work with the one spelled out explicitly below, which is indeed a manifold.

There are four 4-simplices $\braket{a_1 b_1 c_1 d_1 e_1}, \braket{a_2 b_2 c_2 d_2 e_2}, \braket{a_3 b_3 c_3 d_3 e_3}, \braket{a_4 b_4 c_4 d_4 e_4}$ 
that are all $+$-oriented and branched as $a \to b \to c \to d \to e$. 
Then there are ten 3-simplices given by some identifications of the simplices. 
The identifications of 3-simplices must be compatible with the branching structure. 
Those identifications then uniquely fix the identifications of all the lower-dimensional simplices via the ordering induced by the branching structure. 
In total, there are twelve 2-simplices, eight 1-simplices, and three 0-simplices.
This means that there is a sum over $10+12=22$ different anyon labels. 
Below we list all the simplices together with an arbitrary number (in the column $\#$) from $1,\dots,22$ as shorthand for the anyon label we chose.
This list includes the identifications of 3-simplices and 2-simplices, in that each $\{\text{simplex-1}, \cdots, \text{simplex-k}\}$ in braces represents that all simplices in that list are identified.
One can also look at Fig.~\ref{rp4_Step1} which lists all the relevant 15j symbols and also marks the anyon label.

\begin{center}
\begin{equation}
\begin{tabular}{ |c|c||c|c| } 
 \hline
 3-simplices & \# & 2-simplices & \# \\ 
 \hline\hline
 $\{\braket{b_1 c_1 d_1 e_1},\braket{a_2 c_2 d_2 e_2}\}$ & 1 & 
 $\{\braket{b_1 c_1 e_1},\braket{a_2 c_2 e_2},\braket{a_3 d_3 e_3},\braket{b_4 d_4 e_4}\}$ & 2  \\
 $\{\braket{a_2 b_2 c_2 e_2},\braket{a_3 b_3 d_3 e_3}\}$ & 4 & 
 $\{\braket{c_1 d_1 e_1},\braket{c_2 d_2 e_2}\}$ & 3 \\
 $\{\braket{a_3 c_3 d_3 e_3},\braket{b_4 c_4 d_4 e_4}\}$ & 6 & 
 $\{\braket{a_1 b_1 d_1},\braket{a_2 b_2 c_2},\braket{a_3 b_3 d_3},\braket{a_4 b_4 c_4}\}$ & 5 \\
 $\{\braket{a_1 c_1 d_1 e_1},\braket{b_2 c_2 d_2 e_2}\}$ & 8 &
 $\{\braket{a_1 c_1 d_1},\braket{b_2 c_2 d_2},\braket{a_3 c_3 d_3},\braket{b_4 c_4 d_4}\}$ & 7  \\
 $\{\braket{a_1 b_1 d_1 e_1},\braket{a_4 b_4 c_4 e_4}\}$ & 10 &
 $\{\braket{a_1 d_1 e_1},\braket{b_2 d_2 e_2},\braket{b_3 c_3 e_3},\braket{a_4 c_4 e_4}\}$ & 9  \\
 $\{\braket{a_2 b_2 d_2 e_2},\braket{a_3 b_3 c_3 e_3}\}$ & 13 & 
 $\{\braket{a_1 c_1 e_1},\braket{b_2 c_2 e_2},\braket{b_3 d_3 e_3},\braket{a_4 d_4 e_4}\}$ & 11 \\
 $\{\braket{a_1 b_1 c_1 d_1},\braket{a_3 b_3 c_3 d_3}\}$ & 17 &
 $\{\braket{a_2 b_2 e_2},\braket{a_3 b_3 e_3}\}$ & 12 \\
 $\{\braket{a_1 b_1 c_1 e_1},\braket{a_4 b_4 d_4 e_4}\}$ & 18 &
 $\{\braket{a_1 b_1 e_1},\braket{a_4 b_4 e_4}\}$ & 14 \\
 $\{\braket{a_2 b_2 c_2 d_2},\braket{a_4 b_4 c_4 d_4}\}$ & 20 &
 $\{\braket{b_1 d_1 e_1},\braket{a_2 d_2 e_2},\braket{a_3 c_3 e_3},\braket{b_4 c_4 e_4}\}$ & 15 \\
 $\{\braket{b_3 c_3 d_3 e_3},\braket{a_4 c_4 d_4 e_4}\}$ & 21 &
 $\{\braket{b_1 c_1 d_1},\braket{a_2 c_2 d_2},\braket{b_3 c_3 d_3},\braket{a_4 c_4 d_4}\}$ & 16 \\
 & & 
 $\{\braket{a_1 b_1 c_1},\braket{a_2 b_2 d_2},\braket{a_3 b_3 c_3},\braket{a_4 b_4 d_4}\}$ & 19 \\
 & & 
 $\{\braket{c_3 d_3 e_3},\braket{c_4 d_4 e_4}\}$ & 22 \\
 
\hline
\end{tabular}
\end{equation}
\end{center}

For reference, the 1-simplices and 0-simplices are:
\begin{center} 
\begin{equation} \label{0And1SimplicesRP4}
\begin{tabular}{|c||c|}
  \hline
  1-simplices & 0-simplices \\
  \hline\hline
  $\{\braket{c_1 d_1},\braket{c_2 d_2},\braket{c_3 d_3},\braket{c_4 d_4}\}$ & $\{a_1,a_2,a_3,a_4,b_1,b_2,b_3,b_4\}$ \\
  $\{\braket{c_1 e_1},\braket{c_2 e_2},\braket{d_3 e_3},\braket{d_4 e_4}\}$ &
  $\{c_1,c_2,c_3,c_4,d_1,d_2,d_3,d_4\}$ \\
  $\{\braket{c_3 e_3},\braket{c_4 e_4},\braket{d_1 e_1},\braket{d_2 e_2}\}$ &
  $\{e_1,e_2,e_3,e_4\}$ \\
  $\{\braket{a_1 b_1},\braket{a_2 b_2},\braket{a_3 b_3},\braket{a_4 b_4}\}$ &  \\
  $\{\braket{a_1 e_1},\braket{a_4 e_4},\braket{b_2 e_2},\braket{b_3 e_3}\}$ &  \\
  $\{\braket{a_2 e_2},\braket{a_3 e_3},\braket{b_1 e_1},\braket{b_4 e_4}\}$ &  \\
  $\{\braket{a_1 d_1},\braket{a_2 c_2},\braket{a_3 d_3},\braket{a_4 c_4},\braket{b_1 c_1},\braket{b_2 d_2},\braket{b_3 c_3},\braket{b_4 d_4}\}$ &  \\
  $\{\braket{a_1 c_1},\braket{a_2 d_2},\braket{a_3 c_3},\braket{a_4 d_4},\braket{b_1 d_1},\braket{b_2 c_2},\braket{b_3 d_3},\braket{b_4 c_4}\}$ &  \\
  \hline
\end{tabular}
\end{equation}
\end{center}

Inputting this cellulation into the program Regina \cite{regina} shows that this cellulation is a manifold whose double-cover is simply-connected with $H^2(-,\Z) = 0$, which means that its double-cover is homeomorphic to $S^4$ by the topological Poincaré conjecture in $d=4$. As such, this indicates that the original manifold is homeomorphic to $\mathbb{RP}^4$.

\subsection{ Representatives of $w_2$ and $\text{pin}^+$ structure, $w_1$ and Gauge Fields, and $[f] \neq 0$ and $z_c(f)$}

We want to find explicit cochain representatives of the $\text{pin}^+$ structure and of a non-trivial background fermion line $[f] \neq 0$ at hand, for which we will need representatives of the orientation-reversing wall $w_1$ and of $w_2$.

\subsubsection{$w_2$ and $\text{pin}^+$ structure} \label{app:w2AndPinPlusRep}
Recall from the beginning of this section that the cochain representative of $w_2 \in Z^{2}((\mathbb{RP}^4,T_{\star})^{\vee},\Z_2)$ is trivialized by $\xi_{\text{pin}^+}$. On the cochain-level, we can compute the canonical representative of $w_2$ using Eq.~\eqref{YuAnFormula}.

We apply this formula to the above cellulation of $\mathbb{RP}^4$. One can compute $w_2$ on the dual of each 2-simplex, slightly abusing the labels $1,\cdots,22$ from the previous subsection to label the dual 2-cells of the corresponding 2-simplices. The result is
\begin{equation}
\begin{split}
w_2(2)  = w_2(5) = w_2(7) = w_2(9) = w_2(11) = w_2(15) = w_2(16) = w_2(19) &= +1 \\
w_2(3) = w_2(12) = w_2(14) = w_2(22) &= -1 \\
\end{split}
\end{equation}
From here, we want to find a cochain $\xi_{\text{pin}^+} \in Z^1((\mathbb{RP}^4,T_{\star})^\vee,\Z_2) \cong Z_{d-1}((\mathbb{RP}^4,T_{\star}),\Z_2)$ such that $\delta \xi_{\text{pin}^+} = w_2$  (dual to $\partial \xi = w_2$). 
Again slightly abusing the labels of 3-simplices to refer to their dual 1-cells, this can be accomplished by choosing $\xi$ to act non-trivially as:
\begin{equation} \label{spinStructRepRP4_calculation}
\xi_{\text{pin}^+}(4) = \xi_{\text{pin}^+}(6) = \xi_{\text{pin}^+}(8) = \xi_{\text{pin}^+}(10) = \xi_{\text{pin}^+}(17) = \xi_{\text{pin}^+}(20) = -1
\end{equation}
with $\xi_{\text{pin}^+}=+1$ on all other dual 1-cells. Note that the only non-trivial homology class is dual to $w_1$ and consists of the 3-simplices $17,20$. So the other $\text{pin}^+$ structure will have $\xi'_{\text{pin}^+}(17)=\xi'_{\text{pin}^+}(20)=+1$ as opposed to $-1$. 

\subsubsection{$w_1$ and ${\bf T}$ Gauge Fields}
The procedure in~\cite{bulmash2020} also tells us what the gauge fields ${\bf g}_{ij}$ should be on the 1-simplices. All 1-simplices that have a ${\bf T}$-valued gauge field are
\begin{equation}
A_b(\braket{b_1 c_1}) = A_b(\braket{c_1 e_1}) = A_b(\braket{b_2 c_2}) = {\bf T}.
\end{equation}
All 1-simplices $\braket{ij}$ not identified with the above via Eq.~\eqref{0And1SimplicesRP4} will have ${\bf g}_{ij} = {\bf 1}$. 

One can also check that the $f_\infty$ map makes the orientation-reversing wall $w_1$ consist of the two 3-simplices labeled as $17,20$.

\subsubsection{$[f] \neq 0$ and $z_c(f)$} \label{repOfBckdFLines}
We will work with a cochain $f \in C^3((\mathbb{RP}^4,T_{\star}),\Z_2)$ defined with respect to the above 3-simplex labels as
\begin{equation}
\begin{split}
f(17) = f(21) = f(10)       &= 1 \\
f(\text{other 3-simplices}) &= 0 \\
\end{split}
\label{eqn:nontrivialF_RP4}
\end{equation}
One can show that it is closed, which will be more apparent once we write out the 15j symbols in Fig.~\ref{rp4_Step1}. Also, we can see that it corresponds to the non-trivial class in $H^3(\mathbb{RP}^4,\Z_2)$. To see it is non-trivial, note that it is nonzero on exactly one 3-simplex, $20$, from the dual of $w_1$, so $\int_{w_1} f = 1$. 

\begin{figure}[h!]
  \centering
  \includegraphics[width=\linewidth]{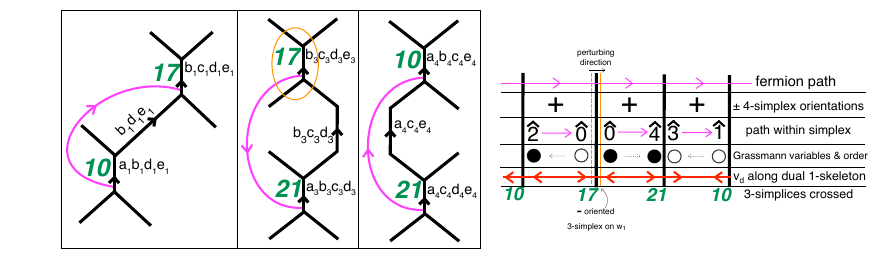}
  \caption{The dual path for the representative $f$ for $[f] \neq 0 \in H^3(\mathbb{RP}^4,\Z_2)$. (Left) How the fermion path looks in the 15j symbols. The arrows on the fermion lines give the loops a direction. The thin blue circle around $17$ on the leftmost diagram indicates that that 3-simplex is part of $w_1$, but it is not perturbed into the 4-simplex $\braket{a_1 b_1 c_1 d_1 e_1}$. (Right) Another depiction of the path of $f$ along with some useful data used to compute $\sigma(f)$. In particular, the orientations of the 4-simplices it passes through, the paths $\hat{i} \to \hat{j}$ it takes within each 4-simplex, the corresponding assignments of Grassmann variables, and the vector $v_d$ along the dual 1-skeleton are all shown. For the Grassmann variables, the solid circles represent $\theta$ while the open circles represent $\overline{\theta}$. The arrows between the variables indicate the order in which they appear in the computation of $\sigma(f)$, as $\text{first} \to \text{second}$ or $\text{second} \leftarrow \text{first}$.}
  \label{f_neq0_fermionPath}
\end{figure}

Now we want to compute $z_c(f) = (-1)^{\int \xi_{\text{pin}^+}(f)} \sigma(f)$. The $\text{pin}^+$ structure part can be quickly computed using Eq.~\eqref{spinStructRepRP4_calculation}. In particular, we will have $(-1)^{\int \xi_{\text{pin}^+}(f)} = +1$ because $\xi_{\text{pin}^+}$ and $f$ share two nonzero 3-simplices, labeled as $10$ and $17$. The $\sigma(f)$ part can be calculated in two ways. First, we can use the winding definition. In this gauge, $f$ encodes a single fermion loop which takes the path shown in Fig.~\ref{f_neq0_fermionPath}. Following the discussion in Sec.~\ref{fInftyAndPerturbationOf_w1} of how the perturbation of the $w_1$ wall is encoded in the 15j symbols and using the winding definition from Eqs.~(\ref{windingGrassmannExpr1},\ref{windingGrassmannExpr2}) will give
\begin{equation*}
  \sigma(f) = -\begin{pmatrix}1 & 0 & 1 & 0\end{pmatrix}
  \underbrace{\begin{pmatrix} -iX & 0  \\ 0   &  iX \end{pmatrix}}_{\substack{\hat{2} \to \hat{0} \text{ on} \\+ \text{ 4-simplex } \\ \braket{a_1 b_1 c_1 d_1 e_1}}}
  \underbrace{\begin{pmatrix}  0  & iY \\ -iY & 0   \end{pmatrix}}_{\substack{\text{crossing } w_1 \text{ in the } \\ \text{perturbing direction on } \\ - \text{ 3-simplex } 17}}
  \underbrace{\begin{pmatrix}  iX & 0  \\ 0   & -iX \end{pmatrix}}_{\substack{\hat{0} \to \hat{4} \text{ on} \\+ \text{ 4-simplex } \\ \braket{a_3 b_3 c_3 d_3 e_3}}}
  \underbrace{\begin{pmatrix}  -iX & 0  \\ 0   & iX \end{pmatrix}}_{\substack{\hat{3} \to \hat{1} \text{ on} \\+ \text{ 4-simplex } \\ \braket{a_4 b_4 c_4 d_4 e_4}}}
  \begin{pmatrix}  1 \\ 0 \\ 0 \\ 0\end{pmatrix} = -i
\end{equation*}
Alternatively, we can use the Grassmann integral to give the same answer
\begin{equation*}
  \sigma(f) = \int d\theta_{10} d\overline{\theta}_{10} d\theta_{17} d\overline{\theta}_{17} d\theta_{21} d\overline{\theta}_{21} 
  \underbrace{(\overline{\theta}_{17} \theta_{10})}_{\substack{\text{from} \\ \braket{a_1 b_1 c_1 d_1 e_1}}}
  \underbrace{(\theta_{17} \theta_{21})}_{\substack{\text{from} \\ \braket{a_3 b_3 c_3 d_3 e_3}}}
  \underbrace{(\overline{\theta}_{10} \overline{\theta}_{21})}_{\substack{\text{from} \\ \braket{a_4 b_4 c_4 d_4 e_4}}}
  \cdot \underbrace{(-i)}_{\substack{\text{crosssing} \\ - \text{ oriented} \\ 17}}
  = -i
\end{equation*}
As such, we have $z_c(f) = (-1)^{\int \xi_{\text{pin}^+}(f)} \sigma(f) = (+1) \cdot (-i) = -i$.

\subsection{ Diagrammatic calculation} \label{rp4Calculation}

Now, we do the diagrammatic calculation on $(\mathbb{RP}^4,T_{\star})$. We are summing over 22 different labelings of anyon lines.

\begin{figure}[h!]
  \centering
  \includegraphics[width=0.95\linewidth]{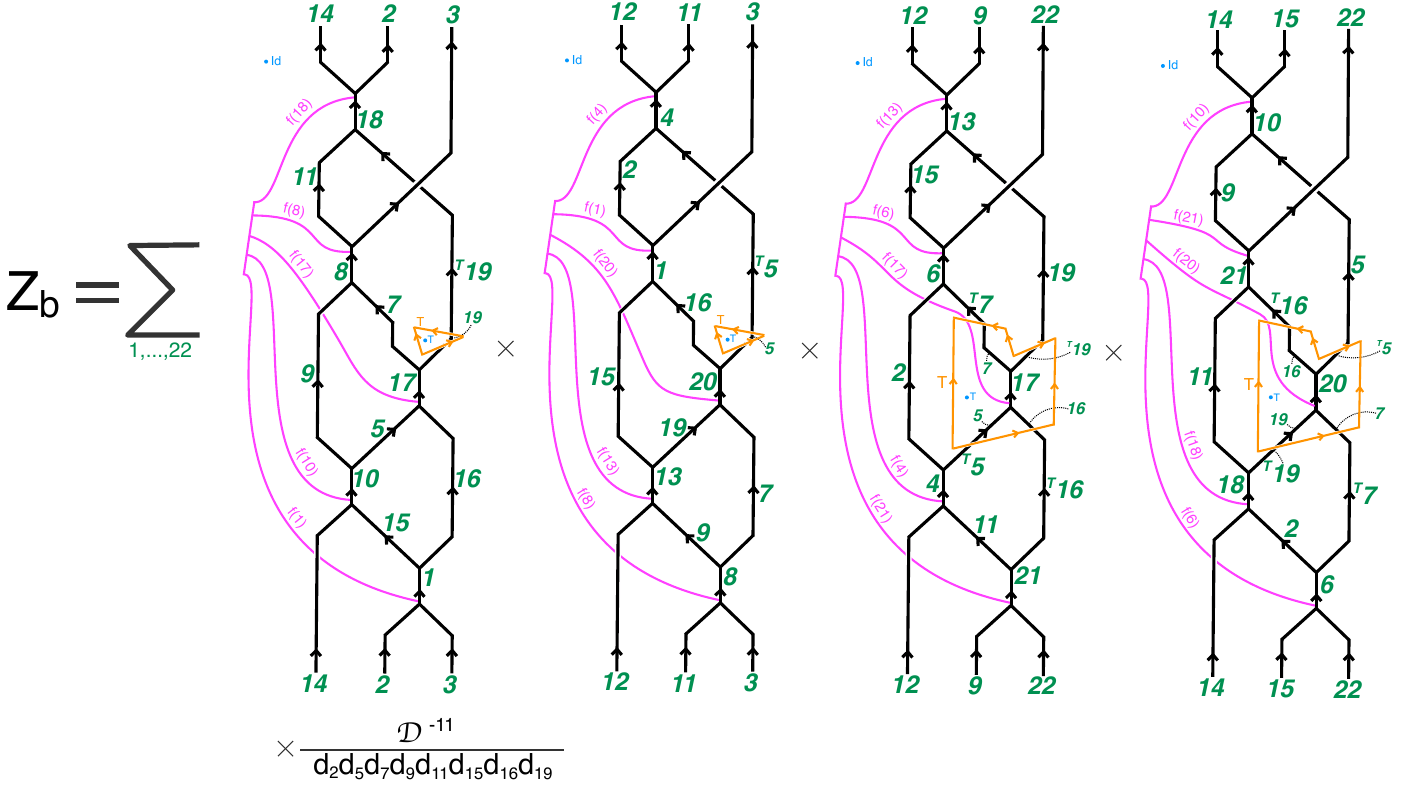}
  \caption{Result of step 1 of $\mathbb{RP}^4$ calculation.}
  \label{rp4_Step1}
\end{figure}

\begin{figure}[h!]
    \centering
    \begin{minipage}{0.47\textwidth}
        \centering
        \includegraphics[width=\linewidth]{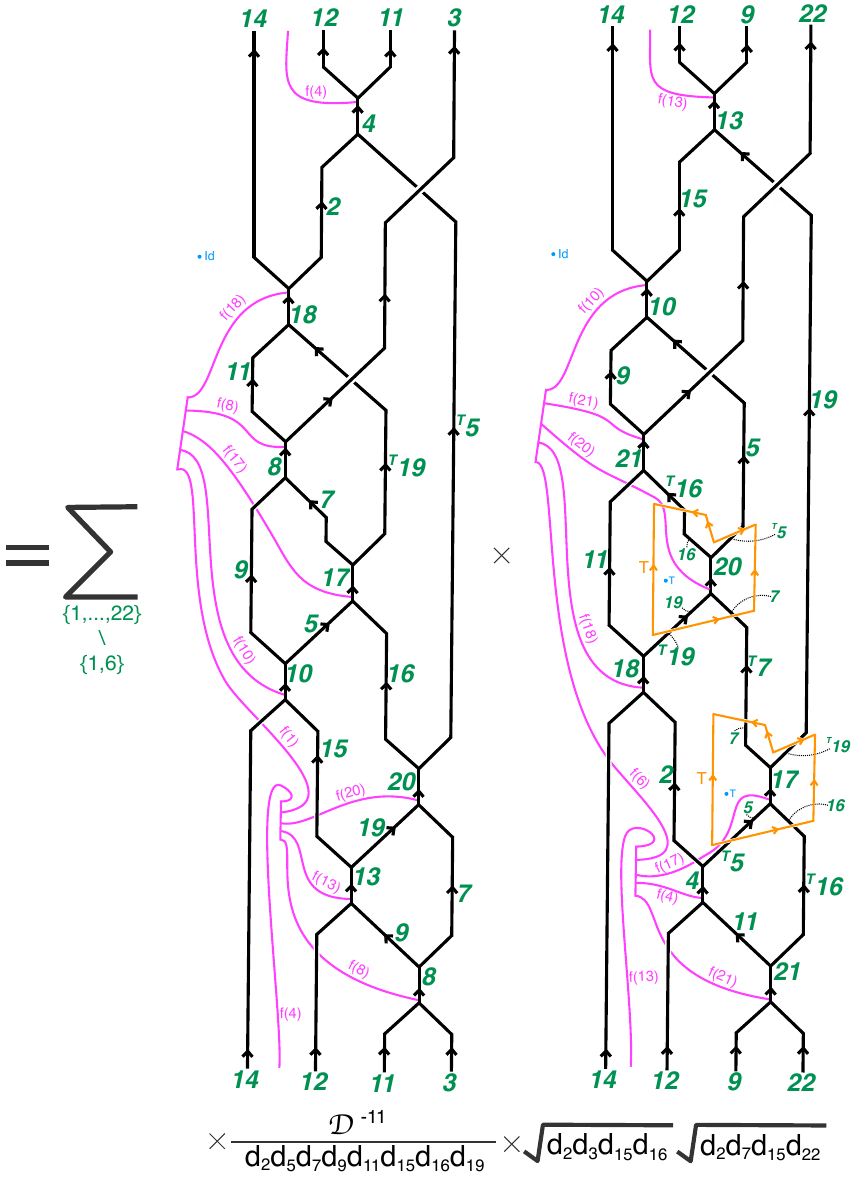}
        \end{minipage} \quad \quad \quad
    \begin{minipage}{0.47\textwidth}
         \centering
         \includegraphics[width=\linewidth]{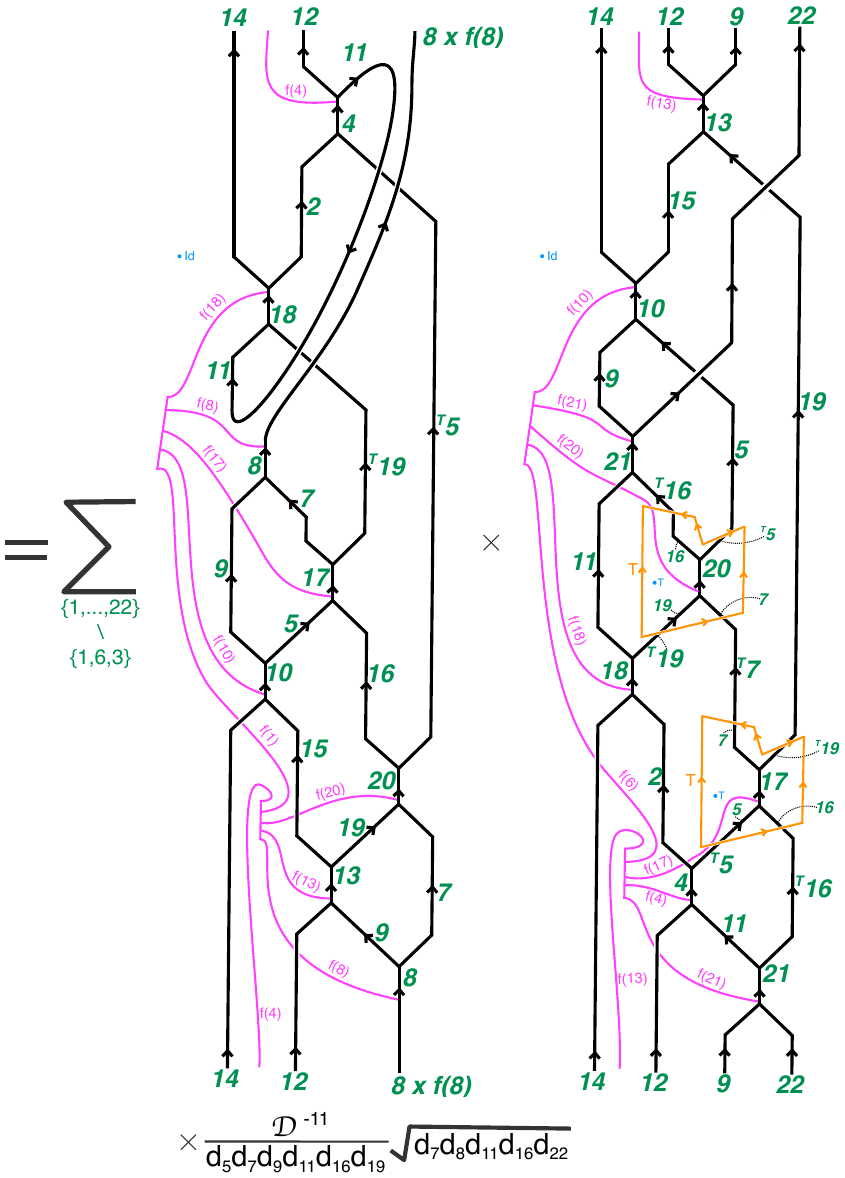}
    \end{minipage}
    \caption{Result of steps 2-3 of $\mathbb{RP}^4$ calculation}
    \label{rp4_Step2to3}
\end{figure}

\begin{figure}[h!]
    \centering
    \begin{minipage}{0.47\textwidth}
        \centering
        \includegraphics[width=\linewidth]{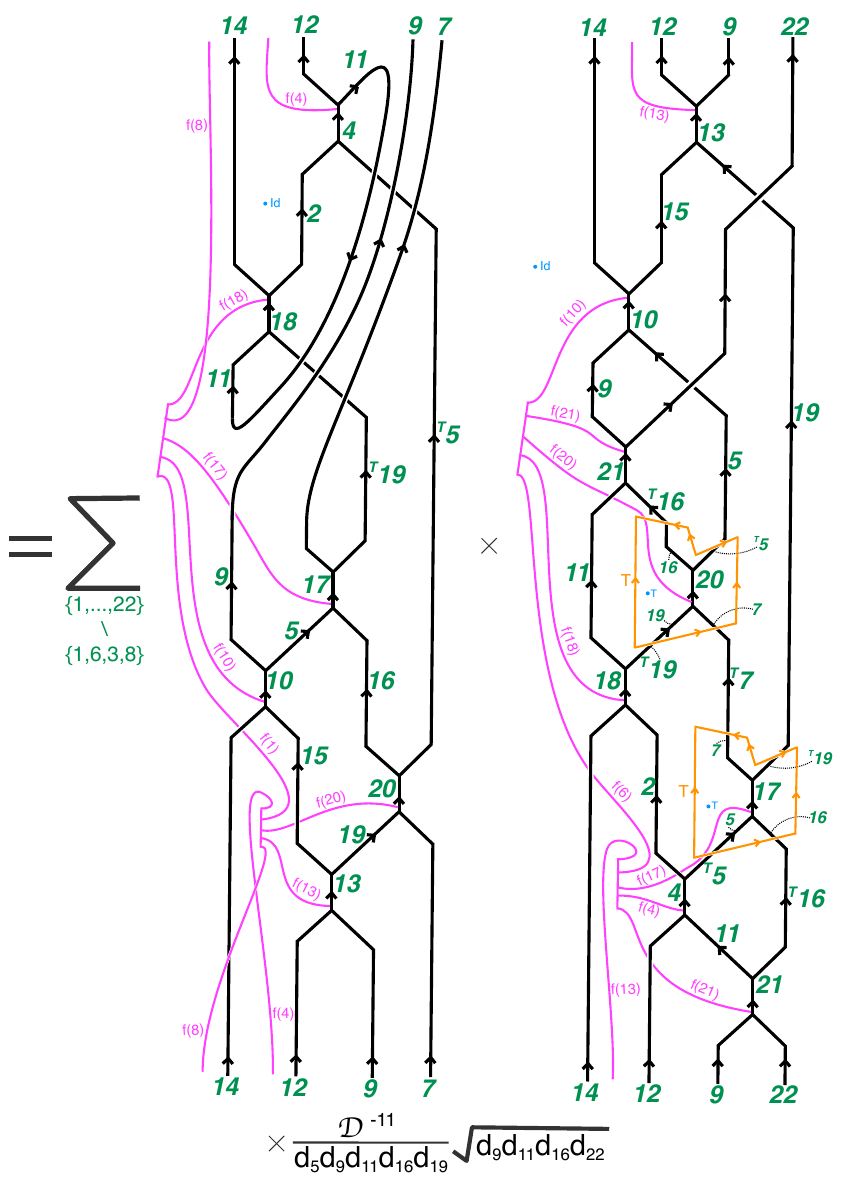}
        \end{minipage} \quad \quad \quad
    \begin{minipage}{0.47\textwidth}
         \centering
         \includegraphics[width=\linewidth]{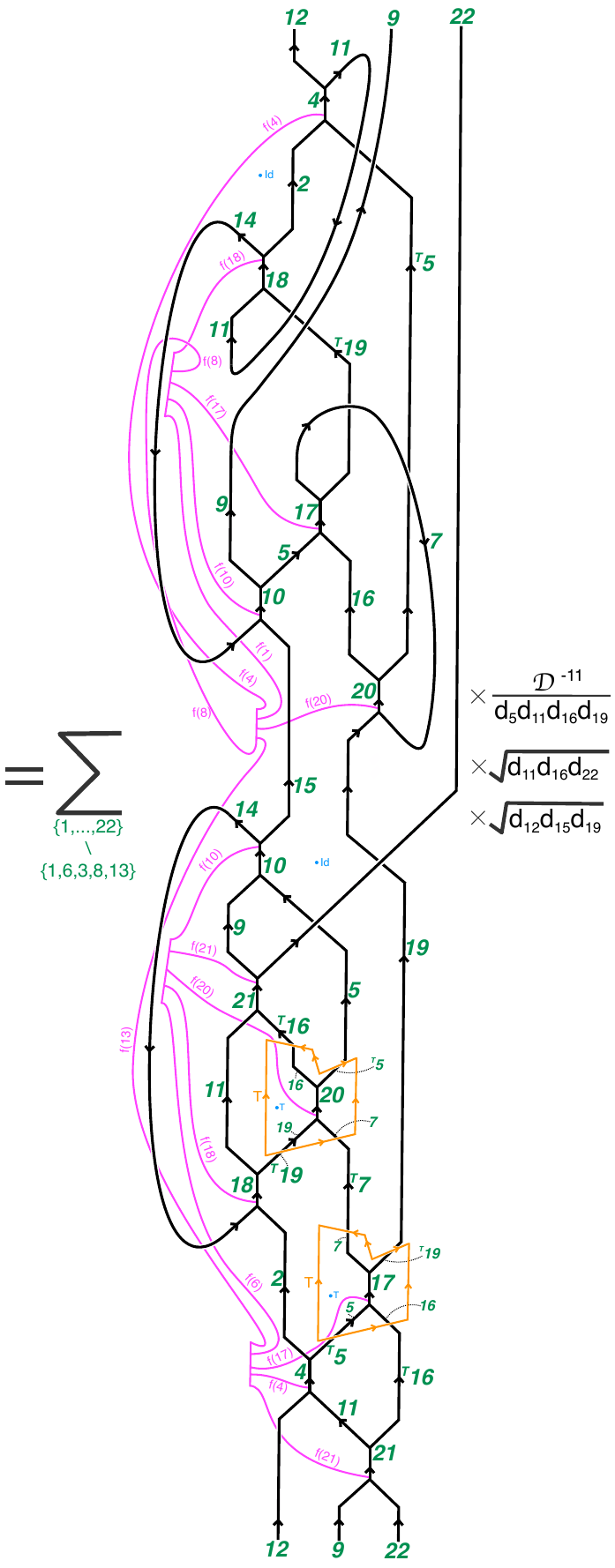}
    \end{minipage}
    \caption{Result of steps 4-5 of $\mathbb{RP}^4$ calculation}
    \label{rp4_Step4to5}
\end{figure}

\begin{figure}[h!]
    \centering
    \begin{minipage}{0.47\textwidth}
        \centering
        \includegraphics[width=\linewidth]{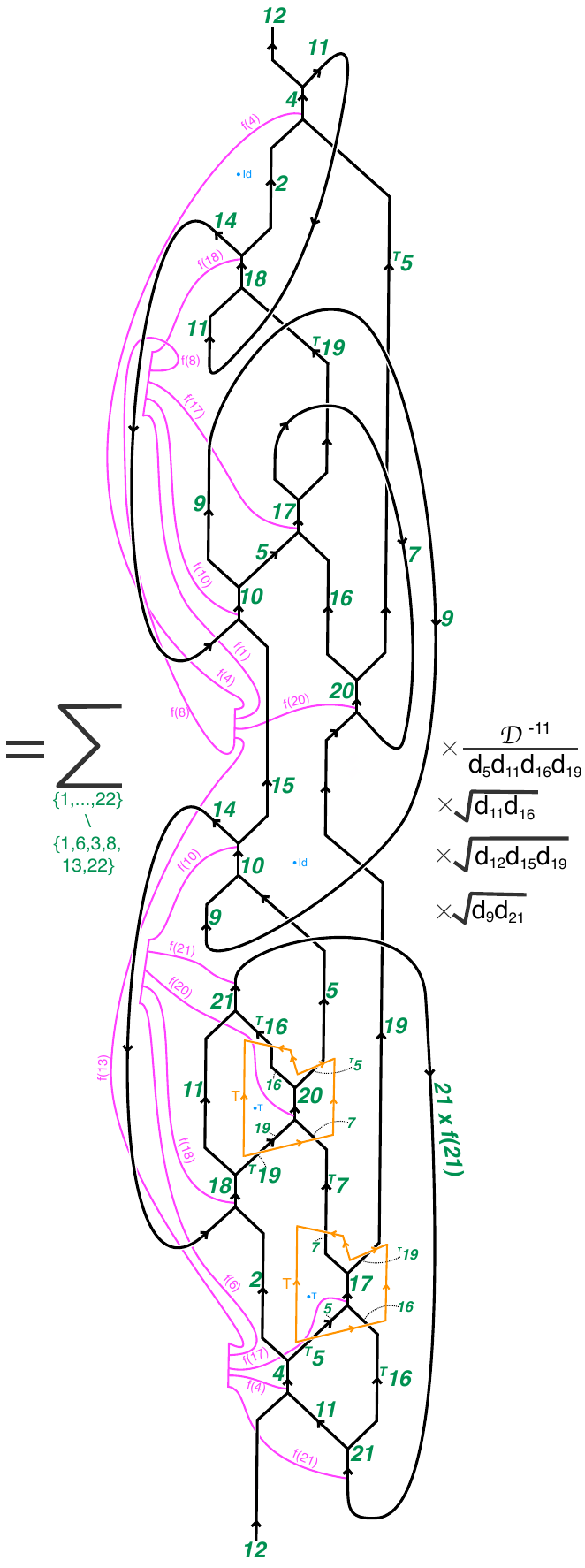}
        \end{minipage} \quad \quad \quad
    \begin{minipage}{0.47\textwidth}
         \centering
         \includegraphics[width=\linewidth]{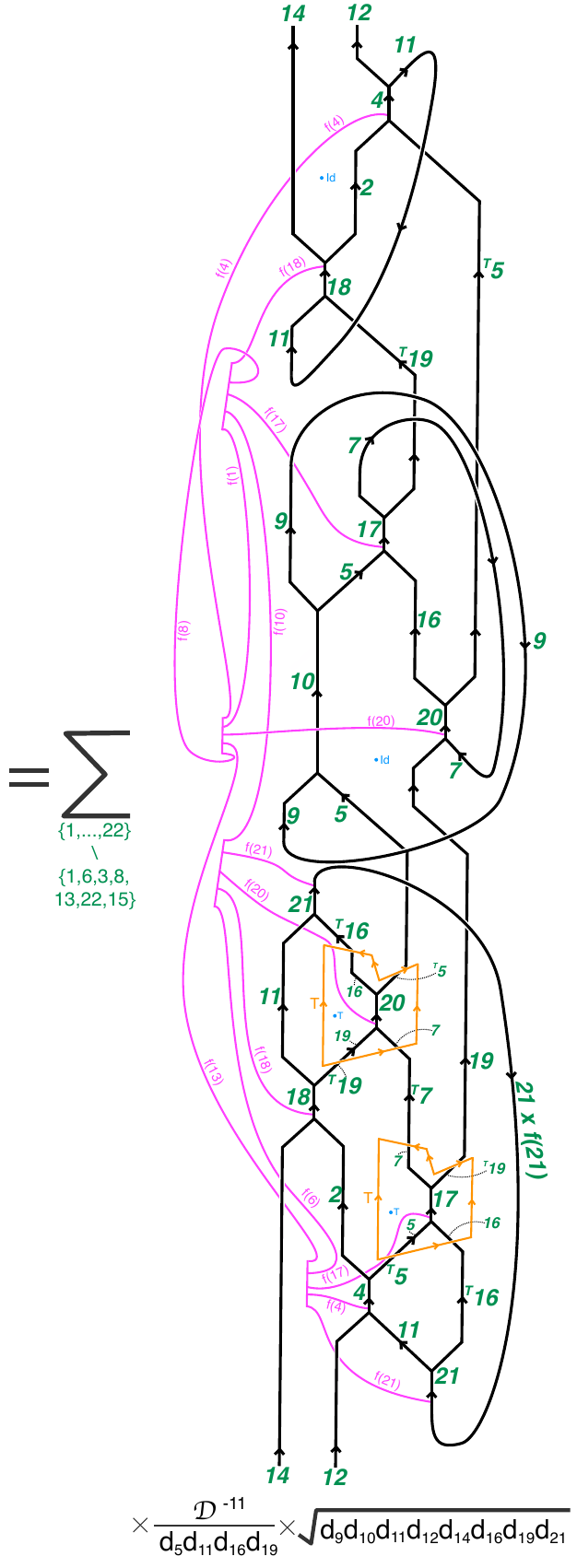}
    \end{minipage}
    \caption{Result of steps 6-7 of $\mathbb{RP}^4$ calculation}
    \label{rp4_Step6to7}
\end{figure}

\begin{figure}[h!]
    \centering
    \begin{minipage}{0.47\textwidth}
        \centering
        \includegraphics[width=\linewidth]{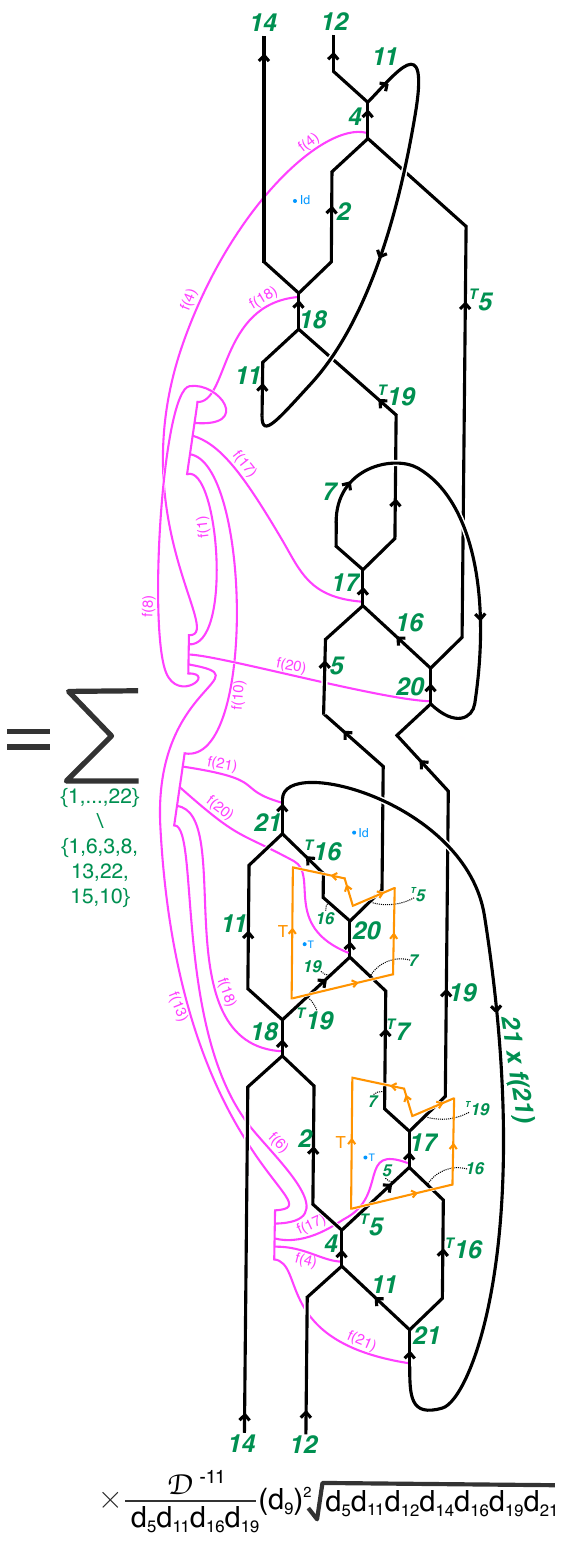}
        \end{minipage} \quad \quad \quad
    \begin{minipage}{0.47\textwidth}
         \centering
         \includegraphics[width=\linewidth]{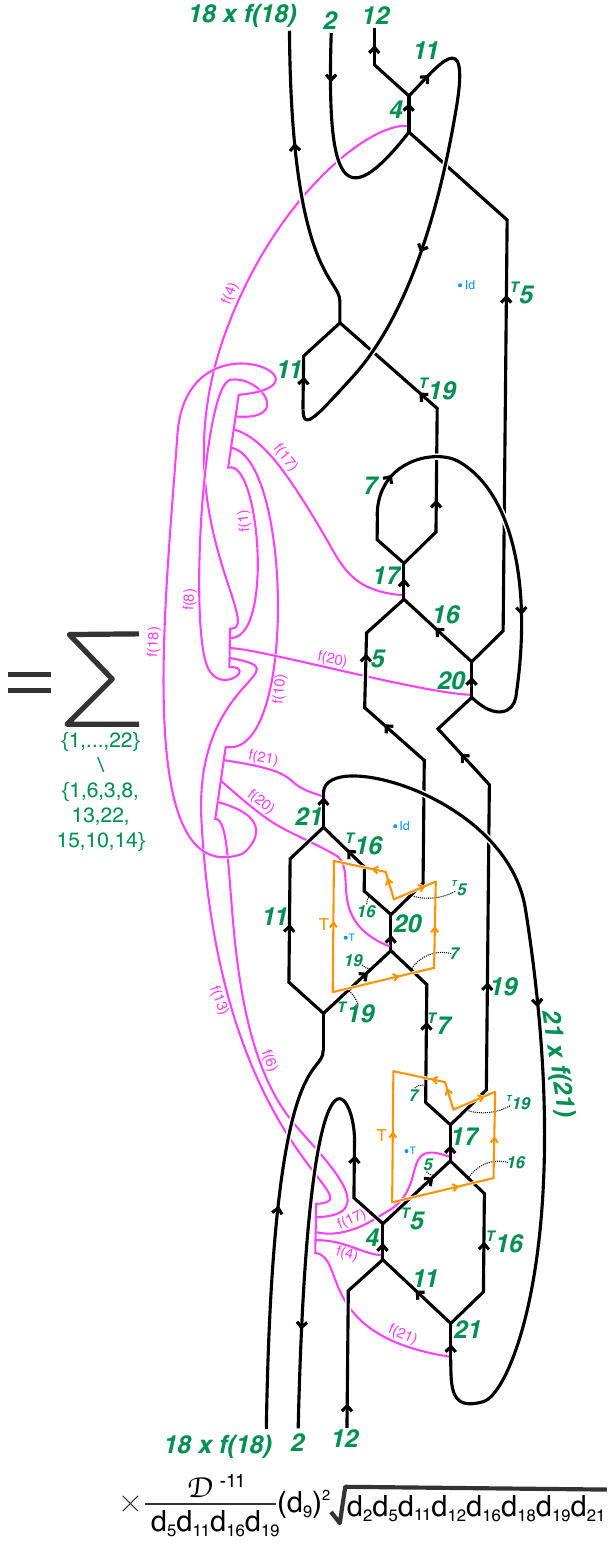}
    \end{minipage}
    \caption{Result of steps 8-9 of $\mathbb{RP}^4$ calculation}
    \label{rp4_Step8to9}
\end{figure}

\begin{figure}[h!]
    \centering
    \begin{minipage}{0.47\textwidth}
        \centering
        \includegraphics[width=\linewidth]{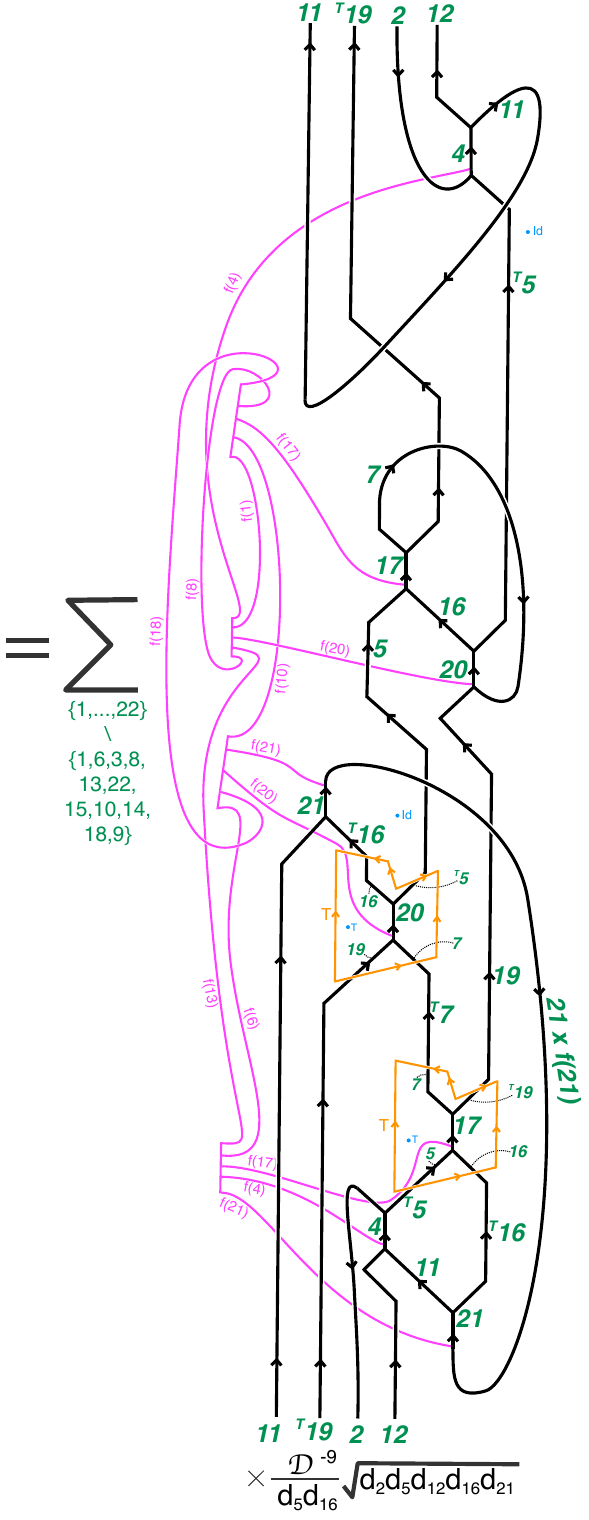}
        \end{minipage} \quad \quad \quad
    \begin{minipage}{0.47\textwidth}
         \centering
         \includegraphics[width=\linewidth]{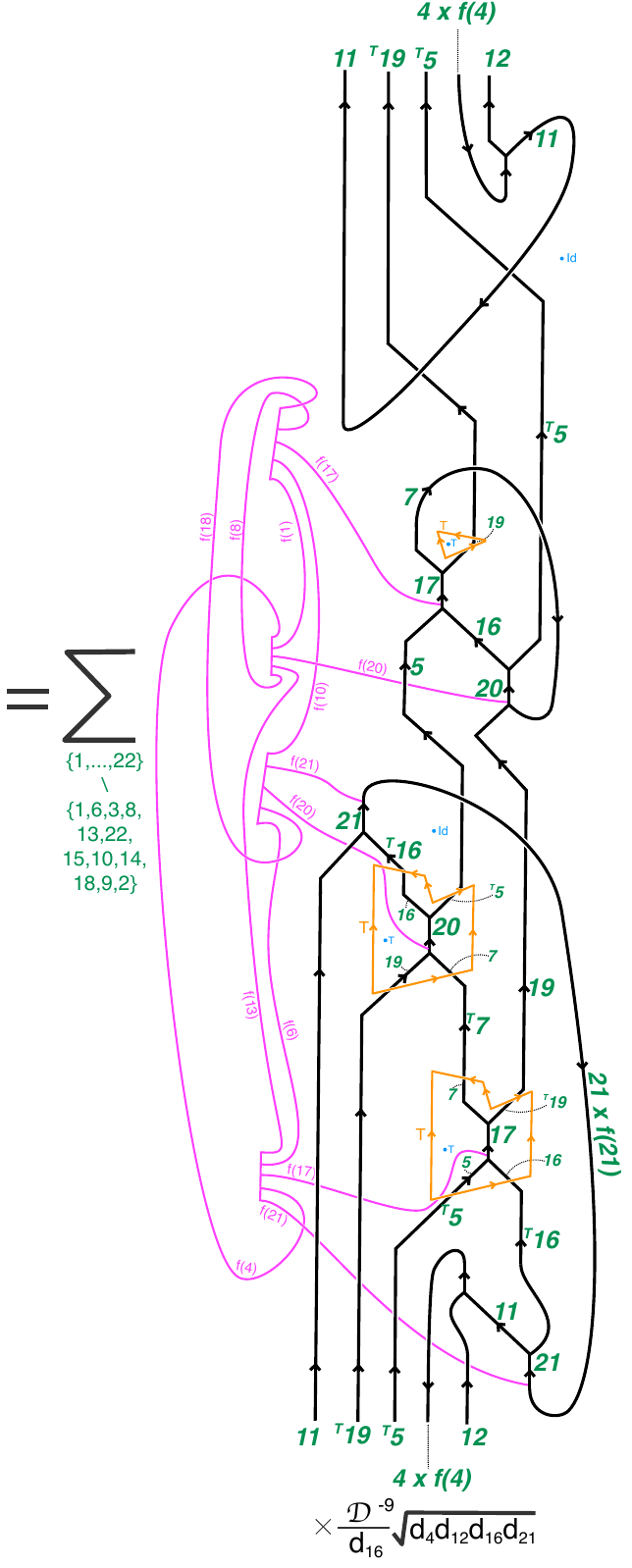}
    \end{minipage}
    \caption{Result of steps 10-11 of $\mathbb{RP}^4$ calculation}
    \label{rp4_Step10to11}
\end{figure}

\begin{figure}[h!]
    \centering
    \begin{minipage}{0.47\textwidth}
        \centering
        \includegraphics[width=\linewidth]{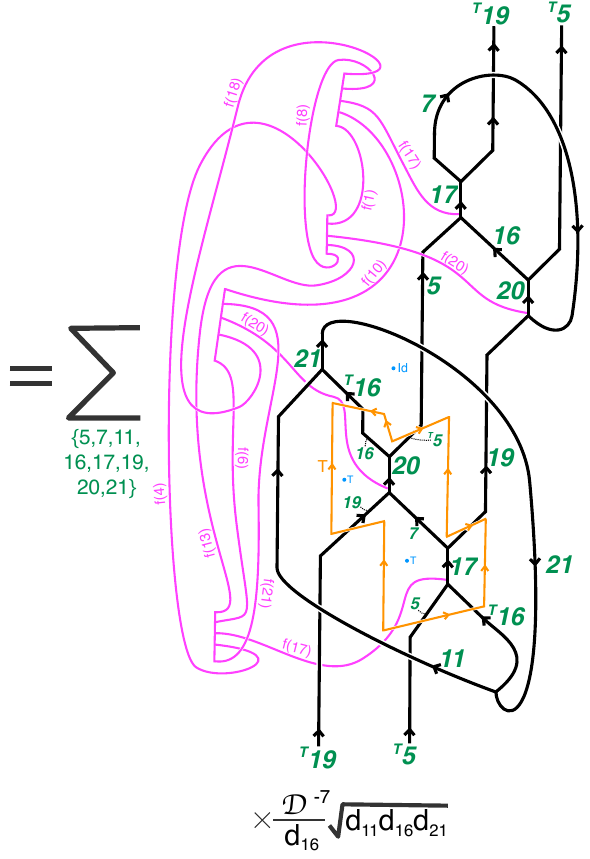}
        \end{minipage} \quad \quad \quad
    \begin{minipage}{0.47\textwidth}
         \centering
         \includegraphics[width=\linewidth]{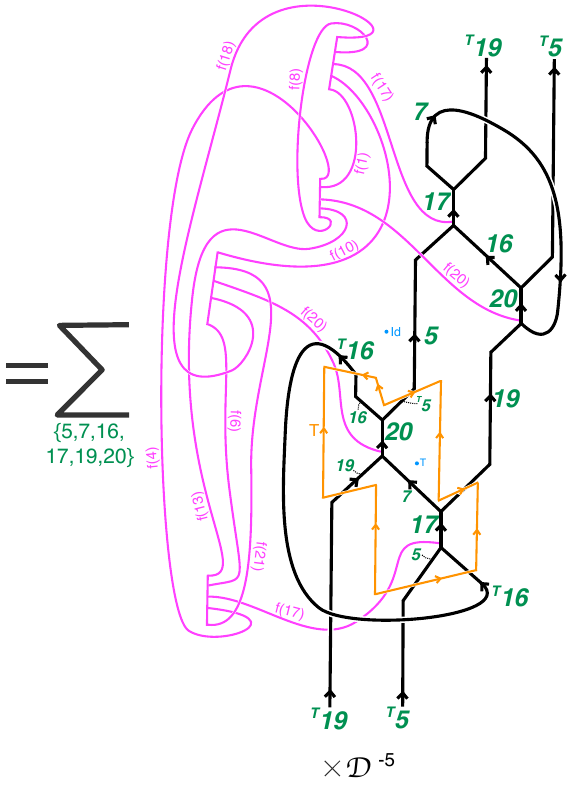}
    \end{minipage}
    \caption{Result of steps 12-13 of $\mathbb{RP}^4$ calculation}
    \label{rp4_Step12to13}
\end{figure}

\begin{figure}[h!]
    \centering
    \begin{minipage}{0.47\textwidth}
        \centering
        \includegraphics[width=\linewidth]{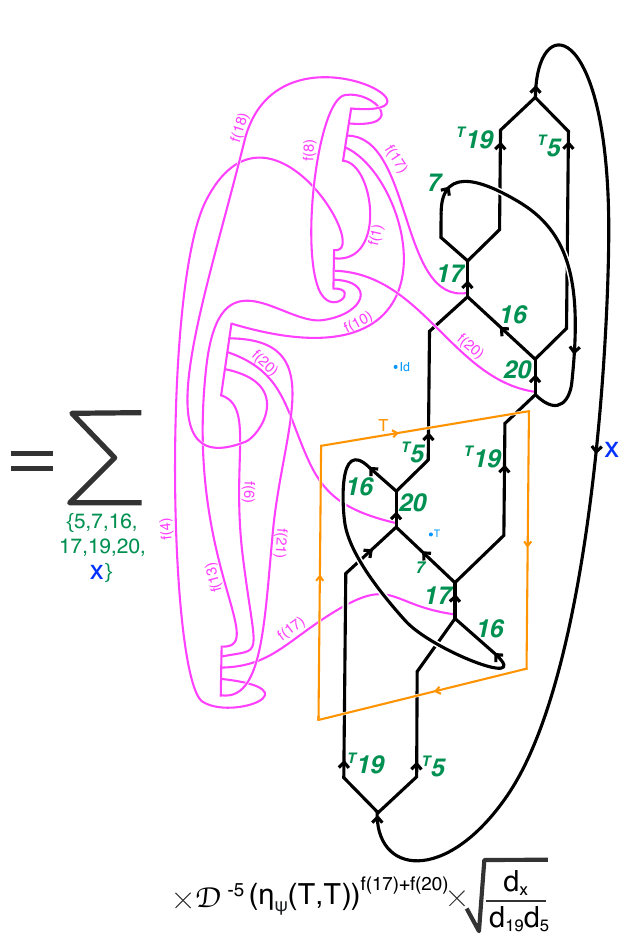}
        \end{minipage} \quad \quad \quad
    \begin{minipage}{0.47\textwidth}
         \centering
         \includegraphics[width=\linewidth]{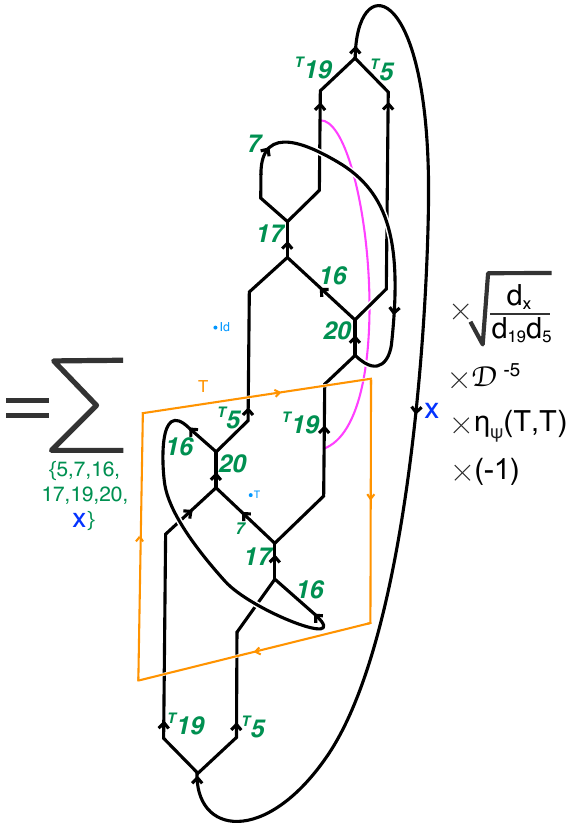}
    \end{minipage}
    \caption{Result of steps 14-15 of $\mathbb{RP}^4$ calculation. NOTE: starting Step 15 we restrict our attention to the specific cochain representative with $f(10)=f(17)=f(21)=1$. The pink line here now refers to the anyon $\psi$.}
    \label{rp4_Step14to15}
\end{figure}

\begin{figure}[h!]
    \centering
    \begin{minipage}{0.47\textwidth}
        \centering
        \includegraphics[width=\linewidth]{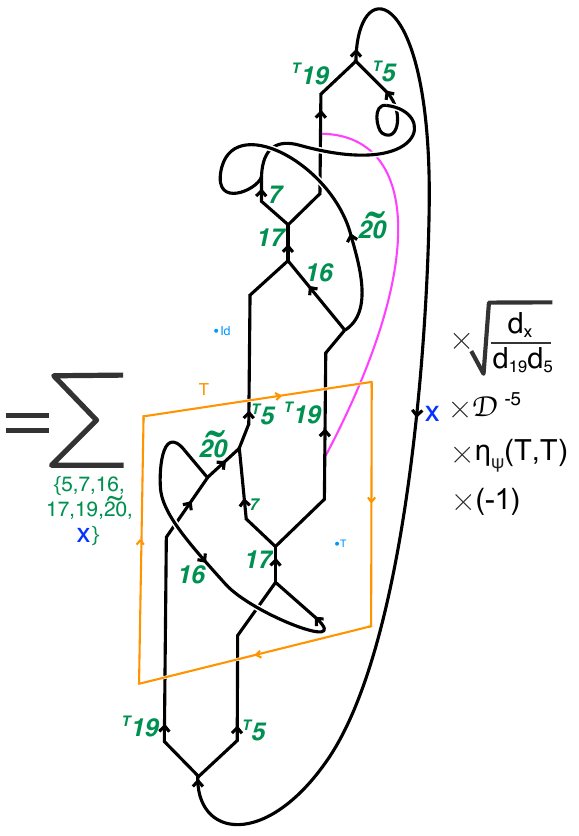}
        \end{minipage} \quad \quad \quad
    \begin{minipage}{0.47\textwidth}
         \centering
         \includegraphics[width=\linewidth]{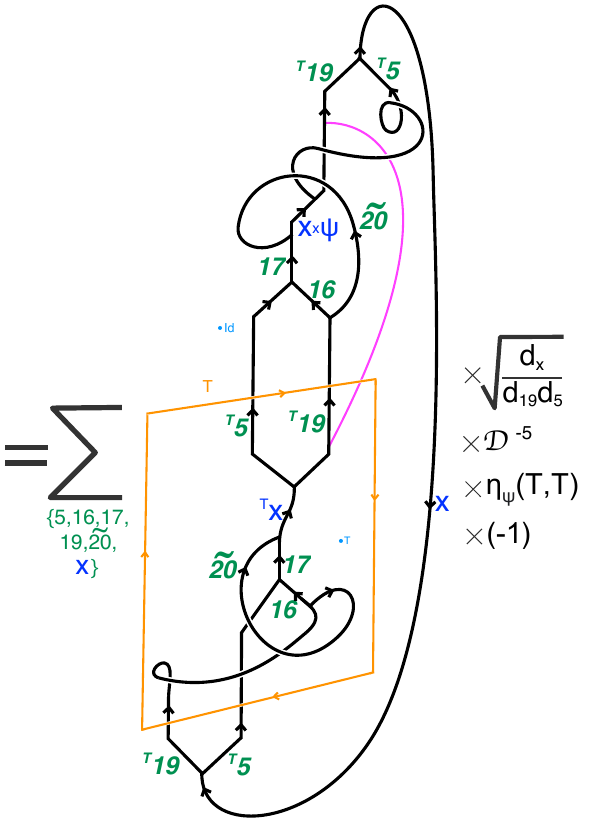}
    \end{minipage}
    \caption{Result of steps 16-17 of $\mathbb{RP}^4$ calculation}
    \label{rp4_Step16to17}
\end{figure}

\begin{figure}[h!]
    \centering
    \begin{minipage}{0.47\textwidth}
        \centering
        \includegraphics[width=\linewidth]{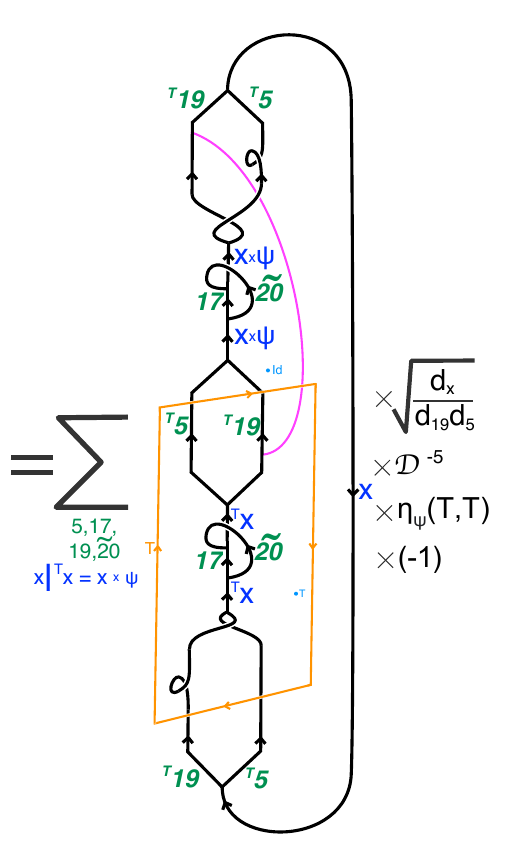}
        \end{minipage} \quad \quad \quad
    \begin{minipage}{0.47\textwidth}
         \centering
         \includegraphics[width=\linewidth]{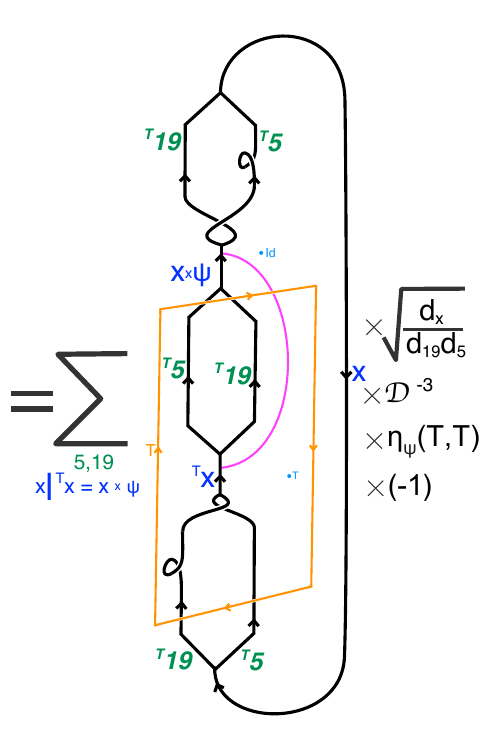}
    \end{minipage}
    \caption{Result of steps 18-19 of $\mathbb{RP}^4$ calculation}
    \label{rp4_Step18to19}
\end{figure}

\begin{figure}[h!]
    \centering
    \begin{minipage}{0.33\textwidth}
        \centering
        \includegraphics[width=\linewidth]{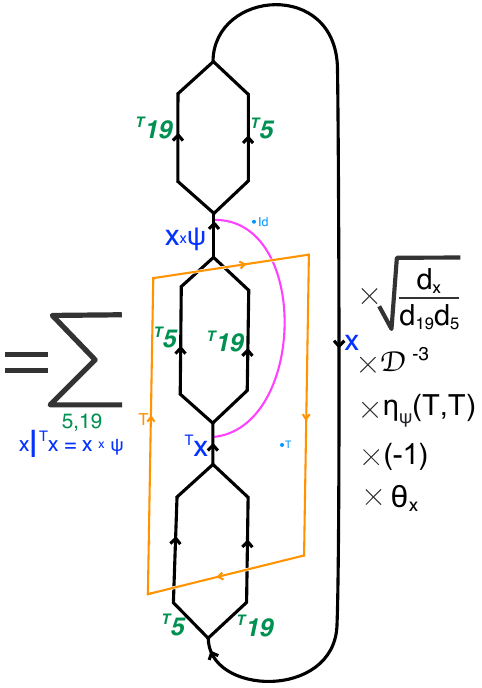}
        \end{minipage} \quad \quad \quad
    \begin{minipage}{0.57\textwidth}
         \centering
         \includegraphics[width=\linewidth]{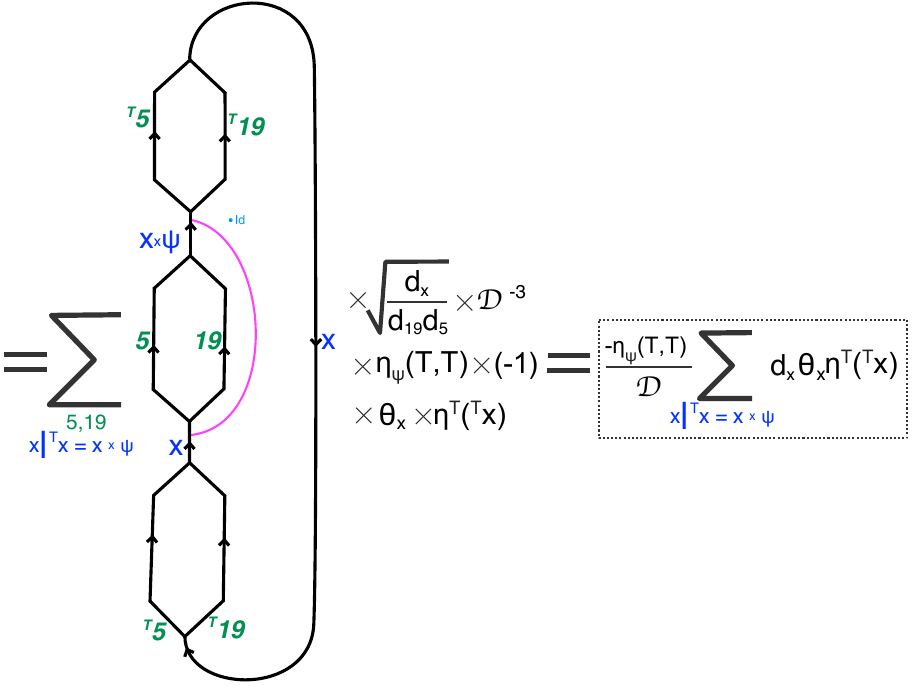}
    \end{minipage}
    \caption{Result of steps 20-22 of $\mathbb{RP}^4$ calculation}
    \label{rp4_Step20to22}
\end{figure}

Note that the links of the 2-simplices $\{2,5,7,9,11,15,16,19\}$ have size 4 whereas the rest of them have size 2. Also, $(N_0 - N_1) = 3-8 = -5$, $\chi(M) = N_0 - N_1 + N_2 - N_3 + N_4 = 3-8+12-10+4=1$. Then the quantum dimensions out in front of the diagrams, as in Eq.~\eqref{quantumDimContribution}, will be
\begin{equation}
\mathcal{N}_\text{q-dim} = \frac{\mathcal{D}^{2(3-8) - 1}}{d_2 d_5 d_7 d_9 d_{11} d_{15} d_{16} d_{20}}
\end{equation}
The rest of the calculation is evaluating the 15j symbols and proceeds as follows. The steps described below correspond to the diagrams of Figs.~\ref{rp4_Step1}-\ref{rp4_Step20to22}.

\begin{enumerate}[start=1,label=\text{Step} {\arabic*}:,leftmargin=1.25cm]
    \item Draw out all the necessary 15j symbols and write quantum dimensions.
    \item Remove the domain walls on the first two 4-simplices since they contribute nothing. Then use Merging Lemma I to sum over anyon labels $1,6$. This adds a factor $\times \sqrt{d_2 d_3 d_{15} d_{16}} \sqrt{d_2 d_7 d_{15} d_{22}}$.
    \item Resolve the identity with sum over anyon $3$. Adds a factor $\sqrt{\frac{d_8 d_{11}}{d_3}}$. Note that we have to use the bending rules in Eqs.~(\ref{eqn:Abend},\ref{eqn:Bbend}). But since the bending factors are unitary, they cancel out. This will happen several more times in the remaining steps although we will not explicitly say it.
    \item Resolve the identity with sum over anyon $8$. Adds a factor $\sqrt{\frac{d_{9} d_{7}}{d_{8}}}$.
    \item Use Merging Lemma I to sum over anyon $13$. Adds a factor $\sqrt{d_{9} d_{12} d_{15} d_{19}}$.
    \item Resolve the identity with sum over anyon $22$. Adds a factor $\sqrt{\frac{d_{9} d_{21}}{d_{22}}}$.
    \item Resolve the identity with sum over anyon $15$. Adds a factor $\sqrt{\frac{d_{14} d_{10}}{d_{15}}}$.
    \item Resolve the identity with sum over anyon $10$ and remove a leftover $d_9$ loop. Together, these add factors $\sqrt{\frac{d_{5} d_{9}}{d_{10}}} \cdot d_9$.
    \item Resolve the identity with sum over anyon $14$. Adds a factor $\sqrt{\frac{d_{2} d_{18}}{d_{14}}}$.
    \item Resolve the identity with sum over anyon $18$. Adds a factor $\sqrt{\frac{d_{11} d_{19}}{d_{18}}}$. Then sum over $9$ gives a factor $\sum_{9} (d_9)^2 = \mathcal{D}^2$.
    \item Resolve the identity with sum over anyon $2$. Adds a factor $\sqrt{\frac{d_{4} d_{5}}{d_{2}}}$.
    \item Resolve the identity with sum over anyon $4$ and gain $d_{12}$ from the leftover $12$ loop. Adds a factor $d_{12} \times \sqrt{\frac{d_{11} d_{12}}{d_{4}}}$. Then sum over $12$ gives a factor $\sum_{12} (d_{12})^2 = \mathcal{D}^2$.
    \item Resolve the identity with sum over anyon $11$ and gain $d_{21}$ from the leftover $21$ loop. Adds a factor $d_{21} \times \sqrt{\frac{d_{16} d_{21}}{d_{11}}}$. Then sum over $21$ gives a factor $\sum_{21} (d_{21})^2 = \mathcal{D}^2$.
    \item Expand the domain walls. Factors of $\eta_{5}({\bf T,T}),\eta_{19}({\bf T,T}),U_{\bf T}({\overline{16}},16)$ each appear twice with opposite exponents and cancel out. Two factors of $\eta_\psi({\bf T,T})$ become  $(\eta_\psi({\bf T,T}))^{f(17)+f(20)}$. Also resolve the $\,^{\bf T}19$ and $\,^{\bf T}5$ lines into a sum over anyons $x$.
\end{enumerate}

The above steps work for any choice of background $f$ lines. The rest of the calculation will be for the particular choice of background $f \in Z^3((\mathbb{RP}^4,T_{\star}),\Z_2)$ from Eq.~\eqref{eqn:nontrivialF_RP4} that corresponds to the nonzero class $[f] \in H^3(\mathbb{RP}^4,\Z_2)$. We will see that this gives the expected formula
\begin{equation*}
\begin{split}
Z_b((\mathbb{RP}^4,T_{\star}),A_b,f) &= -\frac{\eta_\psi({\bf T,T})}{\mathcal{D}} \sum_{x | x = \,^{\bf T}x \times \psi} d_x \theta_x \eta^{\bf T}_{\,^{\bf T}x} \\
& = -\frac{1}{\mathcal{D}} \sum_{x | x = \,^{\bf T}x \times \psi} d_x \theta_x \eta^{\bf T}_x
\end{split}
\end{equation*}
for this path integral. The calculation with $f=0$ is quite similar and slightly simpler; the only difference in the result is a minus sign and that the sum is restricted to $\{x | x = \,{^{\bf T}}x\}$.

A key tool in the rest of the diagrammatics is the fact that there are two `regions' of the diagram containing the same anyon lines but one `half' is inside a time-reversed region labeled by the group element ${\bf T}$. The diagrammatic calculus Fig.~\ref{fig:graphicalCalculus_defs} then states that all $F,R$ moves done inside the $T$-region differ by complex conjugation with respect to the same moves in the identity-group-element region. In particular, unitarity means that these conjugated $F$ and $R$ symbols are the \textit{inverses} of the symbols in the identity-region. This means we can do $F$ and $R$ moves in \textit{pairs}, one in the identity-region and one in the ${\bf T}$-region. Unitary implies that these conjugated $F$,$R$ moves \textit{cancel out} and allow us to diagrammatically simplify things without explicitly needing to write out the respective $F$ and $R$ symbols.

\begin{enumerate}[start=15,label=\text{Step} {\arabic*}:,leftmargin=1.25cm]
    \item Gain a factor of $(-1)$ from the crossing $f(17)=1$ with $f(21)=1$. Cancelling $F$,$R$ moves bring the remaining fermion line from the $17$ 3-simplices onto the $19$ 2-simplices.
    \item Cancelling $F$ moves turn the sum over $20$ lines into a sum over $\widetilde{20}$ lines.
    \item Cancelling $F$ moves turn the sum over $7$ lines into a sum over $\widetilde{7}$ lines. But the $x$ line coming from the bottom necessitate $\widetilde{7} = \,^{\bf T}x$ while the top gives $\widetilde{7} = x \times \psi$. This in turn implies that the sum over $x$ restricts to $\,^{\bf T}x = x \times \psi$.
    \item Cancelling $F$ oves turn the sum over $16$ lines into a sum over $\widetilde{16}$ lines. But charge conservation gives that the only contributing term is $\widetilde{16} = {^T}x = x \times \psi$.
    \item Cancelling $F$,$R$ moves turn the sum over $17$ lines into a sum over $\widetilde{17}$ lines. But $\widetilde{17}$ gets attached to $\widetilde{20}$ as a tadpole so $\widetilde{17}$ must be the identity. Then we get two $\widetilde{20}$ loops whose contraction give a factor of $d_{\widetilde{20}}^2$. The sum over $\widetilde{20}$ then gives $\sum_{\widetilde{20}} (d_{\widetilde{20}})^2 = \mathcal{D}^2$. 
    \item Remove some of the anyon lines' `twists' using $R$ moves to give a factor of $R^{\,{^{\bf T}}5 \,{^{\bf T}}19}_{x}R^{\,{^{\bf T}}19 \,{^{\bf T}5}}_{x} \theta_{\,{^{\bf T}}19} \theta_{\,{^{\bf T}}5}$. Then we can use the ribbon identity Eq.~\eqref{ribbonId} to equate this to a factor of $\theta_x$. Alternatively, one can manipulate the diagram into a twist on the $x$ anyon.  
    \item Contracting the remaining ${\bf T}$-domain-wall bubble gives a factor $(U_{\bf T}(5,19;x \times \psi)U_{\bf T}(5,19;x \times \psi)^{-1}) \cdot \eta^{\bf T}_{\,^{\bf T}x}  \cdot U_{\bf T}(\,^{\bf T}x,\psi;\,^{\bf T}x \times \psi) = \eta_{\,^{\bf T}x}({\bf T,T}) \cdot U_{\bf T}(\,^{\bf T}x,\psi;\,^{\bf T}x \times \psi)$. This is exactly the gauge-invariant quantity $\eta^{\bf T}_{{\,^{\bf T}} x}$ defined in Eq.~\eqref{gaugeInvariant_T_Eta}. The $U$-$\eta$ consistency condition Eq.~\eqref{UUoverU_equals_EtaOverEtaEta} shows this equals the $\eta_\psi({\bf T,T})\eta^{\bf T}_x$ that goes into the sum.
    \item Evaluate the rest of the diagram. First remove the fermion line using canceling $F$-moves. Then evaluate all of the inner products to produce three factors of $\sqrt{\frac{d_{19} d_{5}}{d_x}}$ and $N_{19,5}^{x}$. Finally, evaluate the remaining $x$ loop for another factor of $d_x$. See below for the algebra giving the final simplification.
\end{enumerate}

The sums on $5$ and $19$ can be performed with a little algebra:
\begin{equation}
    \sum_{5,19} d_{19}d_5 N_{19,5}^x = \sum_{5,19} d_{19}d_5 N_{5,\overline{x}}^{\overline{19}} = \sum_{19} d_x d_{19}^2 = \mathcal{D}^2 d_x
\end{equation}

And from here we get the result
\begin{equation*}
Z_b((\mathbb{RP}^4,T_{\star}),A_b,f_\text{non-trivial}) = -\frac{1}{\mathcal{D}} \sum_{x | \,^{\bf T}x = x \times \psi} d_x \theta_x \eta^{\bf T}_x
\end{equation*}
And since $z_c(f_\text{non-trivial}) = -i$ for the representative we chose, we obtain the total partition function exactly corresponding to Eq.~\eqref{totalRP4partitionfunction}.

\clearpage

\bibliography{TI}
        
\end{document}